\documentclass[aps,prl,reprint,groupedaddress]{revtex4-2}

\usepackage{graphicx}
\usepackage{amsmath}
\usepackage{amsfonts}
\usepackage{natbib}
\usepackage{hyperref}
\hypersetup{
    colorlinks = true,
    urlcolor   = blue,
    citecolor  = black,
}

\usepackage{caption}
\usepackage{subcaption}

\usepackage{placeins}
\usepackage{tikz}
\usetikzlibrary{calc,arrows.meta,intersections}

\usepackage{pgffor}

\providecommand\dOne{\delta_*}
\providecommand\dTwo{\theta}
\providecommand\dPe{d \tilde{P}_e}
\providecommand\Rey{\textit{Re}}

\providecommand\myv[1]{\boldsymbol{#1}}

\providecommand\ptau{\Pi^\tau}
\providecommand\ptauo{\Pi^\tau_o}
\providecommand\pU{\Pi^U}
\providecommand\pUo{\Pi^U_o}

\providecommand{\dimensionless}{dimensionless}

\providecommand{\corr}[1]{\textcolor{black}{#1}}

\ifdefined\ShowRevisionMarks
    \colorlet{revisioncolor}{blue}
\else
    \colorlet{revisioncolor}{black}
\fi
\DeclareRobustCommand{\rev}[1]{\textcolor{revisioncolor}{#1}}

\ifdefined\ShowRevisionMarks
\else
    \DeclareRobustCommand{\rev}[1]{\textcolor{black}{#1}}
\fi

\begin{document}

%\title{Unified scaling laws for wall turbulence
%under pressure gradient}

%\title{Unified scaling laws for turbulent boundary layers}
\title{Unified scaling laws for turbulent boundary layers across flow regimes}

%\title{Unified scaling of skin friction and mean velocity in turbulence under
%pressure gradient and separation}
%\title{Unified scaling laws for wall turbulence under pressure gradient } %and
% separation}
%
% separated flows
%\title{Local scaling laws for wall turbulence close and far from equilibrium}
%
\author{Gonzalo Arranz}
\affiliation{California Institute of Technology, Pasadena, CA 91125, USA}
\author{Adri\'an Lozano-Dur\'an }
\email{adrianld@caltech.edu}
\affiliation{California Institute of Technology, Pasadena, CA 91125, USA}
\affiliation{Massachusetts Institute of Technology, Cambridge, MA 02139, USA}

%\aff{1}California Institute of Technology, Pasadena CA 91125, USA
%\aff{2}Massachusetts Institute of Technology, Cambridge MA 02139, USA}

\begin{abstract}
    We discover unified scaling laws for \rev{the mean velocity profile
    and its associated wall shear stress} in
    turbulent boundary layers subject to favorable and adverse mean
    pressure gradients\rev{, across attached, separated, and reattaching
    regimes.} \rev{We approach the problem through the lens of information
    theory, identifying dimensionless groups that encode the information
    needed to characterize the mean velocity profile.} \rev{Three dimensionless variables characterize the
    mean velocity profile, while two suffice to describe the mean
    wall-shear stress.} The scaling laws depend exclusively on variables
    defined at a fixed
    streamwise location, demonstrating that \rev{appropriate combinations}
    of local quantities implicitly encode upstream history
    without requiring global parameters. The results are validated against
    a rich collection of cases and are shown to collapse mean quantities
    across flow regimes previously thought to require distinct treatments.
\end{abstract}

\maketitle

% Intro: importance of TBLs
Wall-bounded turbulent flows are ubiquitous in engineering and nature,
occurring over aircraft wings and turbine blades, in pipelines, and in
ocean currents. In these flows, \rev{the mean velocity profile
characterizes the distribution of momentum across the boundary layer,
while} the skin friction at the wall
determines the energy required to maintain the flow motion, directly
impacting fuel consumption, pumping costs, and overall system
efficiency~\cite{Townsend1976, schlichting1979, Pope2000, Smits2011,
Jimenez2012}. Accordingly, predicting the \rev{mean velocity profile and
mean wall-shear stress} in turbulent boundary layers (TBLs) remains a
central challenge in fluid dynamics, with implications ranging from
fundamental physics to practical design~\cite{rotta1953theory,
    clauser1954turbulent, mellor1966equilibrium, Marusic2010,
Marusic2010Science}.

% ZPG TBLs are annecdotes
While zero-pressure-gradient (ZPG) TBLs over smooth flat plates have
been extensively characterized \cite[e.g.][]{Spalart1988,
DeGraaff2000, Marusic2010, Schlatter2010, sillero2014}, most flows
of practical relevance develop under streamwise-varying pressure
gradients~\citep{vinuesa2017pressuregradient}. Favorable pressure
gradients (FPGs) accelerate the flow and increase skin friction,
whereas adverse pressure gradients (APGs) decelerate it and
\rev{can cause flow separation, degrading performance and increasing drag}~\cite{simpson1989turbulent, perry2002}.  Furthermore,
practical scenarios rarely experience uniform conditions; instead,
boundary layers traverse alternating sequences of FPG, ZPG, and APG
regions, giving rise to complex behaviors, including turbulent
separation bubbles with subsequent
reattachment~\cite{kiya1983structure, alving1996turbulence, Song2004}.

% Non-locality challenge
A longstanding difficulty \rev{is that pressure-gradient TBLs retain
upstream memory. Experiments and simulations document these}
\emph{history effects}: nominally identical local pressure gradients can produce different
flow states depending on the upstream
trajectory~\cite[e.g.][]{bobke2017history, vinuesa2017revisiting,
    virgilio2025pressure, preskett2025effects,
zarei2025decoupling}. This non-universality has motivated decades of
research seeking scaling parameters and similarity frameworks that
capture, at least partially, local and history-dependent
effects~\cite{rotta1953theory, clauser1954turbulent,
    zagarola1998meanflow, castillo2001similarity,
    maciel2006selfsimilarity, gungor2016scaling, kitsios2017direct,
gungor2022energy, gungor2024response, dixit2024generalized}; yet a
unified description applicable across ZPG, FPG, and APG
conditions---including separation and reattachment---has remained
elusive.

% Outline
\corr{In this Letter, we discover unified scaling laws for the mean
    state of TBLs across regimes that have traditionally required
    separate treatments. The central physical advance is twofold. First,
    the same scaling laws describe \rev{the entire mean velocity profile
    across the inner, logarithmic, and outer layers, and its associated
    wall shear stress, across} FPG, APG,
    separated, and reattaching
    regimes.  Second, this unification is achieved using only quantities
    evaluated at the same streamwise location, despite the prevailing
    view that history effects preclude such a local description. \rev{This
    is possible because an appropriately chosen local state encodes the
    relevant information about upstream history.} Building on these findings, we use the discovered
    dimensionless groups to construct \rev{a predictive model for mean
    velocity profiles, with skin friction obtained from its wall derivative,
    that generalizes} to previously
unseen flows.}

%%Recent papers:
%%\cite{mattei2025Signature} Signature of Pressure Gradient History on Wall
% Shear Stress in
%%Turbulent Boundary Layers.
%%%
%%\cite{zarei2025decoupling} Decoupling pressure gradient history effects in
%%turbulent boundary layers through high-Reynolds
%%number experiments.
%%
%%Papers to model the state of the boundary layer:
%%\cite{virgilio2025pressure}
%%\cite{perry2002}
%%\cite{agrawal2024}
%

%%%%%%%%%%%%%%%%%%%%%%%%%%%%%%%%%%%%%%%%%%%%
\section{Methods}
%%%%%%%%%%%%%%%%%%%%%%%%%%%%%%%%%%%%%%%%%%%%

\begin{figure*}
    \centering
    \includegraphics{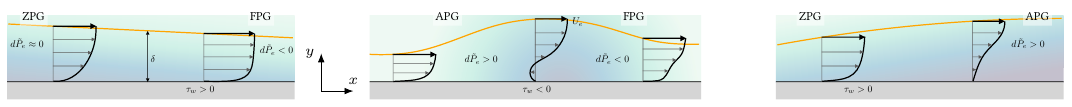}
    \caption{ \corr{\rev{Selected database regimes:} (left) FPG TBL, (center) turbulent
            separation bubble, and (right) APG TBL. The orange line denotes
            $\delta$. \rev{See the main text for the $\Rey_\theta$ ranges.}
            \rev{Canonical training and test cases span pressure-gradient parameters} $-0.028
            \lesssim \gamma_P\beta_* \lesssim 0.045$ \rev{and shape factors}
        $1.26 \lesssim H \lesssim 8.39$.}
    \label{fig:schematic}}
\end{figure*}

% Formulation
We assume that the essential physics controlling the \rev{mean velocity
profile, $U$, and its wall derivative} can be captured
\rev{using flow variables evaluated at the same streamwise location}
$x$ (figure~\ref{fig:schematic}). \rev{These are} the mean streamwise velocity
at the edge of the boundary layer $U_e$; the density-normalized edge
pressure gradient $\dPe = (1/\rho)\,\mathrm{d}P_e/\mathrm{d}x$, where
$\rho$ is the fluid density; the kinematic viscosity $\nu$; the
boundary-layer thickness $\delta$; the displacement thickness $\dOne$;
and the momentum thickness $\dTwo$~\cite{Pope2000}.

% Dimensionless variables
To identify the dimensionless groups governing \rev{$U$}, we
build on the Buckingham--$\Pi$ theorem~\cite{Buckingham1914}, which
states that physical relationships satisfying dimensional homogeneity
(i.e., invariance under a change of units) can be expressed in terms
of dimensionless variables:
\begin{equation}\label{eq:Piexponents}
    \rev{\pU} = \gamma_P \, y^{a_1} U_e^{a_2} \delta^{a_3} \dOne^{a_4} \dTwo^{a_5}
    \nu^{a_6} |\dPe|^{a_7},
\end{equation}
where \rev{$y$ is the wall-normal coordinate and} $a_i$
($i=1,\dots,7$) are unknown exponents. The sign factor
$\gamma_P=\operatorname{sgn}(\dPe)$ is introduced only when the
pressure-gradient term is present; for $a_7=0$ or $\dPe=0$, we set
$\gamma_P=1$. While the Buckingham--$\Pi$ theorem constrains the exponents
$a_i$ to ensure dimensional homogeneity, it admits infinitely many
dimensionless combinations.  Yet, \rev{for determining $U$},
the \rev{predictive information} can vary substantially across different
dimensionless groups.
\corr{ To resolve this non-uniqueness, \rev{we ask whether}
  quantities evaluated at a single streamwise location contain enough
  information to describe the \rev{mean velocity profile.}
  \rev{The} information-theoretic Buckingham--$\Pi$
  theorem~\cite{yuan2025Dimensionless} \rev{selects the dimensionless groups
  that maximize predictive information} for the \rev{dimensionless mean
  velocity}}.
More specifically, given the dimensionless output \rev{$\pUo\equiv U/U_e$}, the most
predictive dimensionless input $\rev{\myv{\Pi}^U=[\pU_1,\cdots,\pU_N]}$ is
obtained by minimizing the information-theoretic \rev{error bound}:
\begin{equation}\label{eq:argminI}
    \rev{\myv{\Pi}^U} = \arg\min_{\rev{\myv{\Pi}^{U\prime}}} \;
    \epsilon_{LB}\!\left(\rev{\pUo},\rev{\myv{\Pi}^{U\prime}}\right),
\end{equation}
where $\epsilon_{LB} = e^{-I(\rev{\pUo};\rev{\myv{\Pi}^{U\prime}})}$ is the normalized
information-theoretic lower bound on the prediction error, and
$I(\rev{\pUo};\rev{\myv{\Pi}^{U\prime}})$ is the mutual information~\citep{shannon1948}
between the \dimensionless~output $\rev{\pUo}$ and the candidate set of
\dimensionless~inputs, $\rev{\myv{\Pi}^{U\prime}}$. An important advantage of the
formulation in Eq.~\eqref{eq:argminI} is that it is independent of the
chosen model class (e.g., linear regression, neural networks).  Hence,
for any model $\rev{\pUo} \approx \rev{f_U'}(\rev{\myv{\Pi}^{U\prime}})$, it holds that $\lVert
\rev{\pUo} - \rev{f_U'}(\rev{\myv{\Pi}^{U\prime}}) \rVert \ge c\,\epsilon_{LB}$, where $c$ is
independent of $\rev{\myv{\Pi}^{U\prime}}$. \rev{Primes label candidate input sets and models.}
The most-informative model is therefore
of the form $\rev{\pUo} \approx \rev{f_U}(\rev{\myv{\Pi}^U})$, with $\rev{\myv{\Pi}^U}$ given by
Eq.~\eqref{eq:argminI}. The number of \dimensionless~inputs required,
$N$, is chosen as the smallest value beyond which $\epsilon_{LB}$ no
longer decreases. \rev{Further theoretical and algorithmic details are provided in the}
Supplementary Material~\cite{suppl} and \citet{yuan2025Dimensionless}.

% Model
Once the optimal dimensionless groups have been identified, we
determine the functional form $\rev{f_U}$ of the optimal model $\rev{\pUo} \approx
\rev{f_U}(\rev{\myv{\Pi}^U})$. To approximate $\rev{f_U}$, we employ a Kolmogorov--Arnold
Network (KAN)~\cite{liu2024kan} with B-spline basis
functions~\cite{deboor2001splines}.
%The architecture consists of 2
%hidden layers with 2 neurons each, using 3 knots per neuron.

% Databases
The approach is applied to a rich database of over 30 high-fidelity
simulations of incompressible TBLs, comprising well-established
reference cases from the literature~\cite{coleman2018numerical,
    coleman2021numerical, gungor2016scaling, gungor2022energy,
gungor2024response, bobke2017history, kitsios2017direct} and 19
simulations performed for this study: 16 TBL simulations and 3
separation-bubble simulations~\cite{suppl}.
\rev{These canonical flows isolate pressure-gradient effects across}
mild FPG, mild-to-strong APG, incipient separation, and separation
bubbles with reattachment.
The Reynolds number based on the momentum thickness $\dTwo$ ranges
from $\Rey_{\dTwo}= U_e\theta/\nu \approx 400$ to $13{,}000$ \corr{for the data used
to train \rev{the model}}.  The dataset comprises more than $70{,}000$
streamwise stations, each with \rev{a mean velocity profile}. We reserve 20\% of that
data for testing \corr{and further augment the test set with
    additional unseen cases reaching up to $\Rey_{\dTwo} \approx
    30{,}000$ for canonical flows and $\Rey_{\dTwo} \approx 180{,}000$
for complex flows}. Details of the architecture, training procedure,
and individual cases are provided in the Supplementary
Material~\cite{suppl}.

%The data are
%split into 80\% for training and 20\% for testing; the training set is
%further divided into 70\% and 30\% for training and validation,
%respectively.
%Details of the individual cases are provided in the
%Supplementary Material~\cite{suppl}.

%%%%%%%%%%%%%%%%%%%%%%%%%%%%%%%%%%%%%%%%%%%%%%%%%%%%%%%%%%%%%%%%%%%
\section{Scaling of the mean velocity profile}
%%%%%%%%%%%%%%%%%%%%%%%%%%%%%%%%%%%%%%%%%%%%%%%%%%%%%%%%%%%%%%%%%%%

% Intro
\rev{We discover} a scaling for the \dimensionless~mean velocity
profile, $\pUo \equiv U/U_e$, a quantity that has received
considerable attention in the
literature~\cite[e.g.][]{kader1978Similarity, skote2002Direct,
nickels2004inner, devenport2022equilibrium}. Since exact
whole-profile self-similarity is only possible in sink
flows~\cite{rotta1962turbulent, perry2002}, most efforts have focused
on developing separate scaling laws for the
inner~\cite{stratford1959Prediction, patel1973, nickels2004inner} and
outer~\cite{zagarola1997scaling, zagarola1998meanflow,
castillo2001similarity, maciel2018outer} regions of wall-bounded
turbulence. Moreover, these studies typically target a single regime
(ZPG, APG, FPG, or separation). Another line of work has sought
mixing-length-based models that are valid across the entire
profile~\cite{cantwell2019, subrahmanyam2022, ma2026}, but only for
ZPG and moderate-APG TBLs. Here, we instead pursue a unified scaling
that remains valid across the entire wall-normal direction $y$ for
ZPG, FPG, and APG cases, including separation and reattachment.

% Pis
We find that three \dimensionless~variables are required to
successfully scale the mean velocity profile:
\begin{align*}
    \pU_1 &= \frac{y^{2/9}\dTwo^{7/9}}{\dOne}, &
    \pU_2 &= \frac{y\,\dOne^{2/3}\nu^{1/8}}{U_e^{1/8}\delta^{9/8}\dTwo^{2/3}}, &
    \pU_3 &= \frac{\dTwo^{4/5}\nu^{1/5}}{U_e^{1/5}\dOne}\rev{,}
\end{align*}
where the exponents have been rounded to nearby simple fractions
without affecting the results (see~\cite{suppl}). Interestingly, only
$\pU_1$ and $\pU_2$ depend on $y$, whereas $\pU_3$ is constant at a given streamwise location. This motivates introducing
characteristic length scales such that
\begin{equation*}
    \pUo \approx f_U\!\left(\frac{y}{\ell_1}, \frac{y}{\ell_2}; \pU_3\right),
\end{equation*}
where $\ell_1=\dOne^{9/2}\dTwo^{-7/2}$ is a purely integral length
scale, and $\ell_2 =
U_e^{1/8}\delta^{9/8}\dTwo^{2/3}\dOne^{-2/3}\nu^{-1/8}$ introduces an
explicit viscous dependence. \rev{These} results can be recast in terms
of classical dimensionless variables as
\begin{align*}
    \ell_1 &= \dOne H^{7/2}, &
    \ell_2 &= \delta H^{-2/3}\Rey_\delta^{1/8}, &
    \pU_3 &= \Rey_\theta^{-1/5}H^{-1},
\end{align*}
where \rev{$H=\delta_*/\theta$ is the shape factor and}
$\Rey_\delta=U_e\delta/\nu$.  Physically, $\ell_1$ can be
interpreted as an \emph{outer} length scale set purely by the integral
structure of the boundary layer: it increases rapidly with the shape
factor and therefore grows under APG conditions (especially near
separation), consistent with an enlarged wake region and a thicker
momentum-deficit layer. In contrast, $\ell_2$ acts as a \emph{mixed
inner--outer} scale: it is proportional to $\delta$ but decreases with
increasing shape factor ($\sim H^{-2/3}$) while retaining a weak
viscous sensitivity. Finally, $\pU_3$ \rev{captures} the combined influence of Reynolds number
and profile fullness.

\begin{figure*}
    \centering
    \includegraphics{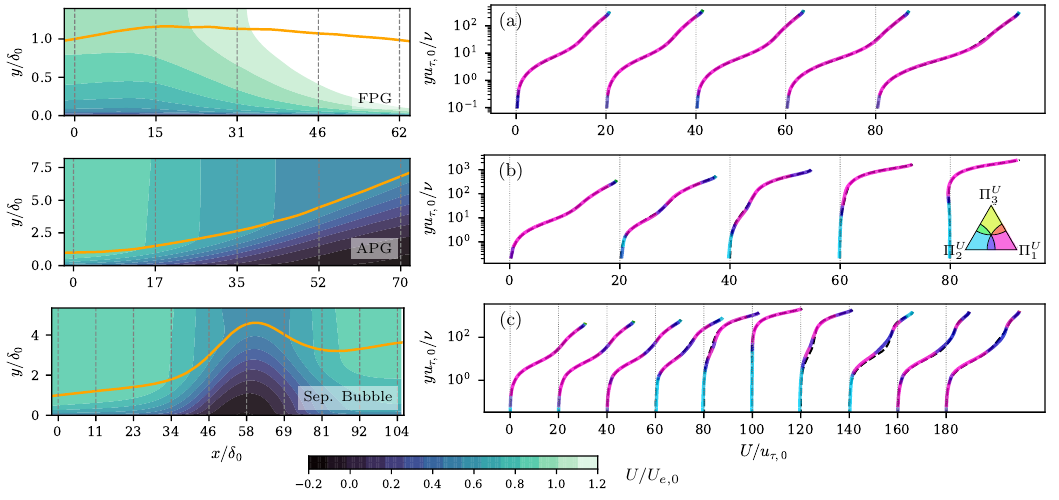}
    \caption{Predicted (colored) and reference (black dashed) mean
        velocity profiles for (a)~an FPG TBL, (b)~an APG TBL, and (c)~a
        turbulent separation bubble~\cite{coleman2018numerical}.  Colors
        identify the dominant \dimensionless~input. Left panels show
        $U/U_{e,0}$ contours, sampled stations (vertical dashed lines),
        and $\delta$ (orange). Here, $u_{\tau,0}=\sqrt{\tau_{w,0}/\rho}$,
        \rev{and} the index $0$ denotes quantities at the first streamwise location\rev{.}
        \rev{Profiles are offset by} $\Delta U/u_{\tau,0} = 20$.
        All results are unseen test data.
        %
        %      Comparison between the model $f_U(\pU_1,\pU_2,\pU_3)$ and
        %        actual mean velocity profiles for (a)~an FPG TBL, (b)~an APG
        %        TBL, and (c)~a turbulent separation
        %        bubble~\cite{coleman2018numerical}. The wall-normal coordinate
        %        is non-dimensionalized as $y^{+,0} \equiv y /
        %        (\nu/\sqrt{\tau_{w,0}/\rho})$, where $\tau_{w,0}$ is the mean
        %        wall-shear stress at the first streamwise location. Black dashed
        %        lines show the actual mean velocity profiles; colored lines show
        %        the model $f_U(\pU_1,\pU_2,\pU_3)$, with color indicating the
        %        dominant \dimensionless~input based on $\partial f_U / \partial
        %        \pU_i$. Left panels show contours of the mean streamwise
        %        velocity, from $-0.2\,U_{e,0}$ (black) to $1.2\,U_{e,0}$ (light
        %        blue) for each case, where $U_{e,0}$ is the edge velocity at the
        %        first location. Vertical dashed lines mark the streamwise
        %        locations of the profile comparisons; the orange line denotes
        %        $\delta$, with $\delta_0$ being the boundary-layer thickness at
        %        the first $x$ location considered. All results shown are for
        %        test (i.e., unseen) data.
    \label{fig:Uprof}}
\end{figure*}

% Model
Figure~\rev{\ref{fig:Uprof}} compares the mean velocity profiles predicted
by the model $\pUo \approx f_U \left( \pU_1, \pU_2, \pU_3 \right)$
against high-fidelity data for three selected cases: an FPG TBL, an
APG TBL approaching separation, and a turbulent separation bubble.
The agreement is within $3\%$ error (figure~\rev{\ref{fig:scaling_tau_2}a}),
with minor differences only in the outer layer downstream of
reattachment (figure~\rev{\ref{fig:Uprof}}c).  \corr{\rev{The} \emph{End Matter}
\rev{compares} the present model with \citet{ma2026}.}

% Sensitivity
\rev{Colors indicate the dominant sensitivity to the standardized
coordinates defined in Sec.~III.2 of the Supplementary Material~\cite{suppl},
using the normalized magnitudes $|\partial F_U/\partial z_i|$. We use}
magenta for $\pU_1$, cyan for $\pU_2$, and yellow for $\pU_3$, with
mixed colors indicating comparable influence. To highlight \rev{strong
sensitivities, normalized gradient components} below $1/\sqrt{3}$
(corresponding to equal contributions) are set to zero.

% Sensitivity: FPG
In FPG regions---including the attached portions of the flow and the
post-reattachment zone in figure~\rev{\ref{fig:Uprof}}(c)---the model is
most sensitive to $\rev{\pU_1}$ throughout the inner and logarithmic
layers, transitioning to $y/\ell_2$ dominance near the boundary-layer
edge. This shift is consistent with the increasing relevance of an
outer length scale as $y\to\delta$~\citep{cal2008Similarity}\rev{.}
The dominance of $\rev{\pU_1}$ in these regions aligns with
classical integral methods: for ZPG and mild pressure gradients, mean
velocity profiles are often described by composite or
equilibrium-profile parameterizations, and their evolution correlates
strongly with integral measures such as the shape factor
$H$~\citep{coles1956wake,mellor1966equilibrium,monty2011parametric};
in some regimes they also admit effective power-law
representations~\citep{buschmann2000powerlaw, afzal2001powerlaw,
dey2024approximate}.

% Sensitivity: APG
In APG regions, $\rev{\pU_1}$ remains dominant over a larger portion of
the boundary layer, extending closer to the edge, consistent with a
strongly modified outer layer under adverse
gradients~\citep{maciel2016coherent, maciel2018outer}.  As the flow
approaches separation (figures~\rev{\ref{fig:Uprof}}b,c), the sensitivity
shifts toward $y/\ell_2$\rev{.} Although the present scaling
does not explicitly include $\dPe$, the integral quantities entering
$\ell_1$ and $\ell_2$ (e.g., $\delta$, $\dOne$, $\dTwo$) are
themselves shaped by the pressure gradient and may therefore encode
its influence implicitly. This transition suggests that the mean
velocity profile is no longer well parameterized by $H$ alone, and the
appearance of viscosity in the dominant scaling parameter may serve as
an indicator of non-equilibrium behavior in the inner
layer~\citep{gungor2016scaling, mellor1966equilibrium}.

%%%%%%%%%%%%%%%%%%%%%%%%%%%%%%%%%%%%%%%%%%%%%%%%%%%%%%%%%%%%
\section{Scaling of the mean wall-shear stress}
%%%%%%%%%%%%%%%%%%%%%%%%%%%%%%%%%%%%%%%%%%%%%%%%%%%%%%%%%%%%

% Note:
% 0.46 ~ 4/9. We know that 2*0.27 + 4/9  = 1 for the velocity, so 0.27 ~ 5/18
% dOne -> 1/3 + 5/18 = 11/18
% dTwo -> 1/3 + 4/9 = 7/9
% Ue -> 1
% nu -> 4/9
% dPe = 5/18
%

\begin{figure*}
    \centering
    \includegraphics{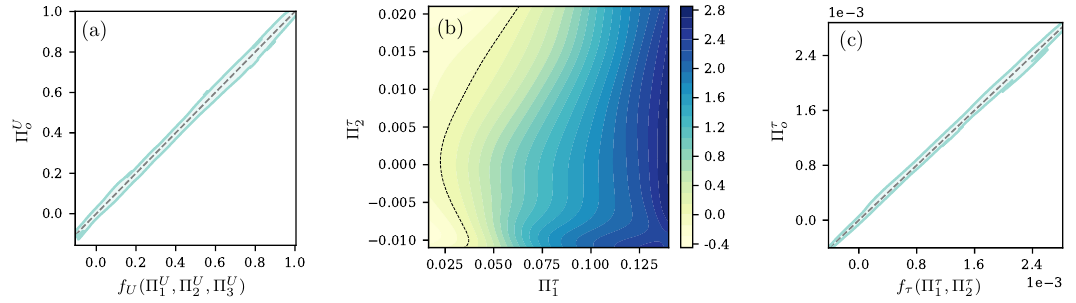}
    \caption{ Discovered scalings on unseen test data.
      \rev{(a) Predicted versus actual $\pUo$.}
      \rev{(b)} Contours of
      $f_\tau(\ptau_1,\ptau_2)$; color \rev{shows predicted}
      $\ptauo\times10^3$, and the dashed contour marks
      $\tau_w=0$. \rev{(c) Predicted versus actual $\ptauo$.}
      Cyan contours enclose 99\% of the joint probability
      mass; diagonal dashed lines mark exact agreement.
        %
        %(a) Optimal scaling law for the dimensionless mean wall
        %shear stress.  Contours represent the KAN model
        %$f_\tau(\ptau_1,\ptau_2)$.  The colorbar shows $\ptauo \times
        %10^3$. (b) Actual $\ptauo$ values and model predictions from
        %$f_\tau(\ptau_1,\ptau_2)$. (c) Optimal scaling law for the
        %dimensionless mean velocity: actual $\pUo$ values and model
        %predictions from $f_U(\pU_1,\pU_2,\pU_3)$. The cyan contour
        %represents 99\% of the joint probability mass; the dashed line
        %indicates exact agreement (zero error). All results shown are
        %for test (i.e., unseen) data.
    \label{fig:scaling_tau_2}}
\end{figure*}

% Pis and model
\rev{Taking the wall derivative of $f_U$ at the wall yields the wall-shear
prediction. This prediction can be expressed in terms of two dimensionless
groups through a reparameterization that preserves predictive information
about wall shear stress. We identify these groups using the approach in
Eq.~\eqref{eq:argminI}, with $\myv{\Pi}^{\tau}$ as the new inputs and the
dimensionless wall shear stress prediction derived from $f_U$ as the output.
This yields the following inputs:}
\begin{align}\label{eq:pitaus}
    \ptau_1 &= \frac{\dTwo^{3/4}\nu^{1/4}}{U_e^{1/4}\dOne}, &
    \ptau_2 &= \gamma_P\,\frac{\dOne^{11/18}\nu^{4/9}|\dPe|^{5/18}}{U_e\,\dTwo^{7/9}}\rev{.}
\end{align}
\rev{Here,} $\ptauo \equiv \tau_w /(\rho U_e^2)$ \rev{with
$\tau_w=\rho\nu\left.\partial U/\partial y\right|_{y=0}$.}
\rev{The wall-shear relation obtained from the reparametrized wall derivative of $f_U$}, $\ptauo \approx
f_\tau(\ptau_1,\ptau_2)$, is shown in figure~\rev{\ref{fig:scaling_tau_2}b}.
The predominantly vertical contours indicate a strong dependence on
$\ptau_1$, while $\ptau_2$ primarily discriminates among flow regimes
through its dependence on the pressure-gradient term
$\dPe$. Figure~\rev{\ref{fig:scaling_tau_2}c} quantifies the model accuracy
by comparing the predicted values $f_\tau(\ptau_1,\ptau_2)$ with the
corresponding $\ptauo$.  The results are shown for the test data,
i.e., cases that were not used to discover the scaling.  The close
agreement (normalized root-mean-square error $\langle[\ptau_o -
    f_\tau]^2\rangle^{1/2}/\sigma_\tau$ below $4\%$, where $\sigma_\tau$
is the standard deviation of $\ptau_o$) confirms that $\ptau_1$ and
$\ptau_2$ capture the flow regimes imposed across different mean
pressure gradients.

% Reinterpretation in terms of Re, H, and beta*
The dimensional variables entering $\ptau_1$ and $\ptau_2$ in
Eq.~\eqref{eq:pitaus} can be regrouped to better expose their physical
content. Multiple rearrangements are possible. Here, we express the
groups in terms of familiar dimensionless parameters:
\begin{align}\label{eq:pitaus1}
    \ptau_1 &= H^{-1}\Rey_\theta^{-1/4}, &
    \ptau_2 &= \gamma_P H^{1/3}\Rey_{\dTwo}^{-4/9}\beta_*^{5/18},
\end{align}
\rev{where} $\beta_* =
|\dPe|\dOne/U_e^2$ is the pressure-gradient parameter.  This
representation facilitates a direct connection with analyses of
turbulent boundary layers under local-equilibrium conditions (i.e.,
ZPG and mild pressure-gradient cases). In that regime, the classical
scaling follows the form $\ptauo\approx
g_\tau(\Rey_\theta,H)$~\citep[p.~590]{schlichting2000boundary}.  The
first group, $\ptau_1$, involves the same dimensionless variables as
the classical formulation, \corr{ while the influence of $\ptau_2$
    becomes subdominant as $\beta_*\to0$. Thus, the discovered scaling
    naturally recovers the classical result in the local-equilibrium
limit}. However, whereas the classical approach treats $\Rey_\theta$
and $H$ as two independent inputs linked by an unspecified function,
our results identify a single combined variable with fixed
exponents. The second group, $\ptau_2$, captures strong
pressure-gradient effects through $\beta_*$, which is absent from the
classical derivation but has been recognized as an important
dimensionless pressure-gradient
parameter~\citep{maciel2006selfsimilarity}.  \corr{\rev{The} \emph{End Matter}
\rev{compares the scaling with} established empirical correlations and a
recent physics-based model, \rev{showing substantially lower errors and
accurate predictions across} attached and separated regimes.}
%
%$-(L_o/U_o)dU_e/dx$~\citep{maciel2006selfsimilarity}, where $L_o = \dOne$
%and $U_o \equiv U_e$.

\providecommand{\myf}[1]{F_{#1}}
% Interpretation of Pis
It is useful to interpret $\ptau_1$ and $\ptau_2$ in terms of the
characteristic viscous force, $\myf{\nu}=\rho\,\nu U_e/\dTwo^2$, and
the free-stream pressure force, $\myf{P}=\rho\,|\dPe|$, together with
the displacement inertia, $\myf{\dOne}= \rho U_e^2/\dOne$, and the
momentum-deficit inertia, $\myf{\dTwo}=\rho U_e^2/\dTwo$, where the
latter two encode information about the upstream history of the TBL.
With these definitions, the dimensionless inputs can be written as
\begin{align*}
    \ptau_1 &= (\myf{\dOne}/\myf{\dTwo})\,(\myf{\nu}/\myf{\dTwo})^{1/4}, \\
    \ptau_2 &= \gamma_P\, (\myf{\dTwo}/\myf{\dOne})^{11/18}
    (\myf{\nu}/\myf{\dTwo})^{4/9} (\myf{P}/\myf{\dTwo})^{5/18}.
\end{align*}
Hence, at a given streamwise location $x$, the TBL response is
governed by the inertia ratio $\myf{\dOne}/\myf{\dTwo}$, adjusted by
ratios involving viscous and pressure forces, with the exponents
\rev{specifying the weighting within each dimensionless group}.  In this
interpretation, $\ptau_1$, which is the primary determinant of
$\ptauo$, reflects the inertia state of the TBL with a \rev{viscous
correction} through $\myf{\nu}/\myf{\dTwo}$. By contrast, $\ptau_2$
introduces \rev{an additional} correction to the inertia ratio by accounting
for both viscous effects, $\myf{\nu}/\myf{\dTwo}$, and pressure
forcing, $\myf{P}/\myf{\dTwo}$, enabling the detection of separation
and reattachment. The condition for incipient separation, $\tau_w=0$,
is given by $f_\tau(\ptau_1,\ptau_2)=0$ (dashed line in
figure~\rev{\ref{fig:scaling_tau_2}b}). \corr{This relation defines a curve
    in the $(\ptau_1,\ptau_2)$ plane, showing that separation does not
    occur at a universal threshold of either input alone.}

%%%%%%%%%%%%%%%%%%%%%%%%%%%%%%%%%%%%%%%%%%%%%%%%%%%
\section{Discussion}
%%%%%%%%%%%%%%%%%%%%%%%%%%%%%%%%%%%%%%%%%%%%%%%%%%%

\begin{figure*}
    \includegraphics{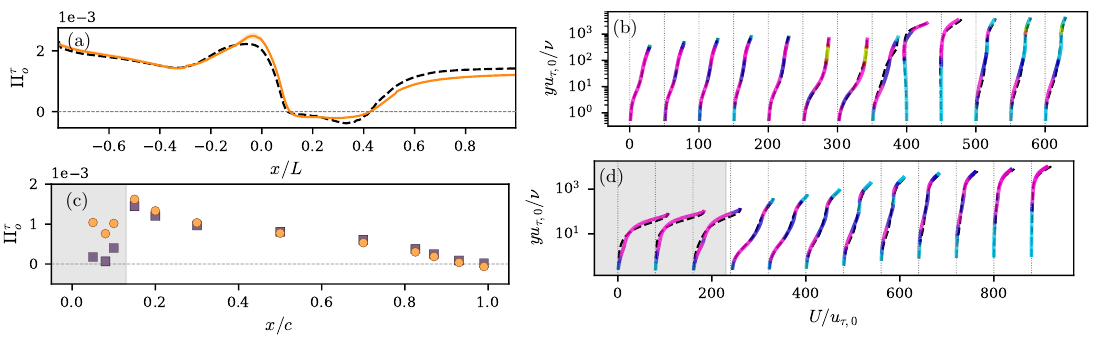}
    \caption{Predictions for unseen complex flows: (a,b)~a
      spanwise-periodic Gaussian bump~\citep{uzun2022High} and (c,d)~a
      near-stall airfoil~\citep{tamaki2023Wall}. \rev{The reference length $L$ is the
      experimental bump width, and $c$ is the
      airfoil chord.} Panels (a,c) compare
      predicted (orange) and reference (black dashed lines or purple
      squares) wall-shear stress; panels (b,d) compare mean velocity
      profiles, colored by the dominant \dimensionless~input as in
      figure~\rev{\ref{fig:Uprof}}. \rev{Shading denotes the airfoil's laminar region.}
      \rev{Profiles in (b,d) are offset by} $\Delta U/u_{\tau,0} = 50$
      and $80$, respectively.
        %
        %      Application of the models based on the discovered
        %        dimensionless groups ($f_\tau$ and $f_U$) to predicting wall
        %        shear stress and mean velocity profiles in complex
        %        configurations.  (a,b)~Spanwise-periodic Gaussian
        %        bump~\citep{uzun2022High}; (c,d)~near-stall
        %        airfoil~\citep{tamaki2023Wall}.  (a,c)~wall-shear stress:
        %        reference data (black dashed/purple squares) and model
        %        prediction (orange).  (b,d)~Mean velocity profiles at selected
        %        streamwise stations; colors indicate the dominant
        %        \dimensionless~input as in figure~\ref{fig:Uprof}.
    \label{fig:generalization}}
\end{figure*}

% @GA-corrections: the maximum reynolds theta are:
%   - 80,000 for tamaki2023 (See fig 11c)
%   - 180,000 for uzun2022 (see fig 6 top)

% Complex cases
To assess whether these conclusions extend beyond idealized configurations,
\rev{we evaluate the predictions on 17 previously unseen flow cases, comprising more than 1,000 mean velocity profiles across different streamwise locations and flow conditions, and find errors ranging from 1\% to 5\%.
Figure}~\ref{fig:generalization} \rev{shows two examples---}a spanwise-periodic Gaussian
bump~\citep{uzun2022High} and a near-stall
airfoil~\citep{tamaki2023Wall}\rev{---with the remaining cases presented in the
Supplementary Material}~\cite{suppl}. These \rev{two}
cases introduce physical effects absent from the training dataset,
including surface curvature, laminar-to-turbulent transition, and
higher Reynolds numbers up to \corr{$\Rey_\theta = 180{,}000$}.
Despite these differences, the information encoded in the
dimensionless groups predicts the \rev{mean velocity profiles and,
through their wall derivative, the wall-shear stress} with reasonable
accuracy across attached, separated, and
reattaching regions. \corr{ The largest discrepancies occur where
    curvature effects are strongest over the bump and in the laminar
    portions of the airfoil boundary layer. \rev{These regions lie primarily
    outside the intended scope of scaling laws discovered exclusively from
    TBL data without strong wall curvature or laminar flow.}}

% Closing
The success of the predictions for cases far from the canonical
dataset suggests that the identified \dimensionless~groups capture
fundamental scaling laws of pressure-gradient boundary layers, rather
than merely fitting specific configurations. A central finding is that
streamwise-local boundary-layer quantities alone---without explicit
knowledge of upstream history---are sufficient to predict \rev{the
mean velocity profile and its wall derivative}, even in the
nonequilibrium flows with separation and reattachment considered here.

It is also noteworthy that, unlike previous approaches for predicting
mean velocity profiles, the present model predicts $U(y)$ directly
from the dimensionless inputs. It therefore does not require
\corr{solving a differential equation by marching in the wall-normal
    direction---as in the mixing-length models of \cite{cantwell2019,
    subrahmanyam2022, ma2026}, which reconstruct $U(y)$ by integrating
    $\mathrm{d}U/\mathrm{d}y$ from the wall---nor does it require the
near-wall velocity gradient \rev{as an input}.}  Consequently, accurate mean velocity
profiles can be obtained without streamwise integration along $x$ or
wall-normal integration \corr{of a modeled governing relation} up to
the location $y$.

This locality provides physical support for turbulence models that
rely solely on local information while accurately representing
wall-bounded flows, including \rev{composite mean velocity
profiles}~\cite{coles1956wake, chauhan2007composite} \rev{and models
for wall-shear stress}~\citep{bose2018}. \rev{Experimentally, one can estimate}
$\tau_w$ \rev{from the modeled wall derivative using}
integral and edge quantities (e.g.,
$\delta_*$, $\theta$, $\delta$, $U_e$) without requiring \corr{direct
    wall-shear stress measurements or \rev{measurements of} the
    near-wall velocity gradient
$\partial U/\partial y$}, both of which are often challenging to
obtain. \corr{While $\delta_*$ and $\theta$ are themselves wall-normal
    integrals of $U$, they are less sensitive to near-wall resolution
    than $\partial U/\partial y$, providing a more robust route for
estimating $\tau_w$.}

\emph{Acknowledgments}---This research was supported by the National
Science Foundation under Grant No. 2140775, NNSA Predictive Science
Academic Alliance Program (PSAAP; grant DE-NA0003993), and by a Seed
Grant from the Center for Sensing to Intelligence at
Caltech. \rev{Computing time was provided by the ASCR Leadership
Computing Challenge (ALCC) program under project TUR152. This research
used resources of the Oak Ridge Leadership Computing Facility, which
is a DOE Office of Science User Facility supported under Contract
DE-AC05-00OR22725.} \corr{The authors thank Julio Soria for sharing the data.}

\emph{Data availability}---The data that support the findings of this
article, together with \rev{the velocity model $f_U$, its derived wall-shear
relation $f_\tau$,} and the code used
to discover the scalings and train \rev{the velocity model}, are available in the
associated GitHub repository~\cite{github}.

%%%%%%%%%%%%%%%%%%%%%%%%%%%%%%%%%%%%%%%%%%%%%%%%%%%%%%%%%%%%%%%%%%%%
\bibliography{references,additional_validation_references}

%apsrev4-2.bst 2019-01-14 (MD) hand-edited version of apsrev4-1.bst
%Control: key (0)
%Control: author (8) initials jnrlst
%Control: editor formatted (1) identically to author
%Control: production of article title (0) allowed
%Control: page (0) single
%Control: year (1) truncated
%Control: production of eprint (0) enabled
\begin{thebibliography}{140}%
\makeatletter
\providecommand \@ifxundefined [1]{%
 \@ifx{#1\undefined}
}%
\providecommand \@ifnum [1]{%
 \ifnum #1\expandafter \@firstoftwo
 \else \expandafter \@secondoftwo
 \fi
}%
\providecommand \@ifx [1]{%
 \ifx #1\expandafter \@firstoftwo
 \else \expandafter \@secondoftwo
 \fi
}%
\providecommand \natexlab [1]{#1}%
\providecommand \enquote  [1]{``#1''}%
\providecommand \bibnamefont  [1]{#1}%
\providecommand \bibfnamefont [1]{#1}%
\providecommand \citenamefont [1]{#1}%
\providecommand \href@noop [0]{\@secondoftwo}%
\providecommand \href [0]{\begingroup \@sanitize@url \@href}%
\providecommand \@href[1]{\@@startlink{#1}\@@href}%
\providecommand \@@href[1]{\endgroup#1\@@endlink}%
\providecommand \@sanitize@url [0]{\catcode `\\12\catcode `\$12\catcode
  `\&12\catcode `\#12\catcode `\^12\catcode `\_12\catcode `\%12\relax}%
\providecommand \@@startlink[1]{}%
\providecommand \@@endlink[0]{}%
\providecommand \url  [0]{\begingroup\@sanitize@url \@url }%
\providecommand \@url [1]{\endgroup\@href {#1}{\urlprefix }}%
\providecommand \urlprefix  [0]{URL }%
\providecommand \Eprint [0]{\href }%
\providecommand \doibase [0]{https://doi.org/}%
\providecommand \selectlanguage [0]{\@gobble}%
\providecommand \bibinfo  [0]{\@secondoftwo}%
\providecommand \bibfield  [0]{\@secondoftwo}%
\providecommand \translation [1]{[#1]}%
\providecommand \BibitemOpen [0]{}%
\providecommand \bibitemStop [0]{}%
\providecommand \bibitemNoStop [0]{.\EOS\space}%
\providecommand \EOS [0]{\spacefactor3000\relax}%
\providecommand \BibitemShut  [1]{\csname bibitem#1\endcsname}%
\let\auto@bib@innerbib\@empty
%</preamble>
\bibitem [{\citenamefont {Townsend}(1976)}]{Townsend1976}%
  \BibitemOpen
  \bibfield  {author} {\bibinfo {author} {\bibfnamefont {A.~A.}\ \bibnamefont
  {Townsend}},\ }\href@noop {} {\emph {\bibinfo {title} {The Structure of
  Turbulent Shear Flow}}},\ \bibinfo {edition} {2nd}\ ed.\ (\bibinfo
  {publisher} {Cambridge University Press},\ \bibinfo {address} {Cambridge},\
  \bibinfo {year} {1976})\BibitemShut {NoStop}%
\bibitem [{\citenamefont {Schlichting}(1979)}]{schlichting1979}%
  \BibitemOpen
  \bibfield  {author} {\bibinfo {author} {\bibfnamefont {H.}~\bibnamefont
  {Schlichting}},\ }\href@noop {} {\emph {\bibinfo {title} {Boundary-Layer
  Theory}}},\ \bibinfo {edition} {7th}\ ed.\ (\bibinfo  {publisher}
  {McGraw-Hill},\ \bibinfo {address} {New York},\ \bibinfo {year}
  {1979})\BibitemShut {NoStop}%
\bibitem [{\citenamefont {Pope}(2000)}]{Pope2000}%
  \BibitemOpen
  \bibfield  {author} {\bibinfo {author} {\bibfnamefont {S.~B.}\ \bibnamefont
  {Pope}},\ }\href {https://doi.org/10.1017/CBO9781316179475} {\emph {\bibinfo
  {title} {Turbulent Flows}}}\ (\bibinfo  {publisher} {Cambridge University
  Press},\ \bibinfo {address} {Cambridge},\ \bibinfo {year} {2000})\BibitemShut
  {NoStop}%
\bibitem [{\citenamefont {Smits}\ \emph {et~al.}(2011)\citenamefont {Smits},
  \citenamefont {McKeon},\ and\ \citenamefont {Marusic}}]{Smits2011}%
  \BibitemOpen
  \bibfield  {author} {\bibinfo {author} {\bibfnamefont {A.~J.}\ \bibnamefont
  {Smits}}, \bibinfo {author} {\bibfnamefont {B.~J.}\ \bibnamefont {McKeon}},\
  and\ \bibinfo {author} {\bibfnamefont {I.}~\bibnamefont {Marusic}},\
  }\bibfield  {title} {\bibinfo {title} {High--{Reynolds} number wall
  turbulence},\ }\href {https://doi.org/10.1146/annurev-fluid-122109-160753}
  {\bibfield  {journal} {\bibinfo  {journal} {Annu. Rev. Fluid Mech.}\ }\textbf
  {\bibinfo {volume} {43}},\ \bibinfo {pages} {353} (\bibinfo {year}
  {2011})}\BibitemShut {NoStop}%
\bibitem [{\citenamefont {Jim{\'e}nez}(2012)}]{Jimenez2012}%
  \BibitemOpen
  \bibfield  {author} {\bibinfo {author} {\bibfnamefont {J.}~\bibnamefont
  {Jim{\'e}nez}},\ }\bibfield  {title} {\bibinfo {title} {Cascades in
  wall-bounded turbulence},\ }\href
  {https://doi.org/10.1146/annurev-fluid-120710-101039} {\bibfield  {journal}
  {\bibinfo  {journal} {Annu. Rev. Fluid Mech.}\ }\textbf {\bibinfo {volume}
  {44}},\ \bibinfo {pages} {27} (\bibinfo {year} {2012})}\BibitemShut {NoStop}%
\bibitem [{\citenamefont {Rotta}(1953)}]{rotta1953theory}%
  \BibitemOpen
  \bibfield  {author} {\bibinfo {author} {\bibfnamefont {J.~C.}\ \bibnamefont
  {Rotta}},\ }\href@noop {} {\emph {\bibinfo {title} {On the Theory of the
  Turbulent Boundary Layer}}},\ \bibinfo {type} {NACA Technical Memorandum}\
  \bibinfo {number} {1344}\ (\bibinfo  {institution} {National Advisory
  Committee for Aeronautics},\ \bibinfo {year} {1953})\ \bibinfo {note}
  {translation of ``{\"U}ber die Theorie der turbulenten Grenzschichten''
  (1950)}\BibitemShut {NoStop}%
\bibitem [{\citenamefont {Clauser}(1954)}]{clauser1954turbulent}%
  \BibitemOpen
  \bibfield  {author} {\bibinfo {author} {\bibfnamefont {F.~H.}\ \bibnamefont
  {Clauser}},\ }\bibfield  {title} {\bibinfo {title} {Turbulent boundary layers
  in adverse pressure gradients},\ }\href {https://doi.org/10.2514/8.2938}
  {\bibfield  {journal} {\bibinfo  {journal} {J. Aeronaut. Sci.}\ }\textbf
  {\bibinfo {volume} {21}},\ \bibinfo {pages} {91} (\bibinfo {year}
  {1954})}\BibitemShut {NoStop}%
\bibitem [{\citenamefont {Mellor}\ and\ \citenamefont
  {Gibson}(1966)}]{mellor1966equilibrium}%
  \BibitemOpen
  \bibfield  {author} {\bibinfo {author} {\bibfnamefont {G.~L.}\ \bibnamefont
  {Mellor}}\ and\ \bibinfo {author} {\bibfnamefont {D.~M.}\ \bibnamefont
  {Gibson}},\ }\bibfield  {title} {\bibinfo {title} {Equilibrium turbulent
  boundary layers},\ }\href {https://doi.org/10.1017/S0022112066000612}
  {\bibfield  {journal} {\bibinfo  {journal} {J. Fluid Mech.}\ }\textbf
  {\bibinfo {volume} {24}},\ \bibinfo {pages} {225} (\bibinfo {year}
  {1966})}\BibitemShut {NoStop}%
\bibitem [{\citenamefont {Marusic}\ \emph
  {et~al.}(2010{\natexlab{a}})\citenamefont {Marusic}, \citenamefont {McKeon},
  \citenamefont {Monkewitz}, \citenamefont {Nagib}, \citenamefont {Smits},\
  and\ \citenamefont {Sreenivasan}}]{Marusic2010}%
  \BibitemOpen
  \bibfield  {author} {\bibinfo {author} {\bibfnamefont {I.}~\bibnamefont
  {Marusic}}, \bibinfo {author} {\bibfnamefont {B.~J.}\ \bibnamefont {McKeon}},
  \bibinfo {author} {\bibfnamefont {P.~A.}\ \bibnamefont {Monkewitz}}, \bibinfo
  {author} {\bibfnamefont {H.~M.}\ \bibnamefont {Nagib}}, \bibinfo {author}
  {\bibfnamefont {A.~J.}\ \bibnamefont {Smits}},\ and\ \bibinfo {author}
  {\bibfnamefont {K.~R.}\ \bibnamefont {Sreenivasan}},\ }\bibfield  {title}
  {\bibinfo {title} {Wall-bounded turbulent flows at high {Reynolds} numbers:
  {Recent} advances and key issues},\ }\href
  {https://doi.org/10.1063/1.3453711} {\bibfield  {journal} {\bibinfo
  {journal} {Phys. Fluids}\ }\textbf {\bibinfo {volume} {22}},\ \bibinfo
  {pages} {065103} (\bibinfo {year} {2010}{\natexlab{a}})}\BibitemShut
  {NoStop}%
\bibitem [{\citenamefont {Marusic}\ \emph
  {et~al.}(2010{\natexlab{b}})\citenamefont {Marusic}, \citenamefont {Mathis},\
  and\ \citenamefont {Hutchins}}]{Marusic2010Science}%
  \BibitemOpen
  \bibfield  {author} {\bibinfo {author} {\bibfnamefont {I.}~\bibnamefont
  {Marusic}}, \bibinfo {author} {\bibfnamefont {R.}~\bibnamefont {Mathis}},\
  and\ \bibinfo {author} {\bibfnamefont {N.}~\bibnamefont {Hutchins}},\
  }\bibfield  {title} {\bibinfo {title} {Predictive model for wall-bounded
  turbulent flow},\ }\href {https://doi.org/10.1126/science.1188765} {\bibfield
   {journal} {\bibinfo  {journal} {Science}\ }\textbf {\bibinfo {volume}
  {329}},\ \bibinfo {pages} {193} (\bibinfo {year}
  {2010}{\natexlab{b}})}\BibitemShut {NoStop}%
\bibitem [{\citenamefont {Spalart}(1988)}]{Spalart1988}%
  \BibitemOpen
  \bibfield  {author} {\bibinfo {author} {\bibfnamefont {P.~R.}\ \bibnamefont
  {Spalart}},\ }\bibfield  {title} {\bibinfo {title} {Direct simulation of a
  turbulent boundary layer up to {{\rev{$R_\theta = 1410$}}}},\ }\href
  {https://doi.org/10.1017/S0022112088000345} {\bibfield  {journal} {\bibinfo
  {journal} {J. Fluid Mech.}\ }\textbf {\bibinfo {volume} {187}},\ \bibinfo
  {pages} {61} (\bibinfo {year} {1988})}\BibitemShut {NoStop}%
\bibitem [{\citenamefont {De~Graaff}\ and\ \citenamefont
  {Eaton}(2000)}]{DeGraaff2000}%
  \BibitemOpen
  \bibfield  {author} {\bibinfo {author} {\bibfnamefont {D.~B.}\ \bibnamefont
  {De~Graaff}}\ and\ \bibinfo {author} {\bibfnamefont {J.~K.}\ \bibnamefont
  {Eaton}},\ }\bibfield  {title} {\bibinfo {title} {{Reynolds}-number scaling
  of the flat-plate turbulent boundary layer},\ }\href
  {https://doi.org/10.1017/S0022112000001713} {\bibfield  {journal} {\bibinfo
  {journal} {J. Fluid Mech.}\ }\textbf {\bibinfo {volume} {422}},\ \bibinfo
  {pages} {319} (\bibinfo {year} {2000})}\BibitemShut {NoStop}%
\bibitem [{\citenamefont {Schlatter}\ and\ \citenamefont
  {{\"O}rl{\"u}}(2010)}]{Schlatter2010}%
  \BibitemOpen
  \bibfield  {author} {\bibinfo {author} {\bibfnamefont {P.}~\bibnamefont
  {Schlatter}}\ and\ \bibinfo {author} {\bibfnamefont {R.}~\bibnamefont
  {{\"O}rl{\"u}}},\ }\bibfield  {title} {\bibinfo {title} {Assessment of direct
  numerical simulation data of turbulent boundary layers},\ }\href
  {https://doi.org/10.1017/S0022112010003113} {\bibfield  {journal} {\bibinfo
  {journal} {J. Fluid Mech.}\ }\textbf {\bibinfo {volume} {659}},\ \bibinfo
  {pages} {116} (\bibinfo {year} {2010})}\BibitemShut {NoStop}%
\bibitem [{\citenamefont {Sillero}\ \emph {et~al.}(2014)\citenamefont
  {Sillero}, \citenamefont {Jim{\'e}nez},\ and\ \citenamefont
  {Moser}}]{sillero2014}%
  \BibitemOpen
  \bibfield  {author} {\bibinfo {author} {\bibfnamefont {J.~A.}\ \bibnamefont
  {Sillero}}, \bibinfo {author} {\bibfnamefont {J.}~\bibnamefont
  {Jim{\'e}nez}},\ and\ \bibinfo {author} {\bibfnamefont {R.~D.}\ \bibnamefont
  {Moser}},\ }\bibfield  {title} {\bibinfo {title} {Two-point statistics for
  turbulent boundary layers and channels at {Reynolds} numbers up to $\delta^+
  \approx 2000$},\ }\href {https://doi.org/10.1063/1.4899259} {\bibfield
  {journal} {\bibinfo  {journal} {Phys. Fluids}\ }\textbf {\bibinfo {volume}
  {26}},\ \bibinfo {pages} {105109} (\bibinfo {year} {2014})}\BibitemShut
  {NoStop}%
\bibitem [{\citenamefont {Vinuesa}\ \emph
  {et~al.}(2017{\natexlab{a}})\citenamefont {Vinuesa}, \citenamefont
  {Hosseini}, \citenamefont {Hanifi}, \citenamefont {Henningson},\ and\
  \citenamefont {Schlatter}}]{vinuesa2017pressuregradient}%
  \BibitemOpen
  \bibfield  {author} {\bibinfo {author} {\bibfnamefont {R.}~\bibnamefont
  {Vinuesa}}, \bibinfo {author} {\bibfnamefont {S.~M.}\ \bibnamefont
  {Hosseini}}, \bibinfo {author} {\bibfnamefont {A.}~\bibnamefont {Hanifi}},
  \bibinfo {author} {\bibfnamefont {D.~S.}\ \bibnamefont {Henningson}},\ and\
  \bibinfo {author} {\bibfnamefont {P.}~\bibnamefont {Schlatter}},\ }\bibfield
  {title} {\bibinfo {title} {Pressure-gradient turbulent boundary layers
  developing around a wing section},\ }\href
  {https://doi.org/10.1007/s10494-017-9840-z} {\bibfield  {journal} {\bibinfo
  {journal} {Flow Turbul. Combust.}\ }\textbf {\bibinfo {volume} {99}},\
  \bibinfo {pages} {613} (\bibinfo {year} {2017}{\natexlab{a}})}\BibitemShut
  {NoStop}%
\bibitem [{\citenamefont {Simpson}(1989)}]{simpson1989turbulent}%
  \BibitemOpen
  \bibfield  {author} {\bibinfo {author} {\bibfnamefont {R.~L.}\ \bibnamefont
  {Simpson}},\ }\bibfield  {title} {\bibinfo {title} {Turbulent boundary-layer
  separation},\ }\href {https://doi.org/10.1146/annurev.fl.21.010189.001225}
  {\bibfield  {journal} {\bibinfo  {journal} {Annu. Rev. Fluid Mech.}\ }\textbf
  {\bibinfo {volume} {21}},\ \bibinfo {pages} {205} (\bibinfo {year}
  {1989})}\BibitemShut {NoStop}%
\bibitem [{\citenamefont {Perry}\ \emph {et~al.}(2002)\citenamefont {Perry},
  \citenamefont {Marusic},\ and\ \citenamefont {Jones}}]{perry2002}%
  \BibitemOpen
  \bibfield  {author} {\bibinfo {author} {\bibfnamefont {A.~E.}\ \bibnamefont
  {Perry}}, \bibinfo {author} {\bibfnamefont {I.}~\bibnamefont {Marusic}},\
  and\ \bibinfo {author} {\bibfnamefont {M.~B.}\ \bibnamefont {Jones}},\
  }\bibfield  {title} {\bibinfo {title} {On the streamwise evolution of
  turbulent boundary layers in arbitrary pressure gradients},\ }\href
  {https://doi.org/10.1017/S002211200200825X} {\bibfield  {journal} {\bibinfo
  {journal} {J. Fluid Mech.}\ }\textbf {\bibinfo {volume} {461}},\ \bibinfo
  {pages} {61} (\bibinfo {year} {2002})}\BibitemShut {NoStop}%
\bibitem [{\citenamefont {Kiya}\ and\ \citenamefont
  {Sasaki}(1983)}]{kiya1983structure}%
  \BibitemOpen
  \bibfield  {author} {\bibinfo {author} {\bibfnamefont {M.}~\bibnamefont
  {Kiya}}\ and\ \bibinfo {author} {\bibfnamefont {K.}~\bibnamefont {Sasaki}},\
  }\bibfield  {title} {\bibinfo {title} {Structure of a turbulent separation
  bubble},\ }\href {https://doi.org/10.1017/S002211208300230X} {\bibfield
  {journal} {\bibinfo  {journal} {J. Fluid Mech.}\ }\textbf {\bibinfo {volume}
  {137}},\ \bibinfo {pages} {83} (\bibinfo {year} {1983})}\BibitemShut
  {NoStop}%
\bibitem [{\citenamefont {Alving}\ and\ \citenamefont
  {Fernholz}(1996)}]{alving1996turbulence}%
  \BibitemOpen
  \bibfield  {author} {\bibinfo {author} {\bibfnamefont {A.~E.}\ \bibnamefont
  {Alving}}\ and\ \bibinfo {author} {\bibfnamefont {H.~H.}\ \bibnamefont
  {Fernholz}},\ }\bibfield  {title} {\bibinfo {title} {Turbulence measurements
  around a mild separation bubble and downstream of reattachment},\ }\href
  {https://doi.org/10.1017/S0022112096002807} {\bibfield  {journal} {\bibinfo
  {journal} {J. Fluid Mech.}\ }\textbf {\bibinfo {volume} {322}},\ \bibinfo
  {pages} {297} (\bibinfo {year} {1996})}\BibitemShut {NoStop}%
\bibitem [{\citenamefont {Song}\ and\ \citenamefont {Eaton}(2004)}]{Song2004}%
  \BibitemOpen
  \bibfield  {author} {\bibinfo {author} {\bibfnamefont {S.}~\bibnamefont
  {Song}}\ and\ \bibinfo {author} {\bibfnamefont {J.~K.}\ \bibnamefont
  {Eaton}},\ }\bibfield  {title} {\bibinfo {title} {{Reynolds} number effects
  on a turbulent boundary layer with separation, reattachment, and recovery},\
  }\href {https://doi.org/10.1007/s00348-003-0696-8} {\bibfield  {journal}
  {\bibinfo  {journal} {Exp. Fluids}\ }\textbf {\bibinfo {volume} {36}},\
  \bibinfo {pages} {246} (\bibinfo {year} {2004})}\BibitemShut {NoStop}%
\bibitem [{\citenamefont {Bobke}\ \emph {et~al.}(2017)\citenamefont {Bobke},
  \citenamefont {Vinuesa}, \citenamefont {{\"O}rl{\"u}},\ and\ \citenamefont
  {Schlatter}}]{bobke2017history}%
  \BibitemOpen
  \bibfield  {author} {\bibinfo {author} {\bibfnamefont {A.}~\bibnamefont
  {Bobke}}, \bibinfo {author} {\bibfnamefont {R.}~\bibnamefont {Vinuesa}},
  \bibinfo {author} {\bibfnamefont {R.}~\bibnamefont {{\"O}rl{\"u}}},\ and\
  \bibinfo {author} {\bibfnamefont {P.}~\bibnamefont {Schlatter}},\ }\bibfield
  {title} {\bibinfo {title} {History effects and near equilibrium in
  adverse-pressure-gradient turbulent boundary layers},\ }\href
  {https://doi.org/10.1017/jfm.2017.236} {\bibfield  {journal} {\bibinfo
  {journal} {J. Fluid Mech.}\ }\textbf {\bibinfo {volume} {820}},\ \bibinfo
  {pages} {667} (\bibinfo {year} {2017})}\BibitemShut {NoStop}%
\bibitem [{\citenamefont {Vinuesa}\ \emph
  {et~al.}(2017{\natexlab{b}})\citenamefont {Vinuesa}, \citenamefont
  {{\"O}rl{\"u}}, \citenamefont {Sanmiguel~Vila}, \citenamefont {Ianiro},
  \citenamefont {Discetti},\ and\ \citenamefont
  {Schlatter}}]{vinuesa2017revisiting}%
  \BibitemOpen
  \bibfield  {author} {\bibinfo {author} {\bibfnamefont {R.}~\bibnamefont
  {Vinuesa}}, \bibinfo {author} {\bibfnamefont {R.}~\bibnamefont
  {{\"O}rl{\"u}}}, \bibinfo {author} {\bibfnamefont {C.}~\bibnamefont
  {Sanmiguel~Vila}}, \bibinfo {author} {\bibfnamefont {A.}~\bibnamefont
  {Ianiro}}, \bibinfo {author} {\bibfnamefont {S.}~\bibnamefont {Discetti}},\
  and\ \bibinfo {author} {\bibfnamefont {P.}~\bibnamefont {Schlatter}},\
  }\bibfield  {title} {\bibinfo {title} {Revisiting history effects in
  adverse-pressure-gradient turbulent boundary layers},\ }\href
  {https://doi.org/10.1007/s10494-017-9845-7} {\bibfield  {journal} {\bibinfo
  {journal} {Flow Turbul. Combust.}\ }\textbf {\bibinfo {volume} {99}},\
  \bibinfo {pages} {565} (\bibinfo {year} {2017}{\natexlab{b}})}\BibitemShut
  {NoStop}%
\bibitem [{\citenamefont {Virgilio}\ \emph {et~al.}(2025)\citenamefont
  {Virgilio}, \citenamefont {Preskett}, \citenamefont {Jaiswal},\ and\
  \citenamefont {Ganapathisubramani}}]{virgilio2025pressure}%
  \BibitemOpen
  \bibfield  {author} {\bibinfo {author} {\bibfnamefont {M.}~\bibnamefont
  {Virgilio}}, \bibinfo {author} {\bibfnamefont {T.}~\bibnamefont {Preskett}},
  \bibinfo {author} {\bibfnamefont {P.}~\bibnamefont {Jaiswal}},\ and\ \bibinfo
  {author} {\bibfnamefont {B.}~\bibnamefont {Ganapathisubramani}},\ }\bibfield
  {title} {\bibinfo {title} {Pressure gradient history effects on integral
  quantities of turbulent boundary layers: Experiments and data-driven
  models},\ }\href {https://doi.org/10.1017/jfm.2025.10245} {\bibfield
  {journal} {\bibinfo  {journal} {J. Fluid Mech.}\ }\textbf {\bibinfo {volume}
  {1014}},\ \bibinfo {pages} {A2} (\bibinfo {year} {2025})}\BibitemShut
  {NoStop}%
\bibitem [{\citenamefont {Preskett}\ \emph {et~al.}(2025)\citenamefont
  {Preskett}, \citenamefont {Virgilio}, \citenamefont {Jaiswal},\ and\
  \citenamefont {Ganapathisubramani}}]{preskett2025effects}%
  \BibitemOpen
  \bibfield  {author} {\bibinfo {author} {\bibfnamefont {T.}~\bibnamefont
  {Preskett}}, \bibinfo {author} {\bibfnamefont {M.}~\bibnamefont {Virgilio}},
  \bibinfo {author} {\bibfnamefont {P.}~\bibnamefont {Jaiswal}},\ and\ \bibinfo
  {author} {\bibfnamefont {B.}~\bibnamefont {Ganapathisubramani}},\ }\bibfield
  {title} {\bibinfo {title} {Effects of pressure-gradient histories on skin
  friction and mean flow of high {Reynolds} number turbulent boundary layers
  over smooth and rough walls},\ }\href {https://doi.org/10.1017/jfm.2025.320}
  {\bibfield  {journal} {\bibinfo  {journal} {J. Fluid Mech.}\ }\textbf
  {\bibinfo {volume} {1010}},\ \bibinfo {pages} {A30} (\bibinfo {year}
  {2025})}\BibitemShut {NoStop}%
\bibitem [{\citenamefont {Zarei}\ \emph {et~al.}(2026)\citenamefont {Zarei},
  \citenamefont {Lozier}, \citenamefont {Deshpande},\ and\ \citenamefont
  {Marusic}}]{zarei2025decoupling}%
  \BibitemOpen
  \bibfield  {author} {\bibinfo {author} {\bibfnamefont {A.}~\bibnamefont
  {Zarei}}, \bibinfo {author} {\bibfnamefont {M.}~\bibnamefont {Lozier}},
  \bibinfo {author} {\bibfnamefont {R.}~\bibnamefont {Deshpande}},\ and\
  \bibinfo {author} {\bibfnamefont {I.}~\bibnamefont {Marusic}},\ }\bibfield
  {title} {\bibinfo {title} {{\rev{Turbulent boundary layers under adverse
  pressure gradients. Part 1. Decoupling the influences of local and upstream
  pressure gradients through high-Reynolds-number experiments}}},\ }\href
  {https://doi.org/10.1017/jfm.2026.12090} {\bibfield  {journal} {\bibinfo
  {journal} {\rev{J. Fluid Mech.}}\ }\textbf {\bibinfo {volume} {\rev{1045}}},\
  \bibinfo {pages} {\rev{A7}} (\bibinfo {year} {\rev{2026}})}\BibitemShut
  {NoStop}%
\bibitem [{\citenamefont {Zagarola}\ and\ \citenamefont
  {Smits}(1998)}]{zagarola1998meanflow}%
  \BibitemOpen
  \bibfield  {author} {\bibinfo {author} {\bibfnamefont {M.~V.}\ \bibnamefont
  {Zagarola}}\ and\ \bibinfo {author} {\bibfnamefont {A.~J.}\ \bibnamefont
  {Smits}},\ }\bibfield  {title} {\bibinfo {title} {Mean-flow scaling of
  turbulent pipe flow},\ }\href {https://doi.org/10.1017/S0022112098002419}
  {\bibfield  {journal} {\bibinfo  {journal} {J. Fluid Mech.}\ }\textbf
  {\bibinfo {volume} {373}},\ \bibinfo {pages} {33} (\bibinfo {year}
  {1998})}\BibitemShut {NoStop}%
\bibitem [{\citenamefont {Castillo}\ and\ \citenamefont
  {George}(2001)}]{castillo2001similarity}%
  \BibitemOpen
  \bibfield  {author} {\bibinfo {author} {\bibfnamefont {L.}~\bibnamefont
  {Castillo}}\ and\ \bibinfo {author} {\bibfnamefont {W.~K.}\ \bibnamefont
  {George}},\ }\bibfield  {title} {\bibinfo {title} {Similarity analysis for
  turbulent boundary layer with pressure gradient: Outer flow},\ }\href
  {https://doi.org/10.2514/2.1300} {\bibfield  {journal} {\bibinfo  {journal}
  {AIAA J.}\ }\textbf {\bibinfo {volume} {39}},\ \bibinfo {pages} {41}
  (\bibinfo {year} {2001})}\BibitemShut {NoStop}%
\bibitem [{\citenamefont {Maciel}\ \emph {et~al.}(2006)\citenamefont {Maciel},
  \citenamefont {Rossignol},\ and\ \citenamefont
  {Lemay}}]{maciel2006selfsimilarity}%
  \BibitemOpen
  \bibfield  {author} {\bibinfo {author} {\bibfnamefont {Y.}~\bibnamefont
  {Maciel}}, \bibinfo {author} {\bibfnamefont {K.-S.}\ \bibnamefont
  {Rossignol}},\ and\ \bibinfo {author} {\bibfnamefont {J.}~\bibnamefont
  {Lemay}},\ }\bibfield  {title} {\bibinfo {title} {Self-similarity in the
  outer region of adverse-pressure-gradient turbulent boundary layers},\ }\href
  {https://doi.org/10.2514/1.19234} {\bibfield  {journal} {\bibinfo  {journal}
  {AIAA J.}\ }\textbf {\bibinfo {volume} {44}},\ \bibinfo {pages} {2450}
  (\bibinfo {year} {2006})}\BibitemShut {NoStop}%
\bibitem [{\citenamefont {Gungor}\ \emph {et~al.}(2016)\citenamefont {Gungor},
  \citenamefont {Maciel}, \citenamefont {Simens},\ and\ \citenamefont
  {Soria}}]{gungor2016scaling}%
  \BibitemOpen
  \bibfield  {author} {\bibinfo {author} {\bibfnamefont {A.~G.}\ \bibnamefont
  {Gungor}}, \bibinfo {author} {\bibfnamefont {Y.}~\bibnamefont {Maciel}},
  \bibinfo {author} {\bibfnamefont {M.~P.}\ \bibnamefont {Simens}},\ and\
  \bibinfo {author} {\bibfnamefont {J.}~\bibnamefont {Soria}},\ }\bibfield
  {title} {\bibinfo {title} {Scaling and statistics of large-defect adverse
  pressure gradient turbulent boundary layers},\ }\href
  {https://doi.org/10.1016/j.ijheatfluidflow.2016.03.004} {\bibfield  {journal}
  {\bibinfo  {journal} {Int. J. Heat Fluid Flow}\ }\textbf {\bibinfo {volume}
  {59}},\ \bibinfo {pages} {109} (\bibinfo {year} {2016})}\BibitemShut
  {NoStop}%
\bibitem [{\citenamefont {Kitsios}\ \emph {et~al.}(2017)\citenamefont
  {Kitsios}, \citenamefont {Sekimoto}, \citenamefont {Atkinson}, \citenamefont
  {Sillero}, \citenamefont {Borrell}, \citenamefont {Gungor}, \citenamefont
  {Jim{\'e}nez},\ and\ \citenamefont {Soria}}]{kitsios2017direct}%
  \BibitemOpen
  \bibfield  {author} {\bibinfo {author} {\bibfnamefont {V.}~\bibnamefont
  {Kitsios}}, \bibinfo {author} {\bibfnamefont {A.}~\bibnamefont {Sekimoto}},
  \bibinfo {author} {\bibfnamefont {C.}~\bibnamefont {Atkinson}}, \bibinfo
  {author} {\bibfnamefont {J.~A.}\ \bibnamefont {Sillero}}, \bibinfo {author}
  {\bibfnamefont {G.}~\bibnamefont {Borrell}}, \bibinfo {author} {\bibfnamefont
  {A.~G.}\ \bibnamefont {Gungor}}, \bibinfo {author} {\bibfnamefont
  {J.}~\bibnamefont {Jim{\'e}nez}},\ and\ \bibinfo {author} {\bibfnamefont
  {J.}~\bibnamefont {Soria}},\ }\bibfield  {title} {\bibinfo {title} {Direct
  numerical simulation of a self-similar adverse pressure gradient turbulent
  boundary layer at the verge of separation},\ }\href
  {https://doi.org/10.1017/jfm.2017.549} {\bibfield  {journal} {\bibinfo
  {journal} {J. Fluid Mech.}\ }\textbf {\bibinfo {volume} {829}},\ \bibinfo
  {pages} {392} (\bibinfo {year} {2017})}\BibitemShut {NoStop}%
\bibitem [{\citenamefont {Gungor}\ \emph {et~al.}(2022)\citenamefont {Gungor},
  \citenamefont {Maciel},\ and\ \citenamefont {Gungor}}]{gungor2022energy}%
  \BibitemOpen
  \bibfield  {author} {\bibinfo {author} {\bibfnamefont {T.~R.}\ \bibnamefont
  {Gungor}}, \bibinfo {author} {\bibfnamefont {Y.}~\bibnamefont {Maciel}},\
  and\ \bibinfo {author} {\bibfnamefont {A.~G.}\ \bibnamefont {Gungor}},\
  }\bibfield  {title} {\bibinfo {title} {Energy transfer mechanisms in adverse
  pressure gradient turbulent boundary layers: production and inter-component
  redistribution},\ }\href {https://doi.org/10.1017/jfm.2022.679} {\bibfield
  {journal} {\bibinfo  {journal} {J. Fluid Mech.}\ }\textbf {\bibinfo {volume}
  {948}},\ \bibinfo {pages} {A5} (\bibinfo {year} {2022})}\BibitemShut
  {NoStop}%
\bibitem [{\citenamefont {Gungor}\ \emph {et~al.}(2024)\citenamefont {Gungor},
  \citenamefont {Gungor},\ and\ \citenamefont {Maciel}}]{gungor2024response}%
  \BibitemOpen
  \bibfield  {author} {\bibinfo {author} {\bibfnamefont {T.~R.}\ \bibnamefont
  {Gungor}}, \bibinfo {author} {\bibfnamefont {A.~G.}\ \bibnamefont {Gungor}},\
  and\ \bibinfo {author} {\bibfnamefont {Y.}~\bibnamefont {Maciel}},\
  }\bibfield  {title} {\bibinfo {title} {Turbulent boundary layer response to
  uniform changes of the pressure force contribution},\ }\href
  {https://doi.org/10.1017/jfm.2024.579} {\bibfield  {journal} {\bibinfo
  {journal} {J. Fluid Mech.}\ }\textbf {\bibinfo {volume} {997}},\ \bibinfo
  {pages} {A75} (\bibinfo {year} {2024})}\BibitemShut {NoStop}%
\bibitem [{\citenamefont {Dixit}\ \emph {et~al.}(2024)\citenamefont {Dixit},
  \citenamefont {Gupta}, \citenamefont {Choudhary},\ and\ \citenamefont
  {Prabhakaran}}]{dixit2024generalized}%
  \BibitemOpen
  \bibfield  {author} {\bibinfo {author} {\bibfnamefont {S.~A.}\ \bibnamefont
  {Dixit}}, \bibinfo {author} {\bibfnamefont {A.}~\bibnamefont {Gupta}},
  \bibinfo {author} {\bibfnamefont {H.}~\bibnamefont {Choudhary}},\ and\
  \bibinfo {author} {\bibfnamefont {T.}~\bibnamefont {Prabhakaran}},\
  }\bibfield  {title} {\bibinfo {title} {Generalized scaling and model for
  friction in wall turbulence},\ }\href
  {https://doi.org/10.1103/PhysRevLett.132.014001} {\bibfield  {journal}
  {\bibinfo  {journal} {Phys. Rev. Lett.}\ }\textbf {\bibinfo {volume} {132}},\
  \bibinfo {pages} {014001} (\bibinfo {year} {2024})}\BibitemShut {NoStop}%
\bibitem [{\citenamefont {Buckingham}(1914)}]{Buckingham1914}%
  \BibitemOpen
  \bibfield  {author} {\bibinfo {author} {\bibfnamefont {E.}~\bibnamefont
  {Buckingham}},\ }\bibfield  {title} {\bibinfo {title} {On physically similar
  systems; illustrations of the use of dimensional equations},\ }\href
  {https://doi.org/10.1103/PhysRev.4.345} {\bibfield  {journal} {\bibinfo
  {journal} {Phys. Rev.}\ }\textbf {\bibinfo {volume} {4}},\ \bibinfo {pages}
  {345} (\bibinfo {year} {1914})}\BibitemShut {NoStop}%
\bibitem [{\citenamefont {Yuan}\ and\ \citenamefont
  {Lozano-Dur{\'a}n}(2025)}]{yuan2025Dimensionless}%
  \BibitemOpen
  \bibfield  {author} {\bibinfo {author} {\bibfnamefont {Y.}~\bibnamefont
  {Yuan}}\ and\ \bibinfo {author} {\bibfnamefont {A.}~\bibnamefont
  {Lozano-Dur{\'a}n}},\ }\bibfield  {title} {\bibinfo {title} {Dimensionless
  learning based on information},\ }\href
  {https://doi.org/10.1038/s41467-025-64425-8} {\bibfield  {journal} {\bibinfo
  {journal} {Nat. Commun.}\ }\textbf {\bibinfo {volume} {16}},\ \bibinfo
  {pages} {9171} (\bibinfo {year} {2025})}\BibitemShut {NoStop}%
\bibitem [{\citenamefont {Shannon}(1948)}]{shannon1948}%
  \BibitemOpen
  \bibfield  {author} {\bibinfo {author} {\bibfnamefont {C.~E.}\ \bibnamefont
  {Shannon}},\ }\bibfield  {title} {\bibinfo {title} {A mathematical theory of
  communication},\ }\href {https://doi.org/10.1002/j.1538-7305.1948.tb01338.x}
  {\bibfield  {journal} {\bibinfo  {journal} {Bell Syst. Tech. J.}\ }\textbf
  {\bibinfo {volume} {27}},\ \bibinfo {pages} {379} (\bibinfo {year} {1948})},\
  \bibinfo {note} {published in two parts; the second part has DOI
  10.1002/j.1538-7305.1948.tb00917.x}\BibitemShut {NoStop}%
\bibitem [{sup()}]{suppl}%
  \BibitemOpen
  \href@noop {} {}\bibinfo {note} {See Supplemental Material at [URL to be
  inserted by the publisher] for details of the databases, dimensional
  analysis, scaling sensitivity, model construction, additional validation
  cases, and input-space coverage, which includes Refs.~[70--140].}\BibitemShut
  {Stop}%
\bibitem [{\citenamefont {Liu}\ \emph {et~al.}(2025)\citenamefont {Liu},
  \citenamefont {Wang}, \citenamefont {Vaidya}, \citenamefont {Ruehle},
  \citenamefont {Halverson}, \citenamefont {Solja{\v{c}}i{\'c}}, \citenamefont
  {Hou},\ and\ \citenamefont {Tegmark}}]{liu2024kan}%
  \BibitemOpen
  \bibfield  {author} {\bibinfo {author} {\bibfnamefont {Z.}~\bibnamefont
  {Liu}}, \bibinfo {author} {\bibfnamefont {Y.}~\bibnamefont {Wang}}, \bibinfo
  {author} {\bibfnamefont {S.}~\bibnamefont {Vaidya}}, \bibinfo {author}
  {\bibfnamefont {F.}~\bibnamefont {Ruehle}}, \bibinfo {author} {\bibfnamefont
  {J.}~\bibnamefont {Halverson}}, \bibinfo {author} {\bibfnamefont
  {M.}~\bibnamefont {Solja{\v{c}}i{\'c}}}, \bibinfo {author} {\bibfnamefont
  {T.~Y.}\ \bibnamefont {Hou}},\ and\ \bibinfo {author} {\bibfnamefont
  {M.}~\bibnamefont {Tegmark}},\ }\bibfield  {title} {\bibinfo {title} {{KAN}:
  {Kolmogorov--Arnold} networks},\ }in\ \href
  {https://arxiv.org/abs/2404.19756} {\emph {\bibinfo {booktitle}
  {International Conference on Learning Representations}}}\ (\bibinfo {year}
  {2025})\ \Eprint {https://arxiv.org/abs/2404.19756} {arXiv:2404.19756
  [cs.LG]} \BibitemShut {NoStop}%
\bibitem [{\citenamefont {de~Boor}(2001)}]{deboor2001splines}%
  \BibitemOpen
  \bibfield  {author} {\bibinfo {author} {\bibfnamefont {C.}~\bibnamefont
  {de~Boor}},\ }\href {https://link.springer.com/book/9780387953663} {\emph
  {\bibinfo {title} {A Practical Guide to Splines}}},\ \bibinfo {edition}
  {revised}\ ed.,\ \bibinfo {series} {Applied Mathematical Sciences},
  Vol.~\bibinfo {volume} {27}\ (\bibinfo  {publisher} {Springer},\ \bibinfo
  {address} {New York},\ \bibinfo {year} {2001})\BibitemShut {NoStop}%
\bibitem [{\citenamefont {Coleman}\ \emph {et~al.}(2018)\citenamefont
  {Coleman}, \citenamefont {Rumsey},\ and\ \citenamefont
  {Spalart}}]{coleman2018numerical}%
  \BibitemOpen
  \bibfield  {author} {\bibinfo {author} {\bibfnamefont {G.~N.}\ \bibnamefont
  {Coleman}}, \bibinfo {author} {\bibfnamefont {C.~L.}\ \bibnamefont
  {Rumsey}},\ and\ \bibinfo {author} {\bibfnamefont {P.~R.}\ \bibnamefont
  {Spalart}},\ }\bibfield  {title} {\bibinfo {title} {Numerical study of
  turbulent separation bubbles with varying pressure gradient and {Reynolds}
  number},\ }\href {https://doi.org/10.1017/jfm.2018.257} {\bibfield  {journal}
  {\bibinfo  {journal} {J. Fluid Mech.}\ }\textbf {\bibinfo {volume} {847}},\
  \bibinfo {pages} {28} (\bibinfo {year} {2018})}\BibitemShut {NoStop}%
\bibitem [{\citenamefont {Coleman}(2021)}]{coleman2021numerical}%
  \BibitemOpen
  \bibfield  {author} {\bibinfo {author} {\bibfnamefont {G.~N.}\ \bibnamefont
  {Coleman}},\ }\href@noop {} {\emph {\bibinfo {title} {Numerical Simulation of
  Pressure-Induced Separation of Turbulent Flat-Plate Boundary Layers:
  Definition and Overview of New Cases with Suction-Only Transpiration and a
  Step in {Reynolds} Number}}},\ \bibinfo {type} {NASA Technical Memorandum}\
  \bibinfo {number} {NASA/TM-20210020762}\ (\bibinfo  {institution} {NASA
  Langley Research Center},\ \bibinfo {address} {Hampton, VA},\ \bibinfo {year}
  {2021})\ \bibinfo {note} {errata issued January 2022}\BibitemShut {NoStop}%
\bibitem [{\citenamefont {Kader}\ and\ \citenamefont
  {Yaglom}(1978)}]{kader1978Similarity}%
  \BibitemOpen
  \bibfield  {author} {\bibinfo {author} {\bibfnamefont {B.~A.}\ \bibnamefont
  {Kader}}\ and\ \bibinfo {author} {\bibfnamefont {A.~M.}\ \bibnamefont
  {Yaglom}},\ }\bibfield  {title} {\bibinfo {title} {Similarity treatment of
  moving-equilibrium turbulent boundary layers in adverse pressure gradients},\
  }\href {https://doi.org/10.1017/S0022112078002621} {\bibfield  {journal}
  {\bibinfo  {journal} {J. Fluid Mech.}\ }\textbf {\bibinfo {volume} {89}},\
  \bibinfo {pages} {305} (\bibinfo {year} {1978})}\BibitemShut {NoStop}%
\bibitem [{\citenamefont {Skote}\ and\ \citenamefont
  {Henningson}(2002)}]{skote2002Direct}%
  \BibitemOpen
  \bibfield  {author} {\bibinfo {author} {\bibfnamefont {M.}~\bibnamefont
  {Skote}}\ and\ \bibinfo {author} {\bibfnamefont {D.~S.}\ \bibnamefont
  {Henningson}},\ }\bibfield  {title} {\bibinfo {title} {Direct numerical
  simulation of a separated turbulent boundary layer},\ }\href
  {https://doi.org/10.1017/S0022112002002173} {\bibfield  {journal} {\bibinfo
  {journal} {J. Fluid Mech.}\ }\textbf {\bibinfo {volume} {471}},\ \bibinfo
  {pages} {107} (\bibinfo {year} {2002})}\BibitemShut {NoStop}%
\bibitem [{\citenamefont {Nickels}(2004)}]{nickels2004inner}%
  \BibitemOpen
  \bibfield  {author} {\bibinfo {author} {\bibfnamefont {T.~B.}\ \bibnamefont
  {Nickels}},\ }\bibfield  {title} {\bibinfo {title} {Inner scaling for
  wall-bounded flows subject to large pressure gradients},\ }\href
  {https://doi.org/10.1017/S0022112004001788} {\bibfield  {journal} {\bibinfo
  {journal} {J. Fluid Mech.}\ }\textbf {\bibinfo {volume} {521}},\ \bibinfo
  {pages} {217} (\bibinfo {year} {2004})}\BibitemShut {NoStop}%
\bibitem [{\citenamefont {Devenport}\ and\ \citenamefont
  {Lowe}(2022)}]{devenport2022equilibrium}%
  \BibitemOpen
  \bibfield  {author} {\bibinfo {author} {\bibfnamefont {W.~J.}\ \bibnamefont
  {Devenport}}\ and\ \bibinfo {author} {\bibfnamefont {K.~T.}\ \bibnamefont
  {Lowe}},\ }\bibfield  {title} {\bibinfo {title} {Equilibrium and
  non-equilibrium turbulent boundary layers},\ }\href
  {https://doi.org/10.1016/j.paerosci.2022.100807} {\bibfield  {journal}
  {\bibinfo  {journal} {Prog. Aerosp. Sci.}\ }\textbf {\bibinfo {volume}
  {131}},\ \bibinfo {pages} {100807} (\bibinfo {year} {2022})}\BibitemShut
  {NoStop}%
\bibitem [{\citenamefont {Rotta}(1962)}]{rotta1962turbulent}%
  \BibitemOpen
  \bibfield  {author} {\bibinfo {author} {\bibfnamefont {J.~C.}\ \bibnamefont
  {Rotta}},\ }\bibfield  {title} {\bibinfo {title} {Turbulent boundary layers
  in incompressible flow},\ }\href
  {https://doi.org/10.1016/0376-0421(62)90014-3} {\bibfield  {journal}
  {\bibinfo  {journal} {Prog. Aerosp. Sci.}\ }\textbf {\bibinfo {volume} {2}},\
  \bibinfo {pages} {1} (\bibinfo {year} {1962})}\BibitemShut {NoStop}%
\bibitem [{\citenamefont {Stratford}(1959)}]{stratford1959Prediction}%
  \BibitemOpen
  \bibfield  {author} {\bibinfo {author} {\bibfnamefont {B.~S.}\ \bibnamefont
  {Stratford}},\ }\bibfield  {title} {\bibinfo {title} {The prediction of
  separation of the turbulent boundary layer},\ }\href
  {https://doi.org/10.1017/S0022112059000015} {\bibfield  {journal} {\bibinfo
  {journal} {J. Fluid Mech.}\ }\textbf {\bibinfo {volume} {5}},\ \bibinfo
  {pages} {1} (\bibinfo {year} {1959})}\BibitemShut {NoStop}%
\bibitem [{\citenamefont {Patel}(1973)}]{patel1973}%
  \BibitemOpen
  \bibfield  {author} {\bibinfo {author} {\bibfnamefont {V.~C.}\ \bibnamefont
  {Patel}},\ }\bibfield  {title} {\bibinfo {title} {A unified view of the law
  of the wall using mixing-length theory},\ }\href
  {https://doi.org/10.1017/S0001925900006429} {\bibfield  {journal} {\bibinfo
  {journal} {Aeronaut. Q.}\ }\textbf {\bibinfo {volume} {24}},\ \bibinfo
  {pages} {55} (\bibinfo {year} {1973})}\BibitemShut {NoStop}%
\bibitem [{\citenamefont {Zagarola}\ and\ \citenamefont
  {Smits}(1997)}]{zagarola1997scaling}%
  \BibitemOpen
  \bibfield  {author} {\bibinfo {author} {\bibfnamefont {M.~V.}\ \bibnamefont
  {Zagarola}}\ and\ \bibinfo {author} {\bibfnamefont {A.~J.}\ \bibnamefont
  {Smits}},\ }\bibfield  {title} {\bibinfo {title} {Scaling of the mean
  velocity profile for turbulent pipe flow},\ }\href
  {https://doi.org/10.1103/PhysRevLett.78.239} {\bibfield  {journal} {\bibinfo
  {journal} {Phys. Rev. Lett.}\ }\textbf {\bibinfo {volume} {78}},\ \bibinfo
  {pages} {239} (\bibinfo {year} {1997})}\BibitemShut {NoStop}%
\bibitem [{\citenamefont {Maciel}\ \emph {et~al.}(2018)\citenamefont {Maciel},
  \citenamefont {Wei}, \citenamefont {Gungor},\ and\ \citenamefont
  {Simens}}]{maciel2018outer}%
  \BibitemOpen
  \bibfield  {author} {\bibinfo {author} {\bibfnamefont {Y.}~\bibnamefont
  {Maciel}}, \bibinfo {author} {\bibfnamefont {T.}~\bibnamefont {Wei}},
  \bibinfo {author} {\bibfnamefont {A.~G.}\ \bibnamefont {Gungor}},\ and\
  \bibinfo {author} {\bibfnamefont {M.~P.}\ \bibnamefont {Simens}},\ }\bibfield
   {title} {\bibinfo {title} {Outer scales and parameters of
  adverse-pressure-gradient turbulent boundary layers},\ }\href
  {https://doi.org/10.1017/jfm.2018.193} {\bibfield  {journal} {\bibinfo
  {journal} {J. Fluid Mech.}\ }\textbf {\bibinfo {volume} {844}},\ \bibinfo
  {pages} {5} (\bibinfo {year} {2018})}\BibitemShut {NoStop}%
\bibitem [{\citenamefont {Cantwell}(2019)}]{cantwell2019}%
  \BibitemOpen
  \bibfield  {author} {\bibinfo {author} {\bibfnamefont {B.~J.}\ \bibnamefont
  {Cantwell}},\ }\bibfield  {title} {\bibinfo {title} {A universal velocity
  profile for smooth wall pipe flow},\ }\href
  {https://doi.org/10.1017/jfm.2019.669} {\bibfield  {journal} {\bibinfo
  {journal} {J. Fluid Mech.}\ }\textbf {\bibinfo {volume} {878}},\ \bibinfo
  {pages} {834} (\bibinfo {year} {2019})}\BibitemShut {NoStop}%
\bibitem [{\citenamefont {Subrahmanyam}\ \emph {et~al.}(2022)\citenamefont
  {Subrahmanyam}, \citenamefont {Cantwell},\ and\ \citenamefont
  {Alonso}}]{subrahmanyam2022}%
  \BibitemOpen
  \bibfield  {author} {\bibinfo {author} {\bibfnamefont {M.~A.}\ \bibnamefont
  {Subrahmanyam}}, \bibinfo {author} {\bibfnamefont {B.~J.}\ \bibnamefont
  {Cantwell}},\ and\ \bibinfo {author} {\bibfnamefont {J.~J.}\ \bibnamefont
  {Alonso}},\ }\bibfield  {title} {\bibinfo {title} {A universal velocity
  profile for turbulent wall flows including adverse pressure gradient boundary
  layers},\ }\href {https://doi.org/10.1017/jfm.2021.998} {\bibfield  {journal}
  {\bibinfo  {journal} {J. Fluid Mech.}\ }\textbf {\bibinfo {volume} {933}},\
  \bibinfo {pages} {A16} (\bibinfo {year} {2022})}\BibitemShut {NoStop}%
\bibitem [{\citenamefont {Ma}\ \emph {et~al.}(2026)\citenamefont {Ma},
  \citenamefont {Shi}, \citenamefont {Zhu}, \citenamefont {Han},\ and\
  \citenamefont {Chen}}]{ma2026}%
  \BibitemOpen
  \bibfield  {author} {\bibinfo {author} {\bibfnamefont {M.}~\bibnamefont
  {Ma}}, \bibinfo {author} {\bibfnamefont {Y.}~\bibnamefont {Shi}}, \bibinfo
  {author} {\bibfnamefont {Y.}~\bibnamefont {Zhu}}, \bibinfo {author}
  {\bibfnamefont {A.}~\bibnamefont {Han}},\ and\ \bibinfo {author}
  {\bibfnamefont {X.}~\bibnamefont {Chen}},\ }\bibfield  {title} {\bibinfo
  {title} {A generalized mixing length model with adverse-pressure-gradient
  effects},\ }\href {https://doi.org/10.3390/sym18010105} {\bibfield  {journal}
  {\bibinfo  {journal} {Symmetry}\ }\textbf {\bibinfo {volume} {18}},\ \bibinfo
  {pages} {105} (\bibinfo {year} {2026})}\BibitemShut {NoStop}%
\bibitem [{\citenamefont {Cal}\ and\ \citenamefont
  {Castillo}(2008)}]{cal2008Similarity}%
  \BibitemOpen
  \bibfield  {author} {\bibinfo {author} {\bibfnamefont {R.~B.}\ \bibnamefont
  {Cal}}\ and\ \bibinfo {author} {\bibfnamefont {L.}~\bibnamefont {Castillo}},\
  }\bibfield  {title} {\bibinfo {title} {Similarity analysis of favorable
  pressure gradient turbulent boundary layers with eventual
  quasilaminarization},\ }\href {https://doi.org/10.1063/1.2991433} {\bibfield
  {journal} {\bibinfo  {journal} {Phys. Fluids}\ }\textbf {\bibinfo {volume}
  {20}},\ \bibinfo {pages} {105106} (\bibinfo {year} {2008})}\BibitemShut
  {NoStop}%
\bibitem [{\citenamefont {Coles}(1956)}]{coles1956wake}%
  \BibitemOpen
  \bibfield  {author} {\bibinfo {author} {\bibfnamefont {D.}~\bibnamefont
  {Coles}},\ }\bibfield  {title} {\bibinfo {title} {The law of the wake in the
  turbulent boundary layer},\ }\href
  {https://doi.org/10.1017/S0022112056000135} {\bibfield  {journal} {\bibinfo
  {journal} {J. Fluid Mech.}\ }\textbf {\bibinfo {volume} {1}},\ \bibinfo
  {pages} {191} (\bibinfo {year} {1956})}\BibitemShut {NoStop}%
\bibitem [{\citenamefont {Monty}\ \emph {et~al.}(2011)\citenamefont {Monty},
  \citenamefont {Harun},\ and\ \citenamefont {Marusic}}]{monty2011parametric}%
  \BibitemOpen
  \bibfield  {author} {\bibinfo {author} {\bibfnamefont {J.~P.}\ \bibnamefont
  {Monty}}, \bibinfo {author} {\bibfnamefont {Z.}~\bibnamefont {Harun}},\ and\
  \bibinfo {author} {\bibfnamefont {I.}~\bibnamefont {Marusic}},\ }\bibfield
  {title} {\bibinfo {title} {A parametric study of adverse pressure gradient
  turbulent boundary layers},\ }\href
  {https://doi.org/10.1016/j.ijheatfluidflow.2011.03.004} {\bibfield  {journal}
  {\bibinfo  {journal} {Int. J. Heat Fluid Flow}\ }\textbf {\bibinfo {volume}
  {32}},\ \bibinfo {pages} {575} (\bibinfo {year} {2011})}\BibitemShut
  {NoStop}%
\bibitem [{\citenamefont {Buschmann}(2000)}]{buschmann2000powerlaw}%
  \BibitemOpen
  \bibfield  {author} {\bibinfo {author} {\bibfnamefont {M.}~\bibnamefont
  {Buschmann}},\ }\bibfield  {title} {\bibinfo {title} {Power law or
  logarithmic law? --- a data analysis for zero pressure gradient turbulent
  boundary layers with low {$Re_{\delta_2}$}},\ }\href
  {https://doi.org/10.1007/s11630-000-0041-y} {\bibfield  {journal} {\bibinfo
  {journal} {J. Therm. Sci.}\ }\textbf {\bibinfo {volume} {9}},\ \bibinfo
  {pages} {23} (\bibinfo {year} {2000})}\BibitemShut {NoStop}%
\bibitem [{\citenamefont {Afzal}(2001)}]{afzal2001powerlaw}%
  \BibitemOpen
  \bibfield  {author} {\bibinfo {author} {\bibfnamefont {N.}~\bibnamefont
  {Afzal}},\ }\bibfield  {title} {\bibinfo {title} {Power law and log law
  velocity profiles in turbulent boundary-layer flow: equivalent relations at
  large {Reynolds} numbers},\ }\href {https://doi.org/10.1007/BF01246918}
  {\bibfield  {journal} {\bibinfo  {journal} {Acta Mech.}\ }\textbf {\bibinfo
  {volume} {151}},\ \bibinfo {pages} {195} (\bibinfo {year}
  {2001})}\BibitemShut {NoStop}%
\bibitem [{\citenamefont {Dey}(2024)}]{dey2024approximate}%
  \BibitemOpen
  \bibfield  {author} {\bibinfo {author} {\bibfnamefont {J.}~\bibnamefont
  {Dey}},\ }\bibfield  {title} {\bibinfo {title} {Approximate derivation of the
  power law for the mean streamwise velocity in a turbulent boundary layer
  under zero-pressure gradient},\ }\href
  {https://doi.org/10.1103/PhysRevFluids.9.084601} {\bibfield  {journal}
  {\bibinfo  {journal} {Phys. Rev. Fluids}\ }\textbf {\bibinfo {volume} {9}},\
  \bibinfo {pages} {084601} (\bibinfo {year} {2024})}\BibitemShut {NoStop}%
\bibitem [{\citenamefont {Maciel}\ \emph {et~al.}(2017)\citenamefont {Maciel},
  \citenamefont {Simens},\ and\ \citenamefont {Gungor}}]{maciel2016coherent}%
  \BibitemOpen
  \bibfield  {author} {\bibinfo {author} {\bibfnamefont {Y.}~\bibnamefont
  {Maciel}}, \bibinfo {author} {\bibfnamefont {M.~P.}\ \bibnamefont {Simens}},\
  and\ \bibinfo {author} {\bibfnamefont {A.~G.}\ \bibnamefont {Gungor}},\
  }\bibfield  {title} {\bibinfo {title} {Coherent structures in a
  non-equilibrium large-velocity-defect turbulent boundary layer},\ }\href
  {https://doi.org/10.1007/s10494-016-9737-2} {\bibfield  {journal} {\bibinfo
  {journal} {Flow Turbul. Combust.}\ }\textbf {\bibinfo {volume} {98}},\
  \bibinfo {pages} {1} (\bibinfo {year} {2017})}\BibitemShut {NoStop}%
\bibitem [{\citenamefont {Schlichting}\ and\ \citenamefont
  {Gersten}(2000)}]{schlichting2000boundary}%
  \BibitemOpen
  \bibfield  {author} {\bibinfo {author} {\bibfnamefont {H.}~\bibnamefont
  {Schlichting}}\ and\ \bibinfo {author} {\bibfnamefont {K.}~\bibnamefont
  {Gersten}},\ }\href {https://doi.org/10.1007/978-3-642-85829-1} {\emph
  {\bibinfo {title} {Boundary-Layer Theory}}},\ \bibinfo {edition} {8th}\ ed.\
  (\bibinfo  {publisher} {Springer},\ \bibinfo {address} {Berlin, Heidelberg},\
  \bibinfo {year} {2000})\BibitemShut {NoStop}%
\bibitem [{\citenamefont {Uzun}\ and\ \citenamefont
  {Malik}(2022)}]{uzun2022High}%
  \BibitemOpen
  \bibfield  {author} {\bibinfo {author} {\bibfnamefont {A.}~\bibnamefont
  {Uzun}}\ and\ \bibinfo {author} {\bibfnamefont {M.~R.}\ \bibnamefont
  {Malik}},\ }\bibfield  {title} {\bibinfo {title} {High-fidelity simulation of
  turbulent flow {{\rev{past Gaussian bump}}}},\ }\href
  {https://doi.org/10.2514/1.J060760} {\bibfield  {journal} {\bibinfo
  {journal} {AIAA J.}\ }\textbf {\bibinfo {volume} {60}},\ \bibinfo {pages}
  {2130} (\bibinfo {year} {2022})}\BibitemShut {NoStop}%
\bibitem [{\citenamefont {Tamaki}\ and\ \citenamefont
  {Kawai}(2023)}]{tamaki2023Wall}%
  \BibitemOpen
  \bibfield  {author} {\bibinfo {author} {\bibfnamefont {Y.}~\bibnamefont
  {Tamaki}}\ and\ \bibinfo {author} {\bibfnamefont {S.}~\bibnamefont {Kawai}},\
  }\bibfield  {title} {\bibinfo {title} {Wall-resolved large-eddy simulation of
  near-stall airfoil flow at ${Re}_c = 10^7$},\ }\href
  {https://doi.org/10.2514/1.J062066} {\bibfield  {journal} {\bibinfo
  {journal} {AIAA J.}\ }\textbf {\bibinfo {volume} {61}},\ \bibinfo {pages}
  {698} (\bibinfo {year} {2023})}\BibitemShut {NoStop}%
\bibitem [{\citenamefont {Chauhan}\ \emph {et~al.}(2007)\citenamefont
  {Chauhan}, \citenamefont {Nagib},\ and\ \citenamefont
  {Monkewitz}}]{chauhan2007composite}%
  \BibitemOpen
  \bibfield  {author} {\bibinfo {author} {\bibfnamefont {K.~A.}\ \bibnamefont
  {Chauhan}}, \bibinfo {author} {\bibfnamefont {H.~M.}\ \bibnamefont {Nagib}},\
  and\ \bibinfo {author} {\bibfnamefont {P.~A.}\ \bibnamefont {Monkewitz}},\
  }\bibfield  {title} {\bibinfo {title} {On the composite logarithmic profile
  in zero pressure gradient turbulent boundary layers},\ }in\ \href
  {https://doi.org/10.2514/6.2007-532} {\emph {\bibinfo {booktitle} {45th
  {AIAA} Aerospace Sciences Meeting and Exhibit}}}\ (\bibinfo {year}
  {2007})\BibitemShut {NoStop}%
\bibitem [{\citenamefont {Bose}\ and\ \citenamefont {Park}(2018)}]{bose2018}%
  \BibitemOpen
  \bibfield  {author} {\bibinfo {author} {\bibfnamefont {S.~T.}\ \bibnamefont
  {Bose}}\ and\ \bibinfo {author} {\bibfnamefont {G.~I.}\ \bibnamefont
  {Park}},\ }\bibfield  {title} {\bibinfo {title} {Wall-modeled large-eddy
  simulation for complex turbulent flows},\ }\href
  {https://doi.org/10.1146/annurev-fluid-122316-045241} {\bibfield  {journal}
  {\bibinfo  {journal} {Annu. Rev. Fluid Mech.}\ }\textbf {\bibinfo {volume}
  {50}},\ \bibinfo {pages} {535} (\bibinfo {year} {2018})}\BibitemShut
  {NoStop}%
\bibitem [{\citenamefont {Arranz}\ and\ \citenamefont
  {Lozano-Dur{\'a}n}(2026)}]{github}%
  \BibitemOpen
  \bibfield  {author} {\bibinfo {author} {\bibfnamefont {G.}~\bibnamefont
  {Arranz}}\ and\ \bibinfo {author} {\bibfnamefont {A.}~\bibnamefont
  {Lozano-Dur{\'a}n}},\ }\href {https://github.com/ALD-Lab/TBLscaling}
  {\bibinfo {title} {{TBLscaling}: Unified scaling laws for turbulent boundary
  layers}},\ \bibinfo {howpublished} {GitHub repository} (\bibinfo {year}
  {2026})\BibitemShut {NoStop}%
\bibitem [{\citenamefont {Ludwieg}\ and\ \citenamefont
  {Tillmann}(1949)}]{ludwieg1949}%
  \BibitemOpen
  \bibfield  {author} {\bibinfo {author} {\bibfnamefont {H.}~\bibnamefont
  {Ludwieg}}\ and\ \bibinfo {author} {\bibfnamefont {W.}~\bibnamefont
  {Tillmann}},\ }\bibfield  {title} {\bibinfo {title} {{Untersuchungen}
  {\"u}ber die {Wandschubspannung} in turbulenten {Reibungsschichten}},\ }\href
  {https://doi.org/10.1007/BF00538855} {\bibfield  {journal} {\bibinfo
  {journal} {Ing. Arch.}\ }\textbf {\bibinfo {volume} {17}},\ \bibinfo {pages}
  {288} (\bibinfo {year} {1949})}\BibitemShut {NoStop}%
\bibitem [{\citenamefont {White}(1974)}]{White1974}%
  \BibitemOpen
  \bibfield  {author} {\bibinfo {author} {\bibfnamefont {F.~M.}\ \bibnamefont
  {White}},\ }\href@noop {} {\emph {\bibinfo {title} {Viscous Fluid Flow}}}\
  (\bibinfo  {publisher} {McGraw-Hill},\ \bibinfo {address} {New York},\
  \bibinfo {year} {1974})\BibitemShut {NoStop}%
\bibitem [{\citenamefont {Barr}(1980)}]{Barr1980}%
  \BibitemOpen
  \bibfield  {author} {\bibinfo {author} {\bibfnamefont {P.~K.}\ \bibnamefont
  {Barr}},\ }\href@noop {} {\emph {\bibinfo {title} {Calculation of
  Skin-Friction Coefficients for Low-{Reynolds}-Number Turbulent Boundary Layer
  Flows}}},\ \bibinfo {type} {Technical Report}\ \bibinfo {number}
  {SAND-79-8062}\ (\bibinfo  {institution} {Sandia National Laboratories},\
  \bibinfo {address} {Albuquerque, NM},\ \bibinfo {year} {1980})\ \bibinfo
  {note} {also issued as NASA-CR-163062; NASA NTRS ID 19800017127}\BibitemShut
  {NoStop}%
\bibitem [{\citenamefont {Orlandi}(2000)}]{Orlandi2000}%
  \BibitemOpen
  \bibinfo {editor} {\bibfnamefont {P.}~\bibnamefont {Orlandi}},\ ed.,\ \href
  {https://doi.org/10.1007/978-94-011-4281-6} {\emph {\bibinfo {title} {Fluid
  Flow Phenomena: A Numerical Toolkit}}},\ \bibinfo {series} {Fluid Mechanics
  and Its Applications}, Vol.~\bibinfo {volume} {55}\ (\bibinfo  {publisher}
  {Springer},\ \bibinfo {address} {Dordrecht},\ \bibinfo {year}
  {2000})\BibitemShut {NoStop}%
\bibitem [{\citenamefont {Wray}(1990)}]{Wray1990}%
  \BibitemOpen
  \bibfield  {author} {\bibinfo {author} {\bibfnamefont {A.~A.}\ \bibnamefont
  {Wray}},\ }\href@noop {} {\emph {\bibinfo {title} {Minimal Storage
  Time-Advancement Schemes for Spectral Methods}}},\ \bibinfo {type} {Technical
  Report}\ (\bibinfo  {institution} {NASA Ames Research Center},\ \bibinfo
  {address} {Moffett Field, CA},\ \bibinfo {year} {1990})\BibitemShut {NoStop}%
\bibitem [{\citenamefont {Kim}\ and\ \citenamefont {Moin}(1985)}]{Kim1985}%
  \BibitemOpen
  \bibfield  {author} {\bibinfo {author} {\bibfnamefont {J.}~\bibnamefont
  {Kim}}\ and\ \bibinfo {author} {\bibfnamefont {P.}~\bibnamefont {Moin}},\
  }\bibfield  {title} {\bibinfo {title} {Application of a fractional-step
  method to incompressible {Navier--Stokes} equations},\ }\href
  {https://doi.org/10.1016/0021-9991(85)90148-2} {\bibfield  {journal}
  {\bibinfo  {journal} {J. Comput. Phys.}\ }\textbf {\bibinfo {volume} {59}},\
  \bibinfo {pages} {308} (\bibinfo {year} {1985})}\BibitemShut {NoStop}%
\bibitem [{\citenamefont {Bae}\ \emph {et~al.}(2018)\citenamefont {Bae},
  \citenamefont {Lozano-Dur{\'a}n}, \citenamefont {Bose},\ and\ \citenamefont
  {Moin}}]{Bae2018}%
  \BibitemOpen
  \bibfield  {author} {\bibinfo {author} {\bibfnamefont {H.~J.}\ \bibnamefont
  {Bae}}, \bibinfo {author} {\bibfnamefont {A.}~\bibnamefont
  {Lozano-Dur{\'a}n}}, \bibinfo {author} {\bibfnamefont {S.~T.}\ \bibnamefont
  {Bose}},\ and\ \bibinfo {author} {\bibfnamefont {P.}~\bibnamefont {Moin}},\
  }\bibfield  {title} {\bibinfo {title} {Turbulence intensities in large-eddy
  simulation of wall-bounded flows},\ }\href
  {https://doi.org/10.1103/PhysRevFluids.3.014610} {\bibfield  {journal}
  {\bibinfo  {journal} {Phys. Rev. Fluids}\ }\textbf {\bibinfo {volume} {3}},\
  \bibinfo {pages} {014610} (\bibinfo {year} {2018})}\BibitemShut {NoStop}%
\bibitem [{\citenamefont {Lozano-Dur{\'a}n}\ and\ \citenamefont
  {Bae}(2019)}]{Lozano2019d}%
  \BibitemOpen
  \bibfield  {author} {\bibinfo {author} {\bibfnamefont {A.}~\bibnamefont
  {Lozano-Dur{\'a}n}}\ and\ \bibinfo {author} {\bibfnamefont {H.~J.}\
  \bibnamefont {Bae}},\ }\bibfield  {title} {\bibinfo {title} {Characteristic
  scales of {Townsend}'s wall-attached eddies},\ }\href
  {https://doi.org/10.1017/jfm.2019.209} {\bibfield  {journal} {\bibinfo
  {journal} {J. Fluid Mech.}\ }\textbf {\bibinfo {volume} {868}},\ \bibinfo
  {pages} {698} (\bibinfo {year} {2019})}\BibitemShut {NoStop}%
\bibitem [{\citenamefont {Lozano-Dur{\'a}n}\ \emph {et~al.}(2018)\citenamefont
  {Lozano-Dur{\'a}n}, \citenamefont {Hack},\ and\ \citenamefont
  {Moin}}]{Lozano2018}%
  \BibitemOpen
  \bibfield  {author} {\bibinfo {author} {\bibfnamefont {A.}~\bibnamefont
  {Lozano-Dur{\'a}n}}, \bibinfo {author} {\bibfnamefont {M.~J.~P.}\
  \bibnamefont {Hack}},\ and\ \bibinfo {author} {\bibfnamefont
  {P.}~\bibnamefont {Moin}},\ }\bibfield  {title} {\bibinfo {title} {Modeling
  boundary-layer transition in direct and large-eddy simulations using
  parabolized stability equations},\ }\href
  {https://doi.org/10.1103/PhysRevFluids.3.023901} {\bibfield  {journal}
  {\bibinfo  {journal} {Phys. Rev. Fluids}\ }\textbf {\bibinfo {volume} {3}},\
  \bibinfo {pages} {023901} (\bibinfo {year} {2018})}\BibitemShut {NoStop}%
\bibitem [{\citenamefont {Towne}\ \emph {et~al.}(2023)\citenamefont {Towne},
  \citenamefont {Dawson}, \citenamefont {Br{\`e}s}, \citenamefont
  {Lozano-Dur{\'a}n}, \citenamefont {Saxton-Fox}, \citenamefont
  {Parthasarathy}, \citenamefont {Jones}, \citenamefont {Biler}, \citenamefont
  {Yeh}, \citenamefont {Patel},\ and\ \citenamefont {Taira}}]{Towne2023}%
  \BibitemOpen
  \bibfield  {author} {\bibinfo {author} {\bibfnamefont {A.}~\bibnamefont
  {Towne}}, \bibinfo {author} {\bibfnamefont {S.~T.~M.}\ \bibnamefont
  {Dawson}}, \bibinfo {author} {\bibfnamefont {G.~A.}\ \bibnamefont
  {Br{\`e}s}}, \bibinfo {author} {\bibfnamefont {A.}~\bibnamefont
  {Lozano-Dur{\'a}n}}, \bibinfo {author} {\bibfnamefont {T.}~\bibnamefont
  {Saxton-Fox}}, \bibinfo {author} {\bibfnamefont {A.}~\bibnamefont
  {Parthasarathy}}, \bibinfo {author} {\bibfnamefont {A.~R.}\ \bibnamefont
  {Jones}}, \bibinfo {author} {\bibfnamefont {H.}~\bibnamefont {Biler}},
  \bibinfo {author} {\bibfnamefont {C.-A.}\ \bibnamefont {Yeh}}, \bibinfo
  {author} {\bibfnamefont {H.~D.}\ \bibnamefont {Patel}},\ and\ \bibinfo
  {author} {\bibfnamefont {K.}~\bibnamefont {Taira}},\ }\bibfield  {title}
  {\bibinfo {title} {A database for reduced-complexity modeling of fluid
  flows},\ }\href {https://doi.org/10.2514/1.J062203} {\bibfield  {journal}
  {\bibinfo  {journal} {AIAA J.}\ }\textbf {\bibinfo {volume} {61}},\ \bibinfo
  {pages} {2867} (\bibinfo {year} {2023})}\BibitemShut {NoStop}%
\bibitem [{\citenamefont {Pauley}\ \emph {et~al.}(1990)\citenamefont {Pauley},
  \citenamefont {Moin},\ and\ \citenamefont {Reynolds}}]{Pauley1990}%
  \BibitemOpen
  \bibfield  {author} {\bibinfo {author} {\bibfnamefont {L.~L.}\ \bibnamefont
  {Pauley}}, \bibinfo {author} {\bibfnamefont {P.}~\bibnamefont {Moin}},\ and\
  \bibinfo {author} {\bibfnamefont {W.~C.}\ \bibnamefont {Reynolds}},\
  }\bibfield  {title} {\bibinfo {title} {The structure of two-dimensional
  separation},\ }\href {https://doi.org/10.1017/S0022112090003317} {\bibfield
  {journal} {\bibinfo  {journal} {J. Fluid Mech.}\ }\textbf {\bibinfo {volume}
  {220}},\ \bibinfo {pages} {397} (\bibinfo {year} {1990})}\BibitemShut
  {NoStop}%
\bibitem [{\citenamefont {Lund}\ \emph {et~al.}(1998)\citenamefont {Lund},
  \citenamefont {Wu},\ and\ \citenamefont {Squires}}]{lund1998Generation}%
  \BibitemOpen
  \bibfield  {author} {\bibinfo {author} {\bibfnamefont {T.~S.}\ \bibnamefont
  {Lund}}, \bibinfo {author} {\bibfnamefont {X.}~\bibnamefont {Wu}},\ and\
  \bibinfo {author} {\bibfnamefont {K.~D.}\ \bibnamefont {Squires}},\
  }\bibfield  {title} {\bibinfo {title} {Generation of turbulent inflow data
  for spatially-developing boundary layer simulations},\ }\href
  {https://doi.org/10.1006/jcph.1998.5882} {\bibfield  {journal} {\bibinfo
  {journal} {J. Comput. Phys.}\ }\textbf {\bibinfo {volume} {140}},\ \bibinfo
  {pages} {233} (\bibinfo {year} {1998})}\BibitemShut {NoStop}%
\bibitem [{\citenamefont {Morgan}\ \emph {et~al.}(2011)\citenamefont {Morgan},
  \citenamefont {Larsson}, \citenamefont {Kawai},\ and\ \citenamefont
  {Lele}}]{morgan2011Improving}%
  \BibitemOpen
  \bibfield  {author} {\bibinfo {author} {\bibfnamefont {B.}~\bibnamefont
  {Morgan}}, \bibinfo {author} {\bibfnamefont {J.}~\bibnamefont {Larsson}},
  \bibinfo {author} {\bibfnamefont {S.}~\bibnamefont {Kawai}},\ and\ \bibinfo
  {author} {\bibfnamefont {S.~K.}\ \bibnamefont {Lele}},\ }\bibfield  {title}
  {\bibinfo {title} {Improving low-frequency characteristics of
  recycling/rescaling inflow turbulence generation},\ }\href
  {https://doi.org/10.2514/1.J050705} {\bibfield  {journal} {\bibinfo
  {journal} {AIAA J.}\ }\textbf {\bibinfo {volume} {49}},\ \bibinfo {pages}
  {582} (\bibinfo {year} {2011})}\BibitemShut {NoStop}%
\bibitem [{\citenamefont {Griffin}\ \emph
  {et~al.}(2021{\natexlab{a}})\citenamefont {Griffin}, \citenamefont {Fu},\
  and\ \citenamefont {Moin}}]{griffin2021general}%
  \BibitemOpen
  \bibfield  {author} {\bibinfo {author} {\bibfnamefont {K.~P.}\ \bibnamefont
  {Griffin}}, \bibinfo {author} {\bibfnamefont {L.}~\bibnamefont {Fu}},\ and\
  \bibinfo {author} {\bibfnamefont {P.}~\bibnamefont {Moin}},\ }\bibfield
  {title} {\bibinfo {title} {General method for determining the boundary layer
  thickness in nonequilibrium flows},\ }\href
  {https://doi.org/10.1103/PhysRevFluids.6.024608} {\bibfield  {journal}
  {\bibinfo  {journal} {Phys. Rev. Fluids}\ }\textbf {\bibinfo {volume} {6}},\
  \bibinfo {pages} {024608} (\bibinfo {year} {2021}{\natexlab{a}})}\BibitemShut
  {NoStop}%
\bibitem [{\citenamefont {Torroja}(2025)}]{torroja}%
  \BibitemOpen
  \bibfield  {author} {\bibinfo {author} {\bibnamefont {Torroja}},\ }\href
  {https://torroja.dmt.upm.es/turbdata/} {\bibinfo {title} {Turbulent flow
  databases}},\ \bibinfo {howpublished} {Online database} (\bibinfo {year}
  {2025})\BibitemShut {NoStop}%
\bibitem [{\citenamefont {Wu}\ \emph {et~al.}(2020)\citenamefont {Wu},
  \citenamefont {Meneveau},\ and\ \citenamefont
  {Mittal}}]{wu2019spatiotemporal}%
  \BibitemOpen
  \bibfield  {author} {\bibinfo {author} {\bibfnamefont {W.}~\bibnamefont
  {Wu}}, \bibinfo {author} {\bibfnamefont {C.}~\bibnamefont {Meneveau}},\ and\
  \bibinfo {author} {\bibfnamefont {R.}~\bibnamefont {Mittal}},\ }\bibfield
  {title} {\bibinfo {title} {Spatio-temporal dynamics of turbulent separation
  bubbles},\ }\href {https://doi.org/10.1017/jfm.2019.911} {\bibfield
  {journal} {\bibinfo  {journal} {J. Fluid Mech.}\ }\textbf {\bibinfo {volume}
  {883}},\ \bibinfo {pages} {A45} (\bibinfo {year} {2020})}\BibitemShut
  {NoStop}%
\bibitem [{\citenamefont {Cover}\ and\ \citenamefont
  {Thomas}(2006)}]{cover2006}%
  \BibitemOpen
  \bibfield  {author} {\bibinfo {author} {\bibfnamefont {T.~M.}\ \bibnamefont
  {Cover}}\ and\ \bibinfo {author} {\bibfnamefont {J.~A.}\ \bibnamefont
  {Thomas}},\ }\href {https://doi.org/10.1002/047174882X} {\emph {\bibinfo
  {title} {Elements of Information Theory}}},\ \bibinfo {edition} {2nd}\ ed.\
  (\bibinfo  {publisher} {John Wiley \& Sons},\ \bibinfo {address} {Hoboken,
  NJ},\ \bibinfo {year} {2006})\BibitemShut {NoStop}%
\bibitem [{\citenamefont {Wei}\ \emph {et~al.}(2024)\citenamefont {Wei},
  \citenamefont {Li},\ and\ \citenamefont {Wang}}]{wei2024new}%
  \BibitemOpen
  \bibfield  {author} {\bibinfo {author} {\bibfnamefont {T.}~\bibnamefont
  {Wei}}, \bibinfo {author} {\bibfnamefont {Z.}~\bibnamefont {Li}},\ and\
  \bibinfo {author} {\bibfnamefont {Y.}~\bibnamefont {Wang}},\ }\bibfield
  {title} {\bibinfo {title} {New momentum integral equation applicable to
  boundary layer flows under arbitrary pressure gradients},\ }\href
  {https://doi.org/10.1017/jfm.2024.207} {\bibfield  {journal} {\bibinfo
  {journal} {J. Fluid Mech.}\ }\textbf {\bibinfo {volume} {984}},\ \bibinfo
  {pages} {A64} (\bibinfo {year} {2024})}\BibitemShut {NoStop}%
\bibitem [{\citenamefont {Han}\ \emph {et~al.}(2025)\citenamefont {Han},
  \citenamefont {Ma},\ and\ \citenamefont {Yan}}]{han2025integral}%
  \BibitemOpen
  \bibfield  {author} {\bibinfo {author} {\bibfnamefont {M.}~\bibnamefont
  {Han}}, \bibinfo {author} {\bibfnamefont {M.}~\bibnamefont {Ma}},\ and\
  \bibinfo {author} {\bibfnamefont {C.}~\bibnamefont {Yan}},\ }\bibfield
  {title} {\bibinfo {title} {Integral analysis of adverse pressure gradient
  turbulent boundary layers},\ }\href {https://doi.org/10.1017/jfm.2025.10395}
  {\bibfield  {journal} {\bibinfo  {journal} {J. Fluid Mech.}\ }\textbf
  {\bibinfo {volume} {1019}},\ \bibinfo {pages} {A46} (\bibinfo {year}
  {2025})}\BibitemShut {NoStop}%
\bibitem [{\citenamefont {Sobol'}(2001)}]{sobol2001}%
  \BibitemOpen
  \bibfield  {author} {\bibinfo {author} {\bibfnamefont {I.~M.}\ \bibnamefont
  {Sobol'}},\ }\bibfield  {title} {\bibinfo {title} {Global sensitivity indices
  for nonlinear mathematical models and their {Monte Carlo} estimates},\ }\href
  {https://doi.org/10.1016/S0378-4754(00)00270-6} {\bibfield  {journal}
  {\bibinfo  {journal} {Math. Comput. Simul.}\ }\textbf {\bibinfo {volume}
  {55}},\ \bibinfo {pages} {271} (\bibinfo {year} {2001})}\BibitemShut
  {NoStop}%
\bibitem [{\citenamefont {Saltelli}(2002)}]{saltelli2002}%
  \BibitemOpen
  \bibfield  {author} {\bibinfo {author} {\bibfnamefont {A.}~\bibnamefont
  {Saltelli}},\ }\bibfield  {title} {\bibinfo {title} {Sensitivity analysis for
  importance assessment},\ }\href {https://doi.org/10.1111/0272-4332.00040}
  {\bibfield  {journal} {\bibinfo  {journal} {Risk Anal.}\ }\textbf {\bibinfo
  {volume} {22}},\ \bibinfo {pages} {579} (\bibinfo {year} {2002})}\BibitemShut
  {NoStop}%
\bibitem [{\citenamefont {Vinuesa}\ \emph {et~al.}(2016)\citenamefont
  {Vinuesa}, \citenamefont {Bobke}, \citenamefont {{\"O}rl{\"u}},\ and\
  \citenamefont {Schlatter}}]{vinuesa2016determining}%
  \BibitemOpen
  \bibfield  {author} {\bibinfo {author} {\bibfnamefont {R.}~\bibnamefont
  {Vinuesa}}, \bibinfo {author} {\bibfnamefont {A.}~\bibnamefont {Bobke}},
  \bibinfo {author} {\bibfnamefont {R.}~\bibnamefont {{\"O}rl{\"u}}},\ and\
  \bibinfo {author} {\bibfnamefont {P.}~\bibnamefont {Schlatter}},\ }\bibfield
  {title} {\bibinfo {title} {On determining characteristic length scales in
  pressure-gradient turbulent boundary layers},\ }\href
  {https://doi.org/10.1063/1.4947532} {\bibfield  {journal} {\bibinfo
  {journal} {Phys. Fluids}\ }\textbf {\bibinfo {volume} {28}},\ \bibinfo
  {pages} {055101} (\bibinfo {year} {2016})}\BibitemShut {NoStop}%
\bibitem [{\citenamefont {Wei}\ and\ \citenamefont
  {Knopp}(2023)}]{wei2023outer}%
  \BibitemOpen
  \bibfield  {author} {\bibinfo {author} {\bibfnamefont {T.}~\bibnamefont
  {Wei}}\ and\ \bibinfo {author} {\bibfnamefont {T.}~\bibnamefont {Knopp}},\
  }\bibfield  {title} {\bibinfo {title} {Outer scaling of the mean momentum
  equation for turbulent boundary layers under adverse pressure gradient},\
  }\href {https://doi.org/10.1017/jfm.2023.72} {\bibfield  {journal} {\bibinfo
  {journal} {J. Fluid Mech.}\ }\textbf {\bibinfo {volume} {958}},\ \bibinfo
  {pages} {A9} (\bibinfo {year} {2023})}\BibitemShut {NoStop}%
\bibitem [{\citenamefont {Lozier}\ \emph {et~al.}(2025)\citenamefont {Lozier},
  \citenamefont {Deshpande}, \citenamefont {Zarei}, \citenamefont {Lindi{\'c}},
  \citenamefont {Abu~Rowin},\ and\ \citenamefont
  {Marusic}}]{lozier2025Defining}%
  \BibitemOpen
  \bibfield  {author} {\bibinfo {author} {\bibfnamefont {M.}~\bibnamefont
  {Lozier}}, \bibinfo {author} {\bibfnamefont {R.}~\bibnamefont {Deshpande}},
  \bibinfo {author} {\bibfnamefont {A.}~\bibnamefont {Zarei}}, \bibinfo
  {author} {\bibfnamefont {L.}~\bibnamefont {Lindi{\'c}}}, \bibinfo {author}
  {\bibfnamefont {W.}~\bibnamefont {Abu~Rowin}},\ and\ \bibinfo {author}
  {\bibfnamefont {I.}~\bibnamefont {Marusic}},\ }\bibfield  {title} {\bibinfo
  {title} {Defining the mean turbulent boundary layer thickness based on
  streamwise velocity skewness},\ }\href
  {https://doi.org/10.1017/jfm.2025.10711} {\bibfield  {journal} {\bibinfo
  {journal} {J. Fluid Mech.}\ }\textbf {\bibinfo {volume} {1021}},\ \bibinfo
  {pages} {A19} (\bibinfo {year} {2025})}\BibitemShut {NoStop}%
\bibitem [{\citenamefont {Mart{\'i}nez-S{\'a}nchez}\ \emph
  {et~al.}(2024)\citenamefont {Mart{\'i}nez-S{\'a}nchez}, \citenamefont
  {Arranz},\ and\ \citenamefont {Lozano-Dur{\'a}n}}]{martinez2024surd}%
  \BibitemOpen
  \bibfield  {author} {\bibinfo {author} {\bibfnamefont {A.}~\bibnamefont
  {Mart{\'i}nez-S{\'a}nchez}}, \bibinfo {author} {\bibfnamefont
  {G.}~\bibnamefont {Arranz}},\ and\ \bibinfo {author} {\bibfnamefont
  {A.}~\bibnamefont {Lozano-Dur{\'a}n}},\ }\bibfield  {title} {\bibinfo {title}
  {Decomposing causality into its synergistic, unique, and redundant
  components},\ }\href {https://doi.org/10.1038/s41467-024-53373-4} {\bibfield
  {journal} {\bibinfo  {journal} {Nat. Commun.}\ }\textbf {\bibinfo {volume}
  {15}},\ \bibinfo {pages} {9296} (\bibinfo {year} {2024})}\BibitemShut
  {NoStop}%
\bibitem [{\citenamefont {Mart{\'i}nez-S{\'a}nchez}\ and\ \citenamefont
  {Lozano-Dur{\'a}n}(2026)}]{martinez2026}%
  \BibitemOpen
  \bibfield  {author} {\bibinfo {author} {\bibfnamefont {{\'A}.}~\bibnamefont
  {Mart{\'i}nez-S{\'a}nchez}}\ and\ \bibinfo {author} {\bibfnamefont
  {A.}~\bibnamefont {Lozano-Dur{\'a}n}},\ }\bibfield  {title} {\bibinfo {title}
  {Cause-and-effect approach to turbulence forecasting},\ }\href
  {https://doi.org/10.1108/HFF-09-2025-0715} {\bibfield  {journal} {\bibinfo
  {journal} {Int. J. Numer. Methods Heat Fluid Flow}\ }\textbf {\bibinfo
  {volume} {36}},\ \bibinfo {pages} {2539} (\bibinfo {year}
  {2026})}\BibitemShut {NoStop}%
\bibitem [{\citenamefont {Bradshaw}(1987)}]{bradshaw1987}%
  \BibitemOpen
  \bibfield  {author} {\bibinfo {author} {\bibfnamefont {P.}~\bibnamefont
  {Bradshaw}},\ }\bibfield  {title} {\bibinfo {title} {Turbulent secondary
  flows},\ }\href {https://doi.org/10.1146/annurev.fl.19.010187.000413}
  {\bibfield  {journal} {\bibinfo  {journal} {Annu. Rev. Fluid Mech.}\ }\textbf
  {\bibinfo {volume} {19}},\ \bibinfo {pages} {53} (\bibinfo {year}
  {1987})}\BibitemShut {NoStop}%
\bibitem [{\citenamefont {Lozano-Dur{\'a}n}\ \emph {et~al.}(2020)\citenamefont
  {Lozano-Dur{\'a}n}, \citenamefont {Giometto}, \citenamefont {Park},\ and\
  \citenamefont {Moin}}]{lozano-duran2020}%
  \BibitemOpen
  \bibfield  {author} {\bibinfo {author} {\bibfnamefont {A.}~\bibnamefont
  {Lozano-Dur{\'a}n}}, \bibinfo {author} {\bibfnamefont {M.~G.}\ \bibnamefont
  {Giometto}}, \bibinfo {author} {\bibfnamefont {G.~I.}\ \bibnamefont {Park}},\
  and\ \bibinfo {author} {\bibfnamefont {P.}~\bibnamefont {Moin}},\ }\bibfield
  {title} {\bibinfo {title} {Non-equilibrium three-dimensional boundary layers
  at moderate {Reynolds} numbers},\ }\href
  {https://doi.org/10.1017/jfm.2019.869} {\bibfield  {journal} {\bibinfo
  {journal} {J. Fluid Mech.}\ }\textbf {\bibinfo {volume} {883}},\ \bibinfo
  {pages} {A20} (\bibinfo {year} {2020})}\BibitemShut {NoStop}%
\bibitem [{\citenamefont {Driver}\ and\ \citenamefont
  {Hebbar}(1987)}]{driver1987}%
  \BibitemOpen
  \bibfield  {author} {\bibinfo {author} {\bibfnamefont {D.~M.}\ \bibnamefont
  {Driver}}\ and\ \bibinfo {author} {\bibfnamefont {S.~K.}\ \bibnamefont
  {Hebbar}},\ }\bibfield  {title} {\bibinfo {title} {Experimental study of a
  three-dimensional, shear-driven, turbulent boundary layer},\ }\href
  {https://doi.org/10.2514/3.9575} {\bibfield  {journal} {\bibinfo  {journal}
  {AIAA J.}\ }\textbf {\bibinfo {volume} {25}},\ \bibinfo {pages} {35}
  (\bibinfo {year} {1987})}\BibitemShut {NoStop}%
\bibitem [{\citenamefont {Eaton}(1995)}]{eaton1995}%
  \BibitemOpen
  \bibfield  {author} {\bibinfo {author} {\bibfnamefont {J.~K.}\ \bibnamefont
  {Eaton}},\ }\bibfield  {title} {\bibinfo {title} {Effects of mean flow three
  dimensionality on turbulent boundary-layer structure},\ }\href
  {https://doi.org/10.2514/3.12942} {\bibfield  {journal} {\bibinfo  {journal}
  {AIAA J.}\ }\textbf {\bibinfo {volume} {33}},\ \bibinfo {pages} {2020}
  (\bibinfo {year} {1995})}\BibitemShut {NoStop}%
\bibitem [{\citenamefont {Johnston}(1960)}]{johnston1960}%
  \BibitemOpen
  \bibfield  {author} {\bibinfo {author} {\bibfnamefont {J.~P.}\ \bibnamefont
  {Johnston}},\ }\bibfield  {title} {\bibinfo {title} {On the three-dimensional
  turbulent boundary layer generated by secondary flow},\ }\href
  {https://doi.org/10.1115/1.3662539} {\bibfield  {journal} {\bibinfo
  {journal} {J. Basic Eng.}\ }\textbf {\bibinfo {volume} {82}},\ \bibinfo
  {pages} {233} (\bibinfo {year} {1960})}\BibitemShut {NoStop}%
\bibitem [{\citenamefont {Abe}(2020)}]{abe2020}%
  \BibitemOpen
  \bibfield  {author} {\bibinfo {author} {\bibfnamefont {H.}~\bibnamefont
  {Abe}},\ }\bibfield  {title} {\bibinfo {title} {Direct numerical simulation
  of a non-equilibrium three-dimensional turbulent boundary layer over a flat
  plate},\ }\href {https://doi.org/10.1017/jfm.2020.488} {\bibfield  {journal}
  {\bibinfo  {journal} {J. Fluid Mech.}\ }\textbf {\bibinfo {volume} {902}},\
  \bibinfo {pages} {A20} (\bibinfo {year} {2020})}\BibitemShut {NoStop}%
\bibitem [{\citenamefont {van~den Berg}\ \emph {et~al.}(1975)\citenamefont
  {van~den Berg}, \citenamefont {Elsenaar}, \citenamefont {Lindhout},\ and\
  \citenamefont {Wesseling}}]{vandenberg1975}%
  \BibitemOpen
  \bibfield  {author} {\bibinfo {author} {\bibfnamefont {B.}~\bibnamefont
  {van~den Berg}}, \bibinfo {author} {\bibfnamefont {A.}~\bibnamefont
  {Elsenaar}}, \bibinfo {author} {\bibfnamefont {J.~P.~F.}\ \bibnamefont
  {Lindhout}},\ and\ \bibinfo {author} {\bibfnamefont {P.}~\bibnamefont
  {Wesseling}},\ }\bibfield  {title} {\bibinfo {title} {Measurements in an
  incompressible three-dimensional turbulent boundary layer, under infinite
  swept-wing conditions, and comparison with theory},\ }\href
  {https://doi.org/10.1017/S0022112075001930} {\bibfield  {journal} {\bibinfo
  {journal} {J. Fluid Mech.}\ }\textbf {\bibinfo {volume} {70}},\ \bibinfo
  {pages} {127} (\bibinfo {year} {1975})}\BibitemShut {NoStop}%
\bibitem [{\citenamefont {Coleman}\ \emph {et~al.}(2019)\citenamefont
  {Coleman}, \citenamefont {Rumsey},\ and\ \citenamefont
  {Spalart}}]{coleman2019}%
  \BibitemOpen
  \bibfield  {author} {\bibinfo {author} {\bibfnamefont {G.~N.}\ \bibnamefont
  {Coleman}}, \bibinfo {author} {\bibfnamefont {C.~L.}\ \bibnamefont
  {Rumsey}},\ and\ \bibinfo {author} {\bibfnamefont {P.~R.}\ \bibnamefont
  {Spalart}},\ }\bibfield  {title} {\bibinfo {title} {Numerical study of a
  turbulent separation bubble with sweep},\ }\href
  {https://doi.org/10.1017/jfm.2019.736} {\bibfield  {journal} {\bibinfo
  {journal} {J. Fluid Mech.}\ }\textbf {\bibinfo {volume} {880}},\ \bibinfo
  {pages} {684} (\bibinfo {year} {2019})}\BibitemShut {NoStop}%
\bibitem [{\citenamefont {Jeans}\ and\ \citenamefont
  {Johnston}(1982)}]{jeans1982}%
  \BibitemOpen
  \bibfield  {author} {\bibinfo {author} {\bibfnamefont {A.~H.}\ \bibnamefont
  {Jeans}}\ and\ \bibinfo {author} {\bibfnamefont {J.~P.}\ \bibnamefont
  {Johnston}},\ }\href@noop {} {\emph {\bibinfo {title} {The Effects of
  Streamwise Concave Curvature on Turbulent Boundary Layer Structure}}},\
  \bibinfo {type} {Tech. Rep.}\ \bibinfo {number} {MD-40}\ (\bibinfo
  {institution} {Thermosciences Division, Department of Mechanical Engineering,
  Stanford University},\ \bibinfo {year} {1982})\BibitemShut {NoStop}%
\bibitem [{\citenamefont {Bradshaw}(1973)}]{bradshaw1973}%
  \BibitemOpen
  \bibfield  {author} {\bibinfo {author} {\bibfnamefont {P.}~\bibnamefont
  {Bradshaw}},\ }\href@noop {} {\emph {\bibinfo {title} {Effects of Streamline
  Curvature on Turbulent Flow}}},\ \bibinfo {type} {AGARDograph}\ \bibinfo
  {number} {169}\ (\bibinfo  {institution} {Advisory Group for Aerospace
  Research and Development},\ \bibinfo {year} {1973})\BibitemShut {NoStop}%
\bibitem [{\citenamefont {Ellis}\ and\ \citenamefont
  {Joubert}(1974)}]{ellis1974}%
  \BibitemOpen
  \bibfield  {author} {\bibinfo {author} {\bibfnamefont {L.~B.}\ \bibnamefont
  {Ellis}}\ and\ \bibinfo {author} {\bibfnamefont {P.~N.}\ \bibnamefont
  {Joubert}},\ }\bibfield  {title} {\bibinfo {title} {Turbulent shear flow in a
  curved duct},\ }\href {https://doi.org/10.1017/S0022112074000589} {\bibfield
  {journal} {\bibinfo  {journal} {J. Fluid Mech.}\ }\textbf {\bibinfo {volume}
  {62}},\ \bibinfo {pages} {65} (\bibinfo {year} {1974})}\BibitemShut {NoStop}%
\bibitem [{\citenamefont {Smits}\ \emph {et~al.}(1979)\citenamefont {Smits},
  \citenamefont {Young},\ and\ \citenamefont {Bradshaw}}]{smits1979}%
  \BibitemOpen
  \bibfield  {author} {\bibinfo {author} {\bibfnamefont {A.~J.}\ \bibnamefont
  {Smits}}, \bibinfo {author} {\bibfnamefont {S.~T.~B.}\ \bibnamefont
  {Young}},\ and\ \bibinfo {author} {\bibfnamefont {P.}~\bibnamefont
  {Bradshaw}},\ }\bibfield  {title} {\bibinfo {title} {The effect of short
  regions of high surface curvature on turbulent boundary layers},\ }\href
  {https://doi.org/10.1017/S0022112079001002} {\bibfield  {journal} {\bibinfo
  {journal} {J. Fluid Mech.}\ }\textbf {\bibinfo {volume} {94}},\ \bibinfo
  {pages} {209} (\bibinfo {year} {1979})}\BibitemShut {NoStop}%
\bibitem [{\citenamefont {Barlow}\ and\ \citenamefont
  {Johnston}(1988)}]{barlow1988}%
  \BibitemOpen
  \bibfield  {author} {\bibinfo {author} {\bibfnamefont {R.~S.}\ \bibnamefont
  {Barlow}}\ and\ \bibinfo {author} {\bibfnamefont {J.~P.}\ \bibnamefont
  {Johnston}},\ }\bibfield  {title} {\bibinfo {title} {Structure of a turbulent
  boundary layer on a concave surface},\ }\href
  {https://doi.org/10.1017/S0022112088001545} {\bibfield  {journal} {\bibinfo
  {journal} {J. Fluid Mech.}\ }\textbf {\bibinfo {volume} {191}},\ \bibinfo
  {pages} {137} (\bibinfo {year} {1988})}\BibitemShut {NoStop}%
\bibitem [{\citenamefont {Appelbaum}\ \emph {et~al.}(2025)\citenamefont
  {Appelbaum}, \citenamefont {Kloker},\ and\ \citenamefont
  {Wenzel}}]{appelbaum2025}%
  \BibitemOpen
  \bibfield  {author} {\bibinfo {author} {\bibfnamefont {J.}~\bibnamefont
  {Appelbaum}}, \bibinfo {author} {\bibfnamefont {M.}~\bibnamefont {Kloker}},\
  and\ \bibinfo {author} {\bibfnamefont {C.}~\bibnamefont {Wenzel}},\
  }\bibfield  {title} {\bibinfo {title} {A systematic {DNS} approach to isolate
  wall-curvature effects in spatially developing boundary layers},\ }\href
  {https://doi.org/10.1007/s00162-024-00729-7} {\bibfield  {journal} {\bibinfo
  {journal} {Theor. Comput. Fluid Dyn.}\ }\textbf {\bibinfo {volume} {39}},\
  \bibinfo {pages} {10} (\bibinfo {year} {2025})}\BibitemShut {NoStop}%
\bibitem [{\citenamefont {Pargal}\ \emph {et~al.}(2022)\citenamefont {Pargal},
  \citenamefont {Wu}, \citenamefont {Yuan},\ and\ \citenamefont
  {Moreau}}]{pargal2022}%
  \BibitemOpen
  \bibfield  {author} {\bibinfo {author} {\bibfnamefont {S.}~\bibnamefont
  {Pargal}}, \bibinfo {author} {\bibfnamefont {H.}~\bibnamefont {Wu}}, \bibinfo
  {author} {\bibfnamefont {J.}~\bibnamefont {Yuan}},\ and\ \bibinfo {author}
  {\bibfnamefont {S.}~\bibnamefont {Moreau}},\ }\bibfield  {title} {\bibinfo
  {title} {Adverse-pressure-gradient turbulent boundary layer on convex wall},\
  }\href {https://doi.org/10.1063/5.0083919} {\bibfield  {journal} {\bibinfo
  {journal} {Phys. Fluids}\ }\textbf {\bibinfo {volume} {34}},\ \bibinfo
  {pages} {035107} (\bibinfo {year} {2022})}\BibitemShut {NoStop}%
\bibitem [{\citenamefont {Lagha}\ \emph {et~al.}(2011)\citenamefont {Lagha},
  \citenamefont {Kim}, \citenamefont {Eldredge},\ and\ \citenamefont
  {Zhong}}]{lagha2011}%
  \BibitemOpen
  \bibfield  {author} {\bibinfo {author} {\bibfnamefont {M.}~\bibnamefont
  {Lagha}}, \bibinfo {author} {\bibfnamefont {J.}~\bibnamefont {Kim}}, \bibinfo
  {author} {\bibfnamefont {J.~D.}\ \bibnamefont {Eldredge}},\ and\ \bibinfo
  {author} {\bibfnamefont {X.}~\bibnamefont {Zhong}},\ }\bibfield  {title}
  {\bibinfo {title} {A numerical study of compressible turbulent boundary
  layers},\ }\href {https://doi.org/10.1063/1.3541841} {\bibfield  {journal}
  {\bibinfo  {journal} {Phys. Fluids}\ }\textbf {\bibinfo {volume} {23}},\
  \bibinfo {pages} {015106} (\bibinfo {year} {2011})}\BibitemShut {NoStop}%
\bibitem [{\citenamefont {Spina}\ \emph {et~al.}(1994)\citenamefont {Spina},
  \citenamefont {Smits},\ and\ \citenamefont {Robinson}}]{spina1994}%
  \BibitemOpen
  \bibfield  {author} {\bibinfo {author} {\bibfnamefont {E.~F.}\ \bibnamefont
  {Spina}}, \bibinfo {author} {\bibfnamefont {A.~J.}\ \bibnamefont {Smits}},\
  and\ \bibinfo {author} {\bibfnamefont {S.~K.}\ \bibnamefont {Robinson}},\
  }\bibfield  {title} {\bibinfo {title} {The physics of supersonic turbulent
  boundary layers},\ }\href
  {https://doi.org/10.1146/annurev.fl.26.010194.001443} {\bibfield  {journal}
  {\bibinfo  {journal} {Annu. Rev. Fluid Mech.}\ }\textbf {\bibinfo {volume}
  {26}},\ \bibinfo {pages} {287} (\bibinfo {year} {1994})}\BibitemShut
  {NoStop}%
\bibitem [{\citenamefont {Smits}\ and\ \citenamefont
  {Dussauge}(2006)}]{smits2006}%
  \BibitemOpen
  \bibfield  {author} {\bibinfo {author} {\bibfnamefont {A.~J.}\ \bibnamefont
  {Smits}}\ and\ \bibinfo {author} {\bibfnamefont {J.-P.}\ \bibnamefont
  {Dussauge}},\ }\href {https://doi.org/10.1007/b137383} {\emph {\bibinfo
  {title} {Turbulent Shear Layers in Supersonic Flow}}},\ \bibinfo {edition}
  {2nd}\ ed.\ (\bibinfo  {publisher} {Springer},\ \bibinfo {address} {New
  York},\ \bibinfo {year} {2006})\BibitemShut {NoStop}%
\bibitem [{\citenamefont {Morkovin}(1962)}]{morkovin1962}%
  \BibitemOpen
  \bibfield  {author} {\bibinfo {author} {\bibfnamefont {M.~V.}\ \bibnamefont
  {Morkovin}},\ }\bibfield  {title} {\bibinfo {title} {Effects of
  compressibility on turbulent flows},\ }in\ \href@noop {} {\emph {\bibinfo
  {booktitle} {M{\'e}canique de la Turbulence}}},\ \bibinfo {editor} {edited
  by\ \bibinfo {editor} {\bibfnamefont {A.}~\bibnamefont {Favre}}}\ (\bibinfo
  {publisher} {CNRS},\ \bibinfo {address} {Paris},\ \bibinfo {year} {1962})\
  pp.\ \bibinfo {pages} {367--380}\BibitemShut {NoStop}%
\bibitem [{\citenamefont {Van~Driest}(1951)}]{vandriest1951}%
  \BibitemOpen
  \bibfield  {author} {\bibinfo {author} {\bibfnamefont {E.~R.}\ \bibnamefont
  {Van~Driest}},\ }\bibfield  {title} {\bibinfo {title} {Turbulent boundary
  layer in compressible fluids},\ }\href {https://doi.org/10.2514/8.1895}
  {\bibfield  {journal} {\bibinfo  {journal} {J. Aeronaut. Sci.}\ }\textbf
  {\bibinfo {volume} {18}},\ \bibinfo {pages} {145} (\bibinfo {year}
  {1951})}\BibitemShut {NoStop}%
\bibitem [{\citenamefont {Huang}\ \emph {et~al.}(1995)\citenamefont {Huang},
  \citenamefont {Coleman},\ and\ \citenamefont {Bradshaw}}]{huang1995}%
  \BibitemOpen
  \bibfield  {author} {\bibinfo {author} {\bibfnamefont {P.~G.}\ \bibnamefont
  {Huang}}, \bibinfo {author} {\bibfnamefont {G.~N.}\ \bibnamefont {Coleman}},\
  and\ \bibinfo {author} {\bibfnamefont {P.}~\bibnamefont {Bradshaw}},\
  }\bibfield  {title} {\bibinfo {title} {Compressible turbulent channel flows:
  {DNS} results and modelling},\ }\href
  {https://doi.org/10.1017/S0022112095004599} {\bibfield  {journal} {\bibinfo
  {journal} {J. Fluid Mech.}\ }\textbf {\bibinfo {volume} {305}},\ \bibinfo
  {pages} {185} (\bibinfo {year} {1995})}\BibitemShut {NoStop}%
\bibitem [{\citenamefont {Volpiani}\ \emph {et~al.}(2020)\citenamefont
  {Volpiani}, \citenamefont {Iyer}, \citenamefont {Pirozzoli},\ and\
  \citenamefont {Larsson}}]{volpiani2020}%
  \BibitemOpen
  \bibfield  {author} {\bibinfo {author} {\bibfnamefont {P.~S.}\ \bibnamefont
  {Volpiani}}, \bibinfo {author} {\bibfnamefont {P.~S.}\ \bibnamefont {Iyer}},
  \bibinfo {author} {\bibfnamefont {S.}~\bibnamefont {Pirozzoli}},\ and\
  \bibinfo {author} {\bibfnamefont {J.}~\bibnamefont {Larsson}},\ }\bibfield
  {title} {\bibinfo {title} {Data-driven compressibility transformation for
  turbulent wall layers},\ }\href
  {https://doi.org/10.1103/PhysRevFluids.5.052602} {\bibfield  {journal}
  {\bibinfo  {journal} {Phys. Rev. Fluids}\ }\textbf {\bibinfo {volume} {5}},\
  \bibinfo {pages} {052602(R)} (\bibinfo {year} {2020})}\BibitemShut {NoStop}%
\bibitem [{\citenamefont {Griffin}\ \emph
  {et~al.}(2021{\natexlab{b}})\citenamefont {Griffin}, \citenamefont {Fu},\
  and\ \citenamefont {Moin}}]{griffin2021}%
  \BibitemOpen
  \bibfield  {author} {\bibinfo {author} {\bibfnamefont {K.~P.}\ \bibnamefont
  {Griffin}}, \bibinfo {author} {\bibfnamefont {L.}~\bibnamefont {Fu}},\ and\
  \bibinfo {author} {\bibfnamefont {P.}~\bibnamefont {Moin}},\ }\bibfield
  {title} {\bibinfo {title} {Velocity transformation for compressible
  wall-bounded turbulent flows with and without heat transfer},\ }\href
  {https://doi.org/10.1073/pnas.2111144118} {\bibfield  {journal} {\bibinfo
  {journal} {Proc. Natl. Acad. Sci. U.S.A.}\ }\textbf {\bibinfo {volume}
  {118}},\ \bibinfo {pages} {e2111144118} (\bibinfo {year}
  {2021}{\natexlab{b}})}\BibitemShut {NoStop}%
\bibitem [{\citenamefont {Coles}(1964)}]{coles1964}%
  \BibitemOpen
  \bibfield  {author} {\bibinfo {author} {\bibfnamefont {D.~E.}\ \bibnamefont
  {Coles}},\ }\bibfield  {title} {\bibinfo {title} {The turbulent boundary
  layer in a compressible fluid},\ }\href {https://doi.org/10.1063/1.1711395}
  {\bibfield  {journal} {\bibinfo  {journal} {Phys. Fluids}\ }\textbf {\bibinfo
  {volume} {7}},\ \bibinfo {pages} {1403} (\bibinfo {year} {1964})}\BibitemShut
  {NoStop}%
\bibitem [{\citenamefont {Bons}\ \emph {et~al.}(2001)\citenamefont {Bons},
  \citenamefont {Taylor}, \citenamefont {McClain},\ and\ \citenamefont
  {Rivir}}]{bons2001}%
  \BibitemOpen
  \bibfield  {author} {\bibinfo {author} {\bibfnamefont {J.~P.}\ \bibnamefont
  {Bons}}, \bibinfo {author} {\bibfnamefont {R.~P.}\ \bibnamefont {Taylor}},
  \bibinfo {author} {\bibfnamefont {S.~T.}\ \bibnamefont {McClain}},\ and\
  \bibinfo {author} {\bibfnamefont {R.~B.}\ \bibnamefont {Rivir}},\ }\bibfield
  {title} {\bibinfo {title} {The many faces of turbine surface roughness},\
  }\href {https://doi.org/10.1115/1.1400115} {\bibfield  {journal} {\bibinfo
  {journal} {J. Turbomach.}\ }\textbf {\bibinfo {volume} {123}},\ \bibinfo
  {pages} {739} (\bibinfo {year} {2001})}\BibitemShut {NoStop}%
\bibitem [{\citenamefont {Bons}(2010)}]{bons2010}%
  \BibitemOpen
  \bibfield  {author} {\bibinfo {author} {\bibfnamefont {J.~P.}\ \bibnamefont
  {Bons}},\ }\bibfield  {title} {\bibinfo {title} {A review of surface
  roughness effects in gas turbines},\ }\href
  {https://doi.org/10.1115/1.3066315} {\bibfield  {journal} {\bibinfo
  {journal} {J. Turbomach.}\ }\textbf {\bibinfo {volume} {132}},\ \bibinfo
  {pages} {021004} (\bibinfo {year} {2010})}\BibitemShut {NoStop}%
\bibitem [{\citenamefont {Munk}\ \emph {et~al.}(2009)\citenamefont {Munk},
  \citenamefont {Kane},\ and\ \citenamefont {Yebra}}]{munk2009}%
  \BibitemOpen
  \bibfield  {author} {\bibinfo {author} {\bibfnamefont {T.}~\bibnamefont
  {Munk}}, \bibinfo {author} {\bibfnamefont {D.}~\bibnamefont {Kane}},\ and\
  \bibinfo {author} {\bibfnamefont {D.~M.}\ \bibnamefont {Yebra}},\ }\bibfield
  {title} {\bibinfo {title} {The effects of corrosion and fouling on the
  performance of ocean-going vessels: a naval architectural perspective},\ }in\
  \href {https://doi.org/10.1533/9781845696313.1.148} {\emph {\bibinfo
  {booktitle} {Advances in Marine Antifouling Coatings and Technologies}}},\
  \bibinfo {editor} {edited by\ \bibinfo {editor} {\bibfnamefont
  {C.}~\bibnamefont {Hellio}}\ and\ \bibinfo {editor} {\bibfnamefont {D.~M.}\
  \bibnamefont {Yebra}}}\ (\bibinfo  {publisher} {Woodhead Publishing},\
  \bibinfo {address} {Cambridge, UK},\ \bibinfo {year} {2009})\ pp.\ \bibinfo
  {pages} {148--176}\BibitemShut {NoStop}%
\bibitem [{\citenamefont {Kirschner}\ and\ \citenamefont
  {Brennan}(2012)}]{kirschner2012}%
  \BibitemOpen
  \bibfield  {author} {\bibinfo {author} {\bibfnamefont {C.~M.}\ \bibnamefont
  {Kirschner}}\ and\ \bibinfo {author} {\bibfnamefont {A.~B.}\ \bibnamefont
  {Brennan}},\ }\bibfield  {title} {\bibinfo {title} {Bio-inspired antifouling
  strategies},\ }\href {https://doi.org/10.1146/annurev-matsci-070511-155012}
  {\bibfield  {journal} {\bibinfo  {journal} {Annu. Rev. Mater. Res.}\ }\textbf
  {\bibinfo {volume} {42}},\ \bibinfo {pages} {211} (\bibinfo {year}
  {2012})}\BibitemShut {NoStop}%
\bibitem [{\citenamefont {Nikuradse}(1933)}]{nikuradse1933}%
  \BibitemOpen
  \bibfield  {author} {\bibinfo {author} {\bibfnamefont {J.}~\bibnamefont
  {Nikuradse}},\ }\href@noop {} {\emph {\bibinfo {title} {Str{\"o}mungsgesetze
  in rauhen Rohren}}},\ \bibinfo {type} {VDI-Forschungsheft}\ \bibinfo {number}
  {361}\ (\bibinfo  {institution} {Verein Deutscher Ingenieure},\ \bibinfo
  {address} {Berlin},\ \bibinfo {year} {1933})\ \bibinfo {note} {{English}
  translation: {NACA} Technical Memorandum 1292 (1950)}\BibitemShut {NoStop}%
\bibitem [{\citenamefont {Flack}\ and\ \citenamefont
  {Schultz}(2010)}]{flack2010}%
  \BibitemOpen
  \bibfield  {author} {\bibinfo {author} {\bibfnamefont {K.~A.}\ \bibnamefont
  {Flack}}\ and\ \bibinfo {author} {\bibfnamefont {M.~P.}\ \bibnamefont
  {Schultz}},\ }\bibfield  {title} {\bibinfo {title} {Review of hydraulic
  roughness scales in the fully rough regime},\ }\href
  {https://doi.org/10.1115/1.4001492} {\bibfield  {journal} {\bibinfo
  {journal} {J. Fluids Eng.}\ }\textbf {\bibinfo {volume} {132}},\ \bibinfo
  {pages} {041203} (\bibinfo {year} {2010})}\BibitemShut {NoStop}%
\bibitem [{\citenamefont {Ma}\ and\ \citenamefont
  {Lozano-Dur{\'a}n}(2025)}]{ma2025}%
  \BibitemOpen
  \bibfield  {author} {\bibinfo {author} {\bibfnamefont {R.}~\bibnamefont
  {Ma}}\ and\ \bibinfo {author} {\bibfnamefont {A.}~\bibnamefont
  {Lozano-Dur{\'a}n}},\ }\bibfield  {title} {\bibinfo {title} {Machine-learning
  wall-model large-eddy simulation accounting for isotropic roughness under
  local equilibrium},\ }\href {https://doi.org/10.1017/jfm.2025.29} {\bibfield
  {journal} {\bibinfo  {journal} {J. Fluid Mech.}\ }\textbf {\bibinfo {volume}
  {1007}},\ \bibinfo {pages} {A17} (\bibinfo {year} {2025})}\BibitemShut
  {NoStop}%
\bibitem [{\citenamefont {Kingma}\ and\ \citenamefont
  {Ba}(2015)}]{kingma2015adam}%
  \BibitemOpen
  \bibfield  {author} {\bibinfo {author} {\bibfnamefont {D.~P.}\ \bibnamefont
  {Kingma}}\ and\ \bibinfo {author} {\bibfnamefont {J.}~\bibnamefont {Ba}},\
  }\bibfield  {title} {\bibinfo {title} {Adam: {A} method for stochastic
  optimization},\ }in\ \href {https://arxiv.org/abs/1412.6980} {\emph {\bibinfo
  {booktitle} {3rd International Conference on Learning Representations
  ({ICLR})}}}\ (\bibinfo {year} {2015})\ \Eprint
  {https://arxiv.org/abs/1412.6980} {arXiv:1412.6980 [cs.LG]} \BibitemShut
  {NoStop}%
\bibitem [{\citenamefont {Eitel-Amor}\ \emph {et~al.}(2014)\citenamefont
  {Eitel-Amor}, \citenamefont {{\"O}rl{\"u}},\ and\ \citenamefont
  {Schlatter}}]{eitelamor2014simulation}%
  \BibitemOpen
  \bibfield  {author} {\bibinfo {author} {\bibfnamefont {G.}~\bibnamefont
  {Eitel-Amor}}, \bibinfo {author} {\bibfnamefont {R.}~\bibnamefont
  {{\"O}rl{\"u}}},\ and\ \bibinfo {author} {\bibfnamefont {P.}~\bibnamefont
  {Schlatter}},\ }\bibfield  {title} {\bibinfo {title} {Simulation and
  validation of a spatially evolving turbulent boundary layer up to
  {$Re_{\theta}=8300$}},\ }\href
  {https://doi.org/10.1016/j.ijheatfluidflow.2014.02.006} {\bibfield  {journal}
  {\bibinfo  {journal} {Int. J. Heat Fluid Flow}\ }\textbf {\bibinfo {volume}
  {47}},\ \bibinfo {pages} {57} (\bibinfo {year} {2014})}\BibitemShut {NoStop}%
\bibitem [{\citenamefont {Spalart}\ \emph {et~al.}(2024)\citenamefont
  {Spalart}, \citenamefont {Jansen},\ and\ \citenamefont
  {Coleman}}]{spalart2024flattened}%
  \BibitemOpen
  \bibfield  {author} {\bibinfo {author} {\bibfnamefont {P.~R.}\ \bibnamefont
  {Spalart}}, \bibinfo {author} {\bibfnamefont {K.~E.}\ \bibnamefont
  {Jansen}},\ and\ \bibinfo {author} {\bibfnamefont {G.~N.}\ \bibnamefont
  {Coleman}},\ }\href {https://doi.org/10.48550/arXiv.2409.00555} {\bibinfo
  {title} {Direct numerical simulation of two boundary layers with the same
  pressure distribution but different surface curvatures}} (\bibinfo {year}
  {2024}),\ \Eprint {https://arxiv.org/abs/2409.00555} {arXiv:2409.00555
  [physics.flu-dyn]} \BibitemShut {NoStop}%
\bibitem [{\citenamefont {{\"O}rl{\"u}}\ and\ \citenamefont
  {Schlatter}(2013)}]{orlu2013dns}%
  \BibitemOpen
  \bibfield  {author} {\bibinfo {author} {\bibfnamefont {R.}~\bibnamefont
  {{\"O}rl{\"u}}}\ and\ \bibinfo {author} {\bibfnamefont {P.}~\bibnamefont
  {Schlatter}},\ }\bibfield  {title} {\bibinfo {title} {Comparison of
  experiments and simulations for zero pressure gradient turbulent boundary
  layers at moderate {Reynolds} numbers},\ }\href
  {https://doi.org/10.1007/s00348-013-1547-x} {\bibfield  {journal} {\bibinfo
  {journal} {Exp. Fluids}\ }\textbf {\bibinfo {volume} {54}},\ \bibinfo {pages}
  {1547} (\bibinfo {year} {2013})}\BibitemShut {NoStop}%
\bibitem [{\citenamefont {Wieghardt}\ and\ \citenamefont
  {Tillmann}(1951)}]{wieghardt1951characteristics}%
  \BibitemOpen
  \bibfield  {author} {\bibinfo {author} {\bibfnamefont {K.}~\bibnamefont
  {Wieghardt}}\ and\ \bibinfo {author} {\bibfnamefont {W.}~\bibnamefont
  {Tillmann}},\ }\href {https://ntrs.nasa.gov/citations/19930093916} {\emph
  {\bibinfo {title} {On the Turbulent Friction Layer for Rising Pressure}}},\
  \bibinfo {type} {Technical Memorandum}\ \bibinfo {number} {1314}\ (\bibinfo
  {institution} {National Advisory Committee for Aeronautics},\ \bibinfo {year}
  {1951})\BibitemShut {NoStop}%
\bibitem [{\citenamefont {Mallor}\ \emph {et~al.}(2025)\citenamefont {Mallor},
  \citenamefont {Sanmiguel~Vila}, \citenamefont {Hajipour}, \citenamefont
  {Vinuesa}, \citenamefont {Schlatter},\ and\ \citenamefont
  {{\"O}rl{\"u}}}]{mallor2025experimental}%
  \BibitemOpen
  \bibfield  {author} {\bibinfo {author} {\bibfnamefont {F.}~\bibnamefont
  {Mallor}}, \bibinfo {author} {\bibfnamefont {C.}~\bibnamefont
  {Sanmiguel~Vila}}, \bibinfo {author} {\bibfnamefont {M.}~\bibnamefont
  {Hajipour}}, \bibinfo {author} {\bibfnamefont {R.}~\bibnamefont {Vinuesa}},
  \bibinfo {author} {\bibfnamefont {P.}~\bibnamefont {Schlatter}},\ and\
  \bibinfo {author} {\bibfnamefont {R.}~\bibnamefont {{\"O}rl{\"u}}},\
  }\bibfield  {title} {\bibinfo {title} {Experimental characterization of
  turbulent boundary layers around a {NACA 4412} wing profile},\ }\href
  {https://doi.org/10.1016/j.expthermflusci.2024.111327} {\bibfield  {journal}
  {\bibinfo  {journal} {Exp. Therm. Fluid Sci.}\ }\textbf {\bibinfo {volume}
  {160}},\ \bibinfo {pages} {111327} (\bibinfo {year} {2025})}\BibitemShut
  {NoStop}%
\bibitem [{\citenamefont {Vinuesa}\ \emph {et~al.}(2018)\citenamefont
  {Vinuesa}, \citenamefont {Negi}, \citenamefont {Atzori}, \citenamefont
  {Hanifi}, \citenamefont {Henningson},\ and\ \citenamefont
  {Schlatter}}]{vinuesa2019wing}%
  \BibitemOpen
  \bibfield  {author} {\bibinfo {author} {\bibfnamefont {R.}~\bibnamefont
  {Vinuesa}}, \bibinfo {author} {\bibfnamefont {P.~S.}\ \bibnamefont {Negi}},
  \bibinfo {author} {\bibfnamefont {M.}~\bibnamefont {Atzori}}, \bibinfo
  {author} {\bibfnamefont {A.}~\bibnamefont {Hanifi}}, \bibinfo {author}
  {\bibfnamefont {D.~S.}\ \bibnamefont {Henningson}},\ and\ \bibinfo {author}
  {\bibfnamefont {P.}~\bibnamefont {Schlatter}},\ }\bibfield  {title} {\bibinfo
  {title} {Turbulent boundary layers around wing sections up to
  {$Re_c=1{,}000{,}000$}},\ }\href
  {https://doi.org/10.1016/j.ijheatfluidflow.2018.04.017} {\bibfield  {journal}
  {\bibinfo  {journal} {Int. J. Heat Fluid Flow}\ }\textbf {\bibinfo {volume}
  {72}},\ \bibinfo {pages} {86} (\bibinfo {year} {2018})}\BibitemShut {NoStop}%
\bibitem [{\citenamefont {Tanarro}\ \emph {et~al.}(2020)\citenamefont
  {Tanarro}, \citenamefont {Vinuesa},\ and\ \citenamefont
  {Schlatter}}]{tanarro2020effect}%
  \BibitemOpen
  \bibfield  {author} {\bibinfo {author} {\bibfnamefont {{\'A}.}~\bibnamefont
  {Tanarro}}, \bibinfo {author} {\bibfnamefont {R.}~\bibnamefont {Vinuesa}},\
  and\ \bibinfo {author} {\bibfnamefont {P.}~\bibnamefont {Schlatter}},\
  }\bibfield  {title} {\bibinfo {title} {Effect of adverse pressure gradients
  on turbulent wing boundary layers},\ }\href
  {https://doi.org/10.1017/jfm.2019.838} {\bibfield  {journal} {\bibinfo
  {journal} {J. Fluid Mech.}\ }\textbf {\bibinfo {volume} {883}},\ \bibinfo
  {pages} {A8} (\bibinfo {year} {2020})}\BibitemShut {NoStop}%
\bibitem [{\citenamefont {Greenblatt}\ \emph {et~al.}(2006)\citenamefont
  {Greenblatt}, \citenamefont {Paschal}, \citenamefont {Yao}, \citenamefont
  {Harris}, \citenamefont {Schaeffler},\ and\ \citenamefont
  {Washburn}}]{greenblatt2006separation}%
  \BibitemOpen
  \bibfield  {author} {\bibinfo {author} {\bibfnamefont {D.}~\bibnamefont
  {Greenblatt}}, \bibinfo {author} {\bibfnamefont {K.~B.}\ \bibnamefont
  {Paschal}}, \bibinfo {author} {\bibfnamefont {C.-S.}\ \bibnamefont {Yao}},
  \bibinfo {author} {\bibfnamefont {J.}~\bibnamefont {Harris}}, \bibinfo
  {author} {\bibfnamefont {N.~W.}\ \bibnamefont {Schaeffler}},\ and\ \bibinfo
  {author} {\bibfnamefont {A.~E.}\ \bibnamefont {Washburn}},\ }\bibfield
  {title} {\bibinfo {title} {Experimental investigation of separation control
  part 1: Baseline and steady suction},\ }\href
  {https://doi.org/10.2514/1.13817} {\bibfield  {journal} {\bibinfo  {journal}
  {AIAA J.}\ }\textbf {\bibinfo {volume} {44}},\ \bibinfo {pages} {2820}
  (\bibinfo {year} {2006})}\BibitemShut {NoStop}%
\bibitem [{\citenamefont {Naughton}\ \emph {et~al.}(2006)\citenamefont
  {Naughton}, \citenamefont {Viken},\ and\ \citenamefont
  {Greenblatt}}]{naughton2006wall}%
  \BibitemOpen
  \bibfield  {author} {\bibinfo {author} {\bibfnamefont {J.~W.}\ \bibnamefont
  {Naughton}}, \bibinfo {author} {\bibfnamefont {S.~A.}\ \bibnamefont
  {Viken}},\ and\ \bibinfo {author} {\bibfnamefont {D.}~\bibnamefont
  {Greenblatt}},\ }\bibfield  {title} {\bibinfo {title} {Skin-friction
  measurements on the {NASA} hump model},\ }\href
  {https://doi.org/10.2514/1.14192} {\bibfield  {journal} {\bibinfo  {journal}
  {AIAA J.}\ }\textbf {\bibinfo {volume} {44}},\ \bibinfo {pages} {1255}
  (\bibinfo {year} {2006})}\BibitemShut {NoStop}%
\bibitem [{\citenamefont {Gray}\ \emph {et~al.}(2023)\citenamefont {Gray},
  \citenamefont {Lakebrink}, \citenamefont {Thomas}, \citenamefont {Corke},
  \citenamefont {Gluzman},\ and\ \citenamefont
  {Straccia}}]{gray2023experimental}%
  \BibitemOpen
  \bibfield  {author} {\bibinfo {author} {\bibfnamefont {P.~D.}\ \bibnamefont
  {Gray}}, \bibinfo {author} {\bibfnamefont {M.~T.}\ \bibnamefont {Lakebrink}},
  \bibinfo {author} {\bibfnamefont {F.~O.}\ \bibnamefont {Thomas}}, \bibinfo
  {author} {\bibfnamefont {T.~C.}\ \bibnamefont {Corke}}, \bibinfo {author}
  {\bibfnamefont {I.}~\bibnamefont {Gluzman}},\ and\ \bibinfo {author}
  {\bibfnamefont {J.}~\bibnamefont {Straccia}},\ }\bibfield  {title} {\bibinfo
  {title} {Experimental and computational evaluation of smooth-body separated
  flow over {Boeing} bump},\ }in\ \href {https://doi.org/10.2514/6.2023-3981}
  {\emph {\bibinfo {booktitle} {{AIAA AVIATION} 2023 Forum}}}\ (\bibinfo
  {publisher} {American Institute of Aeronautics and Astronautics},\ \bibinfo
  {year} {2023})\ \bibinfo {note} {paper 2023-3981}\BibitemShut {NoStop}%
\bibitem [{\citenamefont {Bissonnette}\ and\ \citenamefont
  {Mellor}(1974)}]{bissonnette1974three}%
  \BibitemOpen
  \bibfield  {author} {\bibinfo {author} {\bibfnamefont {L.~R.}\ \bibnamefont
  {Bissonnette}}\ and\ \bibinfo {author} {\bibfnamefont {G.~L.}\ \bibnamefont
  {Mellor}},\ }\bibfield  {title} {\bibinfo {title} {Experiments on the
  behaviour of an axisymmetric turbulent boundary layer with a sudden
  circumferential strain},\ }\href {https://doi.org/10.1017/S0022112074001182}
  {\bibfield  {journal} {\bibinfo  {journal} {J. Fluid Mech.}\ }\textbf
  {\bibinfo {volume} {63}},\ \bibinfo {pages} {369} (\bibinfo {year}
  {1974})}\BibitemShut {NoStop}%
\bibitem [{\citenamefont {Lohmann}(1976)}]{lohmann1976measurement}%
  \BibitemOpen
  \bibfield  {author} {\bibinfo {author} {\bibfnamefont {R.~P.}\ \bibnamefont
  {Lohmann}},\ }\bibfield  {title} {\bibinfo {title} {The response of a
  developed turbulent boundary layer to local transverse surface motion},\
  }\href {https://doi.org/10.1115/1.3448319} {\bibfield  {journal} {\bibinfo
  {journal} {J. Fluids Eng.}\ }\textbf {\bibinfo {volume} {98}},\ \bibinfo
  {pages} {354} (\bibinfo {year} {1976})}\BibitemShut {NoStop}%
\bibitem [{\citenamefont {Rkein}\ and\ \citenamefont
  {Laval}(2023)}]{rkein2023numerical}%
  \BibitemOpen
  \bibfield  {author} {\bibinfo {author} {\bibfnamefont {H.}~\bibnamefont
  {Rkein}}\ and\ \bibinfo {author} {\bibfnamefont {J.-P.}\ \bibnamefont
  {Laval}},\ }\bibfield  {title} {\bibinfo {title} {Direct numerical simulation
  of adverse pressure gradient turbulent boundary layer up to
  {$Re_{\theta}\simeq8000$}},\ }\href
  {https://doi.org/10.1016/j.ijheatfluidflow.2023.109195} {\bibfield  {journal}
  {\bibinfo  {journal} {Int. J. Heat Fluid Flow}\ }\textbf {\bibinfo {volume}
  {103}},\ \bibinfo {pages} {109195} (\bibinfo {year} {2023})}\BibitemShut
  {NoStop}%
\bibitem [{\citenamefont {Spalart}\ and\ \citenamefont
  {Watmuff}(1993)}]{spalart1993experimental}%
  \BibitemOpen
  \bibfield  {author} {\bibinfo {author} {\bibfnamefont {P.~R.}\ \bibnamefont
  {Spalart}}\ and\ \bibinfo {author} {\bibfnamefont {J.~H.}\ \bibnamefont
  {Watmuff}},\ }\bibfield  {title} {\bibinfo {title} {Experimental and
  numerical study of a turbulent boundary layer with pressure gradients},\
  }\href {https://doi.org/10.1017/S002211209300120X} {\bibfield  {journal}
  {\bibinfo  {journal} {J. Fluid Mech.}\ }\textbf {\bibinfo {volume} {249}},\
  \bibinfo {pages} {337} (\bibinfo {year} {1993})}\BibitemShut {NoStop}%
\bibitem [{\citenamefont {Fritsch}\ \emph {et~al.}(2022)\citenamefont
  {Fritsch}, \citenamefont {Vishwanathan}, \citenamefont {Lowe},\ and\
  \citenamefont {Devenport}}]{fritsch2022effects}%
  \BibitemOpen
  \bibfield  {author} {\bibinfo {author} {\bibfnamefont {D.~J.}\ \bibnamefont
  {Fritsch}}, \bibinfo {author} {\bibfnamefont {V.}~\bibnamefont
  {Vishwanathan}}, \bibinfo {author} {\bibfnamefont {K.~T.}\ \bibnamefont
  {Lowe}},\ and\ \bibinfo {author} {\bibfnamefont {W.~J.}\ \bibnamefont
  {Devenport}},\ }\bibfield  {title} {\bibinfo {title} {Fluctuating pressure
  beneath smooth wall boundary layers in nonequilibrium pressure gradients},\
  }\href {https://doi.org/10.2514/1.J061431} {\bibfield  {journal} {\bibinfo
  {journal} {AIAA J.}\ }\textbf {\bibinfo {volume} {60}},\ \bibinfo {pages}
  {4725} (\bibinfo {year} {2022})}\BibitemShut {NoStop}%
\bibitem [{\citenamefont {Vishwanathan}\ \emph {et~al.}(2023)\citenamefont
  {Vishwanathan}, \citenamefont {Fritsch}, \citenamefont {Lowe},\ and\
  \citenamefont {Devenport}}]{vishwanathan2023effects}%
  \BibitemOpen
  \bibfield  {author} {\bibinfo {author} {\bibfnamefont {V.}~\bibnamefont
  {Vishwanathan}}, \bibinfo {author} {\bibfnamefont {D.~J.}\ \bibnamefont
  {Fritsch}}, \bibinfo {author} {\bibfnamefont {K.~T.}\ \bibnamefont {Lowe}},\
  and\ \bibinfo {author} {\bibfnamefont {W.~J.}\ \bibnamefont {Devenport}},\
  }\bibfield  {title} {\bibinfo {title} {History effects and wall-similarity of
  non-equilibrium turbulent boundary layers in varying pressure gradient over
  rough and smooth surfaces},\ }\href
  {https://doi.org/10.1016/j.ijheatfluidflow.2023.109145} {\bibfield  {journal}
  {\bibinfo  {journal} {Int. J. Heat Fluid Flow}\ }\textbf {\bibinfo {volume}
  {102}},\ \bibinfo {pages} {109145} (\bibinfo {year} {2023})}\BibitemShut
  {NoStop}%
\end{thebibliography}%


%apsrev4-2.bst 2019-01-14 (MD) hand-edited version of apsrev4-1.bst
%Control: key (0)
%Control: author (8) initials jnrlst
%Control: editor formatted (1) identically to author
%Control: production of article title (0) allowed
%Control: page (0) single
%Control: year (1) truncated
%Control: production of eprint (0) enabled
\begin{thebibliography}{91}%
\makeatletter
\providecommand \@ifxundefined [1]{%
 \@ifx{#1\undefined}
}%
\providecommand \@ifnum [1]{%
 \ifnum #1\expandafter \@firstoftwo
 \else \expandafter \@secondoftwo
 \fi
}%
\providecommand \@ifx [1]{%
 \ifx #1\expandafter \@firstoftwo
 \else \expandafter \@secondoftwo
 \fi
}%
\providecommand \natexlab [1]{#1}%
\providecommand \enquote  [1]{``#1''}%
\providecommand \bibnamefont  [1]{#1}%
\providecommand \bibfnamefont [1]{#1}%
\providecommand \citenamefont [1]{#1}%
\providecommand \href@noop [0]{\@secondoftwo}%
\providecommand \href [0]{\begingroup \@sanitize@url \@href}%
\providecommand \@href[1]{\@@startlink{#1}\@@href}%
\providecommand \@@href[1]{\endgroup#1\@@endlink}%
\providecommand \@sanitize@url [0]{\catcode `\\12\catcode `\$12\catcode
  `\&12\catcode `\#12\catcode `\^12\catcode `\_12\catcode `\%12\relax}%
\providecommand \@@startlink[1]{}%
\providecommand \@@endlink[0]{}%
\providecommand \url  [0]{\begingroup\@sanitize@url \@url }%
\providecommand \@url [1]{\endgroup\@href {#1}{\urlprefix }}%
\providecommand \urlprefix  [0]{URL }%
\providecommand \Eprint [0]{\href }%
\providecommand \doibase [0]{https://doi.org/}%
\providecommand \selectlanguage [0]{\@gobble}%
\providecommand \bibinfo  [0]{\@secondoftwo}%
\providecommand \bibfield  [0]{\@secondoftwo}%
\providecommand \translation [1]{[#1]}%
\providecommand \BibitemOpen [0]{}%
\providecommand \bibitemStop [0]{}%
\providecommand \bibitemNoStop [0]{.\EOS\space}%
\providecommand \EOS [0]{\spacefactor3000\relax}%
\providecommand \BibitemShut  [1]{\csname bibitem#1\endcsname}%
\let\auto@bib@innerbib\@empty
%</preamble>
\bibitem [{\citenamefont {Orlandi}(2000)}]{Orlandi2000}%
  \BibitemOpen
  \bibinfo {editor} {\bibfnamefont {P.}~\bibnamefont {Orlandi}},\ ed.,\ \href
  {https://doi.org/10.1007/978-94-011-4281-6} {\emph {\bibinfo {title} {Fluid
  Flow Phenomena: A Numerical Toolkit}}},\ \bibinfo {series} {Fluid Mechanics
  and Its Applications}, Vol.~\bibinfo {volume} {55}\ (\bibinfo  {publisher}
  {Springer},\ \bibinfo {address} {Dordrecht},\ \bibinfo {year}
  {2000})\BibitemShut {NoStop}%
\bibitem [{\citenamefont {Wray}(1990)}]{Wray1990}%
  \BibitemOpen
  \bibfield  {author} {\bibinfo {author} {\bibfnamefont {A.~A.}\ \bibnamefont
  {Wray}},\ }\href@noop {} {\emph {\bibinfo {title} {Minimal Storage
  Time-Advancement Schemes for Spectral Methods}}},\ \bibinfo {type} {Technical
  Report}\ (\bibinfo  {institution} {NASA Ames Research Center},\ \bibinfo
  {address} {Moffett Field, CA},\ \bibinfo {year} {1990})\BibitemShut {NoStop}%
\bibitem [{\citenamefont {Kim}\ and\ \citenamefont {Moin}(1985)}]{Kim1985}%
  \BibitemOpen
  \bibfield  {author} {\bibinfo {author} {\bibfnamefont {J.}~\bibnamefont
  {Kim}}\ and\ \bibinfo {author} {\bibfnamefont {P.}~\bibnamefont {Moin}},\
  }\bibfield  {title} {\bibinfo {title} {Application of a fractional-step
  method to incompressible {Navier--Stokes} equations},\ }\href
  {https://doi.org/10.1016/0021-9991(85)90148-2} {\bibfield  {journal}
  {\bibinfo  {journal} {J. Comput. Phys.}\ }\textbf {\bibinfo {volume} {59}},\
  \bibinfo {pages} {308} (\bibinfo {year} {1985})}\BibitemShut {NoStop}%
\bibitem [{\citenamefont {Bae}\ \emph {et~al.}(2018)\citenamefont {Bae},
  \citenamefont {Lozano-Dur{\'a}n}, \citenamefont {Bose},\ and\ \citenamefont
  {Moin}}]{Bae2018}%
  \BibitemOpen
  \bibfield  {author} {\bibinfo {author} {\bibfnamefont {H.~J.}\ \bibnamefont
  {Bae}}, \bibinfo {author} {\bibfnamefont {A.}~\bibnamefont
  {Lozano-Dur{\'a}n}}, \bibinfo {author} {\bibfnamefont {S.~T.}\ \bibnamefont
  {Bose}},\ and\ \bibinfo {author} {\bibfnamefont {P.}~\bibnamefont {Moin}},\
  }\bibfield  {title} {\bibinfo {title} {Turbulence intensities in large-eddy
  simulation of wall-bounded flows},\ }\href
  {https://doi.org/10.1103/PhysRevFluids.3.014610} {\bibfield  {journal}
  {\bibinfo  {journal} {Phys. Rev. Fluids}\ }\textbf {\bibinfo {volume} {3}},\
  \bibinfo {pages} {014610} (\bibinfo {year} {2018})}\BibitemShut {NoStop}%
\bibitem [{\citenamefont {Lozano-Dur{\'a}n}\ and\ \citenamefont
  {Bae}(2019)}]{Lozano2019d}%
  \BibitemOpen
  \bibfield  {author} {\bibinfo {author} {\bibfnamefont {A.}~\bibnamefont
  {Lozano-Dur{\'a}n}}\ and\ \bibinfo {author} {\bibfnamefont {H.~J.}\
  \bibnamefont {Bae}},\ }\bibfield  {title} {\bibinfo {title} {Characteristic
  scales of {Townsend}'s wall-attached eddies},\ }\href
  {https://doi.org/10.1017/jfm.2019.209} {\bibfield  {journal} {\bibinfo
  {journal} {J. Fluid Mech.}\ }\textbf {\bibinfo {volume} {868}},\ \bibinfo
  {pages} {698} (\bibinfo {year} {2019})}\BibitemShut {NoStop}%
\bibitem [{\citenamefont {Lozano-Dur{\'a}n}\ \emph {et~al.}(2018)\citenamefont
  {Lozano-Dur{\'a}n}, \citenamefont {Hack},\ and\ \citenamefont
  {Moin}}]{Lozano2018}%
  \BibitemOpen
  \bibfield  {author} {\bibinfo {author} {\bibfnamefont {A.}~\bibnamefont
  {Lozano-Dur{\'a}n}}, \bibinfo {author} {\bibfnamefont {M.~J.~P.}\
  \bibnamefont {Hack}},\ and\ \bibinfo {author} {\bibfnamefont
  {P.}~\bibnamefont {Moin}},\ }\bibfield  {title} {\bibinfo {title} {Modeling
  boundary-layer transition in direct and large-eddy simulations using
  parabolized stability equations},\ }\href
  {https://doi.org/10.1103/PhysRevFluids.3.023901} {\bibfield  {journal}
  {\bibinfo  {journal} {Phys. Rev. Fluids}\ }\textbf {\bibinfo {volume} {3}},\
  \bibinfo {pages} {023901} (\bibinfo {year} {2018})}\BibitemShut {NoStop}%
\bibitem [{\citenamefont {Towne}\ \emph {et~al.}(2023)\citenamefont {Towne},
  \citenamefont {Dawson}, \citenamefont {Br{\`e}s}, \citenamefont
  {Lozano-Dur{\'a}n}, \citenamefont {Saxton-Fox}, \citenamefont
  {Parthasarathy}, \citenamefont {Jones}, \citenamefont {Biler}, \citenamefont
  {Yeh}, \citenamefont {Patel},\ and\ \citenamefont {Taira}}]{Towne2023}%
  \BibitemOpen
  \bibfield  {author} {\bibinfo {author} {\bibfnamefont {A.}~\bibnamefont
  {Towne}}, \bibinfo {author} {\bibfnamefont {S.~T.~M.}\ \bibnamefont
  {Dawson}}, \bibinfo {author} {\bibfnamefont {G.~A.}\ \bibnamefont
  {Br{\`e}s}}, \bibinfo {author} {\bibfnamefont {A.}~\bibnamefont
  {Lozano-Dur{\'a}n}}, \bibinfo {author} {\bibfnamefont {T.}~\bibnamefont
  {Saxton-Fox}}, \bibinfo {author} {\bibfnamefont {A.}~\bibnamefont
  {Parthasarathy}}, \bibinfo {author} {\bibfnamefont {A.~R.}\ \bibnamefont
  {Jones}}, \bibinfo {author} {\bibfnamefont {H.}~\bibnamefont {Biler}},
  \bibinfo {author} {\bibfnamefont {C.-A.}\ \bibnamefont {Yeh}}, \bibinfo
  {author} {\bibfnamefont {H.~D.}\ \bibnamefont {Patel}},\ and\ \bibinfo
  {author} {\bibfnamefont {K.}~\bibnamefont {Taira}},\ }\bibfield  {title}
  {\bibinfo {title} {A database for reduced-complexity modeling of fluid
  flows},\ }\href {https://doi.org/10.2514/1.J062203} {\bibfield  {journal}
  {\bibinfo  {journal} {AIAA J.}\ }\textbf {\bibinfo {volume} {61}},\ \bibinfo
  {pages} {2867} (\bibinfo {year} {2023})}\BibitemShut {NoStop}%
\bibitem [{\citenamefont {Pauley}\ \emph {et~al.}(1990)\citenamefont {Pauley},
  \citenamefont {Moin},\ and\ \citenamefont {Reynolds}}]{Pauley1990}%
  \BibitemOpen
  \bibfield  {author} {\bibinfo {author} {\bibfnamefont {L.~L.}\ \bibnamefont
  {Pauley}}, \bibinfo {author} {\bibfnamefont {P.}~\bibnamefont {Moin}},\ and\
  \bibinfo {author} {\bibfnamefont {W.~C.}\ \bibnamefont {Reynolds}},\
  }\bibfield  {title} {\bibinfo {title} {The structure of two-dimensional
  separation},\ }\href {https://doi.org/10.1017/S0022112090003317} {\bibfield
  {journal} {\bibinfo  {journal} {J. Fluid Mech.}\ }\textbf {\bibinfo {volume}
  {220}},\ \bibinfo {pages} {397} (\bibinfo {year} {1990})}\BibitemShut
  {NoStop}%
\bibitem [{\citenamefont {Sillero}\ \emph {et~al.}(2014)\citenamefont
  {Sillero}, \citenamefont {Jim{\'e}nez},\ and\ \citenamefont
  {Moser}}]{sillero2014}%
  \BibitemOpen
  \bibfield  {author} {\bibinfo {author} {\bibfnamefont {J.~A.}\ \bibnamefont
  {Sillero}}, \bibinfo {author} {\bibfnamefont {J.}~\bibnamefont
  {Jim{\'e}nez}},\ and\ \bibinfo {author} {\bibfnamefont {R.~D.}\ \bibnamefont
  {Moser}},\ }\bibfield  {title} {\bibinfo {title} {Two-point statistics for
  turbulent boundary layers and channels at {Reynolds} numbers up to $\delta^+
  \approx 2000$},\ }\href {https://doi.org/10.1063/1.4899259} {\bibfield
  {journal} {\bibinfo  {journal} {Phys. Fluids}\ }\textbf {\bibinfo {volume}
  {26}},\ \bibinfo {pages} {105109} (\bibinfo {year} {2014})}\BibitemShut
  {NoStop}%
\bibitem [{\citenamefont {Lund}\ \emph {et~al.}(1998)\citenamefont {Lund},
  \citenamefont {Wu},\ and\ \citenamefont {Squires}}]{lund1998Generation}%
  \BibitemOpen
  \bibfield  {author} {\bibinfo {author} {\bibfnamefont {T.~S.}\ \bibnamefont
  {Lund}}, \bibinfo {author} {\bibfnamefont {X.}~\bibnamefont {Wu}},\ and\
  \bibinfo {author} {\bibfnamefont {K.~D.}\ \bibnamefont {Squires}},\
  }\bibfield  {title} {\bibinfo {title} {Generation of turbulent inflow data
  for spatially-developing boundary layer simulations},\ }\href
  {https://doi.org/10.1006/jcph.1998.5882} {\bibfield  {journal} {\bibinfo
  {journal} {J. Comput. Phys.}\ }\textbf {\bibinfo {volume} {140}},\ \bibinfo
  {pages} {233} (\bibinfo {year} {1998})}\BibitemShut {NoStop}%
\bibitem [{\citenamefont {Morgan}\ \emph {et~al.}(2011)\citenamefont {Morgan},
  \citenamefont {Larsson}, \citenamefont {Kawai},\ and\ \citenamefont
  {Lele}}]{morgan2011Improving}%
  \BibitemOpen
  \bibfield  {author} {\bibinfo {author} {\bibfnamefont {B.}~\bibnamefont
  {Morgan}}, \bibinfo {author} {\bibfnamefont {J.}~\bibnamefont {Larsson}},
  \bibinfo {author} {\bibfnamefont {S.}~\bibnamefont {Kawai}},\ and\ \bibinfo
  {author} {\bibfnamefont {S.~K.}\ \bibnamefont {Lele}},\ }\bibfield  {title}
  {\bibinfo {title} {Improving low-frequency characteristics of
  recycling/rescaling inflow turbulence generation},\ }\href
  {https://doi.org/10.2514/1.J050705} {\bibfield  {journal} {\bibinfo
  {journal} {AIAA J.}\ }\textbf {\bibinfo {volume} {49}},\ \bibinfo {pages}
  {582} (\bibinfo {year} {2011})}\BibitemShut {NoStop}%
\bibitem [{\citenamefont {Griffin}\ \emph
  {et~al.}(2021{\natexlab{a}})\citenamefont {Griffin}, \citenamefont {Fu},\
  and\ \citenamefont {Moin}}]{griffin2021general}%
  \BibitemOpen
  \bibfield  {author} {\bibinfo {author} {\bibfnamefont {K.~P.}\ \bibnamefont
  {Griffin}}, \bibinfo {author} {\bibfnamefont {L.}~\bibnamefont {Fu}},\ and\
  \bibinfo {author} {\bibfnamefont {P.}~\bibnamefont {Moin}},\ }\bibfield
  {title} {\bibinfo {title} {General method for determining the boundary layer
  thickness in nonequilibrium flows},\ }\href
  {https://doi.org/10.1103/PhysRevFluids.6.024608} {\bibfield  {journal}
  {\bibinfo  {journal} {Phys. Rev. Fluids}\ }\textbf {\bibinfo {volume} {6}},\
  \bibinfo {pages} {024608} (\bibinfo {year} {2021}{\natexlab{a}})}\BibitemShut
  {NoStop}%
\bibitem [{\citenamefont {Coleman}\ \emph {et~al.}(2018)\citenamefont
  {Coleman}, \citenamefont {Rumsey},\ and\ \citenamefont
  {Spalart}}]{coleman2018numerical}%
  \BibitemOpen
  \bibfield  {author} {\bibinfo {author} {\bibfnamefont {G.~N.}\ \bibnamefont
  {Coleman}}, \bibinfo {author} {\bibfnamefont {C.~L.}\ \bibnamefont
  {Rumsey}},\ and\ \bibinfo {author} {\bibfnamefont {P.~R.}\ \bibnamefont
  {Spalart}},\ }\bibfield  {title} {\bibinfo {title} {Numerical study of
  turbulent separation bubbles with varying pressure gradient and {Reynolds}
  number},\ }\href {https://doi.org/10.1017/jfm.2018.257} {\bibfield  {journal}
  {\bibinfo  {journal} {J. Fluid Mech.}\ }\textbf {\bibinfo {volume} {847}},\
  \bibinfo {pages} {28} (\bibinfo {year} {2018})}\BibitemShut {NoStop}%
\bibitem [{\citenamefont {Maciel}\ \emph {et~al.}(2006)\citenamefont {Maciel},
  \citenamefont {Rossignol},\ and\ \citenamefont
  {Lemay}}]{maciel2006selfsimilarity}%
  \BibitemOpen
  \bibfield  {author} {\bibinfo {author} {\bibfnamefont {Y.}~\bibnamefont
  {Maciel}}, \bibinfo {author} {\bibfnamefont {K.-S.}\ \bibnamefont
  {Rossignol}},\ and\ \bibinfo {author} {\bibfnamefont {J.}~\bibnamefont
  {Lemay}},\ }\bibfield  {title} {\bibinfo {title} {Self-similarity in the
  outer region of adverse-pressure-gradient turbulent boundary layers},\ }\href
  {https://doi.org/10.2514/1.19234} {\bibfield  {journal} {\bibinfo  {journal}
  {AIAA J.}\ }\textbf {\bibinfo {volume} {44}},\ \bibinfo {pages} {2450}
  (\bibinfo {year} {2006})}\BibitemShut {NoStop}%
\bibitem [{\citenamefont {Zagarola}\ and\ \citenamefont
  {Smits}(1997)}]{zagarola1997scaling}%
  \BibitemOpen
  \bibfield  {author} {\bibinfo {author} {\bibfnamefont {M.~V.}\ \bibnamefont
  {Zagarola}}\ and\ \bibinfo {author} {\bibfnamefont {A.~J.}\ \bibnamefont
  {Smits}},\ }\bibfield  {title} {\bibinfo {title} {Scaling of the mean
  velocity profile for turbulent pipe flow},\ }\href
  {https://doi.org/10.1103/PhysRevLett.78.239} {\bibfield  {journal} {\bibinfo
  {journal} {Phys. Rev. Lett.}\ }\textbf {\bibinfo {volume} {78}},\ \bibinfo
  {pages} {239} (\bibinfo {year} {1997})}\BibitemShut {NoStop}%
\bibitem [{\citenamefont {Torroja}(2025)}]{torroja}%
  \BibitemOpen
  \bibfield  {author} {\bibinfo {author} {\bibnamefont {Torroja}},\ }\href
  {https://torroja.dmt.upm.es/turbdata/} {\bibinfo {title} {Turbulent flow
  databases}},\ \bibinfo {howpublished} {Online database} (\bibinfo {year}
  {2025})\BibitemShut {NoStop}%
\bibitem [{\citenamefont {Bobke}\ \emph {et~al.}(2017)\citenamefont {Bobke},
  \citenamefont {Vinuesa}, \citenamefont {{\"O}rl{\"u}},\ and\ \citenamefont
  {Schlatter}}]{bobke2017history}%
  \BibitemOpen
  \bibfield  {author} {\bibinfo {author} {\bibfnamefont {A.}~\bibnamefont
  {Bobke}}, \bibinfo {author} {\bibfnamefont {R.}~\bibnamefont {Vinuesa}},
  \bibinfo {author} {\bibfnamefont {R.}~\bibnamefont {{\"O}rl{\"u}}},\ and\
  \bibinfo {author} {\bibfnamefont {P.}~\bibnamefont {Schlatter}},\ }\bibfield
  {title} {\bibinfo {title} {History effects and near equilibrium in
  adverse-pressure-gradient turbulent boundary layers},\ }\href
  {https://doi.org/10.1017/jfm.2017.236} {\bibfield  {journal} {\bibinfo
  {journal} {J. Fluid Mech.}\ }\textbf {\bibinfo {volume} {820}},\ \bibinfo
  {pages} {667} (\bibinfo {year} {2017})}\BibitemShut {NoStop}%
\bibitem [{\citenamefont {Gungor}\ \emph {et~al.}(2016)\citenamefont {Gungor},
  \citenamefont {Maciel}, \citenamefont {Simens},\ and\ \citenamefont
  {Soria}}]{gungor2016scaling}%
  \BibitemOpen
  \bibfield  {author} {\bibinfo {author} {\bibfnamefont {A.~G.}\ \bibnamefont
  {Gungor}}, \bibinfo {author} {\bibfnamefont {Y.}~\bibnamefont {Maciel}},
  \bibinfo {author} {\bibfnamefont {M.~P.}\ \bibnamefont {Simens}},\ and\
  \bibinfo {author} {\bibfnamefont {J.}~\bibnamefont {Soria}},\ }\bibfield
  {title} {\bibinfo {title} {Scaling and statistics of large-defect adverse
  pressure gradient turbulent boundary layers},\ }\href
  {https://doi.org/10.1016/j.ijheatfluidflow.2016.03.004} {\bibfield  {journal}
  {\bibinfo  {journal} {Int. J. Heat Fluid Flow}\ }\textbf {\bibinfo {volume}
  {59}},\ \bibinfo {pages} {109} (\bibinfo {year} {2016})}\BibitemShut
  {NoStop}%
\bibitem [{\citenamefont {Gungor}\ \emph {et~al.}(2022)\citenamefont {Gungor},
  \citenamefont {Maciel},\ and\ \citenamefont {Gungor}}]{gungor2022energy}%
  \BibitemOpen
  \bibfield  {author} {\bibinfo {author} {\bibfnamefont {T.~R.}\ \bibnamefont
  {Gungor}}, \bibinfo {author} {\bibfnamefont {Y.}~\bibnamefont {Maciel}},\
  and\ \bibinfo {author} {\bibfnamefont {A.~G.}\ \bibnamefont {Gungor}},\
  }\bibfield  {title} {\bibinfo {title} {Energy transfer mechanisms in adverse
  pressure gradient turbulent boundary layers: production and inter-component
  redistribution},\ }\href {https://doi.org/10.1017/jfm.2022.679} {\bibfield
  {journal} {\bibinfo  {journal} {J. Fluid Mech.}\ }\textbf {\bibinfo {volume}
  {948}},\ \bibinfo {pages} {A5} (\bibinfo {year} {2022})}\BibitemShut
  {NoStop}%
\bibitem [{\citenamefont {Gungor}\ \emph {et~al.}(2024)\citenamefont {Gungor},
  \citenamefont {Gungor},\ and\ \citenamefont {Maciel}}]{gungor2024response}%
  \BibitemOpen
  \bibfield  {author} {\bibinfo {author} {\bibfnamefont {T.~R.}\ \bibnamefont
  {Gungor}}, \bibinfo {author} {\bibfnamefont {A.~G.}\ \bibnamefont {Gungor}},\
  and\ \bibinfo {author} {\bibfnamefont {Y.}~\bibnamefont {Maciel}},\
  }\bibfield  {title} {\bibinfo {title} {Turbulent boundary layer response to
  uniform changes of the pressure force contribution},\ }\href
  {https://doi.org/10.1017/jfm.2024.579} {\bibfield  {journal} {\bibinfo
  {journal} {J. Fluid Mech.}\ }\textbf {\bibinfo {volume} {997}},\ \bibinfo
  {pages} {A75} (\bibinfo {year} {2024})}\BibitemShut {NoStop}%
\bibitem [{\citenamefont {Coleman}(2021)}]{coleman2021numerical}%
  \BibitemOpen
  \bibfield  {author} {\bibinfo {author} {\bibfnamefont {G.~N.}\ \bibnamefont
  {Coleman}},\ }\href@noop {} {\emph {\bibinfo {title} {Numerical Simulation of
  Pressure-Induced Separation of Turbulent Flat-Plate Boundary Layers:
  Definition and Overview of New Cases with Suction-Only Transpiration and a
  Step in {Reynolds} Number}}},\ \bibinfo {type} {NASA Technical Memorandum}\
  \bibinfo {number} {NASA/TM-20210020762}\ (\bibinfo  {institution} {NASA
  Langley Research Center},\ \bibinfo {address} {Hampton, VA},\ \bibinfo {year}
  {2021})\ \bibinfo {note} {errata issued January 2022}\BibitemShut {NoStop}%
\bibitem [{\citenamefont {Wu}\ \emph {et~al.}(2020)\citenamefont {Wu},
  \citenamefont {Meneveau},\ and\ \citenamefont
  {Mittal}}]{wu2019spatiotemporal}%
  \BibitemOpen
  \bibfield  {author} {\bibinfo {author} {\bibfnamefont {W.}~\bibnamefont
  {Wu}}, \bibinfo {author} {\bibfnamefont {C.}~\bibnamefont {Meneveau}},\ and\
  \bibinfo {author} {\bibfnamefont {R.}~\bibnamefont {Mittal}},\ }\bibfield
  {title} {\bibinfo {title} {Spatio-temporal dynamics of turbulent separation
  bubbles},\ }\href {https://doi.org/10.1017/jfm.2019.911} {\bibfield
  {journal} {\bibinfo  {journal} {J. Fluid Mech.}\ }\textbf {\bibinfo {volume}
  {883}},\ \bibinfo {pages} {A45} (\bibinfo {year} {2020})}\BibitemShut
  {NoStop}%
\bibitem [{\citenamefont {Maciel}\ \emph {et~al.}(2017)\citenamefont {Maciel},
  \citenamefont {Simens},\ and\ \citenamefont {Gungor}}]{maciel2016coherent}%
  \BibitemOpen
  \bibfield  {author} {\bibinfo {author} {\bibfnamefont {Y.}~\bibnamefont
  {Maciel}}, \bibinfo {author} {\bibfnamefont {M.~P.}\ \bibnamefont {Simens}},\
  and\ \bibinfo {author} {\bibfnamefont {A.~G.}\ \bibnamefont {Gungor}},\
  }\bibfield  {title} {\bibinfo {title} {Coherent structures in a
  non-equilibrium large-velocity-defect turbulent boundary layer},\ }\href
  {https://doi.org/10.1007/s10494-016-9737-2} {\bibfield  {journal} {\bibinfo
  {journal} {Flow Turbul. Combust.}\ }\textbf {\bibinfo {volume} {98}},\
  \bibinfo {pages} {1} (\bibinfo {year} {2017})}\BibitemShut {NoStop}%
\bibitem [{\citenamefont {Kitsios}\ \emph {et~al.}(2017)\citenamefont
  {Kitsios}, \citenamefont {Sekimoto}, \citenamefont {Atkinson}, \citenamefont
  {Sillero}, \citenamefont {Borrell}, \citenamefont {Gungor}, \citenamefont
  {Jim{\'e}nez},\ and\ \citenamefont {Soria}}]{kitsios2017direct}%
  \BibitemOpen
  \bibfield  {author} {\bibinfo {author} {\bibfnamefont {V.}~\bibnamefont
  {Kitsios}}, \bibinfo {author} {\bibfnamefont {A.}~\bibnamefont {Sekimoto}},
  \bibinfo {author} {\bibfnamefont {C.}~\bibnamefont {Atkinson}}, \bibinfo
  {author} {\bibfnamefont {J.~A.}\ \bibnamefont {Sillero}}, \bibinfo {author}
  {\bibfnamefont {G.}~\bibnamefont {Borrell}}, \bibinfo {author} {\bibfnamefont
  {A.~G.}\ \bibnamefont {Gungor}}, \bibinfo {author} {\bibfnamefont
  {J.}~\bibnamefont {Jim{\'e}nez}},\ and\ \bibinfo {author} {\bibfnamefont
  {J.}~\bibnamefont {Soria}},\ }\bibfield  {title} {\bibinfo {title} {Direct
  numerical simulation of a self-similar adverse pressure gradient turbulent
  boundary layer at the verge of separation},\ }\href
  {https://doi.org/10.1017/jfm.2017.549} {\bibfield  {journal} {\bibinfo
  {journal} {J. Fluid Mech.}\ }\textbf {\bibinfo {volume} {829}},\ \bibinfo
  {pages} {392} (\bibinfo {year} {2017})}\BibitemShut {NoStop}%
\bibitem [{\citenamefont {Buckingham}(1914)}]{Buckingham1914}%
  \BibitemOpen
  \bibfield  {author} {\bibinfo {author} {\bibfnamefont {E.}~\bibnamefont
  {Buckingham}},\ }\bibfield  {title} {\bibinfo {title} {On physically similar
  systems; illustrations of the use of dimensional equations},\ }\href
  {https://doi.org/10.1103/PhysRev.4.345} {\bibfield  {journal} {\bibinfo
  {journal} {Phys. Rev.}\ }\textbf {\bibinfo {volume} {4}},\ \bibinfo {pages}
  {345} (\bibinfo {year} {1914})}\BibitemShut {NoStop}%
\bibitem [{\citenamefont {Yuan}\ and\ \citenamefont
  {Lozano-Dur{\'a}n}(2025)}]{yuan2025Dimensionless}%
  \BibitemOpen
  \bibfield  {author} {\bibinfo {author} {\bibfnamefont {Y.}~\bibnamefont
  {Yuan}}\ and\ \bibinfo {author} {\bibfnamefont {A.}~\bibnamefont
  {Lozano-Dur{\'a}n}},\ }\bibfield  {title} {\bibinfo {title} {Dimensionless
  learning based on information},\ }\href
  {https://doi.org/10.1038/s41467-025-64425-8} {\bibfield  {journal} {\bibinfo
  {journal} {Nat. Commun.}\ }\textbf {\bibinfo {volume} {16}},\ \bibinfo
  {pages} {9171} (\bibinfo {year} {2025})}\BibitemShut {NoStop}%
\bibitem [{\citenamefont {Cover}\ and\ \citenamefont
  {Thomas}(2006)}]{cover2006}%
  \BibitemOpen
  \bibfield  {author} {\bibinfo {author} {\bibfnamefont {T.~M.}\ \bibnamefont
  {Cover}}\ and\ \bibinfo {author} {\bibfnamefont {J.~A.}\ \bibnamefont
  {Thomas}},\ }\href {https://doi.org/10.1002/047174882X} {\emph {\bibinfo
  {title} {Elements of Information Theory}}},\ \bibinfo {edition} {2nd}\ ed.\
  (\bibinfo  {publisher} {John Wiley \& Sons},\ \bibinfo {address} {Hoboken,
  NJ},\ \bibinfo {year} {2006})\BibitemShut {NoStop}%
\bibitem [{\citenamefont {Wei}\ \emph {et~al.}(2024)\citenamefont {Wei},
  \citenamefont {Li},\ and\ \citenamefont {Wang}}]{wei2024new}%
  \BibitemOpen
  \bibfield  {author} {\bibinfo {author} {\bibfnamefont {T.}~\bibnamefont
  {Wei}}, \bibinfo {author} {\bibfnamefont {Z.}~\bibnamefont {Li}},\ and\
  \bibinfo {author} {\bibfnamefont {Y.}~\bibnamefont {Wang}},\ }\bibfield
  {title} {\bibinfo {title} {New momentum integral equation applicable to
  boundary layer flows under arbitrary pressure gradients},\ }\href
  {https://doi.org/10.1017/jfm.2024.207} {\bibfield  {journal} {\bibinfo
  {journal} {J. Fluid Mech.}\ }\textbf {\bibinfo {volume} {984}},\ \bibinfo
  {pages} {A64} (\bibinfo {year} {2024})}\BibitemShut {NoStop}%
\bibitem [{\citenamefont {Han}\ \emph {et~al.}(2025)\citenamefont {Han},
  \citenamefont {Ma},\ and\ \citenamefont {Yan}}]{han2025integral}%
  \BibitemOpen
  \bibfield  {author} {\bibinfo {author} {\bibfnamefont {M.}~\bibnamefont
  {Han}}, \bibinfo {author} {\bibfnamefont {M.}~\bibnamefont {Ma}},\ and\
  \bibinfo {author} {\bibfnamefont {C.}~\bibnamefont {Yan}},\ }\bibfield
  {title} {\bibinfo {title} {Integral analysis of adverse pressure gradient
  turbulent boundary layers},\ }\href {https://doi.org/10.1017/jfm.2025.10395}
  {\bibfield  {journal} {\bibinfo  {journal} {J. Fluid Mech.}\ }\textbf
  {\bibinfo {volume} {1019}},\ \bibinfo {pages} {A46} (\bibinfo {year}
  {2025})}\BibitemShut {NoStop}%
\bibitem [{\citenamefont {Sobol'}(2001)}]{sobol2001}%
  \BibitemOpen
  \bibfield  {author} {\bibinfo {author} {\bibfnamefont {I.~M.}\ \bibnamefont
  {Sobol'}},\ }\bibfield  {title} {\bibinfo {title} {Global sensitivity indices
  for nonlinear mathematical models and their {Monte Carlo} estimates},\ }\href
  {https://doi.org/10.1016/S0378-4754(00)00270-6} {\bibfield  {journal}
  {\bibinfo  {journal} {Math. Comput. Simul.}\ }\textbf {\bibinfo {volume}
  {55}},\ \bibinfo {pages} {271} (\bibinfo {year} {2001})}\BibitemShut
  {NoStop}%
\bibitem [{\citenamefont {Saltelli}(2002)}]{saltelli2002}%
  \BibitemOpen
  \bibfield  {author} {\bibinfo {author} {\bibfnamefont {A.}~\bibnamefont
  {Saltelli}},\ }\bibfield  {title} {\bibinfo {title} {Sensitivity analysis for
  importance assessment},\ }\href {https://doi.org/10.1111/0272-4332.00040}
  {\bibfield  {journal} {\bibinfo  {journal} {Risk Anal.}\ }\textbf {\bibinfo
  {volume} {22}},\ \bibinfo {pages} {579} (\bibinfo {year} {2002})}\BibitemShut
  {NoStop}%
\bibitem [{\citenamefont {Dixit}\ \emph {et~al.}(2024)\citenamefont {Dixit},
  \citenamefont {Gupta}, \citenamefont {Choudhary},\ and\ \citenamefont
  {Prabhakaran}}]{dixit2024generalized}%
  \BibitemOpen
  \bibfield  {author} {\bibinfo {author} {\bibfnamefont {S.~A.}\ \bibnamefont
  {Dixit}}, \bibinfo {author} {\bibfnamefont {A.}~\bibnamefont {Gupta}},
  \bibinfo {author} {\bibfnamefont {H.}~\bibnamefont {Choudhary}},\ and\
  \bibinfo {author} {\bibfnamefont {T.}~\bibnamefont {Prabhakaran}},\
  }\bibfield  {title} {\bibinfo {title} {Generalized scaling and model for
  friction in wall turbulence},\ }\href
  {https://doi.org/10.1103/PhysRevLett.132.014001} {\bibfield  {journal}
  {\bibinfo  {journal} {Phys. Rev. Lett.}\ }\textbf {\bibinfo {volume} {132}},\
  \bibinfo {pages} {014001} (\bibinfo {year} {2024})}\BibitemShut {NoStop}%
\bibitem [{\citenamefont {Vinuesa}\ \emph {et~al.}(2016)\citenamefont
  {Vinuesa}, \citenamefont {Bobke}, \citenamefont {{\"O}rl{\"u}},\ and\
  \citenamefont {Schlatter}}]{vinuesa2016determining}%
  \BibitemOpen
  \bibfield  {author} {\bibinfo {author} {\bibfnamefont {R.}~\bibnamefont
  {Vinuesa}}, \bibinfo {author} {\bibfnamefont {A.}~\bibnamefont {Bobke}},
  \bibinfo {author} {\bibfnamefont {R.}~\bibnamefont {{\"O}rl{\"u}}},\ and\
  \bibinfo {author} {\bibfnamefont {P.}~\bibnamefont {Schlatter}},\ }\bibfield
  {title} {\bibinfo {title} {On determining characteristic length scales in
  pressure-gradient turbulent boundary layers},\ }\href
  {https://doi.org/10.1063/1.4947532} {\bibfield  {journal} {\bibinfo
  {journal} {Phys. Fluids}\ }\textbf {\bibinfo {volume} {28}},\ \bibinfo
  {pages} {055101} (\bibinfo {year} {2016})}\BibitemShut {NoStop}%
\bibitem [{\citenamefont {Wei}\ and\ \citenamefont
  {Knopp}(2023)}]{wei2023outer}%
  \BibitemOpen
  \bibfield  {author} {\bibinfo {author} {\bibfnamefont {T.}~\bibnamefont
  {Wei}}\ and\ \bibinfo {author} {\bibfnamefont {T.}~\bibnamefont {Knopp}},\
  }\bibfield  {title} {\bibinfo {title} {Outer scaling of the mean momentum
  equation for turbulent boundary layers under adverse pressure gradient},\
  }\href {https://doi.org/10.1017/jfm.2023.72} {\bibfield  {journal} {\bibinfo
  {journal} {J. Fluid Mech.}\ }\textbf {\bibinfo {volume} {958}},\ \bibinfo
  {pages} {A9} (\bibinfo {year} {2023})}\BibitemShut {NoStop}%
\bibitem [{\citenamefont {Lozier}\ \emph {et~al.}(2025)\citenamefont {Lozier},
  \citenamefont {Deshpande}, \citenamefont {Zarei}, \citenamefont {Lindi{\'c}},
  \citenamefont {Abu~Rowin},\ and\ \citenamefont
  {Marusic}}]{lozier2025Defining}%
  \BibitemOpen
  \bibfield  {author} {\bibinfo {author} {\bibfnamefont {M.}~\bibnamefont
  {Lozier}}, \bibinfo {author} {\bibfnamefont {R.}~\bibnamefont {Deshpande}},
  \bibinfo {author} {\bibfnamefont {A.}~\bibnamefont {Zarei}}, \bibinfo
  {author} {\bibfnamefont {L.}~\bibnamefont {Lindi{\'c}}}, \bibinfo {author}
  {\bibfnamefont {W.}~\bibnamefont {Abu~Rowin}},\ and\ \bibinfo {author}
  {\bibfnamefont {I.}~\bibnamefont {Marusic}},\ }\bibfield  {title} {\bibinfo
  {title} {Defining the mean turbulent boundary layer thickness based on
  streamwise velocity skewness},\ }\href
  {https://doi.org/10.1017/jfm.2025.10711} {\bibfield  {journal} {\bibinfo
  {journal} {J. Fluid Mech.}\ }\textbf {\bibinfo {volume} {1021}},\ \bibinfo
  {pages} {A19} (\bibinfo {year} {2025})}\BibitemShut {NoStop}%
\bibitem [{\citenamefont {Mart{\'i}nez-S{\'a}nchez}\ \emph
  {et~al.}(2024)\citenamefont {Mart{\'i}nez-S{\'a}nchez}, \citenamefont
  {Arranz},\ and\ \citenamefont {Lozano-Dur{\'a}n}}]{martinez2024surd}%
  \BibitemOpen
  \bibfield  {author} {\bibinfo {author} {\bibfnamefont {A.}~\bibnamefont
  {Mart{\'i}nez-S{\'a}nchez}}, \bibinfo {author} {\bibfnamefont
  {G.}~\bibnamefont {Arranz}},\ and\ \bibinfo {author} {\bibfnamefont
  {A.}~\bibnamefont {Lozano-Dur{\'a}n}},\ }\bibfield  {title} {\bibinfo {title}
  {Decomposing causality into its synergistic, unique, and redundant
  components},\ }\href {https://doi.org/10.1038/s41467-024-53373-4} {\bibfield
  {journal} {\bibinfo  {journal} {Nat. Commun.}\ }\textbf {\bibinfo {volume}
  {15}},\ \bibinfo {pages} {9296} (\bibinfo {year} {2024})}\BibitemShut
  {NoStop}%
\bibitem [{\citenamefont {Shannon}(1948)}]{shannon1948}%
  \BibitemOpen
  \bibfield  {author} {\bibinfo {author} {\bibfnamefont {C.~E.}\ \bibnamefont
  {Shannon}},\ }\bibfield  {title} {\bibinfo {title} {A mathematical theory of
  communication},\ }\href {https://doi.org/10.1002/j.1538-7305.1948.tb01338.x}
  {\bibfield  {journal} {\bibinfo  {journal} {Bell Syst. Tech. J.}\ }\textbf
  {\bibinfo {volume} {27}},\ \bibinfo {pages} {379} (\bibinfo {year} {1948})},\
  \bibinfo {note} {published in two parts; the second part has DOI
  10.1002/j.1538-7305.1948.tb00917.x}\BibitemShut {NoStop}%
\bibitem [{\citenamefont {Mart{\'i}nez-S{\'a}nchez}\ and\ \citenamefont
  {Lozano-Dur{\'a}n}(2026)}]{martinez2026}%
  \BibitemOpen
  \bibfield  {author} {\bibinfo {author} {\bibfnamefont {{\'A}.}~\bibnamefont
  {Mart{\'i}nez-S{\'a}nchez}}\ and\ \bibinfo {author} {\bibfnamefont
  {A.}~\bibnamefont {Lozano-Dur{\'a}n}},\ }\bibfield  {title} {\bibinfo {title}
  {Cause-and-effect approach to turbulence forecasting},\ }\href
  {https://doi.org/10.1108/HFF-09-2025-0715} {\bibfield  {journal} {\bibinfo
  {journal} {Int. J. Numer. Methods Heat Fluid Flow}\ }\textbf {\bibinfo
  {volume} {36}},\ \bibinfo {pages} {2539} (\bibinfo {year}
  {2026})}\BibitemShut {NoStop}%
\bibitem [{\citenamefont {Bradshaw}(1987)}]{bradshaw1987}%
  \BibitemOpen
  \bibfield  {author} {\bibinfo {author} {\bibfnamefont {P.}~\bibnamefont
  {Bradshaw}},\ }\bibfield  {title} {\bibinfo {title} {Turbulent secondary
  flows},\ }\href {https://doi.org/10.1146/annurev.fl.19.010187.000413}
  {\bibfield  {journal} {\bibinfo  {journal} {Annu. Rev. Fluid Mech.}\ }\textbf
  {\bibinfo {volume} {19}},\ \bibinfo {pages} {53} (\bibinfo {year}
  {1987})}\BibitemShut {NoStop}%
\bibitem [{\citenamefont {Lozano-Dur{\'a}n}\ \emph {et~al.}(2020)\citenamefont
  {Lozano-Dur{\'a}n}, \citenamefont {Giometto}, \citenamefont {Park},\ and\
  \citenamefont {Moin}}]{lozano-duran2020}%
  \BibitemOpen
  \bibfield  {author} {\bibinfo {author} {\bibfnamefont {A.}~\bibnamefont
  {Lozano-Dur{\'a}n}}, \bibinfo {author} {\bibfnamefont {M.~G.}\ \bibnamefont
  {Giometto}}, \bibinfo {author} {\bibfnamefont {G.~I.}\ \bibnamefont {Park}},\
  and\ \bibinfo {author} {\bibfnamefont {P.}~\bibnamefont {Moin}},\ }\bibfield
  {title} {\bibinfo {title} {Non-equilibrium three-dimensional boundary layers
  at moderate {Reynolds} numbers},\ }\href
  {https://doi.org/10.1017/jfm.2019.869} {\bibfield  {journal} {\bibinfo
  {journal} {J. Fluid Mech.}\ }\textbf {\bibinfo {volume} {883}},\ \bibinfo
  {pages} {A20} (\bibinfo {year} {2020})}\BibitemShut {NoStop}%
\bibitem [{\citenamefont {Driver}\ and\ \citenamefont
  {Hebbar}(1987)}]{driver1987}%
  \BibitemOpen
  \bibfield  {author} {\bibinfo {author} {\bibfnamefont {D.~M.}\ \bibnamefont
  {Driver}}\ and\ \bibinfo {author} {\bibfnamefont {S.~K.}\ \bibnamefont
  {Hebbar}},\ }\bibfield  {title} {\bibinfo {title} {Experimental study of a
  three-dimensional, shear-driven, turbulent boundary layer},\ }\href
  {https://doi.org/10.2514/3.9575} {\bibfield  {journal} {\bibinfo  {journal}
  {AIAA J.}\ }\textbf {\bibinfo {volume} {25}},\ \bibinfo {pages} {35}
  (\bibinfo {year} {1987})}\BibitemShut {NoStop}%
\bibitem [{\citenamefont {Eaton}(1995)}]{eaton1995}%
  \BibitemOpen
  \bibfield  {author} {\bibinfo {author} {\bibfnamefont {J.~K.}\ \bibnamefont
  {Eaton}},\ }\bibfield  {title} {\bibinfo {title} {Effects of mean flow three
  dimensionality on turbulent boundary-layer structure},\ }\href
  {https://doi.org/10.2514/3.12942} {\bibfield  {journal} {\bibinfo  {journal}
  {AIAA J.}\ }\textbf {\bibinfo {volume} {33}},\ \bibinfo {pages} {2020}
  (\bibinfo {year} {1995})}\BibitemShut {NoStop}%
\bibitem [{\citenamefont {Johnston}(1960)}]{johnston1960}%
  \BibitemOpen
  \bibfield  {author} {\bibinfo {author} {\bibfnamefont {J.~P.}\ \bibnamefont
  {Johnston}},\ }\bibfield  {title} {\bibinfo {title} {On the three-dimensional
  turbulent boundary layer generated by secondary flow},\ }\href
  {https://doi.org/10.1115/1.3662539} {\bibfield  {journal} {\bibinfo
  {journal} {J. Basic Eng.}\ }\textbf {\bibinfo {volume} {82}},\ \bibinfo
  {pages} {233} (\bibinfo {year} {1960})}\BibitemShut {NoStop}%
\bibitem [{\citenamefont {Abe}(2020)}]{abe2020}%
  \BibitemOpen
  \bibfield  {author} {\bibinfo {author} {\bibfnamefont {H.}~\bibnamefont
  {Abe}},\ }\bibfield  {title} {\bibinfo {title} {Direct numerical simulation
  of a non-equilibrium three-dimensional turbulent boundary layer over a flat
  plate},\ }\href {https://doi.org/10.1017/jfm.2020.488} {\bibfield  {journal}
  {\bibinfo  {journal} {J. Fluid Mech.}\ }\textbf {\bibinfo {volume} {902}},\
  \bibinfo {pages} {A20} (\bibinfo {year} {2020})}\BibitemShut {NoStop}%
\bibitem [{\citenamefont {van~den Berg}\ \emph {et~al.}(1975)\citenamefont
  {van~den Berg}, \citenamefont {Elsenaar}, \citenamefont {Lindhout},\ and\
  \citenamefont {Wesseling}}]{vandenberg1975}%
  \BibitemOpen
  \bibfield  {author} {\bibinfo {author} {\bibfnamefont {B.}~\bibnamefont
  {van~den Berg}}, \bibinfo {author} {\bibfnamefont {A.}~\bibnamefont
  {Elsenaar}}, \bibinfo {author} {\bibfnamefont {J.~P.~F.}\ \bibnamefont
  {Lindhout}},\ and\ \bibinfo {author} {\bibfnamefont {P.}~\bibnamefont
  {Wesseling}},\ }\bibfield  {title} {\bibinfo {title} {Measurements in an
  incompressible three-dimensional turbulent boundary layer, under infinite
  swept-wing conditions, and comparison with theory},\ }\href
  {https://doi.org/10.1017/S0022112075001930} {\bibfield  {journal} {\bibinfo
  {journal} {J. Fluid Mech.}\ }\textbf {\bibinfo {volume} {70}},\ \bibinfo
  {pages} {127} (\bibinfo {year} {1975})}\BibitemShut {NoStop}%
\bibitem [{\citenamefont {Coleman}\ \emph {et~al.}(2019)\citenamefont
  {Coleman}, \citenamefont {Rumsey},\ and\ \citenamefont
  {Spalart}}]{coleman2019}%
  \BibitemOpen
  \bibfield  {author} {\bibinfo {author} {\bibfnamefont {G.~N.}\ \bibnamefont
  {Coleman}}, \bibinfo {author} {\bibfnamefont {C.~L.}\ \bibnamefont
  {Rumsey}},\ and\ \bibinfo {author} {\bibfnamefont {P.~R.}\ \bibnamefont
  {Spalart}},\ }\bibfield  {title} {\bibinfo {title} {Numerical study of a
  turbulent separation bubble with sweep},\ }\href
  {https://doi.org/10.1017/jfm.2019.736} {\bibfield  {journal} {\bibinfo
  {journal} {J. Fluid Mech.}\ }\textbf {\bibinfo {volume} {880}},\ \bibinfo
  {pages} {684} (\bibinfo {year} {2019})}\BibitemShut {NoStop}%
\bibitem [{\citenamefont {Jeans}\ and\ \citenamefont
  {Johnston}(1982)}]{jeans1982}%
  \BibitemOpen
  \bibfield  {author} {\bibinfo {author} {\bibfnamefont {A.~H.}\ \bibnamefont
  {Jeans}}\ and\ \bibinfo {author} {\bibfnamefont {J.~P.}\ \bibnamefont
  {Johnston}},\ }\href@noop {} {\emph {\bibinfo {title} {The Effects of
  Streamwise Concave Curvature on Turbulent Boundary Layer Structure}}},\
  \bibinfo {type} {Tech. Rep.}\ \bibinfo {number} {MD-40}\ (\bibinfo
  {institution} {Thermosciences Division, Department of Mechanical Engineering,
  Stanford University},\ \bibinfo {year} {1982})\BibitemShut {NoStop}%
\bibitem [{\citenamefont {Bradshaw}(1973)}]{bradshaw1973}%
  \BibitemOpen
  \bibfield  {author} {\bibinfo {author} {\bibfnamefont {P.}~\bibnamefont
  {Bradshaw}},\ }\href@noop {} {\emph {\bibinfo {title} {Effects of Streamline
  Curvature on Turbulent Flow}}},\ \bibinfo {type} {AGARDograph}\ \bibinfo
  {number} {169}\ (\bibinfo  {institution} {Advisory Group for Aerospace
  Research and Development},\ \bibinfo {year} {1973})\BibitemShut {NoStop}%
\bibitem [{\citenamefont {Ellis}\ and\ \citenamefont
  {Joubert}(1974)}]{ellis1974}%
  \BibitemOpen
  \bibfield  {author} {\bibinfo {author} {\bibfnamefont {L.~B.}\ \bibnamefont
  {Ellis}}\ and\ \bibinfo {author} {\bibfnamefont {P.~N.}\ \bibnamefont
  {Joubert}},\ }\bibfield  {title} {\bibinfo {title} {Turbulent shear flow in a
  curved duct},\ }\href {https://doi.org/10.1017/S0022112074000589} {\bibfield
  {journal} {\bibinfo  {journal} {J. Fluid Mech.}\ }\textbf {\bibinfo {volume}
  {62}},\ \bibinfo {pages} {65} (\bibinfo {year} {1974})}\BibitemShut {NoStop}%
\bibitem [{\citenamefont {Smits}\ \emph {et~al.}(1979)\citenamefont {Smits},
  \citenamefont {Young},\ and\ \citenamefont {Bradshaw}}]{smits1979}%
  \BibitemOpen
  \bibfield  {author} {\bibinfo {author} {\bibfnamefont {A.~J.}\ \bibnamefont
  {Smits}}, \bibinfo {author} {\bibfnamefont {S.~T.~B.}\ \bibnamefont
  {Young}},\ and\ \bibinfo {author} {\bibfnamefont {P.}~\bibnamefont
  {Bradshaw}},\ }\bibfield  {title} {\bibinfo {title} {The effect of short
  regions of high surface curvature on turbulent boundary layers},\ }\href
  {https://doi.org/10.1017/S0022112079001002} {\bibfield  {journal} {\bibinfo
  {journal} {J. Fluid Mech.}\ }\textbf {\bibinfo {volume} {94}},\ \bibinfo
  {pages} {209} (\bibinfo {year} {1979})}\BibitemShut {NoStop}%
\bibitem [{\citenamefont {Barlow}\ and\ \citenamefont
  {Johnston}(1988)}]{barlow1988}%
  \BibitemOpen
  \bibfield  {author} {\bibinfo {author} {\bibfnamefont {R.~S.}\ \bibnamefont
  {Barlow}}\ and\ \bibinfo {author} {\bibfnamefont {J.~P.}\ \bibnamefont
  {Johnston}},\ }\bibfield  {title} {\bibinfo {title} {Structure of a turbulent
  boundary layer on a concave surface},\ }\href
  {https://doi.org/10.1017/S0022112088001545} {\bibfield  {journal} {\bibinfo
  {journal} {J. Fluid Mech.}\ }\textbf {\bibinfo {volume} {191}},\ \bibinfo
  {pages} {137} (\bibinfo {year} {1988})}\BibitemShut {NoStop}%
\bibitem [{\citenamefont {Appelbaum}\ \emph {et~al.}(2025)\citenamefont
  {Appelbaum}, \citenamefont {Kloker},\ and\ \citenamefont
  {Wenzel}}]{appelbaum2025}%
  \BibitemOpen
  \bibfield  {author} {\bibinfo {author} {\bibfnamefont {J.}~\bibnamefont
  {Appelbaum}}, \bibinfo {author} {\bibfnamefont {M.}~\bibnamefont {Kloker}},\
  and\ \bibinfo {author} {\bibfnamefont {C.}~\bibnamefont {Wenzel}},\
  }\bibfield  {title} {\bibinfo {title} {A systematic {DNS} approach to isolate
  wall-curvature effects in spatially developing boundary layers},\ }\href
  {https://doi.org/10.1007/s00162-024-00729-7} {\bibfield  {journal} {\bibinfo
  {journal} {Theor. Comput. Fluid Dyn.}\ }\textbf {\bibinfo {volume} {39}},\
  \bibinfo {pages} {10} (\bibinfo {year} {2025})}\BibitemShut {NoStop}%
\bibitem [{\citenamefont {Pargal}\ \emph {et~al.}(2022)\citenamefont {Pargal},
  \citenamefont {Wu}, \citenamefont {Yuan},\ and\ \citenamefont
  {Moreau}}]{pargal2022}%
  \BibitemOpen
  \bibfield  {author} {\bibinfo {author} {\bibfnamefont {S.}~\bibnamefont
  {Pargal}}, \bibinfo {author} {\bibfnamefont {H.}~\bibnamefont {Wu}}, \bibinfo
  {author} {\bibfnamefont {J.}~\bibnamefont {Yuan}},\ and\ \bibinfo {author}
  {\bibfnamefont {S.}~\bibnamefont {Moreau}},\ }\bibfield  {title} {\bibinfo
  {title} {Adverse-pressure-gradient turbulent boundary layer on convex wall},\
  }\href {https://doi.org/10.1063/5.0083919} {\bibfield  {journal} {\bibinfo
  {journal} {Phys. Fluids}\ }\textbf {\bibinfo {volume} {34}},\ \bibinfo
  {pages} {035107} (\bibinfo {year} {2022})}\BibitemShut {NoStop}%
\bibitem [{\citenamefont {Lagha}\ \emph {et~al.}(2011)\citenamefont {Lagha},
  \citenamefont {Kim}, \citenamefont {Eldredge},\ and\ \citenamefont
  {Zhong}}]{lagha2011}%
  \BibitemOpen
  \bibfield  {author} {\bibinfo {author} {\bibfnamefont {M.}~\bibnamefont
  {Lagha}}, \bibinfo {author} {\bibfnamefont {J.}~\bibnamefont {Kim}}, \bibinfo
  {author} {\bibfnamefont {J.~D.}\ \bibnamefont {Eldredge}},\ and\ \bibinfo
  {author} {\bibfnamefont {X.}~\bibnamefont {Zhong}},\ }\bibfield  {title}
  {\bibinfo {title} {A numerical study of compressible turbulent boundary
  layers},\ }\href {https://doi.org/10.1063/1.3541841} {\bibfield  {journal}
  {\bibinfo  {journal} {Phys. Fluids}\ }\textbf {\bibinfo {volume} {23}},\
  \bibinfo {pages} {015106} (\bibinfo {year} {2011})}\BibitemShut {NoStop}%
\bibitem [{\citenamefont {Spina}\ \emph {et~al.}(1994)\citenamefont {Spina},
  \citenamefont {Smits},\ and\ \citenamefont {Robinson}}]{spina1994}%
  \BibitemOpen
  \bibfield  {author} {\bibinfo {author} {\bibfnamefont {E.~F.}\ \bibnamefont
  {Spina}}, \bibinfo {author} {\bibfnamefont {A.~J.}\ \bibnamefont {Smits}},\
  and\ \bibinfo {author} {\bibfnamefont {S.~K.}\ \bibnamefont {Robinson}},\
  }\bibfield  {title} {\bibinfo {title} {The physics of supersonic turbulent
  boundary layers},\ }\href
  {https://doi.org/10.1146/annurev.fl.26.010194.001443} {\bibfield  {journal}
  {\bibinfo  {journal} {Annu. Rev. Fluid Mech.}\ }\textbf {\bibinfo {volume}
  {26}},\ \bibinfo {pages} {287} (\bibinfo {year} {1994})}\BibitemShut
  {NoStop}%
\bibitem [{\citenamefont {Smits}\ and\ \citenamefont
  {Dussauge}(2006)}]{smits2006}%
  \BibitemOpen
  \bibfield  {author} {\bibinfo {author} {\bibfnamefont {A.~J.}\ \bibnamefont
  {Smits}}\ and\ \bibinfo {author} {\bibfnamefont {J.-P.}\ \bibnamefont
  {Dussauge}},\ }\href {https://doi.org/10.1007/b137383} {\emph {\bibinfo
  {title} {Turbulent Shear Layers in Supersonic Flow}}},\ \bibinfo {edition}
  {2nd}\ ed.\ (\bibinfo  {publisher} {Springer},\ \bibinfo {address} {New
  York},\ \bibinfo {year} {2006})\BibitemShut {NoStop}%
\bibitem [{\citenamefont {Morkovin}(1962)}]{morkovin1962}%
  \BibitemOpen
  \bibfield  {author} {\bibinfo {author} {\bibfnamefont {M.~V.}\ \bibnamefont
  {Morkovin}},\ }\bibfield  {title} {\bibinfo {title} {Effects of
  compressibility on turbulent flows},\ }in\ \href@noop {} {\emph {\bibinfo
  {booktitle} {M{\'e}canique de la Turbulence}}},\ \bibinfo {editor} {edited
  by\ \bibinfo {editor} {\bibfnamefont {A.}~\bibnamefont {Favre}}}\ (\bibinfo
  {publisher} {CNRS},\ \bibinfo {address} {Paris},\ \bibinfo {year} {1962})\
  pp.\ \bibinfo {pages} {367--380}\BibitemShut {NoStop}%
\bibitem [{\citenamefont {Van~Driest}(1951)}]{vandriest1951}%
  \BibitemOpen
  \bibfield  {author} {\bibinfo {author} {\bibfnamefont {E.~R.}\ \bibnamefont
  {Van~Driest}},\ }\bibfield  {title} {\bibinfo {title} {Turbulent boundary
  layer in compressible fluids},\ }\href {https://doi.org/10.2514/8.1895}
  {\bibfield  {journal} {\bibinfo  {journal} {J. Aeronaut. Sci.}\ }\textbf
  {\bibinfo {volume} {18}},\ \bibinfo {pages} {145} (\bibinfo {year}
  {1951})}\BibitemShut {NoStop}%
\bibitem [{\citenamefont {Huang}\ \emph {et~al.}(1995)\citenamefont {Huang},
  \citenamefont {Coleman},\ and\ \citenamefont {Bradshaw}}]{huang1995}%
  \BibitemOpen
  \bibfield  {author} {\bibinfo {author} {\bibfnamefont {P.~G.}\ \bibnamefont
  {Huang}}, \bibinfo {author} {\bibfnamefont {G.~N.}\ \bibnamefont {Coleman}},\
  and\ \bibinfo {author} {\bibfnamefont {P.}~\bibnamefont {Bradshaw}},\
  }\bibfield  {title} {\bibinfo {title} {Compressible turbulent channel flows:
  {DNS} results and modelling},\ }\href
  {https://doi.org/10.1017/S0022112095004599} {\bibfield  {journal} {\bibinfo
  {journal} {J. Fluid Mech.}\ }\textbf {\bibinfo {volume} {305}},\ \bibinfo
  {pages} {185} (\bibinfo {year} {1995})}\BibitemShut {NoStop}%
\bibitem [{\citenamefont {Volpiani}\ \emph {et~al.}(2020)\citenamefont
  {Volpiani}, \citenamefont {Iyer}, \citenamefont {Pirozzoli},\ and\
  \citenamefont {Larsson}}]{volpiani2020}%
  \BibitemOpen
  \bibfield  {author} {\bibinfo {author} {\bibfnamefont {P.~S.}\ \bibnamefont
  {Volpiani}}, \bibinfo {author} {\bibfnamefont {P.~S.}\ \bibnamefont {Iyer}},
  \bibinfo {author} {\bibfnamefont {S.}~\bibnamefont {Pirozzoli}},\ and\
  \bibinfo {author} {\bibfnamefont {J.}~\bibnamefont {Larsson}},\ }\bibfield
  {title} {\bibinfo {title} {Data-driven compressibility transformation for
  turbulent wall layers},\ }\href
  {https://doi.org/10.1103/PhysRevFluids.5.052602} {\bibfield  {journal}
  {\bibinfo  {journal} {Phys. Rev. Fluids}\ }\textbf {\bibinfo {volume} {5}},\
  \bibinfo {pages} {052602(R)} (\bibinfo {year} {2020})}\BibitemShut {NoStop}%
\bibitem [{\citenamefont {Griffin}\ \emph
  {et~al.}(2021{\natexlab{b}})\citenamefont {Griffin}, \citenamefont {Fu},\
  and\ \citenamefont {Moin}}]{griffin2021}%
  \BibitemOpen
  \bibfield  {author} {\bibinfo {author} {\bibfnamefont {K.~P.}\ \bibnamefont
  {Griffin}}, \bibinfo {author} {\bibfnamefont {L.}~\bibnamefont {Fu}},\ and\
  \bibinfo {author} {\bibfnamefont {P.}~\bibnamefont {Moin}},\ }\bibfield
  {title} {\bibinfo {title} {Velocity transformation for compressible
  wall-bounded turbulent flows with and without heat transfer},\ }\href
  {https://doi.org/10.1073/pnas.2111144118} {\bibfield  {journal} {\bibinfo
  {journal} {Proc. Natl. Acad. Sci. U.S.A.}\ }\textbf {\bibinfo {volume}
  {118}},\ \bibinfo {pages} {e2111144118} (\bibinfo {year}
  {2021}{\natexlab{b}})}\BibitemShut {NoStop}%
\bibitem [{\citenamefont {Coles}(1964)}]{coles1964}%
  \BibitemOpen
  \bibfield  {author} {\bibinfo {author} {\bibfnamefont {D.~E.}\ \bibnamefont
  {Coles}},\ }\bibfield  {title} {\bibinfo {title} {The turbulent boundary
  layer in a compressible fluid},\ }\href {https://doi.org/10.1063/1.1711395}
  {\bibfield  {journal} {\bibinfo  {journal} {Phys. Fluids}\ }\textbf {\bibinfo
  {volume} {7}},\ \bibinfo {pages} {1403} (\bibinfo {year} {1964})}\BibitemShut
  {NoStop}%
\bibitem [{\citenamefont {Bons}\ \emph {et~al.}(2001)\citenamefont {Bons},
  \citenamefont {Taylor}, \citenamefont {McClain},\ and\ \citenamefont
  {Rivir}}]{bons2001}%
  \BibitemOpen
  \bibfield  {author} {\bibinfo {author} {\bibfnamefont {J.~P.}\ \bibnamefont
  {Bons}}, \bibinfo {author} {\bibfnamefont {R.~P.}\ \bibnamefont {Taylor}},
  \bibinfo {author} {\bibfnamefont {S.~T.}\ \bibnamefont {McClain}},\ and\
  \bibinfo {author} {\bibfnamefont {R.~B.}\ \bibnamefont {Rivir}},\ }\bibfield
  {title} {\bibinfo {title} {The many faces of turbine surface roughness},\
  }\href {https://doi.org/10.1115/1.1400115} {\bibfield  {journal} {\bibinfo
  {journal} {J. Turbomach.}\ }\textbf {\bibinfo {volume} {123}},\ \bibinfo
  {pages} {739} (\bibinfo {year} {2001})}\BibitemShut {NoStop}%
\bibitem [{\citenamefont {Bons}(2010)}]{bons2010}%
  \BibitemOpen
  \bibfield  {author} {\bibinfo {author} {\bibfnamefont {J.~P.}\ \bibnamefont
  {Bons}},\ }\bibfield  {title} {\bibinfo {title} {A review of surface
  roughness effects in gas turbines},\ }\href
  {https://doi.org/10.1115/1.3066315} {\bibfield  {journal} {\bibinfo
  {journal} {J. Turbomach.}\ }\textbf {\bibinfo {volume} {132}},\ \bibinfo
  {pages} {021004} (\bibinfo {year} {2010})}\BibitemShut {NoStop}%
\bibitem [{\citenamefont {Munk}\ \emph {et~al.}(2009)\citenamefont {Munk},
  \citenamefont {Kane},\ and\ \citenamefont {Yebra}}]{munk2009}%
  \BibitemOpen
  \bibfield  {author} {\bibinfo {author} {\bibfnamefont {T.}~\bibnamefont
  {Munk}}, \bibinfo {author} {\bibfnamefont {D.}~\bibnamefont {Kane}},\ and\
  \bibinfo {author} {\bibfnamefont {D.~M.}\ \bibnamefont {Yebra}},\ }\bibfield
  {title} {\bibinfo {title} {The effects of corrosion and fouling on the
  performance of ocean-going vessels: a naval architectural perspective},\ }in\
  \href {https://doi.org/10.1533/9781845696313.1.148} {\emph {\bibinfo
  {booktitle} {Advances in Marine Antifouling Coatings and Technologies}}},\
  \bibinfo {editor} {edited by\ \bibinfo {editor} {\bibfnamefont
  {C.}~\bibnamefont {Hellio}}\ and\ \bibinfo {editor} {\bibfnamefont {D.~M.}\
  \bibnamefont {Yebra}}}\ (\bibinfo  {publisher} {Woodhead Publishing},\
  \bibinfo {address} {Cambridge, UK},\ \bibinfo {year} {2009})\ pp.\ \bibinfo
  {pages} {148--176}\BibitemShut {NoStop}%
\bibitem [{\citenamefont {Kirschner}\ and\ \citenamefont
  {Brennan}(2012)}]{kirschner2012}%
  \BibitemOpen
  \bibfield  {author} {\bibinfo {author} {\bibfnamefont {C.~M.}\ \bibnamefont
  {Kirschner}}\ and\ \bibinfo {author} {\bibfnamefont {A.~B.}\ \bibnamefont
  {Brennan}},\ }\bibfield  {title} {\bibinfo {title} {Bio-inspired antifouling
  strategies},\ }\href {https://doi.org/10.1146/annurev-matsci-070511-155012}
  {\bibfield  {journal} {\bibinfo  {journal} {Annu. Rev. Mater. Res.}\ }\textbf
  {\bibinfo {volume} {42}},\ \bibinfo {pages} {211} (\bibinfo {year}
  {2012})}\BibitemShut {NoStop}%
\bibitem [{\citenamefont {Nikuradse}(1933)}]{nikuradse1933}%
  \BibitemOpen
  \bibfield  {author} {\bibinfo {author} {\bibfnamefont {J.}~\bibnamefont
  {Nikuradse}},\ }\href@noop {} {\emph {\bibinfo {title} {Str{\"o}mungsgesetze
  in rauhen Rohren}}},\ \bibinfo {type} {VDI-Forschungsheft}\ \bibinfo {number}
  {361}\ (\bibinfo  {institution} {Verein Deutscher Ingenieure},\ \bibinfo
  {address} {Berlin},\ \bibinfo {year} {1933})\ \bibinfo {note} {{English}
  translation: {NACA} Technical Memorandum 1292 (1950)}\BibitemShut {NoStop}%
\bibitem [{\citenamefont {Flack}\ and\ \citenamefont
  {Schultz}(2010)}]{flack2010}%
  \BibitemOpen
  \bibfield  {author} {\bibinfo {author} {\bibfnamefont {K.~A.}\ \bibnamefont
  {Flack}}\ and\ \bibinfo {author} {\bibfnamefont {M.~P.}\ \bibnamefont
  {Schultz}},\ }\bibfield  {title} {\bibinfo {title} {Review of hydraulic
  roughness scales in the fully rough regime},\ }\href
  {https://doi.org/10.1115/1.4001492} {\bibfield  {journal} {\bibinfo
  {journal} {J. Fluids Eng.}\ }\textbf {\bibinfo {volume} {132}},\ \bibinfo
  {pages} {041203} (\bibinfo {year} {2010})}\BibitemShut {NoStop}%
\bibitem [{\citenamefont {Ma}\ and\ \citenamefont
  {Lozano-Dur{\'a}n}(2025)}]{ma2025}%
  \BibitemOpen
  \bibfield  {author} {\bibinfo {author} {\bibfnamefont {R.}~\bibnamefont
  {Ma}}\ and\ \bibinfo {author} {\bibfnamefont {A.}~\bibnamefont
  {Lozano-Dur{\'a}n}},\ }\bibfield  {title} {\bibinfo {title} {Machine-learning
  wall-model large-eddy simulation accounting for isotropic roughness under
  local equilibrium},\ }\href {https://doi.org/10.1017/jfm.2025.29} {\bibfield
  {journal} {\bibinfo  {journal} {J. Fluid Mech.}\ }\textbf {\bibinfo {volume}
  {1007}},\ \bibinfo {pages} {A17} (\bibinfo {year} {2025})}\BibitemShut
  {NoStop}%
\bibitem [{\citenamefont {Uzun}\ and\ \citenamefont
  {Malik}(2022)}]{uzun2022High}%
  \BibitemOpen
  \bibfield  {author} {\bibinfo {author} {\bibfnamefont {A.}~\bibnamefont
  {Uzun}}\ and\ \bibinfo {author} {\bibfnamefont {M.~R.}\ \bibnamefont
  {Malik}},\ }\bibfield  {title} {\bibinfo {title} {High-fidelity simulation of
  turbulent flow {{\rev{past Gaussian bump}}}},\ }\href
  {https://doi.org/10.2514/1.J060760} {\bibfield  {journal} {\bibinfo
  {journal} {AIAA J.}\ }\textbf {\bibinfo {volume} {60}},\ \bibinfo {pages}
  {2130} (\bibinfo {year} {2022})}\BibitemShut {NoStop}%
\bibitem [{\citenamefont {Tamaki}\ and\ \citenamefont
  {Kawai}(2023)}]{tamaki2023Wall}%
  \BibitemOpen
  \bibfield  {author} {\bibinfo {author} {\bibfnamefont {Y.}~\bibnamefont
  {Tamaki}}\ and\ \bibinfo {author} {\bibfnamefont {S.}~\bibnamefont {Kawai}},\
  }\bibfield  {title} {\bibinfo {title} {Wall-resolved large-eddy simulation of
  near-stall airfoil flow at ${Re}_c = 10^7$},\ }\href
  {https://doi.org/10.2514/1.J062066} {\bibfield  {journal} {\bibinfo
  {journal} {AIAA J.}\ }\textbf {\bibinfo {volume} {61}},\ \bibinfo {pages}
  {698} (\bibinfo {year} {2023})}\BibitemShut {NoStop}%
\bibitem [{\citenamefont {Liu}\ \emph {et~al.}(2025)\citenamefont {Liu},
  \citenamefont {Wang}, \citenamefont {Vaidya}, \citenamefont {Ruehle},
  \citenamefont {Halverson}, \citenamefont {Solja{\v{c}}i{\'c}}, \citenamefont
  {Hou},\ and\ \citenamefont {Tegmark}}]{liu2024kan}%
  \BibitemOpen
  \bibfield  {author} {\bibinfo {author} {\bibfnamefont {Z.}~\bibnamefont
  {Liu}}, \bibinfo {author} {\bibfnamefont {Y.}~\bibnamefont {Wang}}, \bibinfo
  {author} {\bibfnamefont {S.}~\bibnamefont {Vaidya}}, \bibinfo {author}
  {\bibfnamefont {F.}~\bibnamefont {Ruehle}}, \bibinfo {author} {\bibfnamefont
  {J.}~\bibnamefont {Halverson}}, \bibinfo {author} {\bibfnamefont
  {M.}~\bibnamefont {Solja{\v{c}}i{\'c}}}, \bibinfo {author} {\bibfnamefont
  {T.~Y.}\ \bibnamefont {Hou}},\ and\ \bibinfo {author} {\bibfnamefont
  {M.}~\bibnamefont {Tegmark}},\ }\bibfield  {title} {\bibinfo {title} {{KAN}:
  {Kolmogorov--Arnold} networks},\ }in\ \href
  {https://arxiv.org/abs/2404.19756} {\emph {\bibinfo {booktitle}
  {International Conference on Learning Representations}}}\ (\bibinfo {year}
  {2025})\ \Eprint {https://arxiv.org/abs/2404.19756} {arXiv:2404.19756
  [cs.LG]} \BibitemShut {NoStop}%
\bibitem [{\citenamefont {de~Boor}(2001)}]{deboor2001splines}%
  \BibitemOpen
  \bibfield  {author} {\bibinfo {author} {\bibfnamefont {C.}~\bibnamefont
  {de~Boor}},\ }\href {https://link.springer.com/book/9780387953663} {\emph
  {\bibinfo {title} {A Practical Guide to Splines}}},\ \bibinfo {edition}
  {revised}\ ed.,\ \bibinfo {series} {Applied Mathematical Sciences},
  Vol.~\bibinfo {volume} {27}\ (\bibinfo  {publisher} {Springer},\ \bibinfo
  {address} {New York},\ \bibinfo {year} {2001})\BibitemShut {NoStop}%
\bibitem [{\citenamefont {Kingma}\ and\ \citenamefont
  {Ba}(2015)}]{kingma2015adam}%
  \BibitemOpen
  \bibfield  {author} {\bibinfo {author} {\bibfnamefont {D.~P.}\ \bibnamefont
  {Kingma}}\ and\ \bibinfo {author} {\bibfnamefont {J.}~\bibnamefont {Ba}},\
  }\bibfield  {title} {\bibinfo {title} {Adam: {A} method for stochastic
  optimization},\ }in\ \href {https://arxiv.org/abs/1412.6980} {\emph {\bibinfo
  {booktitle} {3rd International Conference on Learning Representations
  ({ICLR})}}}\ (\bibinfo {year} {2015})\ \Eprint
  {https://arxiv.org/abs/1412.6980} {arXiv:1412.6980 [cs.LG]} \BibitemShut
  {NoStop}%
\bibitem [{\citenamefont {Schlatter}\ and\ \citenamefont
  {{\"O}rl{\"u}}(2010)}]{Schlatter2010}%
  \BibitemOpen
  \bibfield  {author} {\bibinfo {author} {\bibfnamefont {P.}~\bibnamefont
  {Schlatter}}\ and\ \bibinfo {author} {\bibfnamefont {R.}~\bibnamefont
  {{\"O}rl{\"u}}},\ }\bibfield  {title} {\bibinfo {title} {Assessment of direct
  numerical simulation data of turbulent boundary layers},\ }\href
  {https://doi.org/10.1017/S0022112010003113} {\bibfield  {journal} {\bibinfo
  {journal} {J. Fluid Mech.}\ }\textbf {\bibinfo {volume} {659}},\ \bibinfo
  {pages} {116} (\bibinfo {year} {2010})}\BibitemShut {NoStop}%
\bibitem [{\citenamefont {Eitel-Amor}\ \emph {et~al.}(2014)\citenamefont
  {Eitel-Amor}, \citenamefont {{\"O}rl{\"u}},\ and\ \citenamefont
  {Schlatter}}]{eitelamor2014simulation}%
  \BibitemOpen
  \bibfield  {author} {\bibinfo {author} {\bibfnamefont {G.}~\bibnamefont
  {Eitel-Amor}}, \bibinfo {author} {\bibfnamefont {R.}~\bibnamefont
  {{\"O}rl{\"u}}},\ and\ \bibinfo {author} {\bibfnamefont {P.}~\bibnamefont
  {Schlatter}},\ }\bibfield  {title} {\bibinfo {title} {Simulation and
  validation of a spatially evolving turbulent boundary layer up to
  {$Re_{\theta}=8300$}},\ }\href
  {https://doi.org/10.1016/j.ijheatfluidflow.2014.02.006} {\bibfield  {journal}
  {\bibinfo  {journal} {Int. J. Heat Fluid Flow}\ }\textbf {\bibinfo {volume}
  {47}},\ \bibinfo {pages} {57} (\bibinfo {year} {2014})}\BibitemShut {NoStop}%
\bibitem [{\citenamefont {Spalart}\ \emph {et~al.}(2024)\citenamefont
  {Spalart}, \citenamefont {Jansen},\ and\ \citenamefont
  {Coleman}}]{spalart2024flattened}%
  \BibitemOpen
  \bibfield  {author} {\bibinfo {author} {\bibfnamefont {P.~R.}\ \bibnamefont
  {Spalart}}, \bibinfo {author} {\bibfnamefont {K.~E.}\ \bibnamefont
  {Jansen}},\ and\ \bibinfo {author} {\bibfnamefont {G.~N.}\ \bibnamefont
  {Coleman}},\ }\href {https://doi.org/10.48550/arXiv.2409.00555} {\bibinfo
  {title} {Direct numerical simulation of two boundary layers with the same
  pressure distribution but different surface curvatures}} (\bibinfo {year}
  {2024}),\ \Eprint {https://arxiv.org/abs/2409.00555} {arXiv:2409.00555
  [physics.flu-dyn]} \BibitemShut {NoStop}%
\bibitem [{\citenamefont {{\"O}rl{\"u}}\ and\ \citenamefont
  {Schlatter}(2013)}]{orlu2013dns}%
  \BibitemOpen
  \bibfield  {author} {\bibinfo {author} {\bibfnamefont {R.}~\bibnamefont
  {{\"O}rl{\"u}}}\ and\ \bibinfo {author} {\bibfnamefont {P.}~\bibnamefont
  {Schlatter}},\ }\bibfield  {title} {\bibinfo {title} {Comparison of
  experiments and simulations for zero pressure gradient turbulent boundary
  layers at moderate {Reynolds} numbers},\ }\href
  {https://doi.org/10.1007/s00348-013-1547-x} {\bibfield  {journal} {\bibinfo
  {journal} {Exp. Fluids}\ }\textbf {\bibinfo {volume} {54}},\ \bibinfo {pages}
  {1547} (\bibinfo {year} {2013})}\BibitemShut {NoStop}%
\bibitem [{\citenamefont {Wieghardt}\ and\ \citenamefont
  {Tillmann}(1951)}]{wieghardt1951characteristics}%
  \BibitemOpen
  \bibfield  {author} {\bibinfo {author} {\bibfnamefont {K.}~\bibnamefont
  {Wieghardt}}\ and\ \bibinfo {author} {\bibfnamefont {W.}~\bibnamefont
  {Tillmann}},\ }\href {https://ntrs.nasa.gov/citations/19930093916} {\emph
  {\bibinfo {title} {On the Turbulent Friction Layer for Rising Pressure}}},\
  \bibinfo {type} {Technical Memorandum}\ \bibinfo {number} {1314}\ (\bibinfo
  {institution} {National Advisory Committee for Aeronautics},\ \bibinfo {year}
  {1951})\BibitemShut {NoStop}%
\bibitem [{\citenamefont {Mallor}\ \emph {et~al.}(2025)\citenamefont {Mallor},
  \citenamefont {Sanmiguel~Vila}, \citenamefont {Hajipour}, \citenamefont
  {Vinuesa}, \citenamefont {Schlatter},\ and\ \citenamefont
  {{\"O}rl{\"u}}}]{mallor2025experimental}%
  \BibitemOpen
  \bibfield  {author} {\bibinfo {author} {\bibfnamefont {F.}~\bibnamefont
  {Mallor}}, \bibinfo {author} {\bibfnamefont {C.}~\bibnamefont
  {Sanmiguel~Vila}}, \bibinfo {author} {\bibfnamefont {M.}~\bibnamefont
  {Hajipour}}, \bibinfo {author} {\bibfnamefont {R.}~\bibnamefont {Vinuesa}},
  \bibinfo {author} {\bibfnamefont {P.}~\bibnamefont {Schlatter}},\ and\
  \bibinfo {author} {\bibfnamefont {R.}~\bibnamefont {{\"O}rl{\"u}}},\
  }\bibfield  {title} {\bibinfo {title} {Experimental characterization of
  turbulent boundary layers around a {NACA 4412} wing profile},\ }\href
  {https://doi.org/10.1016/j.expthermflusci.2024.111327} {\bibfield  {journal}
  {\bibinfo  {journal} {Exp. Therm. Fluid Sci.}\ }\textbf {\bibinfo {volume}
  {160}},\ \bibinfo {pages} {111327} (\bibinfo {year} {2025})}\BibitemShut
  {NoStop}%
\bibitem [{\citenamefont {Vinuesa}\ \emph {et~al.}(2018)\citenamefont
  {Vinuesa}, \citenamefont {Negi}, \citenamefont {Atzori}, \citenamefont
  {Hanifi}, \citenamefont {Henningson},\ and\ \citenamefont
  {Schlatter}}]{vinuesa2019wing}%
  \BibitemOpen
  \bibfield  {author} {\bibinfo {author} {\bibfnamefont {R.}~\bibnamefont
  {Vinuesa}}, \bibinfo {author} {\bibfnamefont {P.~S.}\ \bibnamefont {Negi}},
  \bibinfo {author} {\bibfnamefont {M.}~\bibnamefont {Atzori}}, \bibinfo
  {author} {\bibfnamefont {A.}~\bibnamefont {Hanifi}}, \bibinfo {author}
  {\bibfnamefont {D.~S.}\ \bibnamefont {Henningson}},\ and\ \bibinfo {author}
  {\bibfnamefont {P.}~\bibnamefont {Schlatter}},\ }\bibfield  {title} {\bibinfo
  {title} {Turbulent boundary layers around wing sections up to
  {$Re_c=1{,}000{,}000$}},\ }\href
  {https://doi.org/10.1016/j.ijheatfluidflow.2018.04.017} {\bibfield  {journal}
  {\bibinfo  {journal} {Int. J. Heat Fluid Flow}\ }\textbf {\bibinfo {volume}
  {72}},\ \bibinfo {pages} {86} (\bibinfo {year} {2018})}\BibitemShut {NoStop}%
\bibitem [{\citenamefont {Tanarro}\ \emph {et~al.}(2020)\citenamefont
  {Tanarro}, \citenamefont {Vinuesa},\ and\ \citenamefont
  {Schlatter}}]{tanarro2020effect}%
  \BibitemOpen
  \bibfield  {author} {\bibinfo {author} {\bibfnamefont {{\'A}.}~\bibnamefont
  {Tanarro}}, \bibinfo {author} {\bibfnamefont {R.}~\bibnamefont {Vinuesa}},\
  and\ \bibinfo {author} {\bibfnamefont {P.}~\bibnamefont {Schlatter}},\
  }\bibfield  {title} {\bibinfo {title} {Effect of adverse pressure gradients
  on turbulent wing boundary layers},\ }\href
  {https://doi.org/10.1017/jfm.2019.838} {\bibfield  {journal} {\bibinfo
  {journal} {J. Fluid Mech.}\ }\textbf {\bibinfo {volume} {883}},\ \bibinfo
  {pages} {A8} (\bibinfo {year} {2020})}\BibitemShut {NoStop}%
\bibitem [{\citenamefont {Greenblatt}\ \emph {et~al.}(2006)\citenamefont
  {Greenblatt}, \citenamefont {Paschal}, \citenamefont {Yao}, \citenamefont
  {Harris}, \citenamefont {Schaeffler},\ and\ \citenamefont
  {Washburn}}]{greenblatt2006separation}%
  \BibitemOpen
  \bibfield  {author} {\bibinfo {author} {\bibfnamefont {D.}~\bibnamefont
  {Greenblatt}}, \bibinfo {author} {\bibfnamefont {K.~B.}\ \bibnamefont
  {Paschal}}, \bibinfo {author} {\bibfnamefont {C.-S.}\ \bibnamefont {Yao}},
  \bibinfo {author} {\bibfnamefont {J.}~\bibnamefont {Harris}}, \bibinfo
  {author} {\bibfnamefont {N.~W.}\ \bibnamefont {Schaeffler}},\ and\ \bibinfo
  {author} {\bibfnamefont {A.~E.}\ \bibnamefont {Washburn}},\ }\bibfield
  {title} {\bibinfo {title} {Experimental investigation of separation control
  part 1: Baseline and steady suction},\ }\href
  {https://doi.org/10.2514/1.13817} {\bibfield  {journal} {\bibinfo  {journal}
  {AIAA J.}\ }\textbf {\bibinfo {volume} {44}},\ \bibinfo {pages} {2820}
  (\bibinfo {year} {2006})}\BibitemShut {NoStop}%
\bibitem [{\citenamefont {Naughton}\ \emph {et~al.}(2006)\citenamefont
  {Naughton}, \citenamefont {Viken},\ and\ \citenamefont
  {Greenblatt}}]{naughton2006wall}%
  \BibitemOpen
  \bibfield  {author} {\bibinfo {author} {\bibfnamefont {J.~W.}\ \bibnamefont
  {Naughton}}, \bibinfo {author} {\bibfnamefont {S.~A.}\ \bibnamefont
  {Viken}},\ and\ \bibinfo {author} {\bibfnamefont {D.}~\bibnamefont
  {Greenblatt}},\ }\bibfield  {title} {\bibinfo {title} {Skin-friction
  measurements on the {NASA} hump model},\ }\href
  {https://doi.org/10.2514/1.14192} {\bibfield  {journal} {\bibinfo  {journal}
  {AIAA J.}\ }\textbf {\bibinfo {volume} {44}},\ \bibinfo {pages} {1255}
  (\bibinfo {year} {2006})}\BibitemShut {NoStop}%
\bibitem [{\citenamefont {Gray}\ \emph {et~al.}(2023)\citenamefont {Gray},
  \citenamefont {Lakebrink}, \citenamefont {Thomas}, \citenamefont {Corke},
  \citenamefont {Gluzman},\ and\ \citenamefont
  {Straccia}}]{gray2023experimental}%
  \BibitemOpen
  \bibfield  {author} {\bibinfo {author} {\bibfnamefont {P.~D.}\ \bibnamefont
  {Gray}}, \bibinfo {author} {\bibfnamefont {M.~T.}\ \bibnamefont {Lakebrink}},
  \bibinfo {author} {\bibfnamefont {F.~O.}\ \bibnamefont {Thomas}}, \bibinfo
  {author} {\bibfnamefont {T.~C.}\ \bibnamefont {Corke}}, \bibinfo {author}
  {\bibfnamefont {I.}~\bibnamefont {Gluzman}},\ and\ \bibinfo {author}
  {\bibfnamefont {J.}~\bibnamefont {Straccia}},\ }\bibfield  {title} {\bibinfo
  {title} {Experimental and computational evaluation of smooth-body separated
  flow over {Boeing} bump},\ }in\ \href {https://doi.org/10.2514/6.2023-3981}
  {\emph {\bibinfo {booktitle} {{AIAA AVIATION} 2023 Forum}}}\ (\bibinfo
  {publisher} {American Institute of Aeronautics and Astronautics},\ \bibinfo
  {year} {2023})\ \bibinfo {note} {paper 2023-3981}\BibitemShut {NoStop}%
\bibitem [{\citenamefont {Bissonnette}\ and\ \citenamefont
  {Mellor}(1974)}]{bissonnette1974three}%
  \BibitemOpen
  \bibfield  {author} {\bibinfo {author} {\bibfnamefont {L.~R.}\ \bibnamefont
  {Bissonnette}}\ and\ \bibinfo {author} {\bibfnamefont {G.~L.}\ \bibnamefont
  {Mellor}},\ }\bibfield  {title} {\bibinfo {title} {Experiments on the
  behaviour of an axisymmetric turbulent boundary layer with a sudden
  circumferential strain},\ }\href {https://doi.org/10.1017/S0022112074001182}
  {\bibfield  {journal} {\bibinfo  {journal} {J. Fluid Mech.}\ }\textbf
  {\bibinfo {volume} {63}},\ \bibinfo {pages} {369} (\bibinfo {year}
  {1974})}\BibitemShut {NoStop}%
\bibitem [{\citenamefont {Lohmann}(1976)}]{lohmann1976measurement}%
  \BibitemOpen
  \bibfield  {author} {\bibinfo {author} {\bibfnamefont {R.~P.}\ \bibnamefont
  {Lohmann}},\ }\bibfield  {title} {\bibinfo {title} {The response of a
  developed turbulent boundary layer to local transverse surface motion},\
  }\href {https://doi.org/10.1115/1.3448319} {\bibfield  {journal} {\bibinfo
  {journal} {J. Fluids Eng.}\ }\textbf {\bibinfo {volume} {98}},\ \bibinfo
  {pages} {354} (\bibinfo {year} {1976})}\BibitemShut {NoStop}%
\bibitem [{\citenamefont {Rkein}\ and\ \citenamefont
  {Laval}(2023)}]{rkein2023numerical}%
  \BibitemOpen
  \bibfield  {author} {\bibinfo {author} {\bibfnamefont {H.}~\bibnamefont
  {Rkein}}\ and\ \bibinfo {author} {\bibfnamefont {J.-P.}\ \bibnamefont
  {Laval}},\ }\bibfield  {title} {\bibinfo {title} {Direct numerical simulation
  of adverse pressure gradient turbulent boundary layer up to
  {$Re_{\theta}\simeq8000$}},\ }\href
  {https://doi.org/10.1016/j.ijheatfluidflow.2023.109195} {\bibfield  {journal}
  {\bibinfo  {journal} {Int. J. Heat Fluid Flow}\ }\textbf {\bibinfo {volume}
  {103}},\ \bibinfo {pages} {109195} (\bibinfo {year} {2023})}\BibitemShut
  {NoStop}%
\bibitem [{\citenamefont {Spalart}\ and\ \citenamefont
  {Watmuff}(1993)}]{spalart1993experimental}%
  \BibitemOpen
  \bibfield  {author} {\bibinfo {author} {\bibfnamefont {P.~R.}\ \bibnamefont
  {Spalart}}\ and\ \bibinfo {author} {\bibfnamefont {J.~H.}\ \bibnamefont
  {Watmuff}},\ }\bibfield  {title} {\bibinfo {title} {Experimental and
  numerical study of a turbulent boundary layer with pressure gradients},\
  }\href {https://doi.org/10.1017/S002211209300120X} {\bibfield  {journal}
  {\bibinfo  {journal} {J. Fluid Mech.}\ }\textbf {\bibinfo {volume} {249}},\
  \bibinfo {pages} {337} (\bibinfo {year} {1993})}\BibitemShut {NoStop}%
\bibitem [{\citenamefont {Fritsch}\ \emph {et~al.}(2022)\citenamefont
  {Fritsch}, \citenamefont {Vishwanathan}, \citenamefont {Lowe},\ and\
  \citenamefont {Devenport}}]{fritsch2022effects}%
  \BibitemOpen
  \bibfield  {author} {\bibinfo {author} {\bibfnamefont {D.~J.}\ \bibnamefont
  {Fritsch}}, \bibinfo {author} {\bibfnamefont {V.}~\bibnamefont
  {Vishwanathan}}, \bibinfo {author} {\bibfnamefont {K.~T.}\ \bibnamefont
  {Lowe}},\ and\ \bibinfo {author} {\bibfnamefont {W.~J.}\ \bibnamefont
  {Devenport}},\ }\bibfield  {title} {\bibinfo {title} {Fluctuating pressure
  beneath smooth wall boundary layers in nonequilibrium pressure gradients},\
  }\href {https://doi.org/10.2514/1.J061431} {\bibfield  {journal} {\bibinfo
  {journal} {AIAA J.}\ }\textbf {\bibinfo {volume} {60}},\ \bibinfo {pages}
  {4725} (\bibinfo {year} {2022})}\BibitemShut {NoStop}%
\bibitem [{\citenamefont {Vishwanathan}\ \emph {et~al.}(2023)\citenamefont
  {Vishwanathan}, \citenamefont {Fritsch}, \citenamefont {Lowe},\ and\
  \citenamefont {Devenport}}]{vishwanathan2023effects}%
  \BibitemOpen
  \bibfield  {author} {\bibinfo {author} {\bibfnamefont {V.}~\bibnamefont
  {Vishwanathan}}, \bibinfo {author} {\bibfnamefont {D.~J.}\ \bibnamefont
  {Fritsch}}, \bibinfo {author} {\bibfnamefont {K.~T.}\ \bibnamefont {Lowe}},\
  and\ \bibinfo {author} {\bibfnamefont {W.~J.}\ \bibnamefont {Devenport}},\
  }\bibfield  {title} {\bibinfo {title} {History effects and wall-similarity of
  non-equilibrium turbulent boundary layers in varying pressure gradient over
  rough and smooth surfaces},\ }\href
  {https://doi.org/10.1016/j.ijheatfluidflow.2023.109145} {\bibfield  {journal}
  {\bibinfo  {journal} {Int. J. Heat Fluid Flow}\ }\textbf {\bibinfo {volume}
  {102}},\ \bibinfo {pages} {109145} (\bibinfo {year} {2023})}\BibitemShut
  {NoStop}%
\end{thebibliography}%
%%%%%%%%%%%%%%%%%%%%%%%%%%%%%%%%%%%%%%%%%%%%%%%%%%%%%%%%%%%%%%%%%%%%

%%%%%%%%%%%%%%%%%%%%%%%%%%%%%%%%%%%%%%%%%%%%%%%%%%%%%%%%%%%%%%%%%%%%
\section{End Matter}
%%%%%%%%%%%%%%%%%%%%%%%%%%%%%%%%%%%%%%%%%%%%%%%%%%%%%%%%%%%%%%%%%%%%

%-----------------------------------------------------------------%
\subsection{\corr{Comparison with previous models}}
%-----------------------------------------------------------------%

% Intro to model
\corr{\emph{Mean velocity profiles---}We compare the present model
    $f_U(\pU_1,\pU_2,\pU_3)$ with the
    mixing-length model of~\citet{ma2026}, which was developed to predict
    mean velocity profiles in APG TBLs up to separation onset. \rev{Here,
    $U^+=U/u_\tau$, $y^+=yu_\tau/\nu$, and
    $u_\tau=\sqrt{|\tau_w|/\rho}$.} In that
    approach, the mean velocity is reconstructed from the mean shear as
    \begin{equation}
        U^+(y^+) = \int_0^{y^+} \frac{\partial U^+}{\partial \eta}\,\mathrm{d}\eta,
        \label{eq:ma_integral}
    \end{equation}
    where $\partial U^+/\partial \eta$ is obtained from a modeled
\emph{mixing length}.}

% Ma vs ours
\corr{There are several key distinctions between the model
    of~\citet{ma2026} and the present one. First, our model predicts the
    dimensionless mean velocity $\pUo \equiv U/U_e$ directly and
    non-iteratively as $\pUo \approx f_U(\pU_1,\pU_2,\pU_3)$, rather
    than reconstructing the profile by integrating a modeled mean
    shear. Second, the implementation of~\citet{ma2026} employs two
    parameters, $b$ and $n$, whose values vary with flow type and APG
    strength and are calibrated on a case-by-case basis. Its predictive
    accuracy therefore depends on prior knowledge of these parameters
    for a given flow. By contrast, once $f_U$ has been trained, our
    model requires no case-specific calibration or adjustable parameters
    and is applied unchanged to all cases, including those not seen
    during training.  Finally, the model of~\citet{ma2026} is formulated
    and validated for ZPG and APG boundary layers prior to separation,
    whereas the present approach also extends to FPG conditions and
separated flows.}

\begin{figure*}
    \centering
    \includegraphics{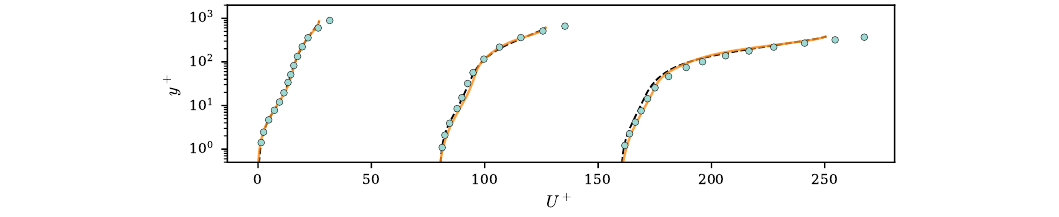}
    \caption{\protect\corr{Comparison of mean velocity profiles from
            the present model $f_U(\pU_1,\pU_2,\pU_3)$ (orange) and the
            mixing-length model of~\citet{ma2026} (cyan circles) against
            the DNS data of \citet{coleman2018numerical} (black dashed) at
            three streamwise stations upstream of separation. The
            wall-normal coordinate and velocity are defined as
            $y^+=yu_\tau/\nu$ and $U^+=U/u_\tau$, respectively. An offset
            of $\Delta U^+ = 80$ is added between the velocity profiles to
    ease the comparison.}}
    \label{fig:velcomp}
\end{figure*}

% Results
\corr{Figure~\rev{\ref{fig:velcomp}} compares the predictions of both models with
    \rev{direct numerical simulation (DNS)} data at three streamwise stations of the separating-flow case of
    \citet{coleman2018numerical}, all located upstream of separation and
    corresponding to figure~8 of~\citet{ma2026}. In the inner layer, the
    model of~\citet{ma2026} agrees reasonably well with the DNS,
    consistent with its accurate near-wall mixing-length behavior.
    However, its accuracy deteriorates in the outer region, especially at
    the most downstream station, where the wake is strongly affected as
    the flow approaches separation. By contrast, our model
    $f_U(\pU_1,\pU_2,\pU_3)$ remains accurate over the full wall-normal
    extent at all three stations, capturing both the inner and outer
layers without case-specific adjustment.}

% Intro
\corr{ \emph{Mean wall-shear stress---}We compare the predictive
    accuracy of existing models for the mean wall-shear stress against
    the present database. Specifically, we consider classical empirical
    correlations~\citep{ludwieg1949, White1974}, which were calibrated
    on turbulent boundary layers under mild pressure gradients, together
    with the recent physics-based framework of
    \citet{dixit2024generalized}, which extends to a broader class of
attached flows.  }

\corr{The classical correlations considered here were developed under
    the assumption of local equilibrium---where boundary layer
    development is gradual and pressure-gradient effects are
    balanced---so that the normalized wall-shear stress depends only on
    the momentum-thickness Reynolds number and the shape
    factor~\citep[p.~590]{schlichting2000boundary}:
    \begin{equation}\label{eq:tauweq}
        \ptauo = g_\tau(\Rey_\theta, H).
    \end{equation}
    A widely used empirical correlation of this form is that
    of~\citet{ludwieg1949}:
    \begin{equation}\label{eq:ludwieg1949}
        \ptauo = 0.123 \cdot 10^{-0.678H} \Rey_\theta^{-0.268}.
    \end{equation}
    We also compare against the correlation given by~\citet{White1974}:
    \begin{equation}\label{eq:white}
        \ptauo =  \frac{0.15 e^{-1.33H}}
        {(\log_{10}{ \Rey_\theta })^{ 1.74 + 0.31H}},
    \end{equation}
    a curve-fitting expression based on Coles' constants for the law
    of the wall~\cite{Barr1980}. Note that, although both
    \rev{\eqref{eq:ludwieg1949}} and \rev{\eqref{eq:white}} are based on the
    variables $H$ and $\Rey_\theta$ identified by classical theory,
    their grouping is suboptimal compared with the combination
$H^{-1}\Rey_\theta^{-1/4}$ identified by the present analysis.}

\corr{ We further compare against the physics-based model
    of~\citet{dixit2024generalized}.  Their approach, termed
    $M$-$\nu$-$G$ scaling, is derived from the wall-normal integral of
    the streamwise mean-flow kinetic energy equation rather than from
    empirical correlations.  Predictions are obtained by solving an
    implicit equation for the friction velocity, \rev{$u_\tau$},
    iteratively.  The scope of the model is
    explicitly restricted to \emph{attached} turbulent boundary layers.
    The authors note that for sufficiently strong adverse pressure
    gradients, the boundary layer may approach separation, at which
    point wall friction ceases to be dynamically relevant and falls
    outside the model's domain of
validity~\citep{dixit2024generalized}.}

\begin{figure*}
    \centering
    \includegraphics{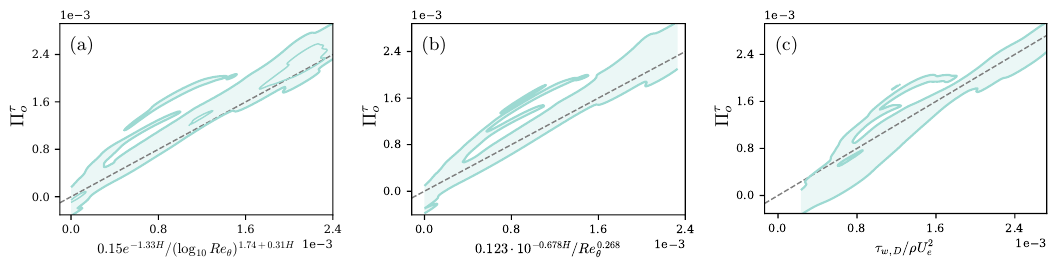}
    \caption{\protect\corr{Comparison of actual $\ptauo$ values with
            predictions from (a) Eq.~\rev{\eqref{eq:white}}, given by
            \citet{White1974}; (b) Eq.~\rev{\eqref{eq:ludwieg1949}}, given by
            \citet{ludwieg1949}; and (c) the iterative approach of
            \citet{dixit2024generalized}. The cyan contours enclose
            99\% of the joint probability mass; the diagonal dashed lines
            indicate perfect agreement (zero error).
    \label{fig:taucomp}}}
\end{figure*}

\corr{Figure~\rev{\ref{fig:taucomp}} compares the predictive accuracy of the
    models against the present database.  Figure~\rev{\ref{fig:taucomp}}a
    shows the corresponding results for Eq.~\rev{\eqref{eq:white}},
    figure~\rev{\ref{fig:taucomp}}b shows the results for
    Eq.~\rev{\eqref{eq:ludwieg1949}}, and figure~\rev{\ref{fig:taucomp}}c shows
    those for the model of~\citet{dixit2024generalized}.  All three
    reference models exhibit substantially larger errors, and none
    is capable of predicting separated flow conditions. In contrast, the
    proposed two-parameter scaling $f_\tau(\ptau_1, \ptau_2)$ accurately
captures both attached and separated flow regimes.}

%-----------------------------------------------------------------%
\subsection{\corr{Scaling laws from governing equations}}
\label{ss:equations_main}
%-----------------------------------------------------------------%

\corr{An interesting question is whether the discovered scaling laws
    could be obtained directly from the governing equations.  A more
    detailed discussion of this point is provided in Sec.~II.2 of the
    Supplemental Materials~\cite{suppl}. Here, we summarize only the
    connection relevant to the mean wall-shear stress.  For
    illustration, consider the classical von K\'arm\'an
    momentum-integral equation for a two-dimensional incompressible
    TBL~\citep{schlichting2000boundary},
    \begin{equation}
        \frac{\mathrm{d}\theta}{\mathrm{d}x}
        =
        \ptauo
        -
        (H+2)\frac{\theta}{U_e}
        \frac{\mathrm{d}U_e}{\mathrm{d}x}.
        \label{eq:vonkarman_main}
    \end{equation}
}

\corr{Using the inviscid relation at the boundary-layer edge,
    $U_e\,\mathrm{d}U_e/\mathrm{d}x=-\dPe$, this balance becomes
    \begin{equation}
        \ptauo
        =
        \frac{\mathrm{d}\theta}{\mathrm{d}x}
        -
        \left(1+\frac{2}{H}\right)\gamma_P\beta_*.
        \label{eq:vonkarman_beta_main}
    \end{equation}
    Equation~\eqref{eq:vonkarman_beta_main} shows that the wall-shear
    stress is constrained by the local evolution of the momentum
    thickness and by the pressure forcing acting on the boundary
layer.}

\corr{The governing balance, however, does not uniquely identify the
    most predictive dimensionless groups. For example, the same equation
    may be rearranged as
    \begin{align}
        \ptauo
        &=
        \frac{\theta}{\Rey_\theta}
        \frac{\mathrm{d}\Rey_\theta}{\mathrm{d}x}
        -
        \left(1+\ptau_1\Rey_\theta^{1/4}\right)
        \gamma_P\beta_*,
        \label{eq:vonkarman_pi1_main}
    \end{align}
    where the equality follows from
    $\ptau_1=H^{-1}\Rey_\theta^{-1/4}$. Thus, $H$, $\beta_*$,
    $\Rey_\theta$, or $\ptau_1$ can be made explicit through algebraic
    rearrangement without changing the underlying physics. The second
    discovered input, $\ptau_2$, likewise incorporates the pressure
    forcing through $\beta_*$, together with viscous and
integral-state contributions [Eq.~\eqref{eq:pitaus1}].}

\corr{This example helps illustrate two key conclusions. First, the
    fact that a dimensionless group appears explicitly in a given form
    of the equations does not imply that it plays a dominant role---or
    even any role at all---in determining the quantity of
    interest. Second, the fact that a dimensionless group does not
    appear explicitly does not imply that it is irrelevant, since its
    influence may be hidden within other terms that depend on it
functionally.}

\corr{The information-theoretic formulation provides the missing
    selection principle: among the admissible dimensionless
    combinations, it identifies \rev{$(\pU_1,\pU_2,\pU_3)$ as the inputs}
    retaining the greatest predictive information about \rev{$\pUo$}. The
    relation $\ptauo\approx f_\tau(\ptau_1,\ptau_2)$\rev{, obtained from
    the wall derivative of $f_U$,} is therefore
    consistent with and constrained by the momentum-integral balance,
    while the particular combination of variables and exponents \rev{in the velocity model} is
    selected by its predictive content rather than by an arbitrary
rearrangement of the governing equation.}

% Include every reference cited in the Supplemental Material, as required by PRL.
% Keep this after the End Matter citations to preserve their reference numbers.
\nocite{%
    Orlandi2000,Wray1990,Kim1985,
    Bae2018,Lozano2019d,Lozano2018,
    Towne2023,Pauley1990,lund1998Generation,
    morgan2011Improving,griffin2021general,torroja,
    wu2019spatiotemporal,cover2006,wei2024new,
    han2025integral,sobol2001,saltelli2002,
    vinuesa2016determining,wei2023outer,lozier2025Defining,
    martinez2024surd,martinez2026,bradshaw1987,
    lozano-duran2020,driver1987,eaton1995,
    johnston1960,abe2020,vandenberg1975,
    coleman2019,jeans1982,bradshaw1973,
    ellis1974,smits1979,barlow1988,
    appelbaum2025,pargal2022,lagha2011,
    spina1994,smits2006,morkovin1962,
    vandriest1951,huang1995,volpiani2020,
    griffin2021,coles1964,bons2001,
    bons2010,munk2009,kirschner2012,
    nikuradse1933,flack2010,ma2025,
    kingma2015adam,eitelamor2014simulation,spalart2024flattened,
    orlu2013dns,wieghardt1951characteristics,mallor2025experimental,
    vinuesa2019wing,tanarro2020effect,greenblatt2006separation,
    naughton2006wall,gray2023experimental,bissonnette1974three,
    lohmann1976measurement,rkein2023numerical,spalart1993experimental,
    fritsch2022effects,vishwanathan2023effects%
}

\end{document}

\title{Supplemental Materials: \\
Unified scaling laws for turbulent boundary layers across flow regimes}

\author{Gonzalo Arranz}
\affiliation{California Institute of Technology, Pasadena, CA 91125, USA}
\author{Adri\'an Lozano-Dur\'an }
\email{adrianld@caltech.edu}
\affiliation{California Institute of Technology, Pasadena, CA 91125, USA}
\affiliation{Massachusetts Institute of Technology, Cambridge, MA 02139, USA}
%\aff{1}California Institute of Technology, Pasadena CA 91125, USA
%\aff{2}Massachusetts Institute of Technology, Cambridge MA 02139, USA}

\maketitle
\tableofcontents

%%%%%%%%%%%$$$$$$$$$$$$$$$$$%%%%%%%%%%%%%%%%%%%%%%%%%%%%%%%%%%%%%%%%%%%
\section{Overview of turbulent boundary layers and separation
bubbles}\label{sec:oursims}
%%%%%%%%%%%%%%%%%%%%%%%%%%%%$$$$$$$$$$$$$$$$$%%%%%%%%%%%%%%%%%%%%%%%%%%

We describe the new pressure-gradient turbulent boundary layers and
separation bubbles simulated for this study and provide an overview
of the other databases used in the analysis.

%--------------------------------------------------------------%
\subsection{Computational setup for current TBLs and separation bubbles}
%--------------------------------------------------------------%

% Numerical methods
We numerically solve the Navier--Stokes equations for an incompressible
fluid:
\begin{align*}
    \nabla\cdot\myv{u} &= 0, \\
    \frac{\partial\myv{u}}{\partial t} + (\myv{u} \cdot \nabla)\myv{u}
    &= -\frac{1}{\rho}\nabla p + \nu \nabla^2 \myv{u},
\end{align*}
where $\myv{u} = (u,v,w)$ is the velocity field, $\rho$ is the fluid
density, $\nu$ is the kinematic viscosity, and $p$ is the pressure.
The solutions are computed by direct numerical simulation (DNS).
Spatial derivatives are discretized on a staggered grid using a
second-order central finite-difference
scheme~\citep{Orlandi2000}. Time advancement is achieved by a
third-order Runge-Kutta scheme~\citep{Wray1990}, combined with the
fractional-step method~\citep{Kim1985}. The code has been validated in
previous studies of turbulent channel
flows~\citep{Bae2018,Lozano2019d}, transitional boundary
layers~\citep{Lozano2018}, and zero-pressure-gradient turbulent
boundary layers~\citep{Towne2023}.

% Domain and BCs
The computational setup is depicted in figure~\ref{fig:domain}. All
simulations are performed in a rectangular domain of size $L_x \times
L_y \times L_z$ in the streamwise, wall-normal, and spanwise
directions, respectively.  The bottom wall is smooth and no-slip,
periodic boundary conditions are imposed in the spanwise direction,
and an advective condition is prescribed at the
outlet~\citep{Pauley1990}.  At the inlet ($x_i$), we impose the mean
velocity profile of a turbulent boundary layer from
\citet{sillero2014} at a fixed $\Rey_{\theta_i}=U_e\dTwo_i/\nu$, where
$\dTwo_i$ is the inlet momentum thickness, and superimpose rescaled
fluctuations taken from a reference plane at $x_0$, following an
approach similar to that of~\citet{lund1998Generation}.  The recycling
plane is placed sufficiently downstream to ensure decorrelation of the
fluctuations, $x_0 - x_i > 11\delta_0$, as recommended by
\citet{morgan2011Improving}, where $\delta_0$ is the boundary layer
thickness at $x_0$.
\rev{For the simulations introduced here,} the boundary-layer thickness $\delta$ is
computed following \citet{griffin2021general}. \rev{The definitions used for
the additional validation cases are specified in \S\ref{sec:additionalvalidation}.} The displacement
thickness, $\dOne$, and momentum thickness, $\dTwo$, are defined as
\begin{equation}
    \delta_* = \int_0^\infty \left(1 - \frac{U}{U_e}\right) \mathrm{d}y,
    \quad
    \theta = \int_0^\infty \frac{U}{U_e}\left(1 - \frac{U}{U_e}\right) \mathrm{d}y,
\end{equation}
where $U$ is the mean streamwise velocity and $U_e$ is the
boundary-layer edge velocity at $y = \delta$.
\begin{figure}[h!]
    \centering
    \begin{tikzpicture}[>={Latex[length=1.5mm]},scale=.9]

    \pgfmathsetmacro{\xo}{2};
    \pgfmathsetmacro{\xc}{3.5};
    \pgfmathsetmacro{\yc}{2.4};
    \pgfmathsetmacro{\Lx}{10};
    \pgfmathsetmacro{\Ly}{1.5};

    \coordinate (edge) at (\Lx,\Ly);
    \coordinate (x0) at (\xo,0);
    \coordinate (xc) at (\xc,\yc);

    \begin{footnotesize}
    \fill[orange!10!white] (0,0) rectangle (\xc,\Ly) node[midway,black,above] {ZPG}
    node[inner sep=1pt, outer sep=1pt,pos=0,black,above left] {$x_i$};
    \fill[blue!10!white] (\xc,0) rectangle (\Lx-1,\Ly) node[midway,black] {APG};
    \fill[red!10!white] (\Lx-1,0) rectangle (\Lx,\Ly) node[midway,black]
    {outflow};
    \draw[thin,dashed] (0,0) rectangle (\Lx,\Ly);
    \draw[red!40!black,densely dotted] (\xc*.7,0) --+ (0,\Ly) 
        node[inner sep=1pt, outer sep=1pt,text=black,pos=0,above left] {$x_0$};

    \draw[<-,thin,red!40!black] (0,.3*\Ly) to [bend left=20] +(\xc*.7,0);
    \node[inner sep=1pt, outer sep=1pt,fill=orange!10!white,text=red!40!black] 
        at (\xc*.35,.3*\Ly) {\scriptsize recycle};

    \draw[dashed] (xc) --+ (4,0);
    \draw ($(xc)+(3,0)$) arc(0:5.8:3) node[pos=0.6, anchor=west] {$\alpha >
        0^\circ$};

    \draw[thick,blue!50!black] (-1,\yc) -- (xc) -- (\Lx+1,\yc+.8);
    \draw[thick,blue!50!black] (-1,0) -- (\Lx+1,0);

    \draw[<->] (-.8,0) --++ (0,\yc) node[midway,left] {$y_c$};

    \draw[<->] (\Lx+.8,0) --++ (0,\Ly) node[midway,right] {$L_y$};
    \draw (\Lx+.3,\Ly) -- (\Lx+1,\Ly); 

    \node[above left,inner sep=1pt, outer sep=1pt] at (\xc,0) {$x_c$};

    \draw[<->] (0,-.6) --+ (\Lx,0) node[midway,above] {$L_x$};
    \draw (0,-.1-.6) -- (0,-.2); 
    \draw (\Lx,-.1-.6) -- (\Lx,-.2); 

    \end{footnotesize}

    \node[above] at (.8*\Lx,\Ly) {\scriptsize computational domain};
    \node[above] at (.1*\Lx,\yc) {\scriptsize inviscid top wall};

\end{tikzpicture}
    \caption{Computational setup for the APG/FPG TBL cases. The effect of the
        inviscid upper boundary is modeled through a prescribed
        wall-normal velocity profile given by Eq.~\eqref{eq:topramp}. FPG and APG
        TBL cases correspond to $\alpha < 0^\circ$ and $\alpha >
    0^\circ$, respectively.\label{fig:domain}}
\end{figure}

% APG/FPG TBLs
The different flow regimes (APG/FPG TBLs and separation bubbles) are
achieved by prescribing distinct wall-normal velocity profiles at the
top boundary, $v(x,L_y,z)$.  For the APG/FPG cases, we impose
\begin{align}\label{eq:topramp}
    v(x,L_y,z) &=
    \begin{cases}
        \dfrac{ K\left(y_c^2 \cos^2 \alpha+ L_y^2
        \sin^2 \alpha\right)}{ [ x - x_c + y_c/\tan
        \alpha ]^2 + L_y^2}, & \text{if } x > x_c, \\[2em]
        0, & \text{if } x \leq x_c,
    \end{cases}
\end{align}
where $K = 2 (L_y/y_c) U_\infty \sin(2\alpha)$, and $\alpha$, $x_c >
x_0$, and $y_c \geq L_y$ are parameters that approximate the potential
flow induced by a ramp of angle $\alpha$ located at $(x_c, y_c)$, with
$\alpha > 0^\circ$ corresponding to adverse pressure gradients (APG)
and $\alpha < 0^\circ$ to favorable pressure gradients (FPG).  The top
boundary conditions for the streamwise and spanwise velocity
components are obtained from the irrotational flow condition:
\begin{align}\label{eq:bcramp}
    \left. \frac{\partial u}{\partial y } \right|_{y=L_y} &=
    \left. \frac{\partial v}{\partial x } \right|_{y=L_y},  \quad
    \left. \frac{\partial w}{\partial y}\right|_{y=L_y} = 0.
\end{align}
For the APG cases with mean-flow separation, we impose a vertical
blowing velocity at $95\%$ of $L_x$ to promote reattachment of the
boundary layer. The influence of this velocity on the flow statistics
is quantified in \S\ref{app:validationdomain}.

% Summary of TBL cases
We consider a total of 16 APG/FPG TBLs and 3 turbulent
separation-bubble simulations. For the TBLs, we consider eight ramp
angles, $\alpha=-4^\circ$, $-3^\circ$, $-2^\circ$, $-1^\circ$,
$5^\circ$, $10^\circ$, $15^\circ$, and $20^\circ$, at each of two
inlet momentum-thickness Reynolds numbers, $\Rey_{\theta_i}=300$ and
$670$. At the recycling plane $x_0$, which marks the beginning of the
domain used to collect statistics, $\Rey_{\theta_0}$ is approximately
400 and 800 for the two inlet Reynolds numbers, respectively, as shown
in Table~\ref{tab:rampcases}. The table also reports the domain
dimensions and grid resolution. An assessment of numerical grid
resolution is presented in \S\ref{app:validationgrid}.
\begin{table}
    \begin{ruledtabular}
        \begin{tabular}{rcccc}
            $\alpha$ & $\Rey_{\theta_i}$ & $\Rey_{\theta_0}$ & $[L_x \times L_y
            \times L_z]/\delta_0$ & $N_x \times N_y \times N_z$ \\[3pt]
            \hline
            $-4^\circ$ & 300 & 398 & $105 \times ~2 \times 11$ &
            $8192\times350\times1536$ \\
            $-3^\circ$ & 300 & 396 & $106 \times ~2 \times 11$ &
            $6144\times320\times1024$ \\
            $-2^\circ$ & 300 & 401 & $107 \times ~2 \times 11$ &
            $5376\times300\times1024$ \\
            $-1^\circ$ & 300 & 398 & $106 \times ~2 \times 11$ &
            $4096\times280\times~768$ \\
            $~5^\circ$ & 300 & 468 & $125 \times 18 \times 18$ &
            $2048\times144\times1024$ \\
            $10^\circ$ & 300 & 461 & $126 \times 18 \times 18$ &
            $2048\times144\times1024$ \\
            $15^\circ$ & 300 & 455 & $130 \times 19 \times 19$ &
            $2048\times144\times1024$ \\
            $20^\circ$ & 300 & 449 & $133 \times 19 \times 19$ &
            $2048\times144\times1024$ \\
            $-4^\circ$ & 670 & 803 & $105 \times ~2 \times 11$ &
            $8192\times350\times1536$ \\
            $-3^\circ$ & 670 & 807 & $103 \times ~2 \times 10$ &
            $6144\times320\times1024$ \\
            $-2^\circ$ & 670 & 808 & $104 \times ~2 \times 10$ &
            $5376\times300\times1024$ \\
            $-1^\circ$ & 670 & 811 & $104 \times ~2 \times 10$ &
            $4096\times280\times~768$ \\
            $~5^\circ$ & 670 & 840 & $~84 \times 12 \times 12$ &
            $4096\times320\times1536$ \\
            $10^\circ$ & 670 & 832 & $134 \times 19 \times 19$ &
            $4096\times320\times1536$ \\
            $15^\circ$ & 670 & 821 & $142 \times 20 \times 20$ &
            $4096\times320\times1536$ \\
            $20^\circ$ & 670 & 812 & $148 \times 21 \times 21$ & $4096\times320\times1536$
        \end{tabular}
        \caption{Simulation parameters for all APG/FPG TBL cases. The
          table reports the ramp angle $\alpha$ and the Reynolds
          numbers $\Rey_{\theta_i}$ and $\Rey_{\theta_0}$, based on
          the momentum thicknesses at $x_i$ and $x_0$,
          respectively. The domain dimensions are nondimensionalized
          by $\delta_0$. The last column reports the numbers of grid
          points in the streamwise ($N_x$), wall-normal ($N_y$), and
          spanwise ($N_z$) directions.
          \label{tab:rampcases}}
    \end{ruledtabular}
\end{table}

\begin{figure}[h!]
    \centering
    \begin{tikzpicture}[>={Latex[length=.15cm]},scale=.9]

    \pgfmathsetmacro{\xL}{12}
    \pgfmathsetmacro{\yL}{2}
    \pgfmathsetmacro{\xO}{7}
    \pgfmathsetmacro{\sg}{2.}
    \pgfmathsetmacro{\pang}{130}

    \fill[orange!10!white] (0,0) rectangle (\xO*.6,\yL) node[midway,black,above] {ZPG}
    node[inner sep=1pt, outer sep=1pt,pos=0,black,above left] {$x_i$};
    \fill[blue!10!white] (\xO*.6,0) rectangle (\xO,\yL) node[midway,black,above] {APG};
    \fill[red!10!white] (\xO,0) rectangle ++(.4*\xO,\yL) node[midway,black,above] {FPG};
    \fill[orange!10!white] (1.4*\xO,0) rectangle (\xL,\yL) 
    node[midway,black,above] {ZPG};

    \draw[thin,dashed] (0,0) rectangle (\xL,\yL);

    \draw[fill=cyan!10!white] (0,\yL) --++(\xO-\sg,0) .. controls ++(0:1) 
    and ++(\pang:1) .. (\xO,\yL) .. controls ++(\pang:-1) and ++(0:-1) .. (\xO+\sg,\yL)
    -- (\xL,\yL) -- cycle; 

    \node[anchor=south west,cyan] at (\xO,\yL) {$v(x,L_y,z)$};
    \draw[dotted,black!50] (\xO,0) -- (\xO,\yL) 
    node[black,pos=0,above left] {$x_p$};

    \draw[red!40!black,densely dotted] (\xO*.4,0) --+ (0,\yL) 
        node[inner sep=1pt, outer sep=1pt,text=black,pos=0,above left] {$x_0$};

    \draw[<-,thin,red!40!black] (0,.3*\yL) to [bend left=20] +(\xO*.4,0);
    \node[inner sep=1pt, outer sep=1pt,fill=orange!10!white,text=red!40!black] 
        at (\xO*.2,.3*\yL) {\scriptsize recycle};

    %\draw (0,0) .. controls ++(10:.5) and ++(270:.3) .. (.5,.3)
    %-- (.5,\yL) node [pos=1,anchor=south] {$U_\infty$};
    %\foreach \i in {.15,.3,...,1} {
    %    \draw[thin,-{Latex[length=.1cm]}] (0,\yL*\i) --++(.5,0);
    %}

    \draw[black,<->] (\xL+.4,0) --++(0,\yL) node[pos=0.5,anchor=west] {$L_y$};
    \draw[black,-] (\xL+.2,0) --++ (.3,0);
    \draw[black,-] (\xL+.2,\yL) --++ (.3,0);

    %\draw[thick,blue] 
    %(1,.2).. controls ++(0:4) and ++(30:-.2) .. 
    %(\xO-1,.8) .. controls ++(30:1) and ++(160:.2) .. 
    %(\xO+1,.7) .. controls ++(160:-1) and ++(0:-1) ..
    %(\xL-.5,.5);
    %\draw[thick,blue,dashed] (\xO-.1,.5) ellipse (.6 and .3);
    %\draw[dashed] (1,.2) -- (\xO,.3);

    \draw[thick,blue!50!black] (-1,0) -- (\xL+1,0);
\end{tikzpicture}
    \caption{Computational setup for the separation bubbles. The effect of the
        inviscid upper boundary is modeled through the
        wall-normal velocity profile given by Eq.~\eqref{eq:bcsep}.
    \label{fig:BLdomain}}
\end{figure}

% Separation bubbles
For the turbulent separation bubbles, we adopt the top boundary
condition proposed by \citet{coleman2018numerical}:
\begin{equation}\label{eq:bcsep}
    v(x,L_y,z) =  -\sqrt{2} \, V_{\max} \, \tilde{x}
    \exp \left( \frac{1}{2} - \tilde{x}^2\right) +
    \phi_{\text{top}},
\end{equation}
where $\tilde{x} = (x - x_p)/\sigma$, and $V_{\max}$, $x_p$, and
$\sigma$ control the strength, location, and extent of the
adverse-to-favorable pressure gradient distribution, respectively (see
figure~\ref{fig:BLdomain}).
The term $\phi_{\text{top}} = 0.0034 \, U_\infty$ is a small uniform
transpiration included for numerical
stability~\citep{coleman2018numerical}.
The three separation-bubble simulations use
$(V_{\max}/U_\infty,\sigma/\delta_i)=(0.28,14.5)$, $(0.30,14.5)$, and
$(0.40,11.75)$, where $\delta_i$ is the boundary-layer thickness at
the inlet $x_i$. All three cases exhibit instantaneous flow reversal;
SB2 and SB3 also exhibit mean-flow separation.  The Reynolds numbers
and details of the computational grid and domain are provided in
Table~\ref{tab:sbcases}.
\begin{table}
    \begin{ruledtabular}
        \begin{tabular}{lccccc}
            & $(V_{\max}/U_\infty, \sigma/\delta_i)$ & $\Rey_{\theta_i}$ &
            $\Rey_{\theta_0}$ & $[L_x \times L_y \times L_z]/\delta_i$ &
            $N_x \times N_y \times N_z$ \\[3pt]
            \hline
            SB1 &$(0.28, 14.5)$  & 670 & 1114 & $92 \times 11 \times 12$ &
            $2048\times1024\times512$ \\
            SB2 &$(0.30, 14.5)$  & 670 & 1110 & $92 \times 11 \times 12$ &
            $2048\times1024\times512$ \\
            SB3 &$(0.40, 11.75)$ & 670 & 1090 & $92 \times 11 \times 12$ &
            $2048\times1024\times512$ \\
        \end{tabular}
        \caption{Simulation parameters for all turbulent
          separation-bubble cases. The table reports the dimensionless
          maximum-velocity and width parameters of the upper-boundary
          velocity profile and the Reynolds numbers $\Rey_{\theta_i}$
          and $\Rey_{\theta_0}$, based on the momentum thicknesses at
          $x_i$ and $x_0$, respectively.  The domain dimensions are
          nondimensionalized by $\delta_i$. The last column reports
          the numbers of grid points in the streamwise ($N_x$),
          wall-normal ($N_y$), and spanwise ($N_z$) directions.
          \label{tab:sbcases}}
    \end{ruledtabular}
\end{table}
\begin{figure}
    \begin{tikzpicture}
        \node (f1) at (0,0) {\ig[width=\tw]{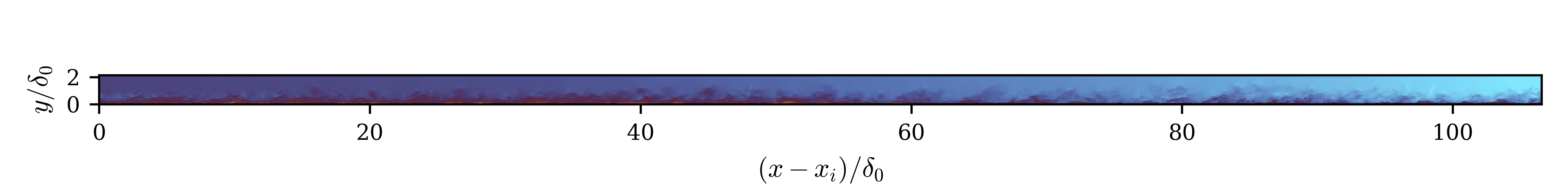}};
        \node[anchor=north west] (f2) at (f1.south west)
        {\ig[width=.5\tw]{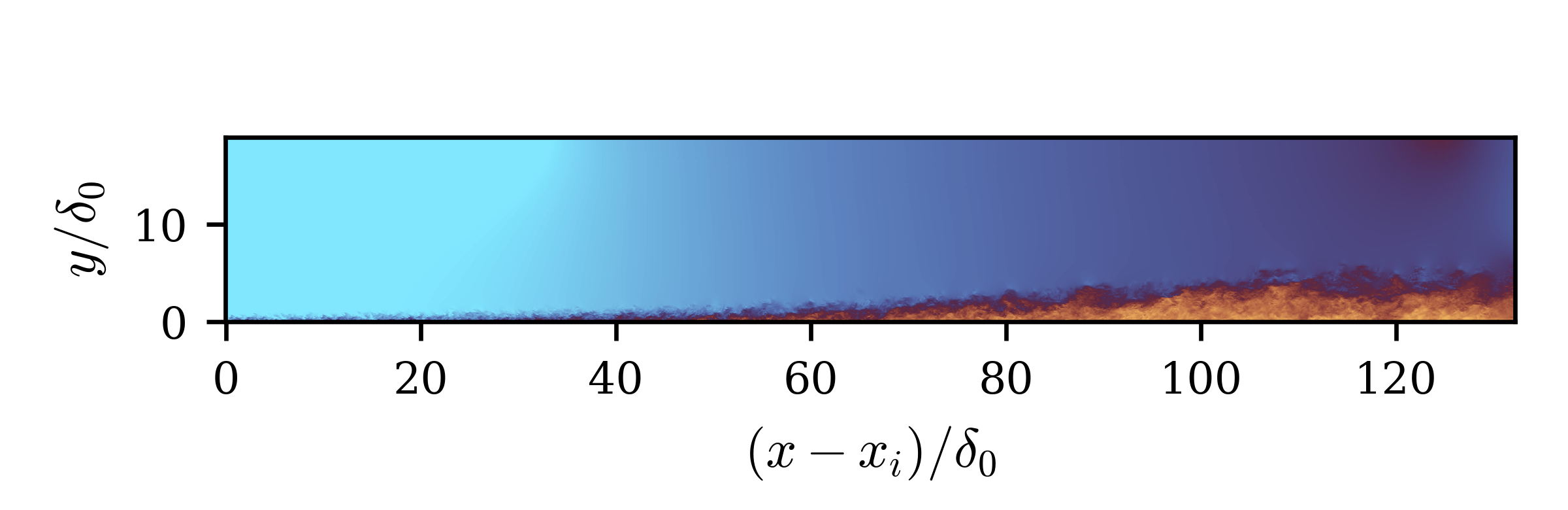}};
        \node[anchor=north east] (f3) at (f1.south east)
        {\ig[width=.5\tw]{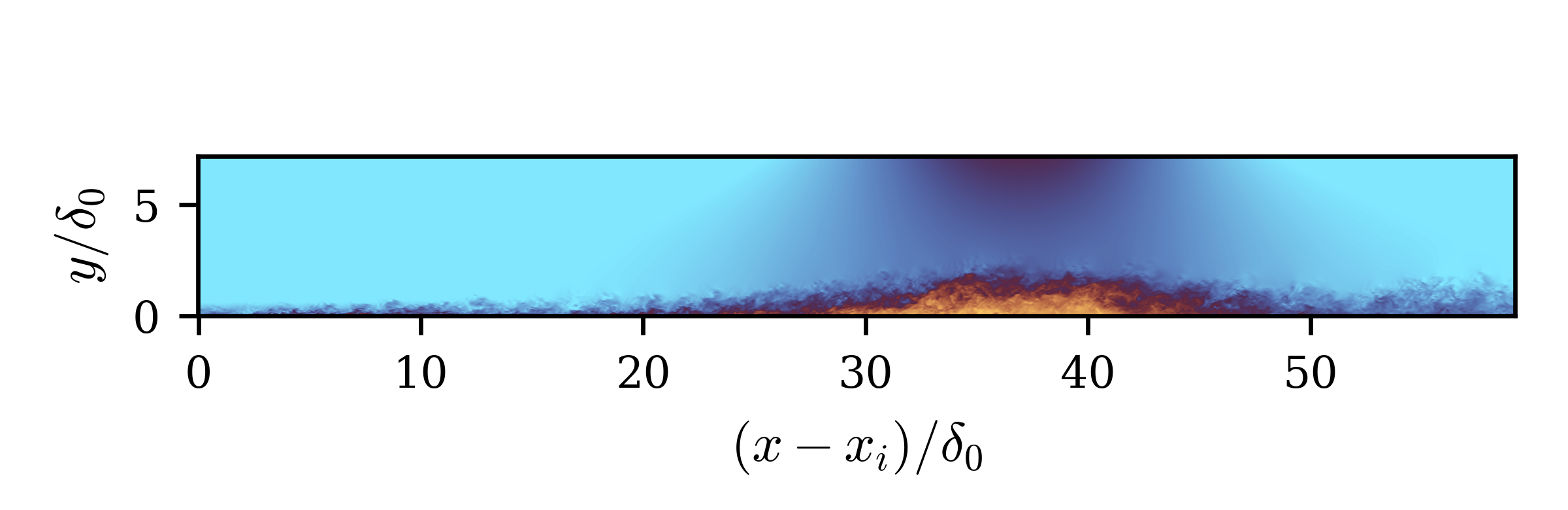}};

        \begin{footnotesize}
            \begin{scope}[shift=(f1.south west),x=(f1.south east),y=(f1.north west)]
                \node at (.05,.9) {(a)};
                \node at (.05,-.2) {(b)};
                \node at (.55,-.2) {(c)};
            \end{scope}

            \node (fc) at (0,-13em)
            {\ig[width=.35\tw]{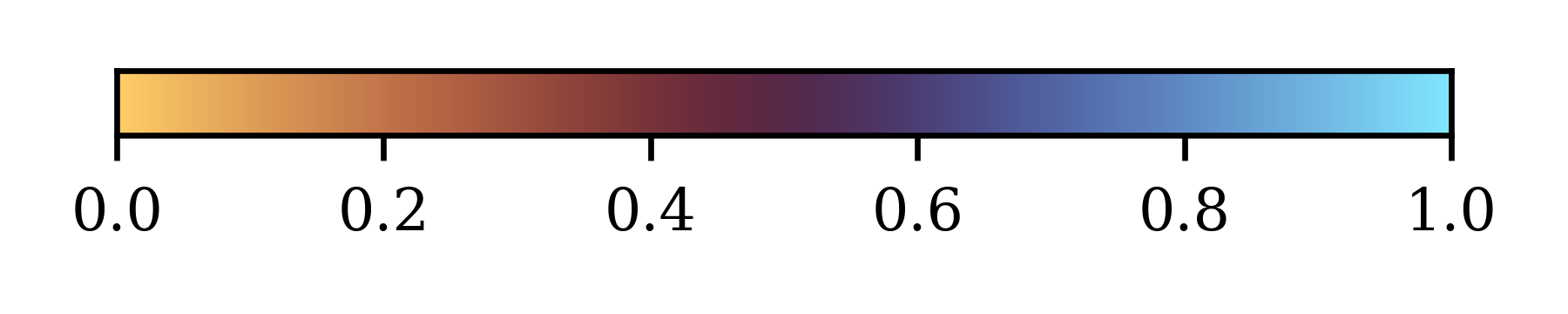}};
            \node at (fc.south) {$u/U_{e,0}$};
        \end{footnotesize}
    \end{tikzpicture}
    \caption{Instantaneous streamwise velocity $u$ in a plane of
      constant $z$ for (a) an FPG case with $\alpha=-3^\circ$ and
      $\Rey_{\theta_i}=300$; (b) an APG case with $\alpha=10^\circ$
      and $\Rey_{\theta_i}=670$; and (c) the separation-bubble case
      SB1. Contour levels range from $u/U_{e,0}=0$ to $1$ in panels
      (b) and (c), and from $0$ to $1.6$ in panel (a).
\label{fig:flowvis}}
\end{figure}
%input_Retheta_300_theta_-4deg_medium.turbb:nxyz = 8194 350 1538 ! (*)
%input_Retheta_300_theta_-3deg_medium.turbb:nxyz = 6146 320 1026 ! (*)
%input_Retheta_300_theta_-2deg_medium.turbb:nxyz = 5378 300 1026 ! (*)
%input_Retheta_300_theta_-1deg_medium.turbb:nxyz = 4098 280 770 ! (*)
%input_Retheta_300_theta_10deg_coarse.turbb:nxyz = 2050 144 1026 ! (*)
%input_Retheta_300_theta_15deg_coarse.turbb:nxyz = 2050 144 1026 ! (*)
%input_Retheta_300_theta_20deg_coarse.turbb:nxyz = 2050 144 1026 ! (*)
%input_Retheta_300_theta_5deg_coarse.turbb:nxyz =  2050 144 1026 ! (*)
%input_Retheta_670_theta_-4deg_medium.turbb:nxyz = 8194 350 1538 ! (*)
%input_Retheta_670_theta_-3deg_medium.turbb:nxyz = 6146 320 1026 ! (*)
%input_Retheta_670_theta_-2deg_medium.turbb:nxyz = 5378 300 1026 ! (*)
%input_Retheta_670_theta_-1deg_medium.turbb:nxyz = 4098 280 770 ! (*)
%input_Retheta_670_theta_10deg_medium.turbb:nxyz = 4098 320 1538 ! (*)
%input_Retheta_670_theta_15deg_medium.turbb:nxyz = 4098 320 1538 ! (*)
%input_Retheta_670_theta_20deg_medium.turbb:nxyz = 4098 320 1538 ! (*)
%input_Retheta_670_theta_5deg_medium.turbb:nxyz =  4098 320 1538 ! (*)

%--------------------------------------------------------------%
\subsection{Statistics of the current TBLs and separation bubbles}
%--------------------------------------------------------------%

\begin{figure}
    \centering
    \begin{subfigure}{.4\tw}
        \centering
        \ig[width=\tw]{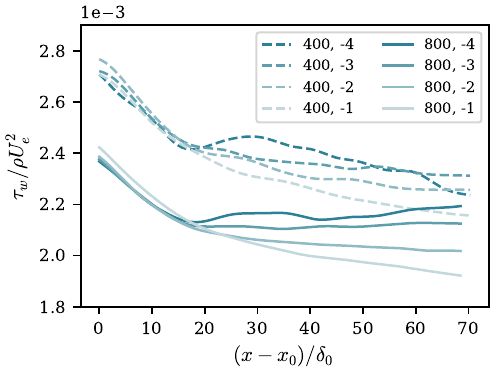}
        \caption{\label{fig:cframpFPG}}
    \end{subfigure}\hfill
    \begin{subfigure}{.4\tw}
        \centering
        \ig[width=\tw]{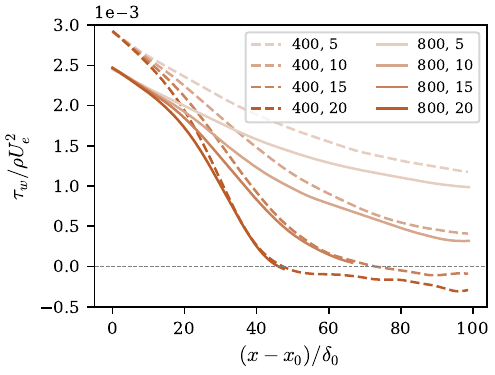}
        \caption{\label{fig:cframpAPG}}
    \end{subfigure}
    \caption{Streamwise evolution of $\ptauo=\tau_w/(\rho U_e^2)$ in
      (a) FPG and (b) APG TBLs.  Dashed and solid lines correspond to
      $\Rey_{\theta_0}\approx400$ and $\Rey_{\theta_0}\approx800$,
      respectively.
\label{fig:Cframp}}
\end{figure}

\begin{figure}
    \centering
    \begin{tikzpicture}
        \node (f1) at (0,0) {\ig[width=\tw]{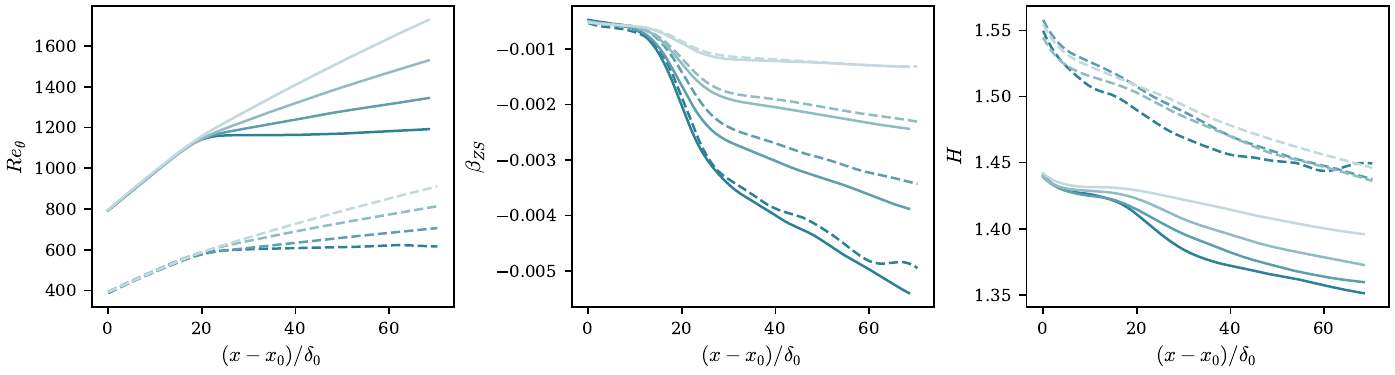}};
        \node[anchor=north west] (f2) at (f1.south west)
        {\ig[width=\tw]{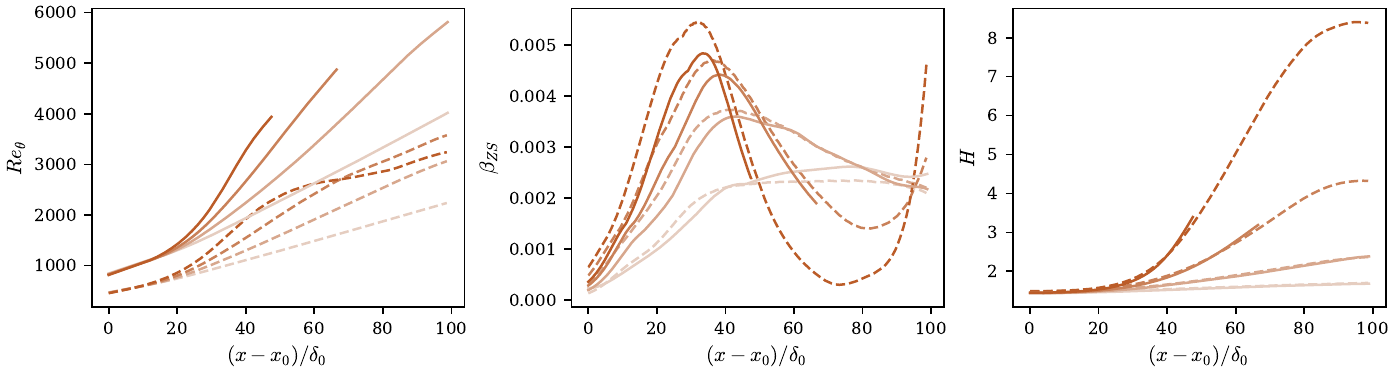}};
        \begin{scope}[shift=(f1.south west),x=(f1.south east),y=(f1.north west)]
            \node at (.10,.87) {\scriptsize(a)};
            \node at (.635,.87) {\scriptsize(b)};
            \node at (.96,.87) {\scriptsize(c)};
        \end{scope}
        \begin{scope}[shift=(f2.south west),x=(f2.south east),y=(f2.north west)]
            \node at (.100,.87) {\scriptsize(d)};
            \node at (.430,.87) {\scriptsize(e)};
            \node at (.745,.87) {\scriptsize(f)};
        \end{scope}
    \end{tikzpicture}
    \caption{Streamwise evolution of (a,d) the momentum-thickness
      Reynolds number, (b,e) the pressure-gradient parameter with
      Zagarola--Smits scaling~\cite{maciel2006selfsimilarity}, and
      (c,f) the shape factor. Panels (a)--(c) show the FPG TBLs, and
      panels (d)--(f) show the APG TBLs. The legend is the same as in
      Figure~\ref{fig:Cframp}.
      \label{fig:Statsramp}}
\end{figure}

% TBLs
Figure~\ref{fig:Cframp} displays the evolution of the mean
dimensionless wall-shear stress, $\ptauo = \tau_w /(\rho U_e^2)$,
along the streamwise direction for the different TBLs.  For all the
FPG cases (figure~\ref{fig:cframpFPG}), $\ptauo$ decreases in the ZPG
region.  For the stronger favorable pressure gradients, this is
followed by an increase in the FPG region, whereas some milder-FPG
cases continue to decrease over the interval shown. The APG cases
(figure~\ref{fig:cframpAPG}) also show this initial decrease, but it
is less noticeable because $\ptauo$ continues to decrease in the APG
region. For both $\Rey_{\theta_0} \approx 400$ and $800$, the flow
separates ($\ptauo \leq 0$) when $\alpha \geq 15^\circ$.
Figure~\ref{fig:Statsramp} displays the evolution of $\Rey_{\theta}$,
the pressure-gradient parameter $\beta_{ZS} =
-(\delta/U_{ZS})\mathrm{d}U_e/\mathrm{d}x$~\cite{maciel2006selfsimilarity},
where $U_{ZS} = U_e \dOne / \delta$ is the Zagarola-Smits
velocity~\cite{zagarola1997scaling}, and the shape factor $H = \dOne /
\dTwo$.

\begin{figure}
    \centering
    \begin{tikzpicture}
        \node (f) at (0,0) {\ig[width=.9\tw]{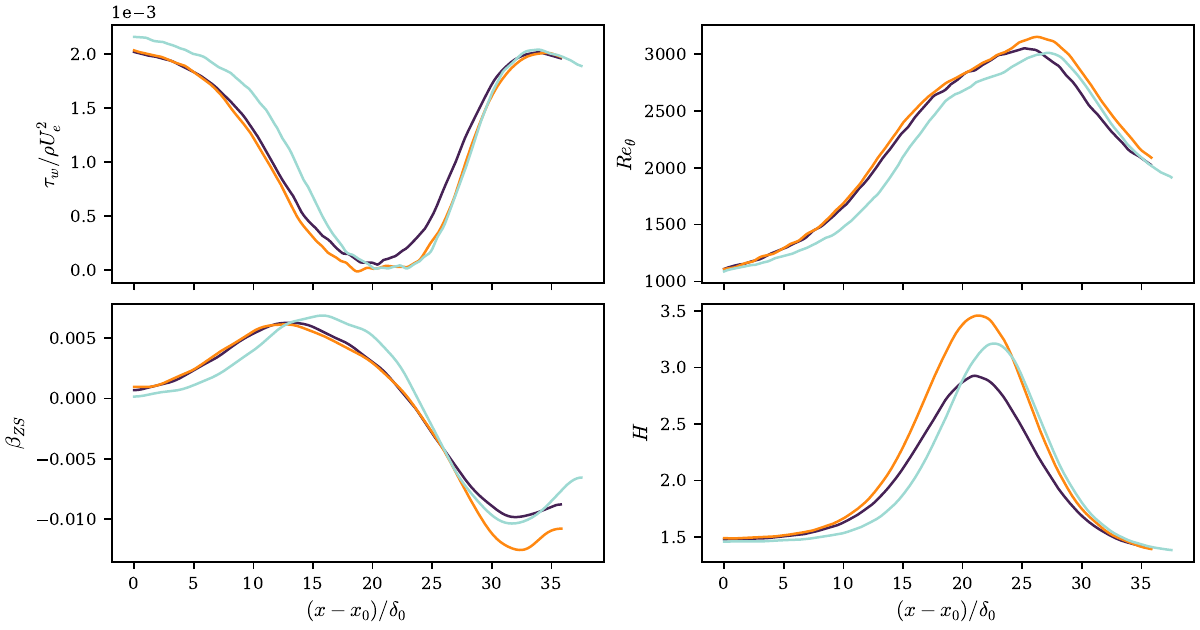}};
        \begin{scriptsize}
            \begin{scope}[shift=(f.south west), x=(f.south east), y=(f.north west)]
                \node at (0.12,.9) {(a)};
                \node at (0.61,.9) {(b)};
                \node at (0.12,.47) {(c)};
                \node at (0.61,.47) {(d)};
            \end{scope}
        \end{scriptsize}
    \end{tikzpicture}
\caption{Streamwise evolution of (a)
    $\ptauo=\tau_w/(\rho U_e^2)$, (b) the momentum-thickness Reynolds
    number, (c) the pressure-gradient parameter with Zagarola--Smits
    scaling, and (d) the shape factor for the turbulent separation
    bubbles. Purple, orange, and cyan lines denote SB1, SB2, and SB3,
    respectively.
\label{fig:sbstats}}
\end{figure}

% Separation bubbles
Figure~\ref{fig:sbstats}a shows the evolution of the dimensionless
wall-shear stress for the turbulent separation bubbles. All cases
exhibit instantaneous flow reversal, whereas SB2 and SB3 also exhibit
mean-flow separation. Panels (b)--(d) show the evolution of the
momentum-thickness Reynolds number, the pressure-gradient parameter
$\beta_{ZS}$, and the shape factor $H$, respectively.

%--------------------------------------------------------------%
\subsection{Effect of computational domain size}
\label{app:validationdomain}
%--------------------------------------------------------------%

% Intro
For APG TBLs with separation ($\alpha \ge 15^\circ$), a downward
velocity component is imposed near the domain exit to ensure that
reattachment occurs within the computational domain. In this section,
we assess the extent to which this boundary treatment influences the
upstream flow.

% Method
To quantify this influence, we performed an additional simulation for
the TBL at $\alpha = 15^\circ$ and $\Rey_{\theta_0} \approx 800$,
which exhibits strong separation. The streamwise domain was
truncated to two-thirds of its original length while maintaining
identical numerical parameters and boundary conditions, except for the
repositioned outflow boundary.

\begin{figure}
    \centering
    \ig[width=0.8\tw,trim=0 .4cm 0 0, clip]{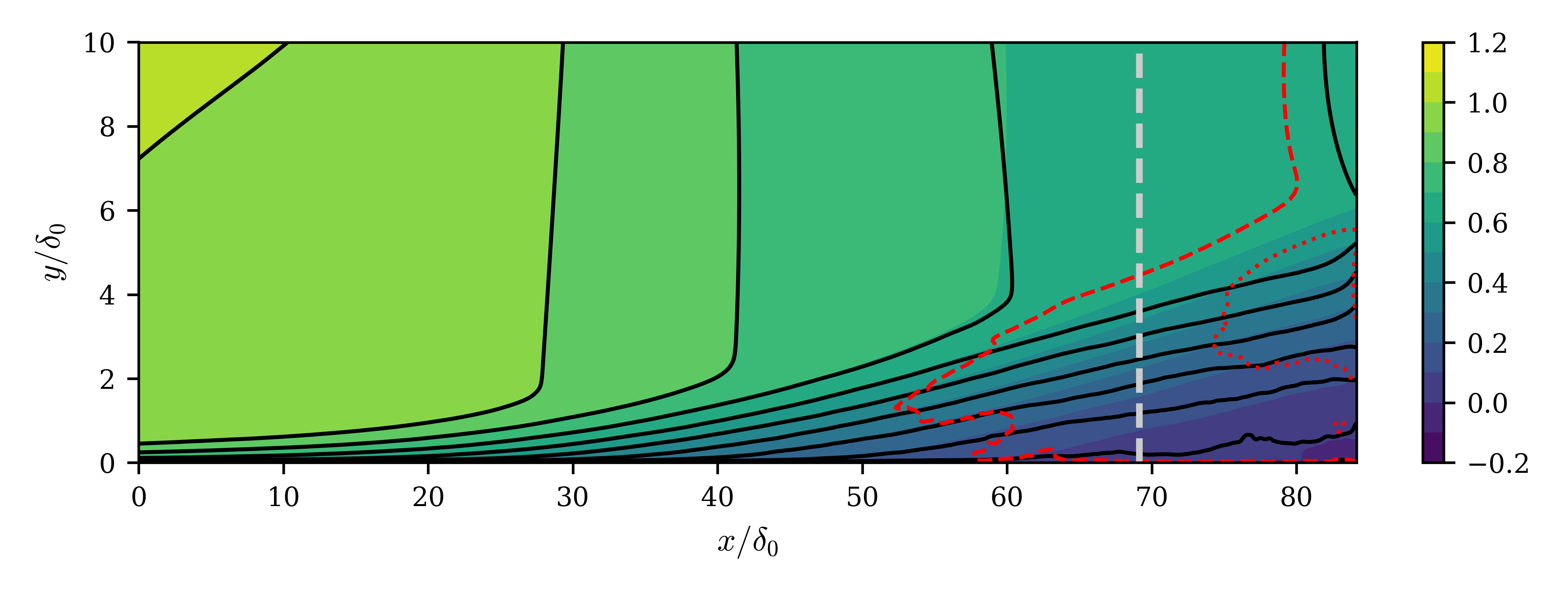}
    \caption{Domain size influence for the $\alpha = 15^\circ$ case at
      $\Rey_{\theta_0} \approx 800$. Background colormap: streamwise
      velocity $U/U_{e,i}$ from the full-length simulation. Black
      contours: corresponding velocity field from the
      two-thirds-domain simulation. Red contours indicate relative
      velocity differences $|U - U_{2/3}|/U_{e,i}$ of 0.1 (dotted
      line) and 0.02 (dashed line). The vertical gray dashed line
      marks the end of the domain used to collect data.
    \label{fig:domaineffect}}
\end{figure}

% Results
Figure~\ref{fig:domaineffect} presents a comparison between the
full-length and truncated-domain simulations. The streamwise velocity
field $U/U_{e,i}$ from the original domain is shown as a colormap
(where $U_{e,i}$ is the edge velocity at $x_i$), with black contour
lines representing the field from the two-thirds-domain simulation,
denoted by $U_{2/3}$. Red contours delineate regions where the
absolute difference $|U - U_{2/3}|$ exceeds a specified threshold:
the dotted line indicates differences greater than $0.1U_{e,i}$, while
the dashed line marks differences exceeding $0.02U_{e,i}$.

% Conclusion
The comparison reveals that significant differences ($> 0.1U_{e,i}$)
are confined to the immediate vicinity of the relocated boundary. The
flow fields remain virtually identical up to $x/\delta_0 \approx 50$,
corresponding to 60\% of the reduced computational domain. Velocity
differences remain below $0.02U_{e,i}$ up to $x/\delta_0 \approx 60$,
or 70\% of the truncated domain length. Based on these findings, we
restrict our analysis of APG cases to $x \leq 0.7L_x$, ensuring that
the extracted flow statistics remain unaffected by the treatment at
the outflow boundary.

%-------------------------------------------------------------------%
\subsection{Assessment of numerical grid resolution}
\label{app:validationgrid}
%-------------------------------------------------------------------%

% Intro
We evaluate the resolution of the computational grid by examining both
the local grid spacing relative to the Kolmogorov scale and the
spectral content of the resolved velocity and enstrophy fields.

% Grid/Kolmogorov
First, we compare the wall-normal and spanwise grid spacings,
$\Delta y$ and $\Delta z$, with the local Kolmogorov length scale,
\begin{equation}
    \eta(x,y)=\left(\frac{\nu^3}{\varepsilon}\right)^{1/4},
\end{equation}
where
$\varepsilon=(\nu/2)\sum_{i,j}
(\partial u_i/\partial x_j+\partial u_j/\partial x_i)^2$
is the kinetic-energy dissipation rate. The largest normalized grid
spacings typically used in DNS of turbulent channel flows and ZPG
TBLs, based on existing databases~\citep{torroja}, are of order
$\Delta y/\eta\approx2$ and $\Delta z/\eta\approx5$.

\begin{figure}
    \centering
    \begin{tikzpicture}
        \node[anchor=south west] (f1) at (0,0)
        {\ig[width=0.95\tw]{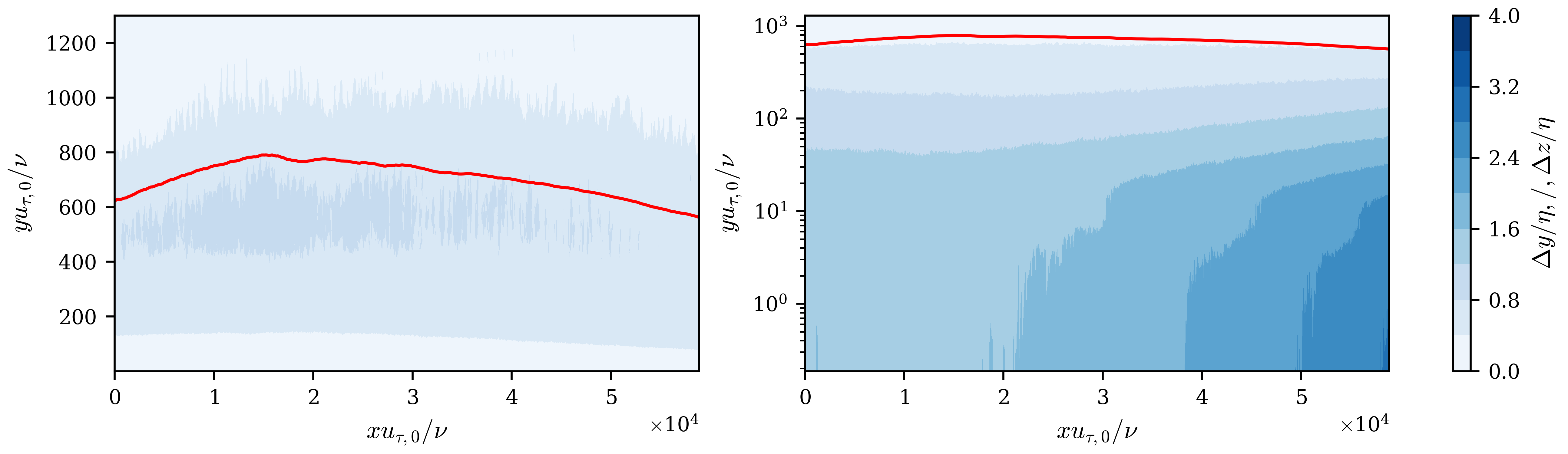}};
        \node[anchor=north west] (f2) at (f1.south west)
        {\ig[width=0.95\tw]{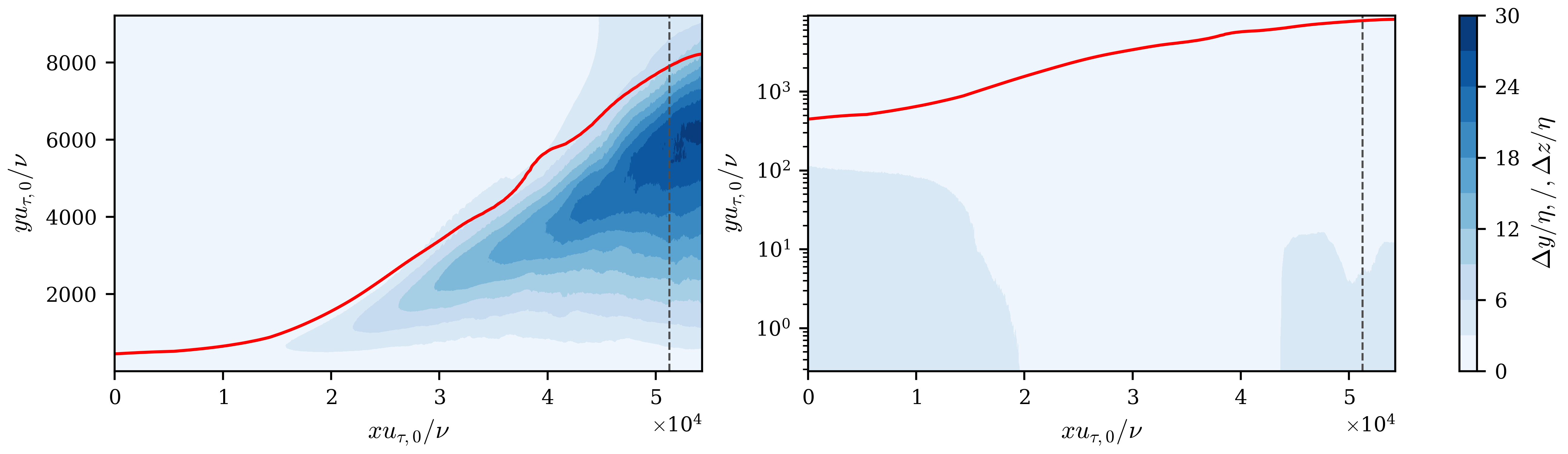}};
        \begin{scope}[shift=(f1.south west), x=(f1.south east), y=(f1.north west)]
            \node at (.105,.88) {(a)};
            \node at (.535,.88) {(b)};
            \node[fill=white] at (.7,.07) {\scriptsize$x u_{\tau,0}/\nu$};
            \node[fill=white] at (.27,.07) {\scriptsize$x u_{\tau,0}/\nu$};
            \node[rotate=90,fill=white] at (.02,.58) {\scriptsize$y u_{\tau,0}/\nu$};
            \node[rotate=90,fill=white,outer sep=0pt, inner sep=0pt]
            at (.465,.58) {\scriptsize$y u_{\tau,0}/\nu$};
            \node[rotate=90,fill=white] at (.98,.57)
            {\scriptsize$\Delta y/\eta,\Delta z/\eta$};
        \end{scope}
        \begin{scope}[shift=(f2.south west), x=(f2.south east), y=(f2.north west)]
            \node at (.11,.88) {(c)};
            \node at (.54,.88) {(d)};
            \node[fill=white] at (.7,.07) {\scriptsize$x u_{\tau,0}/\nu$};
            \node[fill=white] at (.27,.07) {\scriptsize$x u_{\tau,0}/\nu$};
            \node[rotate=90,fill=white,outer sep=0pt, inner sep=0pt]
            at (.023,.58) {\scriptsize$y u_{\tau,0}/\nu$};
            \node[rotate=90,fill=white,outer sep=0pt, inner sep=0pt]
            at (.47,.58) {\scriptsize$y u_{\tau,0}/\nu$};
            \node[rotate=90,fill=white] at (.98,.57)
            {\scriptsize$\Delta y/\eta,\Delta z/\eta$};
        \end{scope}
    \end{tikzpicture}
\caption{Grid spacings in Kolmogorov units for
    (a,b) $\alpha=-4^\circ$ and (c,d) $\alpha=20^\circ$.
    Panels (a,c) show $\Delta y/\eta$, and panels (b,d) show
    $\Delta z/\eta$. In panels (c,d), the dashed line marks the limit
    of the domain used to discover the scaling
    (see~\S\ref{app:validationdomain}). In every panel, the red line
    denotes $\delta(x)$. The streamwise and wall-normal coordinates
    are nondimensionalized by the viscous length $\nu/u_{\tau,0}$,
    where $u_{\tau,0}=\sqrt{\tau_{w,0}/\rho}$ is the friction
    velocity at $x_0$.
\label{fig:kolmogorov}}
\end{figure}

% Results
Figures~\ref{fig:kolmogorov}(a,b) and \ref{fig:kolmogorov}(c,d) show
the FPG ($\alpha=-4^\circ$) and APG ($\alpha=20^\circ$) cases,
respectively, which have the most stringent grid requirements among
the present cases.  Panels (a,c) report $\Delta y/\eta$, and panels
(b,d) report $\Delta z/\eta$.  The grid resolution for the FPG case
(figures~\ref{fig:kolmogorov}(a,b)) falls within the reference values
reported in the literature. For the APG case
(figures~\ref{fig:kolmogorov}(c,d)), the near-wall grid resolution is
similar, but $\Delta y \approx 30\eta$ in the outer region. However,
dissipation is weak within the separation regions, and $\Delta y/\eta$
may not be a representative indicator of whether the flow motions are
actually resolved by the grid.

\begin{figure}
    \centering
    \begin{tikzpicture}
        \node (f1) at (0,0)
        {\ig[width=0.6\tw]{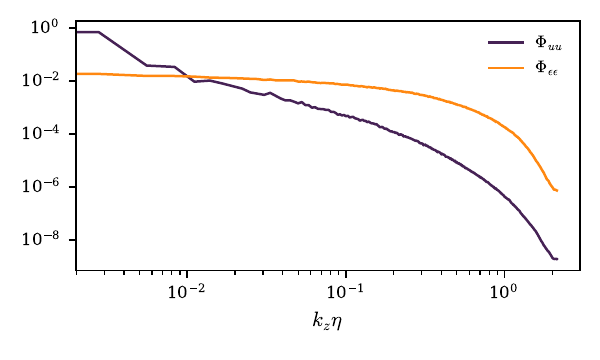}};
        \begin{scope}[shift=(f1.south west), x=(f1.south east), y=(f1.north west)]
            \node[fill=white,anchor=south] at (0.55,.08) {\footnotesize$k_z \eta$};
            \node[fill=white,rotate=90] at (0.0,.60)
            {\footnotesize$\hat{\Phi}_{uu},\,\hat{\Phi}_{\mathcal{E}\mathcal{E}}$};

            \node[fill=white,inner sep=3pt,outer sep=0pt] (ww) at (0.91,.79)
            {\scriptsize$\hat{\Phi}_{\mathcal{E}\mathcal{E}}$};
            \node[fill=white,inner sep=3pt,outer sep=0pt,anchor=south]
            at (ww.north) {\scriptsize$\hat{\Phi}_{uu}$};
        \end{scope}
    \end{tikzpicture}
    \caption{Normalized spectra at the location of maximum $\Delta
        y/\eta$ for the APG case ($\alpha = 20^\circ$): (purple)
        streamwise velocity spectrum $\hat{\Phi}_{uu}(k_z)$ and (orange)
        enstrophy spectrum $\hat{\Phi}_{\mathcal{E}\mathcal{E}}(k_z)$
    as functions of the spanwise wavenumber, $k_z$.\label{fig:enstrophy}}
\end{figure}

% Spectra
To assess whether the coarser resolution in the outer region affects
the accuracy of our simulations, we examine the spanwise power
spectral densities of the streamwise velocity fluctuations,
$\Phi_{uu}(k_z)$, and the enstrophy,
$\Phi_{\mathcal{E}\mathcal{E}}(k_z)$, where
$\mathcal{E}=\frac{1}{2}|\nabla\times\mathbf{u}|^2$ is the enstrophy
density and $k_z$ is the spanwise wavenumber. The spectra are computed
using a Fourier transform in the spanwise direction and normalized
such that $\int_0^{k_{z,\mathrm{max}}}
\hat{\Phi}_{uu}(k_z)\,\mathrm{d}k_z=1$, where
$\hat{\Phi}_{uu}=\Phi_{uu}/\int\Phi_{uu}\,\mathrm{d}k_z$.  The
enstrophy spectrum is normalized analogously. The results are shown in
Figure~\ref{fig:enstrophy}. The exponential decay of the spectra at high
wavenumbers, together with the absence of energy pileup near the
grid-cutoff wavenumber, suggests that the grid resolution is
sufficient for dissipation to remove energy at the smallest resolved
scales.

%--------------------------------------------------------------%
\subsection{Overview and comparison with TBLs from the literature}
\label{app:previous}
%--------------------------------------------------------------%

% Intro
The databases from the literature used in our study are summarized in
Table~\ref{tab:literaturecases}. For each case, the table reports the
ranges of the friction Reynolds number $\Rey_\tau=u_\tau\delta/\nu$,
the momentum-thickness Reynolds number $\Rey_{\dTwo}=U_e\dTwo/\nu$,
the shape factor $H$, the dimensionless wall-shear stress
$\Pi_o^\tau=\tau_w/(\rho U_e^2)$, \corr{and the signed
  pressure-gradient parameter $\gamma_P\beta_*$}.

% Comments
The cases from \citet{bobke2017history} contain TBLs under
mild-to-moderate APG. The TBLs from \citet{gungor2016scaling,
  gungor2022energy, gungor2024response} correspond to highly
nonequilibrium turbulent boundary layers developing under a mean
pressure gradient. The cases from \citet{coleman2018numerical} contain
turbulent separation bubbles generated by applying suction and blowing
at the top boundary of a developing boundary layer. Subsequently,
\citet{coleman2021numerical} considered a similar setup with
suction-only transpiration at the top wall, in line with the
configuration of \citet{wu2019spatiotemporal}.

\begin{sidewaystable}[p]
    \begin{ruledtabular}
        \providecommand\ran[2]{$#1\; - \; #2$}
\begin{tabular}{@{}w{l}{4cm}w{c}{3cm}w{c}{3cm}w{c}{3cm}w{c}{3cm}w{c}{3cm}@{}}
  & $\Rey_\tau$ & $\Rey_{\dTwo}$ & $H$
  & $\gamma_P \beta_{*} \times 100$ & $\ptauo \times 1000$ \\
  \hline

  %1.51 3.12
  %-1.21 2.33
  %1.35 2.17
  \citet{maciel2016coherent}  & \ran{~25}{~444} & \ran{1238}{~3849} &
  \ran{1.50}{3.43} & \corr{\ran{1.51}{3.12}} & \ran{-0.001}{1.79} \\
  \citet{gungor2022energy}  & \ran{336}{~642} & \ran{2066}{~6576} &
  \ran{1.50}{3.13} & \corr{\ran{-1.21}{2.33}} & \ran{~0.063}{1.53} \\
  \citet{gungor2024response}  & \ran{714}{2334} & \ran{2421}{12004} &
  \ran{1.41}{2.88} & \corr{\ran{1.35}{2.17}} & \ran{~0.113}{1.69} \\
  \citet{coleman2018numerical}~18B  & \ran{~~0}{1226} & \ran{~576}{~5346} &
  \ran{1.29}{3.60} & \corr{\ran{-1.78}{1.67}} & \ran{-0.022}{2.45} \\
  \citet{coleman2018numerical}~18C  & \ran{~~0}{2053} & \ran{1106}{10723} &
  \ran{1.26}{3.72} & \corr{\ran{-1.76}{1.62}} & \ran{-0.038}{2.08} \\
  %
  % We did not use 21D for velocity because there is no data.
  %\citet{coleman2021numerical}~21D  & \ran{~51}{1264} & \ran{~697}{13877} &
  %\ran{1.43}{3.47} & \corr{\ran{}{}} & \ran{~0.001}{2.45} \\
  \citet{coleman2021numerical}~21E  & \ran{330}{2651} & \ran{1488}{31532} &
  \ran{1.36}{3.37} & \corr{\ran{0.00}{1.17}} & \ran{~0.007}{2.08} \\

  \citet{bobke2017history} b1n  & \ran{199}{868} & \ran{407}{3186} &
  \ran{1.51}{1.63} & \corr{\ran{0.00}{0.19}} & \ran{1.35}{2.754} \\
  \citet{bobke2017history} b2n  & \ran{195}{896} & \ran{402}{3720} &
  \ran{1.57}{1.72} & \corr{\ran{-0.02}{0.29}} & \ran{1.122}{2.761} \\
  \citet{bobke2017history} m13n & \ran{201}{896} & \ran{415}{3324} &
  \ran{1.51}{1.69} & \corr{\ran{0.00}{0.26}} & \ran{1.325}{2.709} \\
  \citet{bobke2017history} m16n & \ran{197}{916} & \ran{409}{3773} &
  \ran{1.56}{1.79} & \corr{\ran{-0.04}{0.35}} & \ran{1.07}{2.705} \\
  \citet{bobke2017history} m18n & \ran{198}{944} & \ran{402}{4115} &
  \ran{1.57}{1.91} & \corr{\ran{-0.11}{0.43}} & \ran{0.853}{2.767} \\

  \corr{\citet{kitsios2017direct}} $\beta \approx 0$ & \corr{\ran{1251}{1451}} &
  \corr{\ran{3500}{3880}} & \corr{\ran{1.38}{1.39}} & \corr{\ran{0}{0}} &
  \corr{\ran{1.520}{1.565}} \\
  \corr{\citet{kitsios2017direct}} $\beta \approx 1$ & \corr{\ran{781}{868}} &
  \corr{\ran{3100}{3440}} & \corr{\ran{1.57}{1.58}} & \corr{\ran{0.13}{0.13}} &
  \corr{\ran{1.250}{1.274}} \\
  \corr{\citet{kitsios2017direct}} $\beta \approx 39$ & \corr{\ran{667}{711}} &
  \corr{\ran{10000}{12300}} & \corr{\ran{2.25}{2.38}} &
  \corr{\ran{0.65}{0.77}} & \corr{\ran{0.188}{0.264}}
\end{tabular}

    \end{ruledtabular}    
    \caption{Summary of TBLs and separation-bubble databases from the
      literature used in the present study. Reported ranges include
      the friction Reynolds number $\Rey_\tau$, the momentum-thickness
      Reynolds number $\Rey_{\dTwo}$, the shape factor $H$, \corr{the
        pressure-gradient parameter $\gamma_P\beta_*$, where $\beta_*
        = |\dPe|\dOne/U_e^2$ and $\gamma_P = \text{sgn}(\dPe)$,} and
      the dimensionless wall-shear stress $\ptauo$. \corr{The
        parameter $\beta$ used to label the cases of
        \citet{kitsios2017direct} is the Clauser pressure-gradient
        parameter, $\beta = \dOne\,\dPe/u_\tau^2$.}}
    \label{tab:literaturecases}
\end{sidewaystable}

For the sake of comparison we also report in Table~\ref{tab:owncases}
the ranges of the simulations introduced in \S\ref{sec:oursims}.

\begin{sidewaystable}[p]
    \begin{ruledtabular}
        \providecommand\ran[2]{$#1\; - \; #2$}
\begin{tabular}{@{}w{l}{4cm}w{c}{3cm}w{c}{3cm}w{c}{3cm}w{c}{3cm}w{c}{3cm}@{}}
& $\Rey_\tau$ & $\Rey_{\dTwo}$ & $H$ & $\gamma_P \beta_{*} \times 100$ &
$\ptauo \times 1000$ \\
\hline
(300,\,-1) & \ran{236}{412} & \ran{393}{~911} & \ran{1.45}{1.55} & \corr{\ran{-0.064}{-0.022}} & \ran{2.156}{2.707} \\
(300,\,-2) & \ran{240}{407} & \ran{395}{~813} & \ran{1.44}{1.54} & \corr{\ran{-0.093}{-0.021}} & \ran{2.257}{2.767} \\
(300,\,-3) & \ran{235}{388} & \ran{391}{~706} & \ran{1.44}{1.56} & \corr{\ran{-0.116}{-0.021}} & \ran{2.312}{2.721} \\
(300,\,-4) & \ran{239}{377} & \ran{387}{~623} & \ran{1.44}{1.55} & \corr{\ran{-0.136}{-0.020}} & \ran{2.237}{2.708} \\
(670,\,-1) & \ran{454}{737} & \ran{797}{1728} & \ran{1.40}{1.44} & \corr{\ran{-0.056}{-0.017}} & \ran{1.922}{2.423} \\
(670,\,-2) & \ran{450}{738} & \ran{795}{1529} & \ran{1.37}{1.44} & \corr{\ran{-0.083}{-0.017}} & \ran{2.018}{2.386} \\
(670,\,-3) & \ran{454}{737} & \ran{794}{1344} & \ran{1.36}{1.44} & \corr{\ran{-0.106}{-0.017}} & \ran{2.104}{2.377} \\
(670,\,-4) & \ran{444}{715} & \ran{793}{1189} & \ran{1.35}{1.44} & \corr{\ran{-0.124}{-0.017}} & \ran{2.132}{2.368} \\
(300,\,5) &  \ran{292}{535} & \ran{469}{2236} & \ran{1.47}{1.70} & \corr{\ran{0.003}{0.261}} & \ran{1.176}{2.92 } \\
(300,\,10) & \ran{282}{390} & \ran{465}{3062} & \ran{1.48}{2.37} & \corr{\ran{0.009}{0.706}} & \ran{0.41}{2.916} \\
(300,\,15) & \ran{~10}{341} & \ran{460}{3580} & \ran{1.48}{4.33} & \corr{\ran{0.015}{1.945}} & \ran{-0.109}{2.924} \\
(300,\,20) & \ran{~~7}{738} & \ran{456}{3261} & \ran{1.49}{8.39} & \corr{\ran{0.021}{4.539}} & \ran{-0.307}{2.924} \\
(670,\,5) &  \ran{564}{975} & \ran{841}{4024} & \ran{1.44}{1.68} & \corr{\ran{0.003}{0.243}} & \ran{0.986}{2.448} \\
(670,\,10) & \ran{496}{649} & \ran{835}{5811} & \ran{1.44}{2.38} & \corr{\ran{0.005}{0.682}} & \ran{0.317}{2.469} \\
(670,\,15) & \ran{233}{557} & \ran{826}{4870} & \ran{1.44}{3.19} & \corr{\ran{0.008}{0.910}} & \ran{0.048}{2.471} \\
(670,\,20) & \ran{~~9}{493} & \ran{816}{3932} & \ran{1.45}{3.39} & \corr{\ran{0.011}{1.001}} & \ran{-0.03}{2.466} \\
\hline
SB1 & \ran{209}{760} & \ran{1114}{3071} & \ran{1.41}{2.90} & \corr{\ran{-1.696}{2.132}} & \ran{0.048}{2.020}\\
SB2 & \ran{125}{847} & \ran{1111}{3243} & \ran{1.40}{3.45} & \corr{\ran{-2.239}{2.308}} & \ran{-0.016}{2.036}\\
SB3 & \ran{158}{767} & \ran{1090}{3091} & \ran{1.39}{3.25} & \corr{\ran{-2.775}{2.950}} & \ran{0.009}{2.160}
\end{tabular}

% 2.77, 3.25, and 2.98
% 2.9, 3.45, 3.2

    \end{ruledtabular}
\caption{Summary of the new TBLs and separation bubbles described in
  \S\ref{sec:oursims}. The APG/FPG cases are labeled by the pair
  $(\Rey_{\dTwo,i},\alpha)$. Reported ranges include the friction
  Reynolds number $\Rey_\tau$, the momentum-thickness Reynolds number
  $\Rey_{\dTwo}$, the shape factor $H$, the signed pressure-gradient
  parameter $\gamma_P\beta_*$, where $\beta_*=|\dPe|\dOne/U_e^2$ and
  $\gamma_P=\operatorname{sgn}(\dPe)$, and the dimensionless
  wall-shear stress $\ptauo$.
\label{tab:owncases}}
\end{sidewaystable}

%\vspace{1em}
%\begin{longtable}
%      \begin{ruledtabular}
%    \input{literature_table.tex}
%      \end{ruledtabular}
%    \caption{Summary of TBLs and separation-bubble databases from the
%      literature used in the present study. Reported ranges include
%      the friction Reynolds number $\Rey_\tau$, the momentum-thickness
%      Reynolds number $\Rey_\theta$, the shape factor $H$, and the
%      dimensionless wall-shear stress
%      $\ptauo$.} \label{tab:literaturecases}
%  \end{center}
%\end{longtable}

%%%%%%%%%%%%%%%%%%%%%%%%%%%%%%%%%%%%%%%%%%%%%%%%%%%%%%%%%%%%%%%%%%%%%%%%%%%%
\section{Discussion on dimensionally driven and equation-based scaling laws}
%%%%%%%%%%%%%%%%%%%%%%%%%%%%%%%%%%%%%%%%%%%%%%%%%%%%%%%%%%%%%%%%%%%%%%%%%%%%

We address three questions: (1) can the scaling laws discovered here
be obtained from traditional dimensional analysis? (2) is the concept
of history effects justifiable from an information-theoretic
viewpoint? and (3) can the scaling laws be derived from the governing
equations? We argue that the answer to these three questions is, in
general, negative in the strict sense often implied in the
literature. In particular, classical dimensional analysis alone does
not uniquely identify the present local scaling laws; history effects
need not be treated as independent physical ingredients; and the
governing equations do not appear to yield these scaling laws in any
direct or unique way.

%-----------------------------------------------------------------------%
\subsection{Global vs. local dimensional analysis and history effects}
\label{ss:dimensional}
%-----------------------------------------------------------------------%

% Intro
It is useful to distinguish between dimensional analysis based on
\emph{global} parameters that define the problem setup and dimensional
analysis based on \emph{local} variables extracted from the
solution. For example, in the APG/FPG boundary layers considered in
figure~\ref{fig:domain}, global parameters are $[\rho, \nu, x_0,
  \theta_i, \alpha,L_x,L_y,U_\infty,\ldots]$, namely, quantities that
remain fixed within a given case. By contrast, examples of local
variables are $[U_e,\dTwo,\dOne,\delta,\dPe,\ldots]$, which are
obtained from the flow solution at each streamwise location. In some
instances, a given quantity may be interpreted in both ways. For
example, $\rho$ and $\nu$ are global constants for the incompressible
cases considered here, but they also retain a local meaning and can be
used in the local description.

% Global dimensional analysis
If the global parameter list is complete, meaning that there exists a
deterministic mapping from the global parameters to the output,
classical Buckingham--$\Pi$ analysis implies that the problem can be
characterized by $N=N_p-N_u$ independent dimensionless groups, where
$N_p$ is the number of dimensional parameters and $N_u$ is the number
of independent fundamental dimensions, here mass, length, and
time~\cite{Buckingham1914}.  We will denote the corresponding
dimensionless groups by $\myv{\Pi}_g$, where the subscript $g$ stands
for global. For a given family of cases, the solution is then
determined by $\myv{\Pi}_g$, and the dimensionless output of interest
satisfies
\begin{equation}
    \Pi_o = f_g(\myv{\Pi}_g),
\end{equation}
where $\Pi_o$ here represents either $\pUo$ or $\ptauo$ at a given
dimensionless $x$ (and $y$) location, and $f_g$ is a function that, in
essence, contains the Navier--Stokes equations along with boundary and
initial conditions from which $\Pi_o$ can be obtained. From the
information-theoretic viewpoint adopted in this work, this means that
the set $\myv{\Pi}_g$ contains all the information required to predict
$\Pi_o$, and hence the corresponding irreducible error vanishes,
$\epsilon_{\mathrm{LB}}(\Pi_o,\myv{\Pi}_g)=0$~\cite{yuan2025Dimensionless}. If
the diversity of cases is enlarged, for instance by introducing an
additional parameter controlling the wall or ceiling curvature of the
TBLs, then the number of global \dimensionless~groups required to
characterize the family of solutions will generally increase, but the
same argument still applies.

% Implications for history effects
Note that no separate \emph{ad hoc} history effect variable is
required in $\boldsymbol{\Pi}_g$, and these groups are sufficient to
determine $\Pi_o$. Viewed in this way, explicit history
variables are not always necessary for prediction, even in highly
nonequilibrium flows, because their influence may already be encoded
in sufficiently informative global parameters or local variables.
From this perspective, the classical use of upstream integrated
quantities to predict $\tau_w$, or of wall-normal integrated
quantities in a bottom-up manner (from the wall to the freestream) to
predict the mean velocity profile, should be interpreted as a way of
collecting information about the flow state that implicitly encodes
cumulative effects.  For example, quantities integrated over
downstream regions may also correlate with $\tau_w$ at a given point,
although such quantities are not causal predictors in a
forward-modeling sense.  More generally, the success of these integral
forms stems from the fact that they summarize the local flow state
produced by the governing equations.

% Local dimensional analysis
Despite offering a complete characterization of the flow, global
parameters are not particularly attractive for physical insight and
modeling because they are tightly tied to the geometry and boundary
conditions of each specific configuration. Instead, it is far more
desirable to find a minimal set of local variables applicable in a
general setting that still retains enough information to predict
$\pUo$ and $\ptauo$. This has been precisely the objective of the
present work.  However, any low-dimensional local description may be
viewed as a compression of the complete global one.  Since the local
variables are themselves determined by the global parameters, we may
write
\begin{equation}
    \myv{\Pi}_\ell = \boldsymbol{f}_\ell(\myv{\Pi}_g),
\end{equation}
where $\myv{\Pi}_\ell$ denotes a candidate set of local
\dimensionless~groups. Interpreting
$(\Pi_o,\boldsymbol{\Pi}_g,\boldsymbol{\Pi}_\ell)$ as random
variables induced by sampling cases and spatial locations from the
dataset, it follows from the data-processing
inequality~\citep{cover2006} that
\begin{equation}
    I(\Pi_o;\myv{\Pi}_g)\ge I(\Pi_o;\myv{\Pi}_\ell),
\end{equation}
which implies
\begin{equation}
    \epsilon_{\mathrm{LB}}(\Pi_o,\myv{\Pi}_g)\le
    \epsilon_{\mathrm{LB}}(\Pi_o,\myv{\Pi}_\ell).
\end{equation}
Therefore, a local set of variables is not guaranteed \emph{a priori}
to contain sufficient information to predict $\Pi_o$. Moreover,
for a given collection of dimensional local variables, different
admissible \dimensionless~combinations can retain markedly different
amounts of information about the output. This is precisely where the
information-theoretic Buckingham--$\Pi$
theorem~\cite{yuan2025Dimensionless}, leveraged in this work, becomes
essential: it provides a principled way to identify, among the
dimensionally consistent local groups generated from the candidate
variables considered here, those that maximize predictive power for
the target quantity.  \rev{Here, we identify the optimal dimensionless local variables for the
mean velocity profile and obtain wall-shear predictions from its wall
derivative. The low irreducible errors for the velocity inputs and the
two-input wall-shear representation show that these variables encapsulate
most of the predictive information about the corresponding reference
quantities.}

\corr{The optimal dimensionless groups are unique only up to
  information-preserving transformations. Specifically, if
  $\boldsymbol{\Pi}$ is an optimal dimensionless set and
  $\widetilde{\boldsymbol{\Pi}}=g(\boldsymbol{\Pi})$ is a bijective
  transformation, then $I(\Pi_o;\widetilde{\boldsymbol{\Pi}})
  =I(\Pi_o;\boldsymbol{\Pi})$, and the transformed set is also
  optimal. Thus, infinitely many equivalent optimal representations
  exist. This does not imply that arbitrary dimensionless combinations
  are equally valid: there are also infinitely many combinations that
  are not bijectively related to an optimal set and that retain less
  information about $\Pi_o$.  Within the class of optimal
  representations, we select one with simple exponents to facilitate
  physical interpretation.  Hence, the particular representation is
  not unique, but its optimal information content is.
  \rev{For the wall-shear representation, the reduction must preserve
  predictive information about the wall derivative, rather than about
  the entire velocity profile. This does not require a bijective
  transformation of the complete velocity-input set.}}

% Connection with the equations
This also clarifies the sense in which the present method differs from
the classical Buckingham--$\Pi$ theorem and, at the same time,
connects with equation-based scaling arguments. Classical
Buckingham--$\Pi$ analysis relies only on dimensional homogeneity,
regardless of the underlying equations governing the problem. In
contrast, the present approach uses the information content of the
variables, which is ultimately shaped by the way the flow variables
are dynamically coupled through the equations. Hence, the current
method combines the strength of classical dimensional analysis (i.e.,
that no explicit model needs to be postulated in advance) with an
equation-informed notion of optimality that identifies the local
scaling laws most compatible with the underlying flow physics.

%-----------------------------------------------------------------------%
\subsection{Scaling laws from governing equations}
\label{ss:equations}
%-----------------------------------------------------------------------%

% Intro
An interesting question is whether the scaling laws identified above
could have been derived directly from the governing equations.
Although the equations certainly constrain the admissible
relationships, they do not, in general, uniquely determine which
dimensionless groups are the most relevant for predicting a given
quantity. The reason is that the same governing equation can often be
cast in multiple equivalent dimensionless forms, each highlighting a
different set of variables.  Hence, the mere appearance of a
dimensionless group in a given form of the equations does not, by
itself, establish its physical importance.

% Illustration with Karman equation
To illustrate the point, consider the classical von K\'arm\'an
momentum-integral equation for a two-dimensional incompressible
boundary layer,
\begin{equation}
    \frac{\mathrm{d}\theta}{\mathrm{d}x}
    =
    \frac{\tau_w}{\rho U_e^2}
    -
    (H+2)\frac{\theta}{U_e}\frac{\mathrm{d}U_e}{\mathrm{d}x}.
    \label{eq:vonkarman_Ue}
\end{equation}
Although the classical von K\'arm\'an relation is most naturally
interpreted for attached flows and mild pressure gradients, and more
general integral formulations are available for stronger
non-equilibrium conditions \citep{wei2024new,han2025integral}, it is
sufficient here for the present discussion.

% Dimensionless form 1
Using the inviscid relation at the edge of the boundary layer,
$U_e\,\mathrm{d}U_e/\mathrm{d}x =
-(1/\rho)\,\mathrm{d}P_e/\mathrm{d}x$, Eq.~\eqref{eq:vonkarman_Ue} can
be rewritten as
\begin{equation}
    \Pi_o^\tau
    =
    \frac{\mathrm{d}\theta}{\mathrm{d}x}
    -
    \left(1+\frac{2}{H}\right)\gamma_P\beta_*.
    \label{eq:vonkarman_beta}
\end{equation}
Equation~\eqref{eq:vonkarman_beta} involves $H$ and $\beta_*$, but it
contains no explicit dependence on the Reynolds number
$\Rey_\theta$. However, both theory and experiments show that
$\Pi_o^\tau$ depends to some extent on $\Rey_\theta$. Hence, the absence
of $\Rey_\theta$ from a particular representation of the governing
equations does not imply that $\Rey_\theta$ is unimportant; its effect
may simply be embedded implicitly in the dependent variable and in the
streamwise evolution of the integral quantities. For example, it may
be the case that $\mathrm{d}\theta/\mathrm{d}x = f(\Rey_\theta,\cdots)$,
making Eq.~\eqref{eq:vonkarman_beta} sensitive to $\Rey_\theta$. It is
also often the case that the explicit dimensionless groups (i.e., $H$
and $\beta_*$) are themselves not independent of $\Rey_\theta$.

% Dimensionless form 2
Indeed, $\Rey_\theta$ can be made to appear explicitly by rewriting
Eq.~\eqref{eq:vonkarman_beta} as
\begin{equation}
    \Pi_o^\tau
    =
    \frac{\theta}{\Rey_\theta}\frac{\mathrm{d}\Rey_\theta}{\mathrm{d}x}
    -
    \left(1+\frac{1}{H}\right)\gamma_P\beta_*,
    \label{eq:vonkarman_Retheta}
\end{equation}
where now $\Rey_\theta$ appears explicitly, even though the underlying
physics has not changed.

% Dimensionless form 3
The same equation can be rearranged once more to display the
dimensionless group discovered in the present work. Using $\Pi_1^\tau
= H^{-1}\Rey_\theta^{-1/4}$, Eq.~\eqref{eq:vonkarman_Retheta} becomes
\begin{equation}
    \Pi_o^\tau
    =
    \frac{\theta}{\Rey_\theta}\frac{\mathrm{d}\Rey_\theta}{\mathrm{d}x}
    -
    \left(1+\Pi_1^\tau \Rey_\theta^{1/4}\right)\gamma_P\beta_*.
    \label{eq:vonkarman_Pi1}
\end{equation}
Thus, by simple algebraic manipulation, one may produce equivalent
dimensionless forms in which $H$, $\beta_*$, $\Rey_\theta$, or
$\Pi_1^\tau$ appear explicitly.

% Two conclusions
Two conclusions follow from this example. First, the fact that a
dimensionless group appears explicitly in a given form of the
equations does not imply that it plays a dominant role---or even any
role at all---in determining the quantity of interest. Second, the
fact that a dimensionless group does not appear explicitly does not
imply that it is irrelevant, since its influence may be hidden within
other terms that depend on it functionally. For this reason,
equation-based dimensional rearrangements alone are insufficient to
identify the most relevant scaling variables. By contrast, the method
employed in the present work is designed precisely to discover
dimensionless groups according to their predictive power, selecting
those that minimize the information-theoretic irreducible error
\citep{yuan2025Dimensionless}.

% Consequences of our results in terms of equations
Another consequence of the present results concerns what may be termed
the \emph{optimal dimensionless form} of the governing equations.
Consider an equation with $N$ terms,
\begin{equation}
    \ptauo = \sum_{i=1}^N \mathcal{R}_i = f(\ptau_1,\ptau_2) +
    \mathcal{O}(\epsilon_{\mathrm{LB}}).
    \label{eq:Ri_terms}
\end{equation}
In the limit $\epsilon_{\mathrm{LB}}=0$, $\ptauo$ is a deterministic function
of $(\ptau_1,\ptau_2)$.  Let $\Pi_a$ denote an arbitrary additional
dimensionless group. We say that the arrangement of terms in
Eq.~\eqref{eq:Ri_terms} constitutes the optimal dimensionless form
with respect to $(\ptau_1,\ptau_2)$ if
\begin{equation}
    I(\mathcal{R}_i;\Pi_a \mid \ptau_1,\ptau_2)=0,
    \qquad \forall\, \Pi_a,\; i=1,\ldots,N,\label{eq:Ri}
\end{equation}
where $I(\cdot ; \cdot | \cdot)$ is the conditional mutual
information. Therefore, once $(\ptau_1,\ptau_2)$ are specified, no
additional dimensionless group carries further information about any
term $\mathcal{R}_i$.  When $\epsilon_{\mathrm{LB}}>0$ but small, these
quantities may still depend weakly on additional dimensionless groups,
but only in the sense that the extra information they contribute
beyond $(\ptau_1,\ptau_2)$ results in an error smaller than
$\epsilon_{\mathrm{LB}}$.

In the von K\'arm\'an momentum-integral relation, we may express the
equation as
\begin{equation}
    \Pi_o^\tau
    =
    \frac{\mathrm{d}\theta}{\mathrm{d}x}
    -
    (\Pi_1^\tau)^{6/5}(\Pi_2^\tau)^{18/5}\,\widetilde{\Pi}_{3}
    -
    2(\Pi_1^\tau)^{11/5}(\Pi_2^\tau)^{18/5}\,\widetilde{\Pi}_{4},
    \label{eq:vK_Pi_final_U}
\end{equation}
where $\widetilde{\Pi}_{3} \equiv
-\,H\frac{\theta}{U_e}\frac{dU_e}{dx}\left[(\Pi_1^\tau)^{6/5}(\Pi_2^\tau)^{18/5}\right]^{-1},
\qquad \widetilde{\Pi}_{4} \equiv
-\,\frac{\theta}{U_e}\frac{dU_e}{dx}\left[(\Pi_1^\tau)^{11/5}(\Pi_2^\tau)^{18/5}\right]^{-1}$
and we have assumed for simplicity that $\Pi_2^\tau>0$.  Since the
present results indicate that $\ptauo \approx
f_\tau(\ptau_1,\ptau_2)$, we may expect, under the assumption in
Eq.~\eqref{eq:Ri}, that $\mathrm{d}\theta/\mathrm{d}x$,
$\widetilde{\Pi}_{3}$, and $\widetilde{\Pi}_{4}$ are also largely
determined by the same variables, at least within the class of
turbulent boundary layers considered here.  If the condition in
Eq.~\eqref{eq:Ri} is not satisfied, the terms can be regrouped until
it is. In information-theoretic terms, this implies that the
conditional dependence of $\widetilde{\Pi}_{3}$ and
$\widetilde{\Pi}_{4}$ on any additional variables, once
$\left(\ptau_1,\ptau_2\right)$ are specified, should be small.

%%%%%%%%%%%%%%%%%%%%%%%%%%%%%%%%%%%%%%%%%%%%%%%%%%%%%%%%%%%%%%%%%%
\section{Additional analysis of the dimensionless inputs}
%%%%%%%%%%%%%%%%%%%%%%%%%%%%%%%%%%%%%%%%%%%%%%%%%%%%%%%%%%%%%%%%%%

\rev{Throughout the following analyses, $f_U$ denotes the mean-velocity
model after training on velocity data and subsequent fine-tuning using
wall-shear information. The wall-shear predictions are obtained from its
wall derivative. The diagnostic fits reported below are used only to assess the
dimensionless-input representations.}

%-----------------------------------------------------------------------%
\subsection{Two dimensionless inputs for mean velocity profiles}
\label{ss:2vel}
%-----------------------------------------------------------------------%
\providecommand\pUp{\Pi^{U,\prime}}

We assess the impact of reducing the dimensionless input set to two
variables of the form $\pUo \approx f_U^\prime(\pUp_1, \pUp_2)$.
A systematic analysis of the number of dimensionless inputs required
is presented in \S\ref{ss:opt}.

%         y     Ue     delta   d1    d2      nu
%array([[ 0.62, -0.24, -0.41,  0.59, -1.  ,  0.21],
%       [ 0.18,  0.03,  0.04, -1.  ,  0.8 , -0.03]])
For two inputs, the dimensionless groups obtained from the
information-theoretic optimization can be approximated as follows:
\begin{equation}
\Pi_2^{U,\prime}
\approx
\frac{y^{0.2}\theta^{0.8}}{\delta_*},
\qquad
\Pi_1^{U,\prime}
\approx
\frac{y^{0.6}\delta_*^{0.6}\nu^{0.2}}
     {U_e^{0.2}\delta^{0.4}\theta},
\end{equation}
which are qualitatively similar to $\pU_1$ and $\pU_2$ obtained in the
Letter when three inputs are considered.

\begin{figure}
    \centering
    \begin{subfigure}{.4\tw}
        \begin{tikzpicture}
            \node (fig) at (0,0)
            {\ig[height=2in]{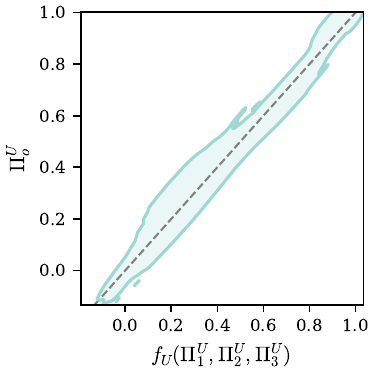}};
            \begin{scope}[shift=(fig.south west), x=(fig.south east), y=(fig.north west)]
                \node at (.30,.88) {\footnotesize(a)};
                \node[fill=white,anchor=south,inner xsep=8pt]
                at (.60,0) {\footnotesize$f_U^\prime(\pUp_1, \pUp_2)$};
                \node[fill=white,rotate=90] at (.04,.60) {$\pUo$};
            \end{scope}
        \end{tikzpicture}
    \end{subfigure}\qquad\qquad
    \begin{subfigure}{.4\tw}
        \centering
        \begin{tikzpicture}
            \node[inner sep=0pt] (f1) {\ig[height=2in]{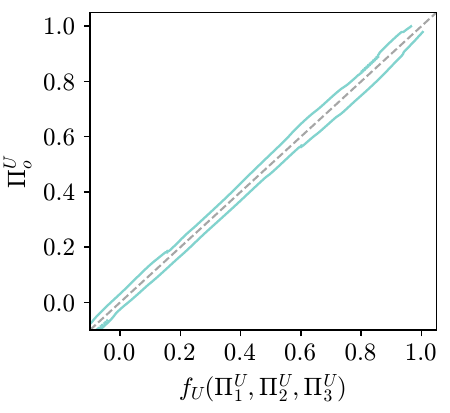}};
            \begin{footnotesize}
                \begin{scope}[shift=(f1.south west), x=(f1.south east), y=(f1.north west)]
                    \node at (.28,.90) {\footnotesize(b)};
                    \node[fill=white] at (.57,.05)
                    {\footnotesize$f_U(\pU_1,\pU_2,\pU_3)$};
                    \node[fill=white,rotate=90] at (.04,.57) {$\pUo$};
                \end{scope}
            \end{footnotesize}
        \end{tikzpicture}
    \end{subfigure}
\caption{Comparison of actual $\pUo$ values with predictions from
    (a) $f_U^\prime(\pUp_1,\pUp_2)$ and
    (b) $f_U(\pU_1,\pU_2,\pU_3)$. \rev{In panel (a), the cyan
    contour encloses 99\% of the joint probability mass. In panel (b),
    it marks the first percentile of the density evaluated at 160,000
    samples; the RMS-normalized prediction error is 1.30\%.
    Panel (b) uses a reconstructed sample subset whose original training
    and testing memberships are unavailable.} The dashed line indicates
    perfect agreement (zero error).
\label{fig:fit2inputs}}   
\end{figure}

\begin{figure}
    \centering
    \begin{tikzpicture}

    \node[anchor=south west,inner sep=0pt, outer sep=0pt] (f1) at (0,0)
    {\ig[width=.49\tw]{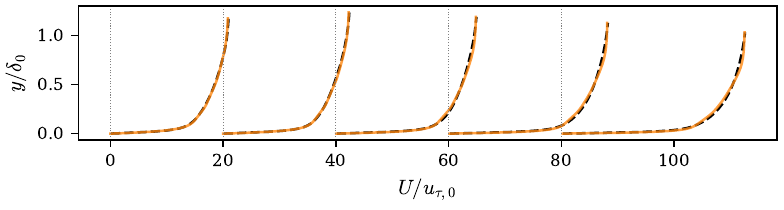}};

    \node[anchor=south west,inner sep=0pt, outer sep=0pt] (f2) at (f1.south east)
    {\ig[width=.49\tw]{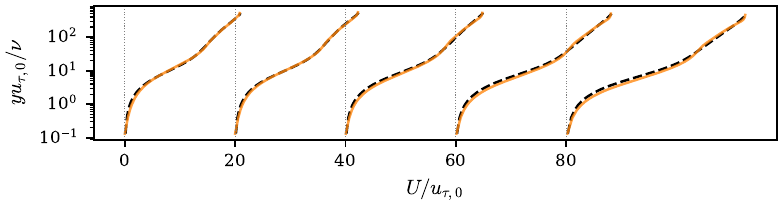}};

    \node[anchor=south west,inner sep=0pt, outer sep=0pt] (f3) at (0,-1.2in)
    {\ig[width=.49\tw]{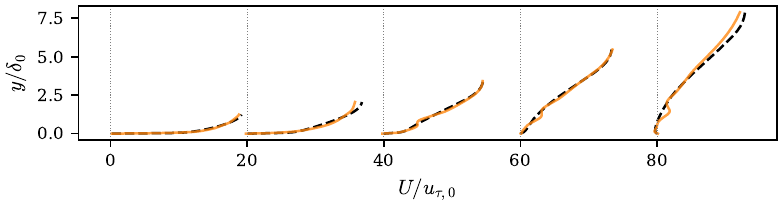}};

    \node[anchor=south west,inner sep=0pt, outer sep=0pt] (f4) at (f3.south east)
    {\ig[width=.49\tw]{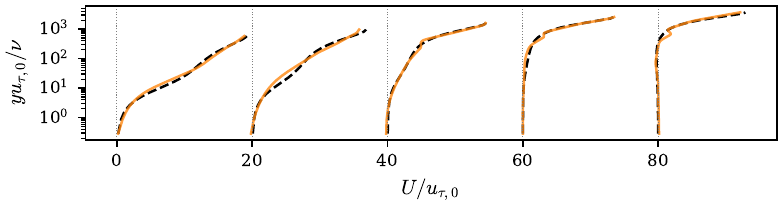}};

    \node[anchor=south west,inner sep=0pt, outer sep=0pt] (f5) at (0,-2.40in)
    {\ig[width=.49\tw]{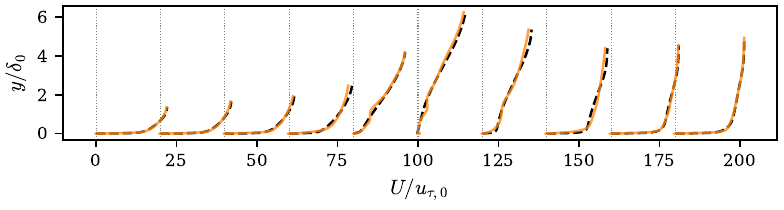}};

    \node[anchor=south west,inner sep=0pt, outer sep=0pt] (f6) at (f5.south east)
    {\ig[width=.49\tw]{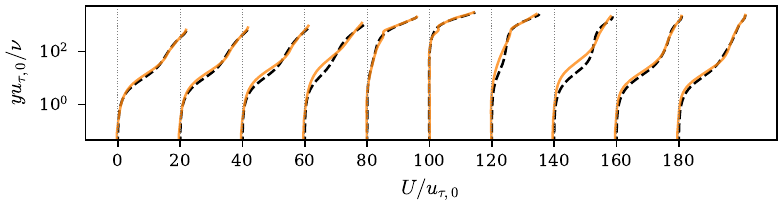}};

    \node[anchor=south west] (fa) at (f1.north west)
    {\ig[width=.31\tw]{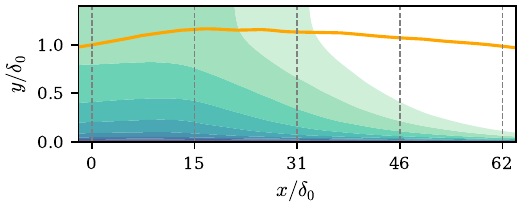}};
    \node[anchor=south west] (fb) at (fa.south east)
    {\ig[width=.31\tw]{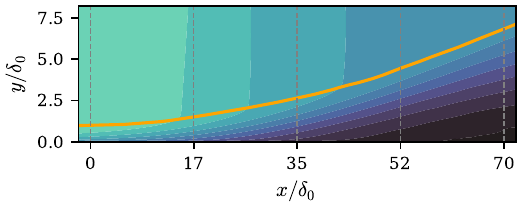}};
    \node[anchor=south west] (fc) at (fb.south east)
    {\ig[width=.31\tw]{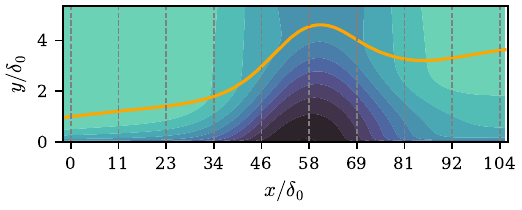}};

    \begin{scriptsize}
        \begin{scope}[shift=(fa.south west), x=(fa.south east), y=(fa.north west)]
            \node[fill=white, fill opacity=.5, text opacity=1., anchor=east] at
            (0.96,.45) {FPG};
        \end{scope}
        \begin{scope}[shift=(fb.south west), x=(fb.south east), y=(fb.north west)]
            \node[fill=white, fill opacity=.5, text opacity=1., anchor=east] at
            (0.96,.45) {APG};
        \end{scope}
        \begin{scope}[shift=(fc.south west), x=(fc.south east), y=(fc.north west)]
            \node[fill=white, fill opacity=.5, text opacity=1., anchor=east] at
            (0.96,.45) {Sep. Bubble};
        \end{scope}
    \end{scriptsize}

    % Parameters with units
    \def\nx{2}
    \def\ny{3}
    \def\dx{.5\tw}  % horizontal spacing
    \def\dy{1.2in}  % vertical spacing

    \foreach \j in {0,...,\numexpr\ny-1} {
        \foreach \i in {0,...,\numexpr\nx-1} {

            \pgfmathtruncatemacro{\idx}{\i + \nx*\j}
            \pgfmathtruncatemacro{\charcode}{97 + \idx} % 97 = 'a'
            \edef\labelchar{\char\charcode}

            % Compute coordinates using TikZ calc syntax (not pure pgfmath)
            \path coordinate (C) at ($(\i*\dx, -\j*\dy)$);

            % Draw figure
            %\draw[thick] (C) rectangle ++(\w,\h);

            % Label
            \node[anchor=north west] at ($(C)+(2.em,0.8in)$) {\footnotesize(\labelchar)};
        }
    }

\end{tikzpicture}
\caption{Comparison of predictions from the model
  $f_U^\prime(\pUp_1,\pUp_2)$ with reference velocity profiles for
  (a,b) an FPG TBL, (c,d) an APG TBL, and (e,f) a turbulent separation
  bubble~\cite{coleman2018numerical}.  Black dashed lines show the
  reference velocity profiles; orange lines show the model predictions
  $f_U^\prime(\pUp_1,\pUp_2)$. An offset of $\Delta U/u_{\tau,0} = 20$
  is added between the velocity profiles to ease the comparison.  The
  upper portions of the panels show contours of the mean streamwise
  velocity, ranging from $-0.2U_{e,0}$ (black) to $1.2U_{e,0}$ (light
  blue), where $U_{e,0}$ is the edge velocity at the first streamwise
  location. Vertical dashed lines mark the locations of the profile
  comparisons; the orange line denotes $\delta$, with $\delta_0$ the
  boundary-layer thickness at the first location.
\label{fig:U2inputs}}
\end{figure}

Figure~\ref{fig:fit2inputs} compares the aggregate prediction accuracy
of the two- and three-input velocity models, while
Figure~\ref{fig:U2inputs} illustrates the corresponding profile-level
differences.
The two-input model yields overall good agreement for the FPG TBL at
$\Rey_{\theta_0} \approx 800$ with $\alpha = -4^\circ$
(figure~\ref{fig:U2inputs}a,b), although a noticeable mismatch develops
in the inner and logarithmic layers, particularly at downstream locations.
The model clearly fails for the APG TBL at $\Rey_{\theta_0} \approx 800$
with $\alpha = 15^\circ$ and for the turbulent separation
bubble~\cite{coleman2018numerical}.
These results demonstrate that $\pU_3$, while not the dominant input in the
three-variable model, plays an essential corrective role that becomes
indispensable under non-equilibrium conditions.
Hence, two dimensionless inputs are insufficient to provide a unified
description across all flow regimes.

%-----------------------------------------------------------------------%
\subsection{Local-variance sensitivity analysis for mean velocity profiles}
\label{ss:local-variance}
%-----------------------------------------------------------------------%

% Intro
We complement the gradient-based sensitivity analysis discussed in the
article with an alternative attribution method based on local variance
decomposition~\cite{sobol2001}. The goal is to ensure that the
conclusions are independent of the method used to assess the role of
each dimensionless input in the prediction of the mean velocity
profiles across flow regimes and wall-normal locations.

% Approach
Local-variance sensitivity analysis is a method for quantifying how
much each input contributes to the variability of a model output
within a local region of the input space. This is achieved by
estimating the fraction of output variance attributable to each
$\pU_i$ when it is perturbed independently over a small neighborhood.
\rev{For numerical evaluation, define
$\mathbf{q}=(\pU_1,(\pU_2)^{8/9},\pU_3)$ and the standardized
coordinates $z_i=(q_i-\mu_i)/s_i$, where $\mu_i$ and $s_i$ are the
input means and standard deviations ordered to match $\mathbf{q}$.
Let $F_U(\mathbf{z})$ denote the standardized network output,
with network input order $(z_3,z_1,z_2)$.
For each coordinate $z_i$ evaluated at a given profile point,}
we introduce a set of artificial perturbations
$\pi\in[-\varepsilon,\varepsilon]$ while holding all other inputs
fixed at their actual values, and then evaluate the resulting spread
in the model output.  Following the variance-decomposition approach of
\citet{sobol2001}, we define
\begingroup
\color{revisioncolor}
\begin{equation}
    V_i = \operatorname{Var}_{\pi \ \in [-\varepsilon,\,\varepsilon]}
    \!\left[
        F_U\!\left( z_1 + \delta_{1i} \pi, \,
            z_2 + \delta_{2i} \pi, \,
            z_3 + \delta_{3i} \pi
        \right)
    \right],
    \label{eq:local_variance}
\end{equation}
\endgroup
where $\delta_{ij}$ is the Kronecker delta, and the variance is
computed over a uniform discrete grid of $N_s = 50$ perturbation
values, with neighborhood size $\varepsilon = 0.1$ in \rev{the standardized
coordinates $\mathbf{z}$, without clipping the perturbations}. The value of $\varepsilon = 0.1$ is chosen to be small enough
to remain local, yet large enough to probe effects beyond the linear
regime. The normalized sensitivity index is then defined as
\begin{equation}
    S_i = \frac{V_i}{\displaystyle\sum_{j} V_j},
    \label{eq:sensitivity_index}
\end{equation}
and is computed independently at each $y$ location along the mean
velocity profile. The index $S_i$ measures the fraction of the summed
local output variance associated with independently perturbing \rev{coordinate
$z_i$} within a finite neighborhood of the current operating
point. It therefore captures nonlinear contributions that may be
missed by first-order gradients~\cite{saltelli2002}.

\begin{figure}
    \centering
    \begin{tikzpicture}
        \node[inner sep=0pt,outer sep=0pt] (f1) at (0,0)
        {\ig[width=.49\tw]{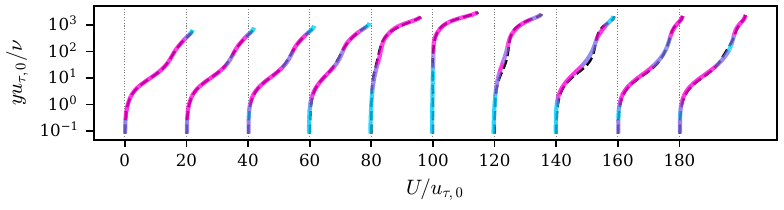}};
        \node[anchor=north west,inner sep=0pt,outer sep=0pt] (f2) at (f1.north east)
        {\ig[width=.49\tw]{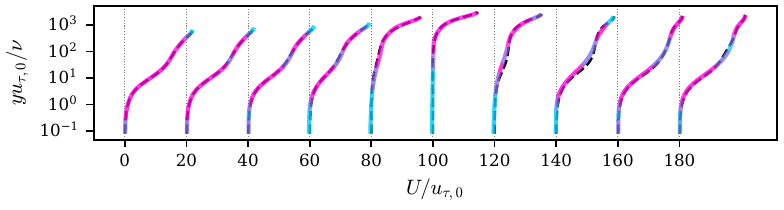}};

        \node[anchor=north west,inner sep=0pt,outer sep=0pt] (f3) at (f1.south west)
        {\ig[width=.49\tw]{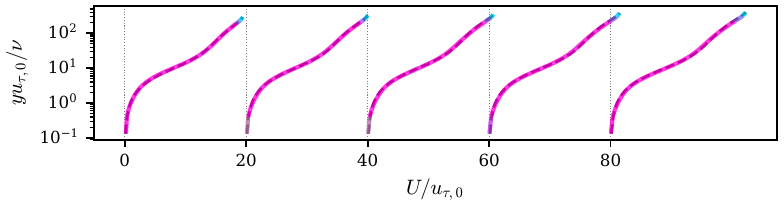}};
        \node[anchor=north west,inner sep=0pt,outer sep=0pt] (f4) at (f3.north east)
        {\ig[width=.49\tw]{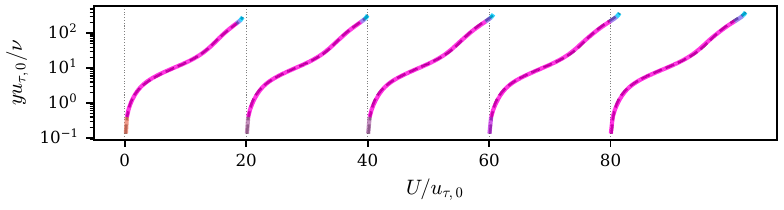}};

        \node[anchor=north west,inner sep=0pt,outer sep=0pt] (f5) at (f3.south west)
        {\ig[width=.49\tw]{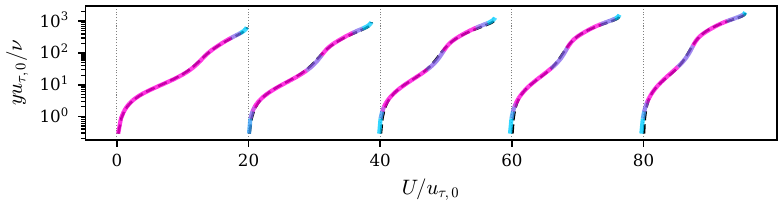}};
        \node[anchor=north west,inner sep=0pt,outer sep=0pt] (f6) at (f5.north east)
        {\ig[width=.49\tw]{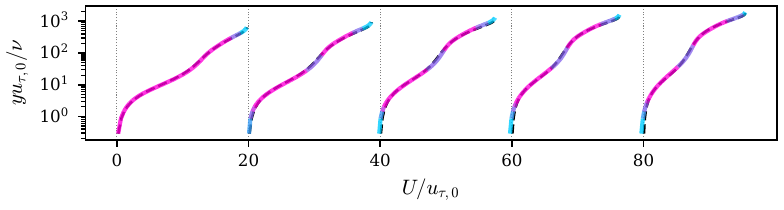}};

        \begin{scriptsize}
            \begin{scope}[shift=(f1.south west), x=(f1.south east), y=(f1.north west)]
                \node at (.15,.85) {(a)};
                \node at (1.15,.85) {(b)};
            \end{scope}
            \begin{scope}[shift=(f3.south west), x=(f3.south east), y=(f3.north west)]
                \node at (.16,.85) {(c)};
                \node at (1.16,.85) {(d)};
            \end{scope}
            \begin{scope}[shift=(f5.south west), x=(f5.south east), y=(f5.north west)]
                \node at (.16,.85) {(e)};
                \node at (1.16,.85) {(f)};
            \end{scope}
        \end{scriptsize}
    \end{tikzpicture}
\caption{Comparison of sensitivity-attribution methods for (a,b) a
  turbulent separation bubble, case 18C, from
  \citet{coleman2018numerical}; (c,d) the FPG TBL
  $(\Rey_{\theta_i},\alpha)=(300,-1^\circ)$; and (e,f) the APG TBL
  $(\Rey_{\theta_i},\alpha)=(670,5^\circ)$. The panels show
    wall-normal profiles of $U/u_{\tau,0}$ at selected streamwise stations,
  colored by the dominant dimensionless input. \rev{Black dashed lines
  show the reference profiles.} Panels (a,c,e) use
  \rev{gradient magnitudes $|\partial F_U/\partial z_i|$ normalized by
  their Euclidean norm}, whereas panels (b,d,f) use the
local-variance sensitivity $S_i$. \rev{Ten separation-bubble profiles
and five per ramp case are selected at approximately uniform source-index
intervals; $u_{\tau,0}$ is the reference friction velocity at the first
available station.} In all plots, an offset of $\Delta U/u_{\tau,0}
= 20$ is added between velocity profiles to ease the comparison.
\label{fig:supp_sensitivity}}   
\end{figure}

% Results
A comparison of the two methods is shown in
figure~\ref{fig:supp_sensitivity} for several validation cases.
\rev{The two approaches identify the same dominant input at more than
99\% of the evaluated points in each of the three cases shown.
For visualization, the gradient weights are
$|\partial F_U/\partial z_i|/\|\nabla_{\mathbf{z}}F_U\|_2$,
whereas the local-variance weights are $\sqrt{S_i}$.
Weights below $1/\sqrt{3}$ are discarded; the remaining weights
blend magenta ($\pU_1$), cyan ($\pU_2$), and yellow ($\pU_3$)
in proportion to their values.}

%-----------------------------------------------------------------------%
\subsection{Linear-scale representation of mean velocity profiles}
%-----------------------------------------------------------------------%

\begin{figure}[h!]
    \centering
    \begin{tikzpicture}
    \node[anchor=south west,inner sep=0pt,outer sep=0pt] (f1) at (0,0)
    {\ig[width=.49\tw]{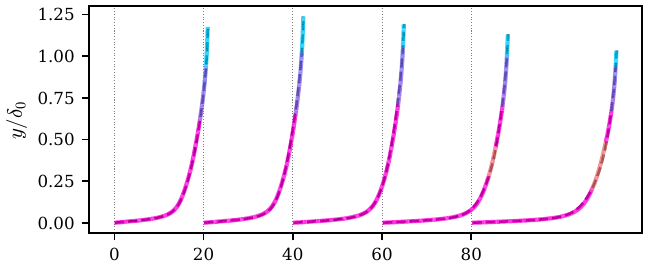}};
    \node[anchor=south west,inner sep=0pt,outer sep=0pt] (f2) at (f1.south east)
    {\ig[width=.49\tw]{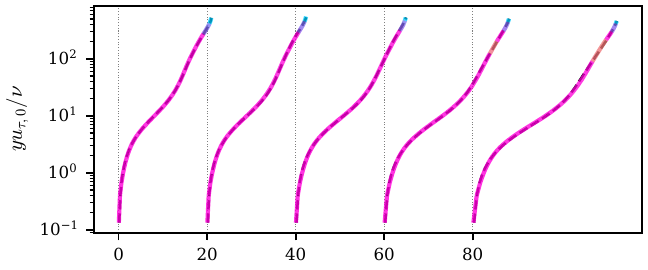}};

    \node[anchor=south west,inner sep=0pt,outer sep=0pt] (f3) at (0,-1.45in)
    {\ig[width=.49\tw]{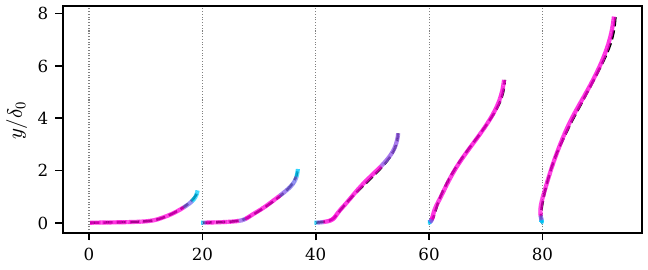}};
    \node[anchor=south west,inner sep=0pt,outer sep=0pt] (f4) at (f3.south east)
    {\ig[width=.49\tw]{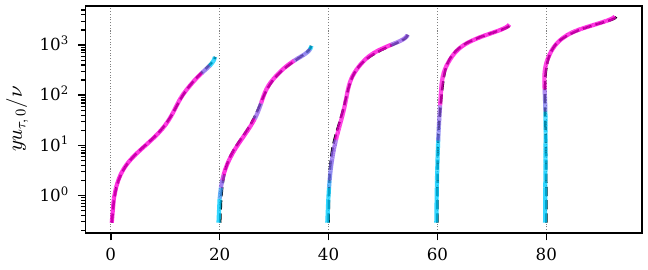}};

    \node[anchor=south west,inner sep=0pt,outer sep=0pt] (f5) at (0,-2.90in)
    {\ig[width=.49\tw]{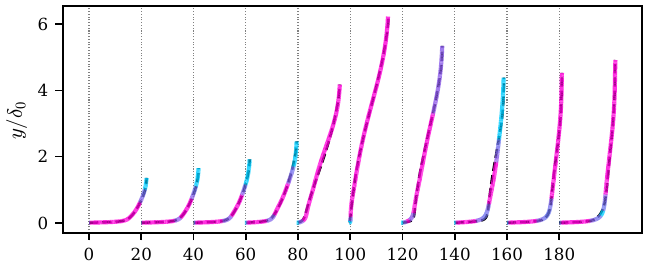}};
    \node[anchor=south west,inner sep=0pt,outer sep=0pt] (f6) at (f5.south east)
    {\ig[width=.49\tw]{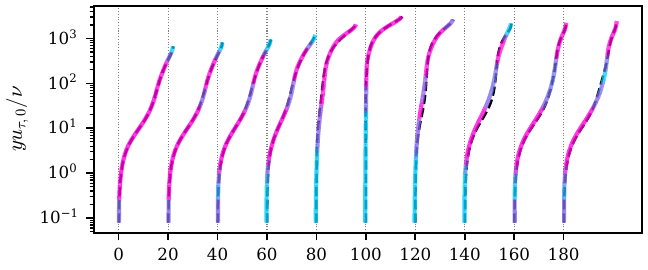}};

    \node[anchor=south west] (fa) at (f1.north west)
    {\ig[width=.31\tw]{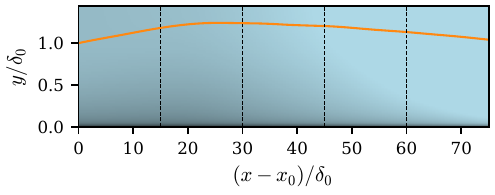}};
    \node[anchor=south west] (fb) at (fa.south east)
    {\ig[width=.31\tw]{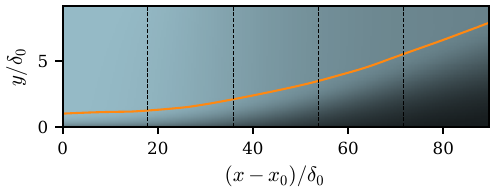}};
    \node[anchor=south west] (fc) at (fb.south east)
    {\ig[width=.31\tw]{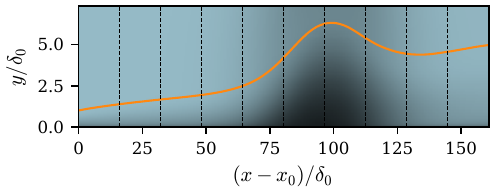}};

\node[anchor=north,scale=.7] at (f5.south) {$U/u_{\tau,0}$};
\node[anchor=north,scale=.7] at (f6.south) {$U/u_{\tau,0}$};
    \begin{scriptsize}
        \begin{scope}[shift=(fa.south west), x=(fa.south east), y=(fa.north west)]
            \node[fill=white, fill opacity=.5, text opacity=1., anchor=east] at
            (0.96,.45) {FPG};
        \end{scope}
        \begin{scope}[shift=(fb.south west), x=(fb.south east), y=(fb.north west)]
            \node[fill=white, fill opacity=.5, text opacity=1., anchor=east] at
            (0.96,.45) {APG};
        \end{scope}
        \begin{scope}[shift=(fc.south west), x=(fc.south east), y=(fc.north west)]
            \node[fill=white, fill opacity=.5, text opacity=1., anchor=east] at
            (0.96,.45) {Sep. Bubble};
        \end{scope}
    \end{scriptsize}
    \node[
        anchor=north west,
        xshift=6em,
        yshift=-1.6em,
        inner sep=0pt,
        outer sep=0pt
    ] (tri) at (f3.north west)
    {\ig[
        width=.065\tw
    ]{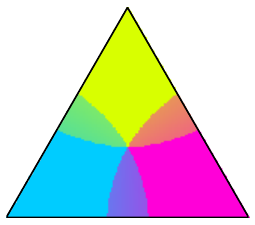}};

    \begin{scope}[
        shift=(tri.south west),
        x=(tri.south east),
        y=(tri.north west)
    ]
        \node[anchor=east,xshift=+0.5em]
            at (0,0) {\scriptsize$\pU_2$};
        \node[anchor=west,xshift=-0.3em]
            at (1,0) {\scriptsize$\pU_1$};
        \node[anchor=south,yshift=-.3em]
            at (.5,1) {\scriptsize$\pU_3$};
    \end{scope}

    \def\nx{2}
    \def\ny{3}
    \def\dx{.5\tw}
    \def\dy{1.45in}

    \foreach \j in {0,...,\numexpr\ny-1} {
        \foreach \i in {0,...,\numexpr\nx-1} {
            \pgfmathtruncatemacro{\idx}{\i + \nx*\j}
            \pgfmathtruncatemacro{\charcode}{97 + \idx}
            \edef\labelchar{\char\charcode}

            \path coordinate (C) at ($(\i*\dx,-\j*\dy)$);
            \node[anchor=north west]
                at ($(C)+(4em,1.25in)$) {(\labelchar)};
        }
    }
\end{tikzpicture}
\caption{Comparison of predictions from the model
    $\pUo=f_U(\pU_1,\pU_2,\pU_3)$ with reference velocity profiles
    for (a,b) an FPG TBL, (c,d) an APG TBL, and (e,f) a turbulent
separation bubble~\cite{coleman2018numerical}. An offset of $\Delta U/u_{\tau,0} =
20$ is added for clarity. Black dashed
    lines show the reference velocity profiles; colored lines show
    the model predictions $f_U(\pU_1,\pU_2,\pU_3)$. \rev{Colors indicate
    the dominant sensitivity to the standardized coordinates defined in
    \S\ref{ss:local-variance}, using the normalized magnitudes
    $|\partial F_U/\partial z_i|$.} The upper portions of the panels
    show contours of the mean streamwise velocity, ranging from
    $-0.2U_{e,0}$ (black) to $1.2U_{e,0}$ (light blue), where
    $U_{e,0}$ is the edge velocity at the first streamwise location.
    Vertical dashed lines mark the locations of the profile
    comparisons; the orange line denotes $\delta$, with $\delta_0$
    the boundary-layer thickness at the first location.
    \rev{Five profiles per ramp case and ten for the separation bubble
    are selected at approximately uniform source-index intervals.
    The scales $U_{e,0}$, $u_{\tau,0}$, and $\delta_0$ are evaluated
    at the first available station.}
\label{fig:Uprof}}
\end{figure}

Figure~\rev{2} of the Letter presents velocity profiles using a
logarithmic wall-normal coordinate, which emphasizes the near-wall
region and facilitates comparison with classical inner-layer
scaling. To complement that representation,
figure~\ref{fig:Uprof} \rev{presents profiles from the same flow cases
using both linear and logarithmic wall-normal coordinates. The linear
representation illustrates the evolution of the boundary-layer thickness
and the outer-layer structure.}

The three representative cases shown are the following: an FPG TBL at
$\Rey_{\theta_0} \approx 800$ with $\alpha = -4^\circ$
(figures~\ref{fig:Uprof}a,b); an APG TBL at $\Rey_{\theta_0} \approx
800$ with $\alpha = 15^\circ$ (figures~\ref{fig:Uprof}c,d); and a
turbulent separation bubble corresponding to case 18C from
\citet{coleman2018numerical} (figures~\ref{fig:Uprof}e,f).
\rev{For the displayed profiles, the RMS errors in $U/U_e$ are
$0.45\%$, $2.03\%$, and $2.45\%$ of the reference RMS, respectively;
local discrepancies remain near the wall.}

%-----------------------------------------------------------------------%
\FloatBarrier
\subsection{\rev{Wall-shear prediction and validation of its two-input representation}}
\label{ss:wall-derivative}
%-----------------------------------------------------------------------%

\rev{The mean wall-shear stress is obtained from the wall-normal derivative
of the final mean-velocity model:}
\begingroup
\color{revisioncolor}
\begin{equation}\label{eq:wall-derivative}
    \widehat{\tau}_w = \rho\nu
    \left.\frac{\partial\widehat{U}}{\partial y}\right|_{y=0},
    \qquad
    \widehat{U}=U_e f_U(\pU_1,\pU_2,\pU_3).
\end{equation}
\endgroup
\rev{The derivative is evaluated at a fixed streamwise location, holding
 the local flow parameters fixed. The resulting dimensionless wall-shear
 prediction is expressed as}
\begingroup
\color{revisioncolor}
\begin{equation}\label{eq:wall-reparameterization}
    \widehat{\Pi}_{o,U}^{\tau}
    \equiv \frac{\widehat{\tau}_w}{\rho U_e^2}
    \approx f_\tau(\ptau_1,\ptau_2),
\end{equation}
\endgroup
\rev{using the two dimensionless groups introduced in the Letter. The
derivative is evaluated using the velocity-model inputs. The
pressure-dependent groups provide an alternative representation of the
resulting predictions over the flow states considered. This
reparameterization retains the predictive information about wall shear
within the flow regimes considered.}

\rev{To validate this reduction, we independently fit a model
$g_\tau^{(2)}(\ptau_1,\ptau_2)$ using wall-shear data and compare its
predictions with $\widehat{\Pi}_{o,U}^{\tau}$. The predictions of
$g_\tau^{(2)}$ and the wall-shear model derived from $f_U$ agree with an
error below $0.01\%$. This agreement confirms the adequacy of the two-input
representation. This fit serves only as a validation of the inputs; the
wall-shear predictions are already supplied by the derivative of $f_U$.}

%-----------------------------------------------------------------------%
\subsection{Single dimensionless input for mean wall-shear stress}
\label{ss:single}
%-----------------------------------------------------------------------%

% Intro
We \rev{assess the need for two inputs in the wall-shear representation
by comparison with a single-input fit},
$\ptauo\approx \rev{g_\tau^{(1)}}(\ptau_1)$.

\begin{figure}
    \centering
    \begin{subfigure}{.45\tw}
        \centering
        \begin{tikzpicture}
            \node (f1) {\ig[width=\tw]{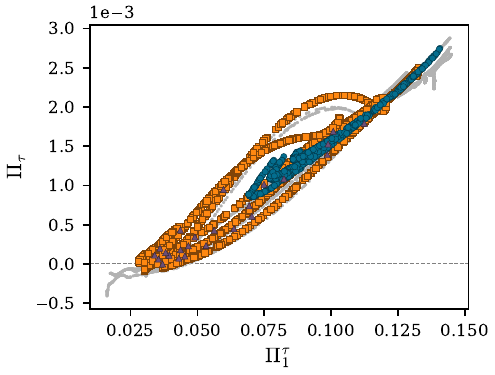}};
            \begin{scope}[shift=(f1.south west), x=(f1.south east), y=(f1.north west)]
                \node[fill=white] at (.58,.05)
                {\footnotesize$\dTwo^{0.75}\nu^{0.25} / U_e^{0.25} \dOne$};
                \node[fill=white,rotate=90] at (.05,.55) {\footnotesize$\ptauo$};
            \end{scope}
        \end{tikzpicture}
        \caption{\label{fig:tau1simple}}
    \end{subfigure}\qquad
    \begin{subfigure}{.45\tw}
        \centering
        \begin{tikzpicture}
            \node (f1) {\ig[width=\tw]{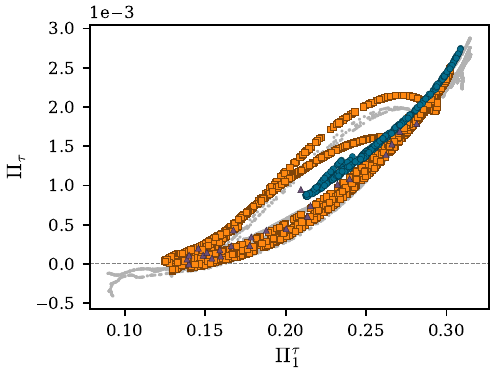}};
            \begin{scope}[shift=(f1.south west), x=(f1.south east), y=(f1.north west)]
                \node[fill=white] at (.58,.05) {\footnotesize$\dTwo^{0.38}\nu^{0.10}
                M^{0.45} / U_e \delta^{0.57} \dOne^{0.36}$};
                \node[fill=white,rotate=90] at (.05,.55) {\footnotesize$\ptauo$};
                %array([-1.   , -0.568, -0.359,  0.377,  0.1  ,  0.   ,  0.45 ])
            \end{scope}
        \end{tikzpicture}
        \caption{\label{fig:tau1full}}
    \end{subfigure}
\caption{Scaling of the dimensionless wall-shear stress $\ptauo$ using
  a single dimensionless input. Panel (a) shows the scaling from
  Eq.~\eqref{eq:tauw1fix}, and panel (b) shows the scaling from
  Eq.~\eqref{eq:tauw1full}. In both panels, gray dots denote the
  present numerical simulations; orange squares denote data from
  \citet{coleman2018numerical} and \citet{coleman2021numerical}; blue
  circles denote data from \citet{bobke2017history}; and purple
  triangles denote data from
  \citet{gungor2016scaling,gungor2022energy} and
  \citet{gungor2024response}.
\label{fig:tau1}}
\end{figure}

% Results
For \rev{this single-input comparison}, the information-theoretic optimization yields
\begin{equation}\label{eq:tauw1fix}
    \ptauo = \frac{\tau_w}{\rho U_e^2} \approx \rev{g_\tau^{(1)}} \left(
    \frac{\dTwo^{0.75} \nu^{0.25}}{U_e^{0.25} \dOne} \right),
\end{equation}
which is exactly the same as the first \dimensionless~input obtained
in Eq.~(4) of the Letter. Figure~\ref{fig:tau1simple}
presents the scaling obtained from Eq.~\eqref{eq:tauw1fix}. The
single-variable scaling achieves a reasonable collapse of the data,
with cases starting from equilibrium conditions exhibiting a
consistent trend along $\ptau_1$. However, two important limitations
are evident. First, a moderate spread persists across different cases.
Second, cases subjected to pressure gradients that induce flow
separation and subsequent reattachment (e.g., the separation bubbles
presented in \citet{coleman2018numerical}) exhibit a characteristic
branching behavior: as the boundary layer approaches separation,
$\ptauo$ decreases along the primary trend, reaching near-zero values
at the separation point before recovering along a distinct branch
during reattachment. This shows that a single-variable scaling, as in
Eq.~\eqref{eq:tauw1fix}, unifies flows across a broad range of
conditions, including favorable and adverse mean pressure gradients
and separated flow, provided that those flows originate from
equilibrium ZPG TBLs. It does not, however, provide a unique collapse
for flows evolving from strongly nonequilibrium initial conditions.
Hence, the presence of history effects manifests as a fundamental
limitation of this single-variable scaling.

% Inclusion of other variables
We also explore additional candidate dimensional variables. In
particular, we include $\delta$ and $M\equiv\int_0^\delta
U^2\,\mathrm{d}y$~\cite{dixit2024generalized}, so that
\[
    \myv{q}=[U_e,\delta,\dOne,\dTwo,\nu,\dPe,M].
    \]
In this case, the optimization process yields the optimal
\dimensionless~input:
\begin{equation}\label{eq:tauw1full}
    \ptauo \approx \rev{g_{\tau,M}^{(1)}}
    \left( \frac{\dTwo^{0.38} \nu^{0.10} M^{0.45}}{U_e \delta^{0.57}\dOne^{0.36}}
    \right).
\end{equation}
\rev{Here, $g_{\tau,M}^{(1)}$ denotes the single-input fit obtained using
the expanded dimensional-variable set.}
As shown in figure~\ref{fig:tau1full}, this scaling slightly improves
the collapse of the main branch compared with Eq.~\eqref{eq:tauw1fix},
but it still fails to collapse the diverging branch. On the other
hand, when two \dimensionless~inputs are allowed, as shown in the main
text, the improvement in the collapse obtained by including $\delta$
and $M$ is marginal. \rev{This indicates that $\delta$ and $M$ provide
little additional predictive information within the flows considered.}
Therefore, we chose to exclude $\delta$ and $M$ from the
\rev{two-input wall-shear representation}. In addition, previous work has
shown that $\delta$ and $M$ are highly sensitive to the exact
definition of the boundary-layer edge~\cite{vinuesa2016determining,
  griffin2021general, wei2023outer, lozier2025Defining}, whereas the
input variables $U_e$, $\dPe$, $\dOne$, and $\dTwo$ are less
sensitive. \rev{Their exclusion simplifies the wall-shear representation
by removing explicit dependence on these quantities. The evaluation of
the wall derivative through $f_U$ still uses $\delta$.}

%-----------------------------------------------------------------------%
\subsection{Optimal number of dimensionless inputs}\label{ss:opt}
%-----------------------------------------------------------------------%

% Intro
The optimization in Eq.~(2) of the main text can be carried out for
any desired number of dimensionless inputs. Since mutual information
is non-decreasing with the inclusion of additional
inputs~\cite{cover2006}, that is, $I(\Pi_o; \Pi_1, \dots, \Pi_n) \geq
I(\Pi_o; \Pi_1, \dots, \Pi_{n-1})$, increasing the number of
dimensionless inputs will either reduce or preserve the irreducible
prediction error, but never increase it. However, beyond a certain
number of inputs, the marginal improvement becomes negligible, while
the increased dimensionality obscures physical interpretation and
makes the subsequent training of models more prone to overfitting.

To determine the minimum number of inputs required, we compute the
information-theoretic lower bound on the prediction error,
$\epsilon_{\mathrm{LB}}$, for different numbers of inputs.

\begin{figure}
    \centering
    \begin{subfigure}{.45\tw}
        \begin{tikzpicture}
            \node (f1) at (0,0) {\ig[width=0.95\tw]{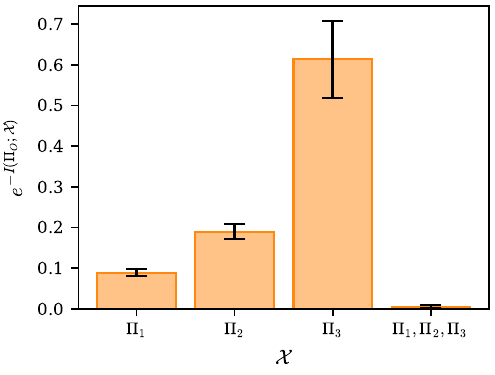}};
            \begin{scriptsize}
                \begin{scope}[shift=(f1.south west),x=(f1.south east),y=(f1.north west)]
                    \node[fill=white] at (.28,.13) {$\pU_1$};
                    \node[fill=white] at (.48,.13) {$\pU_2$};
                    \node[fill=white] at (.67,.13) {$\pU_3$};
                    \node[fill=white] at (.87,.13) {$\pU_1,\pU_2,\pU_3$};

                    \node[fill=white,rotate=90] at
                    (.04,.55) {$e^{-I(\pUo; \mathcal{X})}$};

                \end{scope}
            \end{scriptsize}
        \end{tikzpicture}
        \caption{\label{fig:Uerror}}
    \end{subfigure}\qquad
    \begin{subfigure}{.45\tw}
        \begin{tikzpicture}
            \node (f1) at (0,0) {\ig[width=.95\tw]{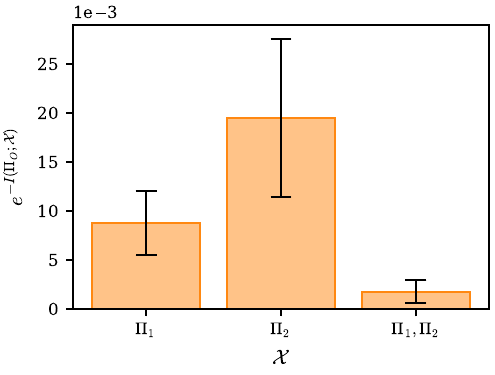}};
            \begin{scriptsize}
                \begin{scope}[shift=(f1.south west),x=(f1.south east),y=(f1.north west)]
                    \node[fill=white] at (.30,.13) {$\ptau_1$};
                    \node[fill=white] at (.56,.13) {$\ptau_2$};
                    \node[fill=white] at (.84,.13) {$\ptau_1,\ptau_2$};

                    \node[fill=white,rotate=90] at
                    (.04,.55) {$e^{-I(\ptauo; \mathcal{X})}$};
                \end{scope}
            \end{scriptsize}
        \end{tikzpicture}
        \caption{\label{fig:tau2error}}
    \end{subfigure}
\caption{Normalized information-theoretic lower bound on the
    prediction error,
    $\epsilon_{\mathrm{LB}}=e^{-I(\Pi_o;\mathcal{X})}$, for different
    combinations of dimensionless inputs: (a) the dimensionless
    mean velocity $\pUo$ and (b) the \rev{reference} dimensionless wall-shear stress
    $\ptauo$. Error bars denote the absolute deviation between the
    full-sample estimate and the mean of 50 half-sample estimates and
    quantify the sensitivity of $\epsilon_{\mathrm{LB}}$ to sample
    size.
\label{fig:bounds}}
\end{figure}

% Results: mean profile
Figure~\ref{fig:Uerror} presents the information-theoretic
error bounds for the mean velocity profile and shows a
reduction when all three inputs are considered jointly.  When the
inputs are considered individually, $\pU_1$ yields the lowest error,
whereas $\pU_3$ yields the largest because it lacks
$y$-dependence. Combining all three dimensionless inputs reduces
$\epsilon_{\mathrm{LB}}$ by a factor of approximately 15 relative to
$\pU_1$ alone.
%

% Results: epsilon tau_w
Figure~\ref{fig:tau2error} presents the results for the \rev{reference
dimensionless wall-shear stress}. When only a single input is considered, $\ptau_1$
achieves the lowest bound, consistent with the findings in
\S\ref{ss:single}. Although $\ptau_2$ alone yields a higher bound than
$\ptau_1$, the joint consideration of $\{\ptau_1, \ptau_2\}$ reduces
$\epsilon_{\mathrm{LB}}$ below the value attainable by either input
individually. This synergy indicates that $\ptau_2$ carries
complementary information about $\ptauo$ that is not captured by
$\ptau_1$ alone. Specifically, as discussed in \S\ref{ss:single} and
the Letter, separation and reattachment can produce distinct branches
associated with different upstream histories, so a single local input
does not provide a unique collapse across all states
(cf.~figure~\ref{fig:tau1}).

% Results: models tau_w
\rev{To assess the wall-shear representation, we compare independently
fitted models with one and two dimensionless inputs against reference
wall-shear data.} The
normalized errors are defined as
\begin{align}\label{eq:errors}
    \mathcal{E}^p_{\ptau_1} &= \left\|\frac{\ptauo -
    \rev{g_\tau^{(1)}}(\ptau_1)}{\sigma_{\ptauo}}\right\|_p, &
    \mathcal{E}^p_{\ptau_1,\ptau_2} &= \left\|\frac{\ptauo -
    \rev{g_\tau^{(2)}}(\ptau_1, \ptau_2)}{\sigma_{\ptauo}}\right\|_p,
\end{align}
where $\|\cdot\|_p$ denotes the $L_p$ norm and $\sigma_{\ptauo}$ is
the standard deviation of the dimensionless output. \rev{Here, $g_\tau^{(m)}$ denotes an independently fitted diagnostic
model with $m$ dimensionless inputs. These fits assess the input
representation; $f_\tau$ denotes the wall-shear relation derived from
$f_U$.}  Normalization by $\sigma_{\ptauo}$ is used instead of a
relative error because $\ptauo$ vanishes at
separation. Table~\ref{tab:errortau} reports the errors for these
models. The two-input model consistently achieves lower errors across
all norms. More notably, as $p$ increases (emphasizing outliers
associated with separated flows) the error for $\rev{g_\tau^{(1)}}(\ptau_1)$
increases from 0.09 to 0.21, whereas $\rev{g_\tau^{(2)}}(\ptau_1, \ptau_2)$
increases only from 0.03 to 0.06. This confirms that the second input
effectively captures the physics of separation and reattachment that
the single-input scaling cannot accommodate.

\corr{For the sake of comparison, we \rev{also consider an independently
  fitted diagnostic model with three dimensionless inputs},
  $\rev{g_\tau^{(3)}}(\rev{\Pi_1^{\tau,(3)}, \Pi_2^{\tau,(3)}, \Pi_3^{\tau,(3)}})$.  In the
  three-input solution, $\rev{\Pi_1^{\tau,(3)}}$ is virtually identical to the
  original $\Pi_1^\tau$ from the Letter, whereas the remaining two
  groups form a different basis for representing the
  pressure-gradient-dependent information:}
\begin{align*}
    \color{revisioncolor}\Pi^{\tau,(3)}_1 &= \color{revisioncolor}\frac{\dTwo^{0.75} \nu^{0.25}}
    {U_e^{0.25} \dOne}, &
    \color{revisioncolor}\Pi^{\tau,(3)}_2 &= \color{revisioncolor}\gamma_P
    \frac{\dTwo^{0.57}\nu^{0.20}|\dPe|^{0.40}}{U_e \dOne^{0.37}}, &
    \color{revisioncolor}\Pi^{\tau,(3)}_3 &= \color{revisioncolor}\gamma_P
    \frac{\dOne^{0.43}\nu^{0.59}|\dPe|^{0.02}}{U_e^{0.63}\dTwo}.
\end{align*}
\corr{ As reported in the last column of Table~\ref{tab:errortau}, the
  model error is essentially saturated once two inputs are
  included. The three-input model yields the same $L_1$ and $L_2$
  errors as the two-input model and only a marginal reduction in the
  $L_3$ error, from 0.06 to 0.05. These results indicate that two
  dimensionless inputs are sufficient to capture the relevant
  dependence of the wall-shear stress. Adding a third group increases
  the dimensionality of the representation, and therefore the
  complexity of the model, without producing a meaningful improvement
  in accuracy. More generally, once the prediction error has nearly
  saturated, additional inputs provide little benefit while reducing
  interpretability and increasing the risk of overfitting during model
  training.}
\begin{table}
    \centering
    \begin{tabular}{@{}w{r}{1.5cm}w{r}{2.cm}w{r}{2.cm}w{r}{3.6cm}@{}}
        $p$-norm & $\mathcal{E}^p_{\ptau_1}$ & $\mathcal{E}^{p}_{\ptau_1,\ptau_2}$
        & $\mathcal{E}^{p}_{\rev{\Pi_1^{\tau,(3)},\Pi_2^{\tau,(3)},\Pi_3^{\tau,(3)}}}$ \\
        \hline
        $1$ & 0.09 & 0.03 & 0.03\\
        $2$ & 0.15 & 0.04 & 0.04\\
        $3$ & 0.21 & 0.06 & 0.05\\
        \hline
    \end{tabular}
    \caption{Normalized \rev{diagnostic-fit error against the reference}
      dimensionless wall-shear stress with one input ($\mathcal{E}^p_{\ptau_1}$), two inputs
      ($\mathcal{E}^{p}_{\ptau_1,\ptau_2}$)\corr{, and three inputs
        ($\mathcal{E}^{p}_{\rev{\Pi_1^{\tau,(3)},\Pi_2^{\tau,(3)},\Pi_3^{\tau,(3)}}}$)}
      for different $L_p$ norms of the error.}\label{tab:errortau}
\end{table}

%-----------------------------------------------------------------------%
\subsection{Fractional-exponent approximation of dimensionless
inputs}\label{sec:approx}
%-----------------------------------------------------------------------%

% Intro
The exponents of the dimensionless groups are first obtained as
decimal values and then approximated by nearby simple fractions.
Here, we assess the effect of this approximation on the resulting
dimensionless inputs.

% Approach
The optimization of Eq.~(2) in the Letter constrains the
exponents of the input variables to the interval $[-2,2]$ (see
\citet{yuan2025Dimensionless} for further details). The resulting
dimensionless inputs are subsequently rescaled so that the largest
exponent magnitude equals unity; this rescaling does not alter the
predictive power of the dimensionless groups themselves. Due to the
non-convex nature of the optimization problem, the procedure is
repeated 30 times with different initial conditions, and the solution
yielding the largest mutual information $I(\Pi_o; \myv{\Pi})$ is
retained. The optimal exponents are then simplified as follows.
First, exponents with magnitude smaller than $0.1$ are set to zero,
effectively removing the corresponding variables from the
dimensionless input. Second, the remaining exponents are approximated
by fractions and adjusted to satisfy dimensional homogeneity.

\begin{figure}
    \centering
    \begin{subfigure}{.48\tw}
        \centering
        \ig[width=\tw]{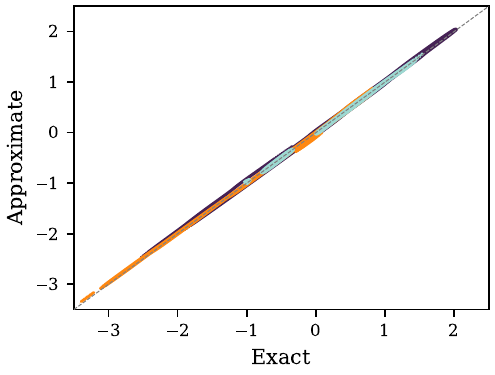}
        \caption{\label{fig:fracU}}
    \end{subfigure}~
    \begin{subfigure}{.48\tw}
        \centering
        \ig[width=\tw]{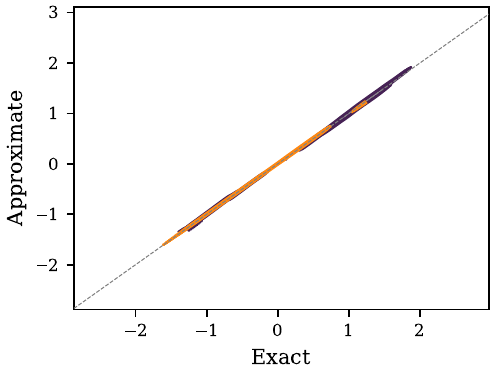}
        \caption{\label{fig:fractau}}
    \end{subfigure}
    \caption{Comparison between the optimal (decimal-valued) inputs
      and their fractional-exponent approximations.  The contours
      enclose 99\% of the joint probability mass.  (a) Velocity-profile
      inputs: $\pU_1$ (orange), $\pU_2$ (purple), and $\pU_3$ (cyan).
      (b) \rev{Inputs of the wall-shear representation}: $\ptau_1$ (purple) and $\ptau_2$ (orange). All quantities are standardized. The dashed line
      denotes exact correspondence.
    \label{fig:frac}}
\end{figure}

% Results for mean profiles
Figure~\ref{fig:fracU} compares the decimal-valued optimal
velocity-profile inputs
\begin{align*}
    \Pi^{U,*}_1 &= \frac{y^{0.21}\delta^{0.04}\dTwo^{0.75}}{\dOne}, &
    \Pi^{U,*}_2 &= \frac{y^{0.87} \dOne^{0.58} \nu^{0.11}}{U_e^{0.11}\delta
    \dTwo^{0.56}}, &
    \Pi^{U,*}_3 &= \frac{\dTwo^{0.80}\delta^{0.02}\nu^{0.18}}{U_e^{0.18}\dOne}
\end{align*}
with their fractional-exponent approximations
\begin{align*}
    \pU_1 &=\frac{y^{2/9}\dTwo^{7/9}}{\dOne}, & \pU_2 &=\frac{y
    \dOne^{2/3} \nu^{1/8}}{U_e^{1/8}\delta^{9/8} \dTwo^{2/3}}, &
    \pU_3 &=\frac{\dTwo^{4/5}\nu^{1/5}}{U_e^{1/5}\dOne}.
\end{align*}
The fractional-exponent approximations closely follow the one-to-one
correspondence line, indicating negligible deviations from the optimal values.

% Results for wall-shear stress
Figure~\ref{fig:fractau} compares the decimal-valued \rev{inputs of the
wall-shear representation}
\begin{align*}
    \Pi^{\tau,*}_1 &= \frac{\dTwo^{0.75} \nu^{0.25}}{U_e^{0.25} \dOne}, &
    \Pi^{\tau,*}_2 &=
    \gamma_P\frac{\dOne^{0.6}\nu^{0.46}|\dPe|^{0.27}}{U_e\dTwo^{0.79}},
\end{align*}
with their fractional-exponent approximations
\begin{align*}
    \ptau_1 &=  \frac{\dTwo^{3/4} \nu^{1/4}}{U_e^{1/4} \dOne}, &
    \ptau_2 &= \gamma_P\frac{\dOne^{11/18}\nu^{4/9}|\dPe|^{5/18}}
    {U_e\dTwo^{7/9}}.
\end{align*}
We can observe that the fractional-exponent approximations closely
follow the one-to-one correspondence line, indicating negligible
deviation from the optimal values.

%-----------------------------------------------------------------------%
\subsection{Redundant, unique, and synergistic contributions of the optimal
dimensionless inputs}
\label{ss:redundant}
%-----------------------------------------------------------------------%

% Method
\rev{We assess the predictive information in the dimensionless inputs
introduced in the Letter and analyzed in \S\ref{ss:opt} using the
normalized information-theoretic error bound,}
\begin{equation}
    \epsilon_{\mathrm{LB}}(\Pi_o,\boldsymbol{\Pi})
    =
    e^{-I(\Pi_o;\boldsymbol{\Pi})},
\end{equation}
where $\Pi_o$ denotes the dimensionless output and
$\boldsymbol{\Pi}=(\Pi_1,\ldots,\Pi_N)$ denotes the candidate set of
dimensionless inputs.  It is possible to determine how the information
that lowers $\epsilon_{\mathrm{LB}}$ is distributed among the optimal
inputs.  To that end, we apply SURD~\cite{martinez2024surd}, which
partitions the mutual information~\cite{shannon1948} between the
output and the inputs into redundant, unique, and synergistic
contributions.  For the two-input case~\citep{martinez2026},
\begin{equation}
    I(\Pi_o;\Pi_1,\Pi_2)
    =
    R_{12}+U_1+U_2+S_{12},
    \qquad
    \epsilon_{\mathrm{LB}}
    =
    e^{-R_{12}}e^{-U_1}e^{-U_2}e^{-S_{12}},
\end{equation}
where $R$, $U$, and $S$ denote redundant, unique and synergistic
contributions of the input variables towards the output, respectively.
In particular, $R_{12}$ measures information about the output that is
available independently in both inputs, $U_i$ denotes information that
is exclusive to input $\Pi_i$, and $S_{12}$ measures information that
becomes accessible only when both inputs are considered jointly.  In
this interpretation, redundancy explains why different inputs can
perform similarly on their own, uniqueness identifies indispensable
inputs, and synergy explains why combining inputs can reduce
$\epsilon_{\mathrm{LB}}$ beyond what any input can achieve separately.
For more than two inputs, the same decomposition extends by including
the corresponding higher-order redundant and synergistic terms.  In
what follows, we report the normalized SURD contributions themselves,
so that all terms add to one. A detailed account of the mathematical
definitions and estimation procedures for the $R$, $U$, and $S$
components is provided in \citet{martinez2024surd}.

\begin{figure}
    \centering
    \begin{subfigure}{.48\tw}
        \centering
        \ig[width=\tw]{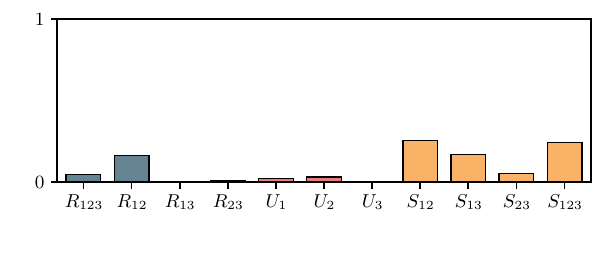}
        \caption{\label{fig:surdU}}
    \end{subfigure}~
    \begin{subfigure}{.48\tw}
        \centering
        \ig[width=\tw]{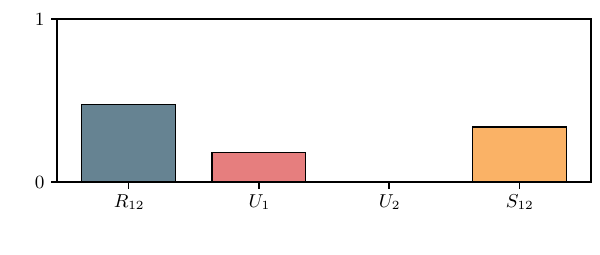}
        \caption{\label{fig:surdtau}}
    \end{subfigure}
    \caption{Synergistic-Unique-Redundant decomposition of the optimal
        dimensionless inputs for (a) $\pUo$ and (b) \rev{the reference} $\ptauo$.  In the plots,
        $R$, $U$, and $S$ denote redundant, unique, and synergistic
        contributions, respectively, and the subscripts indicate the
        corresponding dimensionless inputs.  All values are normalized so
    that their sum is one.\label{fig:surd}}
\end{figure}

% Results for mean velocity
Figure~\ref{fig:surdU} shows the SURD decomposition for $\pUo$ with
respect to $\pU_1$, $\pU_2$, and $\pU_3$.  In this case, the
decomposition contains pairwise and three-way redundant and
synergistic terms (namely $R_{12}$, $R_{13}$, $R_{23}$, $R_{123}$,
$S_{12}$, $S_{13}$, $S_{23}$, and $S_{123}$) in addition to the unique
contributions $U_i$.  The dominant share of the information is
synergistic, the redundant terms provide the second-largest
contribution, and the unique terms are nearly negligible for all three
inputs.

These results are consistent with, and provide a deeper
interpretation of, the scaling analysis for the mean velocity profile
discussed in \S\ref{ss:opt} and \S\ref{ss:2vel}.  The largest single
term is $S_{12}$, indicating that $\pU_1$ and $\pU_2$ jointly encode
most of the information about $\pUo$; this is consistent with these
two variables being the most influential inputs to the mean-velocity
model.  The redundant contribution $R_{12}$ shows that a
non-negligible fraction of this information is independently available
in either input alone, reflecting a structural overlap between $\pU_1$
and $\pU_2$.  Despite the near-vanishing unique contributions, $\pU_3$
remains essential for the complete scaling of $\pUo$: the large values
of $S_{123}$ and $S_{13}$ reveal that $\pU_3$ unlocks information that
is latent in the other inputs but inaccessible without it.

% Results for wall-shear stress
Figure~\ref{fig:surdtau} shows the SURD decomposition for \rev{the reference} $\ptauo$
with respect to $\ptau_1$ and $\ptau_2$.  The dominant contribution is
the redundant term, indicating that a large fraction of the
information about $\ptauo$ is already available in either input alone.
This is expected because $\ptau_1$ and $\ptau_2$ are constructed from
the same set of local dimensional variables and provide redundant
information about the output.  Among the unique contributions, only
$\ptau_1$ carries an appreciable amount of exclusive information,
whereas $\ptau_2$ provides negligible unique information beyond what
is already contained in $\ptau_1$.  The synergistic term, however,
exceeds the unique contribution of $\ptau_1$, showing that the two
inputs combined interact to reveal structure in $\ptauo$ that neither
can expose individually.

This interpretation is consistent with, and complements, the results
in \S\ref{ss:opt}.  The dominant redundancy explains why both
$\ptau_1$ and $\ptau_2$ attain finite lower bounds on the prediction
error when used separately: each input captures the shared
informational content.  The asymmetry in the unique terms is reflected
in the lower error attainable with $\ptau_1$ compared with $\ptau_2$
when each is used as a single predictor.  Finally, the large
synergistic contribution explains the further reduction in
$\epsilon_{\mathrm{LB}}$ when both inputs are used jointly: $\ptau_2$
is valuable mainly as a complementary variable, because part of the
information it carries becomes predictive only when combined with
$\ptau_1$.

%--------------------------------------------------------------%
\subsection{Limitations of the scaling laws}
\label{ss:limitations}
%--------------------------------------------------------------%

% Intro
The scaling laws discovered for the mean velocity profile and mean
wall-shear stress are grounded in data from statistically stationary,
incompressible, turbulent boundary layers that are homogeneous in the
spanwise direction and develop over smooth, flat walls. While these
conditions are representative of many flows encountered in physical
and engineering applications, other flows frequently deviate from this
scenario. In the following, we discuss some of the main physical
mechanisms absent from the current database whose influence on the
scaling laws remains to be assessed.

% 3D effects - seconday flow
% 2 main types: viscous and inviscid
% effect from our point of view: velocity profile not lying on a plane
% coleman2019: in the region of separation we can split between x and z and
% the results are similar to 1-direction (double check)

\emph{TBLs with mean-flow three-dimensionality} (3DTBLs) arise in
geophysical flows, such as tornadoes and river bends, as well as in
engineering applications, including swept wings and turbomachinery, to
name a few~\citep{bradshaw1987}. 3DTBLs can be broadly categorized as
inviscidly induced or viscously induced~\citep{lozano-duran2020}. In
inviscidly induced 3DTBLs, three-dimensionality results from spatially
varying body forces or pressure gradients that skew the mean flow
relative to the free stream. In viscously induced
3DTBLs~\cite{lozano-duran2020,driver1987,eaton1995}, it originates
from viscous effects propagating from the wall, for example through
moving walls, rotating surfaces, or non-inertial reference frames.  In
many practical configurations, both mechanisms may contribute
simultaneously~\cite{lozano-duran2020}. Regardless of their origin,
3DTBLs share key features that distinguish them from their
two-dimensional counterparts.  The mean velocity profile is no longer
contained within a single wall-normal plane; instead, a crossflow
component perpendicular to the boundary-layer edge-velocity vector
$\boldsymbol{U}_e$ develops across the
layer~\cite{johnston1960}. Likewise, the mean wall-shear-stress vector
$\boldsymbol{\tau}_w$ is generally misaligned with $\boldsymbol{U}_e$,
a behavior observed in both viscously induced~\rev{\cite{abe2020}}
and inviscidly induced~\rev{\cite{vandenberg1975,coleman2019}}
flows. Extending the present scaling laws to 3DTBLs requires
\rev{extending the mean-velocity model $f_U$} to predict the wall-normal
distribution of the crossflow component in addition to the streamwise
one. \rev{Its wall derivative must then capture} both the magnitude
$|\boldsymbol{\tau}_w|$ and the orientation of $\boldsymbol{\tau}_w$
relative to $\boldsymbol{U}_e$. In viscously induced flows, the
imposition of \rev{spanwise forcing} can produce a nonmonotonic
response in $|\boldsymbol{\tau}_w|$: it first decreases and
subsequently increases before relaxing toward a new
equilibrium~\cite{lozano-duran2020,abe2020}.
These effects are absent in two-dimensional flows and
must be captured by any predictive model. \rev{This extension} will likely require additional dimensionless inputs
encoding the magnitude and history of the three-dimensional forcing
beyond what is needed to characterize a two-dimensional boundary
layer.

\emph{Streamwise wall curvature} is present in many engineering flows
of interest, including flow over turbine blades and through ducts and
nozzles~\cite{jeans1982}. Curved walls are classified as convex or
concave depending on whether the center of curvature lies inside
(within the solid) or outside (within the fluid) the wall,
respectively~\cite{jeans1982}. Relative to flat-wall flows, curvature
modifies the momentum conservation equations through two additional
acceleration terms: a Coriolis-like contribution in the tangential
(streamwise) direction and a centrifugal contribution in the
wall-normal direction~\cite{bradshaw1973}. These terms alter the mean
velocity profiles~\cite{ellis1974, smits1979, barlow1988} and the mean
wall-shear stress~\cite{appelbaum2025}, regardless of the sign of
curvature. For mild curvatures, several studies
(e.g.,~\citet{pargal2022}) have demonstrated that these additional
accelerations can be effectively absorbed into a modified streamwise
pressure gradient, so that the flat-wall scaling laws remain
applicable without explicitly accounting for curvature. For strong
curvatures, however, the induced accelerations become dynamically
significant and must be accounted for explicitly~\cite{smits1979,
  appelbaum2025}. Within the present work, this may require
constructing additional \dimensionless~inputs that incorporate the
local wall curvature alongside the pressure-gradient and
boundary-layer quantities already used. Incorporating local curvature
as an additional input is a natural direction for further improving
predictive accuracy in these configurations. However, it is worth
noting that the current model, which does not include curvature
explicitly, already yields reasonable agreement for cases with surface
curvature, as shown in the Letter.

\emph{Compressibility effects} become dynamically significant at high
Mach numbers, as encountered in supersonic and hypersonic
vehicles~\cite{lagha2011}. Under these conditions, density and
temperature variations across the boundary layer couple the momentum
and energy equations, modifying the mean velocity and temperature
profiles and the mean wall-shear stress relative to their
incompressible counterparts~\cite{spina1994,smits2006}. Within
Morkovin's hypothesis~\cite{morkovin1962}, one may expect that the
essential dynamics of compressible wall-bounded flows can be accounted
for through variable-density transformations of the mean
velocity~\citep[e.g.,][]{vandriest1951, huang1995, volpiani2020,
griffin2021}, including the prediction of
$\tau_w$~\cite{coles1964}, since compressibility effects are absorbed
by the transformation itself. However, most existing transformations
have been developed for ZPG turbulent boundary layers and are known to
have limitations under strong wall cooling or heating, high Mach
numbers, or significant pressure gradients, where residual
compressibility effects persist~\cite{volpiani2020}. In such cases,
our approach would require additional dimensionless inputs, such as
the edge Mach number $M_e$ and the wall-to-recovery temperature ratio
$T_w/T_r$, to account for residual compressibility effects that
cannot be absorbed by any known transformation.

\emph{Wall roughness} arises from processes such as particle
deposition, erosion, and corrosion on turbomachinery blades and nozzle
surfaces~\cite{bons2001,bons2010}, as well as from biofouling on the
immersed hulls of marine vessels~\cite{munk2009,kirschner2012}, and
can significantly increase skin friction. The primary effect of
roughness on the mean velocity profile in ZPG turbulent boundary
layers is a downward shift of the logarithmic layer, which, in the
fully rough regime, depends only on the equivalent sand-grain
roughness height $k_s$, originally defined by \citet{nikuradse1933}.
In the transitionally rough regime, however, $k_s^+$ is insufficient
to characterize this downward shift, and detailed roughness geometry
may also be required~\cite{flack2010,ma2025}. Extending the present
scaling laws to rough-wall boundary layers would therefore require
additional dimensionless inputs to characterize the geometric and flow
properties of the roughness. \citet{ma2025} suggest that a minimum of
three dimensionless variables is needed to predict the mean wall shear
stress. Whether roughness affects only the inner layer, so that the
present outer-layer scaling remains valid, is still an open question,
particularly for large-scale or strongly anisotropic
roughness~\cite{flack2010}.

\section{Details for complex cases}
%%%%%%%%%%%%%%%%%%%%%%%%%%%%%%%%%%%%%%%%%%%%%%%%%%%

%------------------------------------------------------------------%
\subsection{Spanwise-periodic Gaussian bump at $\Rey_L = 2$ million}
%------------------------------------------------------------------%

% Intro
We evaluate the optimal dimensionless variables from the Letter for
the flow over a spanwise-periodic Gaussian bump at $\Rey_L = U_\infty
L/\nu = 2 \times 10^6$, where $L$ is the \rev{width of the experimental bump
model} and $U_\infty$ is
the free-stream velocity.  The reference data are obtained from
high-fidelity simulations by \citet{uzun2022High}.  This configuration
combines surface curvature (convex for $-0.138 < x/L < 0.138$ and
concave elsewhere, where $x/L = 0$ corresponds to the bump's apex),
which is absent from the training dataset, with strong adverse
pressure gradients and separation with reattachment (at $x/L \approx
0.42$), which are represented in the training data only for canonical
flat-wall configurations.

\begin{figure}[h]
    \centering
    \begin{subfigure}{.9\tw}
        \begin{tikzpicture}
            \node (f1) at (0,0)
{\ig[width=\tw]{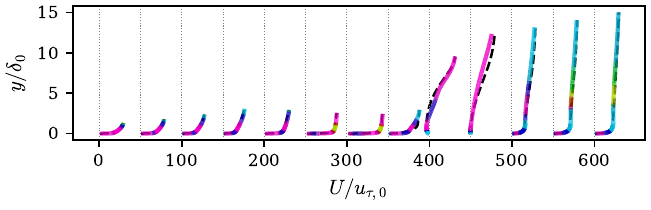}};
        \end{tikzpicture}
        \caption{\label{fig:bumpUnorm}}
    \end{subfigure}
    \begin{subfigure}{.9\tw}
        \begin{tikzpicture}
            \node (f1) at (0,0)
            {\ig[width=\tw]{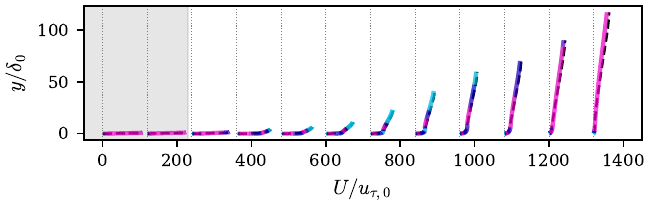}};
        \end{tikzpicture}
        \caption{\label{fig:airfoilUnorm}}
    \end{subfigure}
    \caption{(a) Predictions of the model based on the present scaling
      laws for the flow over a spanwise-periodic Gaussian bump at
      $\Rey_L = 2 \times 10^6$~\citep{uzun2022High}. (b) Predictions
      of the model based on the present scaling laws for the flow over
      a near-stall airfoil at $\Rey_c =
      10^7$~\citep{tamaki2023Wall}. \rev{Colors indicate the dominant
      sensitivity to the standardized coordinates defined in
      \S\ref{ss:local-variance}, using the normalized magnitudes
      $|\partial F_U/\partial z_i|$.} An offset of $\Delta U/u_{\tau,0}
      = 50$ and 120 has been added to the profiles in (a) and (b),
      respectively, for clarity.
    \label{fig:airfoil}}
\end{figure}

% results
The velocity profiles are shown in figure~\ref{fig:bumpUnorm}.  The
trends are consistent with those observed for $\ptauo$: the agreement
is favorable upstream and remains reasonable elsewhere.  Both upstream
and downstream of the apex, the wall-normal distribution of the
dominant parameter resembles that of the separation bubble reported in
Figure~\rev{2}(c) of the Letter (case 18C from
\citet{coleman2018numerical}).  The largest discrepancies occur at
$x/L \approx 0.1$, where the combined effects of convex curvature and
adverse pressure gradient---an interaction not represented in the
training dataset---are most pronounced.

%------------------------------------------------------------------%
\subsection{Near-stall airfoil flow at $\Rey_c = 10$ million}
%------------------------------------------------------------------%

% Intro
We evaluate the optimal dimensionless variables of the Letter for the
flow over an airfoil near stall at $\Rey_c = U_\infty c/\nu = 10^7$,
where $c$ is the chord length and $U_\infty$ is the free-stream
velocity. The angle of attack is $13.3^\circ$, and the free-stream
Mach number is 0.15. The data are obtained from wall-resolved
large-eddy simulations performed by \citet{tamaki2023Wall}. This case
includes effects absent from the training dataset: the airfoil surface
exhibits curvature, and laminar-to-turbulent transition of the
boundary layer occurs at $x/c \approx 0.1$.

% Results
The velocity profiles along the airfoil are shown in
Figure~\ref{fig:airfoilUnorm}. The boundary-layer thickness grows
substantially as the flow approaches separation near the trailing
edge; recall that each profile extends to the local value
$\delta(x)$. The model $f_U(\pU_1,\pU_2,\pU_3)$ accurately captures
the velocity profiles over $0.1<x/c<0.7$, where $x/c=0$ denotes the
leading edge. Discrepancies occur upstream in the laminar region, as
expected because the database used to discover the scaling laws
contains no laminar-flow data. Downstream, the agreement remains
satisfactory, although some deviations occur in the outer layer. At
the most downstream locations, the boundary-layer thickness is
approximately 100 times its value at the initial station.
%

%%%%%%%%%%%$$$$$$$$$$$$$$$$$%%%%%%%%%%%%%%%%%%%%%%%%%%%%%%%%%%%%%%%%%%%
\section{Details of dimensionless-input and model discovery}
%%%%%%%%%%%%%%%%%%%%%%%%%%%%$$$$$$$$$$$$$$$$$%%%%%%%%%%%%%%%%%%%%%%%%%%

\subsection{Details of dimensionless-input discovery}

In order to find the optimal dimensionless inputs \rev{for $U$}, we use data from the current
TBLs and the separation bubbles from \citet{coleman2018numerical}.
\rev{We} use random velocity profiles from $500$
different streamwise locations.
\rev{The} optimization is repeated using several independent
random samples to verify that the discovered scaling and its
fractional-exponent approximation are robust to sample selection
(see~\S\ref{sec:approx}).
\rev{The independent information-theoretic analysis of the wall-shear
representation uses $4{,}000$ random samples of reference wall-shear data
and is likewise repeated to assess sensitivity to sample selection.}

\subsection{Details of Kolmogorov--Arnold networks}\label{sec:kans}

We approximate the mapping from the dimensionless inputs
$\boldsymbol{\Pi}=[\Pi_1,\ldots,\Pi_d]\in\mathbb{R}^d$ to the
dimensionless output $\Pi_o\in\mathbb{R}$ using a Kolmogorov--Arnold
network (KAN)~\cite{liu2024kan}, which places learnable univariate
functions on the edges rather than fixed activation functions at the
nodes, as in conventional neural networks. For the \rev{velocity model $f_U$}, the architecture comprises two hidden layers
with two neurons each, followed by a scalar output layer.

Each KAN layer maps an input $\mathbf{x}\in\mathbb{R}^{d_\mathrm{in}}$
to an output $\mathbf{z}\in\mathbb{R}^{d_\mathrm{out}}$ via
\begin{equation}
    z_j = \sum_{i=1}^{d_\mathrm{in}} \phi_{ji}(x_i),
    \qquad j = 1,\ldots,d_\mathrm{out},
\end{equation}
where each univariate function $\phi_{ji}$ is represented as a linear
combination of B-spline basis functions~\cite{deboor2001splines}
augmented by a residual SiLU path,
\begin{equation}
    \phi_{ji}(x_i)
    = \sum_{k=1}^{N_c} c_{ji}^{(k)}\, B_k^{(p)}(x_i)
    + w_{ji}\,\sigma(x_i).
\end{equation}
Here, $B_k^{(p)}$ denotes the $k$th B-spline basis function of degree
$p$, $c_{ji}^{(k)}$ are learnable spline coefficients, $w_{ji}$ is a
learnable scalar weight, and $\sigma(x)=x/(1+e^{-x})$ is the SiLU
activation. The residual path improves gradient flow during training,
following Liu et al.~\cite{liu2024kan}.

The B-spline basis functions are defined as follows. Let $G$ denote the
number of grid points in a uniform knot sequence over a prescribed
interval $[a,b]$. The grid is extended by $p$ additional knots on each
side to support the basis near the boundaries, yielding an extended
knot sequence $\{t_k\}$ with $G+2p$ total knots. The basis functions
of degree $p$ are evaluated via the Cox--de~Boor
recursion~\cite{deboor2001splines},
\begin{equation}
    B_k^{(0)}(x) = \mathbf{1}[t_k \le x < t_{k+1}],
\end{equation}
\begin{equation}
    B_k^{(\ell)}(x)
    = \frac{x - t_k}{t_{k+\ell} - t_k}\,B_k^{(\ell-1)}(x)
    + \frac{t_{k+\ell+1} - x}{t_{k+\ell+1} - t_{k+1}}\,
    B_{k+1}^{(\ell-1)}(x),
\end{equation}
where zero-over-zero terms are defined as zero. For a spline basis
with $G$ grid points and degree $p$, the number of learnable
coefficients per edge is $N_c = G + p - 1$.
In the present work, we use $G=3$ and $p=3$ (cubic,
$\mathcal{C}^2$-continuous splines), which gives $N_c=5$
coefficients per edge.

\rev{In the first stage, the velocity model $f_U$ is trained using
available mean velocity profiles} by minimizing the mean-squared error
(MSE) of the standardized \rev{velocity targets} using the Adam optimizer~\cite{kingma2015adam} with an initial learning rate of
$10^{-3}$. A reduce-on-plateau scheduler halves the learning rate
whenever the validation loss does not improve for ten consecutive
epochs. The data are split into $80\%$ for training and $20\%$ for
testing; the training set is further divided into $70\%$ and $30\%$
for parameter optimization and validation, respectively. Mini-batches
of size $4096$ are sampled without replacement from the optimization
subset at each epoch. For a dataset
$\{(\boldsymbol{\Pi}^{(n)},\Pi_o^{(n)})\}_{n=1}^N$, each input feature
and the output variable are standardized to zero mean and unit
variance before training.

\rev{In the second stage, the same velocity model is fine-tuned using
wall-shear-stress information from cases for which wall-shear data are
available. This stage constrains the wall-normal derivative of the
predicted velocity profile to improve wall-shear predictions. By construction,
the final velocity prediction is exactly zero at the wall and has a finite
wall-normal derivative there. The
resulting $f_U$ provides both the mean velocity profile and, through its
wall derivative, the mean wall-shear stress.}

The \rev{datasets used for model development and assessment} include the present TBL and
separation-bubble cases, the separation-bubble and APG-TBL cases from
\citet{coleman2018numerical} and \citet{coleman2021numerical}, and the
APG-TBL cases from \citet{bobke2017history}.
\rev{Available mean velocity profiles provide the targets for the first
training stage, while available wall-shear data provide information for
the subsequent fine-tuning and the independent assessment of the
two-input wall-shear representation.}

% Additional validation cases: added in blue; saved full-cohort results.
\clearpage
\begingroup
\color{revisioncolor}
\hypersetup{citecolor=revisioncolor,linkcolor=revisioncolor}
\captionsetup{font={color=revisioncolor},labelfont={color=revisioncolor}}
\section{\rev{Additional validation cases}}\label{sec:additionalvalidation}

This section compares mean-velocity and wall-shear predictions with fifteen
additional benchmark cases that were excluded from model training and evaluated
after training was complete. The comparisons use the edge and integral scales
supplied with each dataset or reconstructed from its reference profiles, so
the predictions are assessed with these quantities provided as inputs.

Table~\ref{tab:additionalvalidation} summarizes the evaluated cases and their
percentage errors, computed using all samples retained for each case and
quantity rather than only the profiles shown in the figures. The errors are
normalized by the reference RMS,
$E_q=100[\sum_n(\widehat q_n-q_n)^2/\sum_n q_n^2]^{1/2}$, where
$q=U/U_e$ for $E_U$ and $q=\tau_w/(\rho U_e^2)$ for $E_\tau$.
For the compressible flat- and concave-wall DNS, $\rho$ is the local edge
density. A dash denotes unavailable data for comparison.

The figures show selected mean velocity profiles plotted against the normalized
wall-normal distance $y/\delta$ on a linear scale. When several profiles are
shown, horizontal offsets $U/U_e+1.3i$ separate them, with $i$ numbering the
displayed profiles from zero, while a single profile is shown without an
offset. Reference profiles are shown as black dashed lines and predictions
as colored solid lines. In wall-shear plots that use symbols, open black and
filled colored symbols show reference values and predictions, respectively,
with different shapes distinguishing the configurations.

When wall-shear data and the required inputs are available, panels (a) and (b)
show the velocity profiles and wall shear, respectively. Otherwise, the
figures show only the velocity profiles for which a comparison is available.
The wall-shear plots use the normalizations given in their captions. For the
zero-pressure-gradient DNS and LES, wall shear is plotted against
$\Rey_\theta$ because streamwise positions are unavailable. In the
zero-pressure-gradient experiments of \"Orl\"u and Schlatter, it is plotted
against edge velocity $U_e$ because all measurements were taken at a single
streamwise position.

\begin{table}[p]
\centering
\renewcommand{\baselinestretch}{1}
\fontsize{10}{12}\selectfont
\setlength{\tabcolsep}{4pt}
\renewcommand{\arraystretch}{1.0}
\caption{\rev{Additional validation cases. Errors are RMS-normalized percentages
computed over all samples included in each comparison. ZPG and APG denote zero
and adverse pressure gradient, respectively. A dash indicates that no
comparison is available.}}
\label{tab:additionalvalidation}
\begin{tabular}{@{}lllcc@{}}
\toprule
\parbox[t]{.16\textwidth}{\raggedright\color{revisioncolor} Case\strut} & \parbox[t]{.29\textwidth}{\raggedright\color{revisioncolor} Brief description\strut} & \parbox[t]{.27\textwidth}{\raggedright\color{revisioncolor} Reference publication\strut} & \parbox[t]{.09\textwidth}{\centering\color{revisioncolor} $E_U$\\(\%)\strut} & \parbox[t]{.09\textwidth}{\centering\color{revisioncolor} $E_\tau$\\(\%)\strut} \\
\midrule
\parbox[t]{.16\textwidth}{\raggedright\color{revisioncolor} ZPG DNS\strut} & \parbox[t]{.29\textwidth}{\raggedright\color{revisioncolor} Planar, zero pressure gradient (ZPG).\strut} & \parbox[t]{.27\textwidth}{\raggedright\color{revisioncolor} \citet{Schlatter2010}\strut} & \parbox[t]{.09\textwidth}{\centering\color{revisioncolor} 1.71\strut} & \parbox[t]{.09\textwidth}{\centering\color{revisioncolor} 3.21\strut} \\
\parbox[t]{.16\textwidth}{\raggedright\color{revisioncolor} ZPG LES\strut} & \parbox[t]{.29\textwidth}{\raggedright\color{revisioncolor} Planar ZPG boundary layer.\strut} & \parbox[t]{.27\textwidth}{\raggedright\color{revisioncolor} \citet{eitelamor2014simulation}\strut} & \parbox[t]{.09\textwidth}{\centering\color{revisioncolor} 1.89\strut} & \parbox[t]{.09\textwidth}{\centering\color{revisioncolor} 1.43\strut} \\
\parbox[t]{.16\textwidth}{\raggedright\color{revisioncolor} Planar bump-pressure DNS\strut} & \parbox[t]{.29\textwidth}{\raggedright\color{revisioncolor} Planar wall with imposed bump-like pressure history.\strut} & \parbox[t]{.27\textwidth}{\raggedright\color{revisioncolor} \citet{spalart2024flattened}\strut} & \parbox[t]{.09\textwidth}{\centering\color{revisioncolor} 2.86\strut} & \parbox[t]{.09\textwidth}{\centering\color{revisioncolor} 1.78\strut} \\
\parbox[t]{.16\textwidth}{\raggedright\color{revisioncolor} ZPG experiment\strut} & \parbox[t]{.29\textwidth}{\raggedright\color{revisioncolor} Hot-wire profiles and oil-film shear calibration.\strut} & \parbox[t]{.27\textwidth}{\raggedright\color{revisioncolor} \citet{orlu2013dns}\strut} & \parbox[t]{.09\textwidth}{\centering\color{revisioncolor} 2.61\strut} & \parbox[t]{.09\textwidth}{\centering\color{revisioncolor} 2.32\strut} \\
\parbox[t]{.16\textwidth}{\raggedright\color{revisioncolor} Historical flat plate\strut} & \parbox[t]{.29\textwidth}{\raggedright\color{revisioncolor} Sparse flat-plate velocity profiles.\strut} & \parbox[t]{.27\textwidth}{\raggedright\color{revisioncolor} \citet{wieghardt1951characteristics}\strut} & \parbox[t]{.09\textwidth}{\centering\color{revisioncolor} 3.18\strut} & \parbox[t]{.09\textwidth}{\centering\color{revisioncolor} ---\strut} \\
\parbox[t]{.16\textwidth}{\raggedright\color{revisioncolor} NACA4412 experiment\strut} & \parbox[t]{.29\textwidth}{\raggedright\color{revisioncolor} Curved-wall hot-wire profiles at three incidences.\strut} & \parbox[t]{.27\textwidth}{\raggedright\color{revisioncolor} \citet{mallor2025experimental}\strut} & \parbox[t]{.09\textwidth}{\centering\color{revisioncolor} 5.36\strut} & \parbox[t]{.09\textwidth}{\centering\color{revisioncolor} ---\strut} \\
\parbox[t]{.16\textwidth}{\raggedright\color{revisioncolor} Wing LES\strut} & \parbox[t]{.29\textwidth}{\raggedright\color{revisioncolor} NACA4412 and NACA0012. Shear for NACA4412 only.\strut} & \parbox[t]{.27\textwidth}{\raggedright\color{revisioncolor} \citet{vinuesa2019wing}, \citet{tanarro2020effect}\strut} & \parbox[t]{.09\textwidth}{\centering\color{revisioncolor} 3.34\strut} & \parbox[t]{.09\textwidth}{\centering\color{revisioncolor} 4.55\strut} \\
\parbox[t]{.16\textwidth}{\raggedright\color{revisioncolor} NASA hump inflow\strut} & \parbox[t]{.29\textwidth}{\raggedright\color{revisioncolor} One flat incoming profile. Hump region excluded.\strut} & \parbox[t]{.27\textwidth}{\raggedright\color{revisioncolor} \citet{greenblatt2006separation}, \citet{naughton2006wall}\strut} & \parbox[t]{.09\textwidth}{\centering\color{revisioncolor} 2.67\strut} & \parbox[t]{.09\textwidth}{\centering\color{revisioncolor} ---\strut} \\
\parbox[t]{.16\textwidth}{\raggedright\color{revisioncolor} Boeing bump inflow\strut} & \parbox[t]{.29\textwidth}{\raggedright\color{revisioncolor} Nearly planar incoming profiles at Mach $0.1$.\strut} & \parbox[t]{.27\textwidth}{\raggedright\color{revisioncolor} \citet{gray2023experimental}\strut} & \parbox[t]{.09\textwidth}{\centering\color{revisioncolor} 1.76\strut} & \parbox[t]{.09\textwidth}{\centering\color{revisioncolor} ---\strut} \\
\parbox[t]{.16\textwidth}{\raggedright\color{revisioncolor} Flat/concave DNS\strut} & \parbox[t]{.29\textwidth}{\raggedright\color{revisioncolor} Matched pressure-gradient pair at Mach $0.5$.\strut} & \parbox[t]{.27\textwidth}{\raggedright\color{revisioncolor} \citet{appelbaum2025}\strut} & \parbox[t]{.09\textwidth}{\centering\color{revisioncolor} 3.30\strut} & \parbox[t]{.09\textwidth}{\centering\color{revisioncolor} 5.17\strut} \\
\parbox[t]{.16\textwidth}{\raggedright\color{revisioncolor} Cylinder inflow\strut} & \parbox[t]{.29\textwidth}{\raggedright\color{revisioncolor} Stationary upstream sections of two cylinder experiments.\strut} & \parbox[t]{.27\textwidth}{\raggedright\color{revisioncolor} \citet{bissonnette1974three}, \citet{lohmann1976measurement}\strut} & \parbox[t]{.09\textwidth}{\centering\color{revisioncolor} 3.14\strut} & \parbox[t]{.09\textwidth}{\centering\color{revisioncolor} ---\strut} \\
\parbox[t]{.16\textwidth}{\raggedright\color{revisioncolor} APG DNS\strut} & \parbox[t]{.29\textwidth}{\raggedright\color{revisioncolor} Planar adverse-pressure-gradient (APG) boundary layer.\strut} & \parbox[t]{.27\textwidth}{\raggedright\color{revisioncolor} \citet{rkein2023numerical}\strut} & \parbox[t]{.09\textwidth}{\centering\color{revisioncolor} 3.11\strut} & \parbox[t]{.09\textwidth}{\centering\color{revisioncolor} ---\strut} \\
\parbox[t]{.16\textwidth}{\raggedright\color{revisioncolor} Pressure-gradient DNS\strut} & \parbox[t]{.29\textwidth}{\raggedright\color{revisioncolor} Planar pressure-gradient boundary layer.\strut} & \parbox[t]{.27\textwidth}{\raggedright\color{revisioncolor} \citet{spalart1993experimental}\strut} & \parbox[t]{.09\textwidth}{\centering\color{revisioncolor} 1.27\strut} & \parbox[t]{.09\textwidth}{\centering\color{revisioncolor} ---\strut} \\
\parbox[t]{.16\textwidth}{\raggedright\color{revisioncolor} Pitot experiment\strut} & \parbox[t]{.29\textwidth}{\raggedright\color{revisioncolor} Smooth-wall profiles with imposed pressure histories.\strut} & \parbox[t]{.27\textwidth}{\raggedright\color{revisioncolor} \citet{fritsch2022effects}\strut} & \parbox[t]{.09\textwidth}{\centering\color{revisioncolor} 2.50\strut} & \parbox[t]{.09\textwidth}{\centering\color{revisioncolor} ---\strut} \\
\parbox[t]{.16\textwidth}{\raggedright\color{revisioncolor} PIV experiment\strut} & \parbox[t]{.29\textwidth}{\raggedright\color{revisioncolor} Companion smooth-wall measurements to the Pitot experiment.\strut} & \parbox[t]{.27\textwidth}{\raggedright\color{revisioncolor} \citet{vishwanathan2023effects}\strut} & \parbox[t]{.09\textwidth}{\centering\color{revisioncolor} 2.81\strut} & \parbox[t]{.09\textwidth}{\centering\color{revisioncolor} ---\strut} \\
\bottomrule
\end{tabular}
\end{table}

\clearpage
\subsection{\rev{Zero-pressure-gradient DNS of Schlatter and \"Orl\"u}}\label{sec:additionalvalidation01}

The direct numerical simulations of Schlatter and \"Orl\"u~\cite{Schlatter2010} provide a canonical zero-pressure-gradient comparison for the mean velocity and wall shear. We evaluate the predictions using ten developed turbulent profiles, comprising 1,585 velocity samples and ten wall-shear values, with selected profiles and the wall-shear results shown in Fig.~\ref{fig:additional_validation_01}. We retain the boundary-layer edge definition used in the original dataset, while recomputing the integral scales over the finite boundary-layer extent and restricting the velocity comparison to that extent. Because the released profile headers do not provide streamwise positions, wall shear is compared against $\Rey_\theta$.

\begin{figure}[htbp]
\centering
\includegraphics[width=\textwidth]{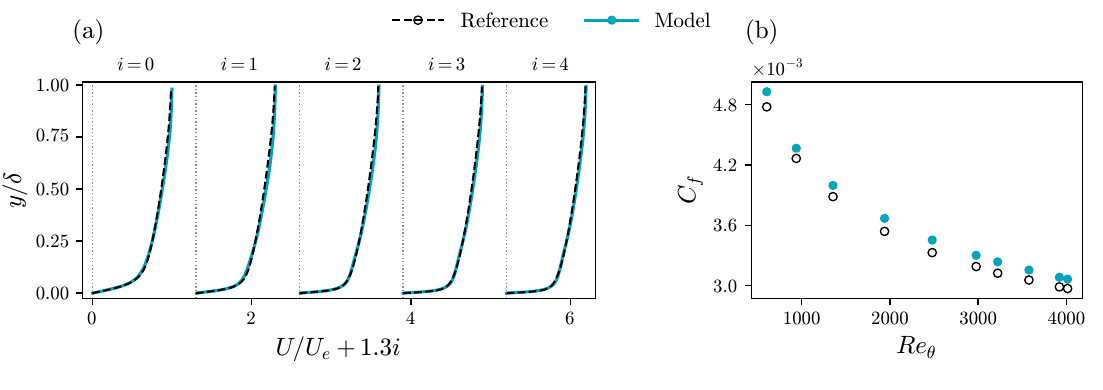}
\caption{\rev{(a) Selected mean velocity profiles, with increasing $i$ corresponding to $\Rey_\theta=605,1355,2480,3575,4014$. (b) Wall shear versus $\Rey_\theta$, with $C_f=2\tau_w/(\rho U_e^2)$. The profiles do not include streamwise positions, so $\Rey_\theta$ is used to distinguish the stations.}}
\label{fig:additional_validation_01}
\end{figure}

\clearpage
\subsection{\rev{Zero-pressure-gradient LES of Eitel-Amor et al.}}\label{sec:additionalvalidation02}

The large-eddy simulations of Eitel-Amor et al.~\cite{eitelamor2014simulation} extend the zero-pressure-gradient comparison to eleven turbulent boundary-layer profiles, comprising 1,737 velocity samples and eleven wall-shear values. We retain the boundary-layer edge definition used in the original dataset, while recomputing the integral scales over the finite boundary-layer extent and evaluating velocity within that extent. Figure~\ref{fig:additional_validation_02} shows selected velocity profiles and the wall-shear comparison, with wall shear plotted against $\Rey_\theta$ because the released profiles do not include streamwise positions.

\begin{figure}[htbp]
\centering
\includegraphics[width=\textwidth]{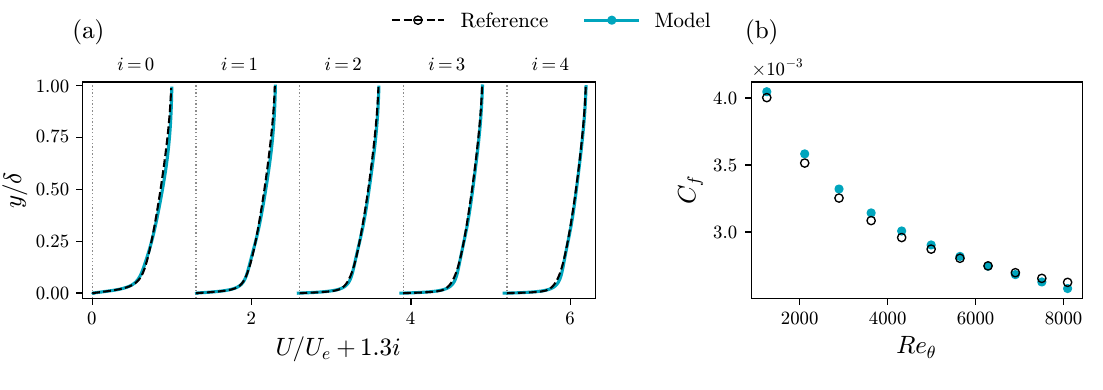}
\caption{\rev{(a) Selected mean velocity profiles, with increasing $i$ corresponding to $\Rey_\theta=1251,2894,4989,6904,8092$. (b) Wall shear versus $\Rey_\theta$, with $C_f=2\tau_w/(\rho U_e^2)$. The profiles do not include streamwise positions, so $\Rey_\theta$ is used to distinguish the stations.}}
\label{fig:additional_validation_02}
\end{figure}

\clearpage
\subsection{\rev{Planar-wall DNS with a speed-bump pressure history}}\label{sec:additionalvalidation07}

Spalart et al.~\cite{spalart2024flattened} simulate an incompressible boundary layer over a planar wall subjected to the pressure history associated with a speed bump. Their Case F isolates the response to acceleration and deceleration without wall curvature, allowing us to compare the predictions at 303 stations over $-0.65\leq x/L\leq0.19$, with 44,004 velocity samples and 303 wall-shear values. For the comparison in Fig.~\ref{fig:additional_validation_07}, edge and integral quantities are reconstructed consistently from the available mean-flow and pressure fields.

\begin{figure}[htbp]
\centering
\includegraphics[width=\textwidth]{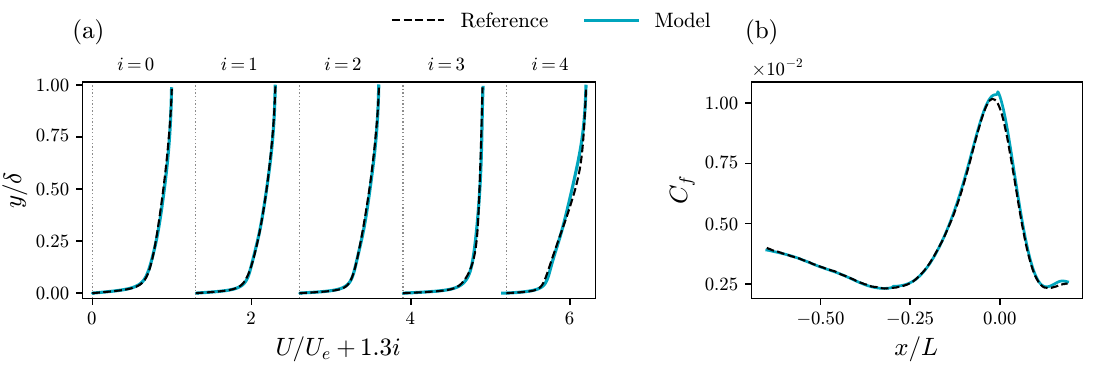}
\caption{\rev{(a) Selected mean velocity profiles, with increasing $i$ corresponding to $x/L=-0.650,-0.439,-0.230,-0.022,0.189$. (b) Streamwise wall shear, with $C_f=2\tau_w/(\rho U_\infty^2)$. Here $L$ is the reference wind-tunnel width, and $x=0$ denotes the apex of the reference bump whose pressure history is imposed on the planar wall.}}
\label{fig:additional_validation_07}
\end{figure}

\clearpage
\subsection{\rev{Zero-pressure-gradient experiments of \"Orl\"u and Schlatter}}\label{sec:additionalvalidation11}

The experiments of \"Orl\"u and Schlatter~\cite{orlu2013dns} provide fourteen zero-pressure-gradient hot-wire profiles obtained with three tripping configurations and a speed sweep at $x=1.625$~m. We retain the near-wall exclusions specified for the original velocity measurements when evaluating the 607 velocity samples and fourteen wall-shear values, with selected profiles and the wall-shear results shown in Fig.~\ref{fig:additional_validation_11}. The wall-shear reference is based on an independent oil-film-interferometry calibration shared by all fourteen values, so their uncertainties are correlated and they do not represent fourteen separate oil-film measurements.

\begin{figure}[htbp]
\centering
\includegraphics[width=\textwidth]{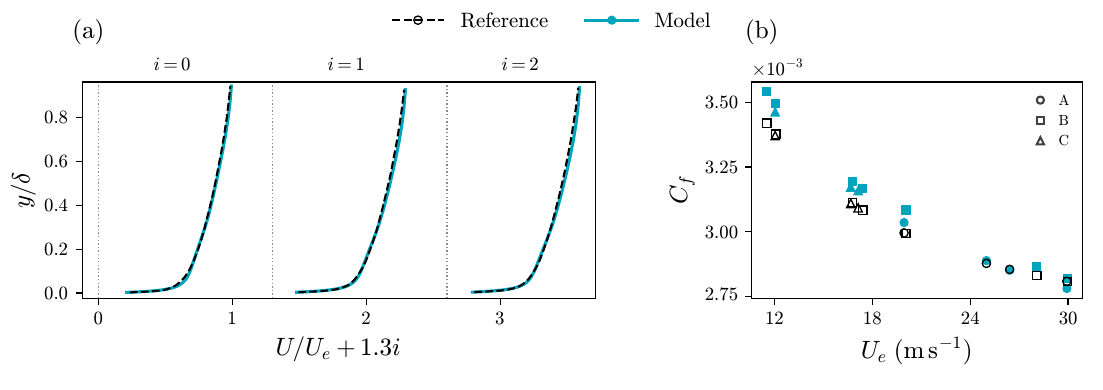}
\caption{\rev{(a) Selected mean velocity profiles for tripping configurations A, B, and C, with $U_e\approx24.99,17.41,16.68$~m/s, respectively, in order of increasing $i$. (b) Wall shear versus edge velocity, with $C_f=2\tau_w/(\rho U_e^2)$. Symbols distinguish the tripping configurations, all measured at $x=1.625$~m.}}
\label{fig:additional_validation_11}
\end{figure}

\clearpage
\subsection{\rev{Historical flat-plate profiles of Wieghardt and Tillmann}}\label{sec:additionalvalidation15}

The historical flat-plate measurements of Wieghardt and Tillmann~\cite{wieghardt1951characteristics} provide a zero-pressure-gradient comparison at twenty-two stations, with 231 velocity samples after excluding the first, potentially transitional profile. We retain the integral scales supplied with the data and recover the coordinates using the transformations specified in the archive. These sparse profiles do not resolve the viscous sublayer, and the comparison in Fig.~\ref{fig:additional_validation_15} is restricted to mean velocity because the archived wall-shear values are inferred rather than independently measured.

\begin{figure}[htbp]
\centering
\includegraphics[width=.60\textwidth]{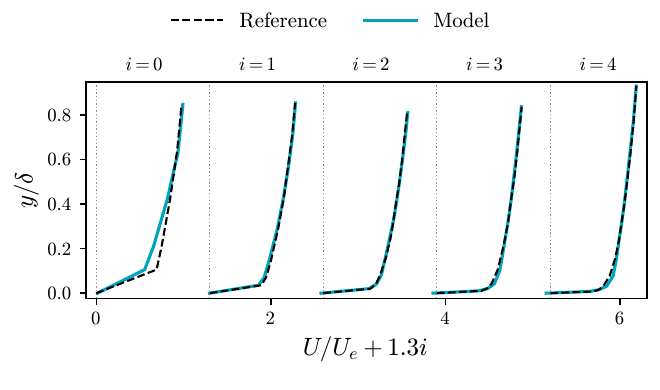}
\caption{\rev{Selected mean velocity profiles, with increasing $i$ corresponding to $x=0.187,0.787,1.687,3.487,4.987$~m.}}
\label{fig:additional_validation_15}
\end{figure}

\clearpage
\subsection{\rev{NACA4412 experiments of Mallor et al.}}\label{sec:additionalvalidation18}

Mallor et al.~\cite{mallor2025experimental} measure boundary-layer profiles on a NACA4412 airfoil at chord Reynolds numbers $4\times10^5$ and $10^6$. The comparison includes twenty-four profiles at angles of attack $\alpha=5^\circ$, $8^\circ$, and $11^\circ$, comprising 674 velocity samples, and excludes the $14^\circ$ profiles. We use the supplied integral scales, which account for the unresolved near-wall portion of the hot-wire profiles. Figure~\ref{fig:additional_validation_18} compares mean velocity only, as the available records do not establish independent wall-shear measurements for each retained station.

\begin{figure}[htbp]
\centering
\includegraphics[width=.60\textwidth]{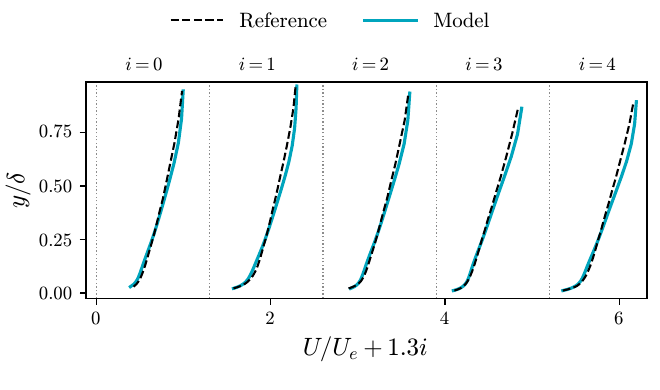}
\caption{\rev{Selected mean velocity profiles, with increasing $i$ corresponding to $(\alpha,\Rey_c)=(5^\circ,10^6),(5^\circ,4\times10^5),(8^\circ,10^6),(11^\circ,10^6),(11^\circ,4\times10^5)$, all at $x/c=0.6$.}}
\label{fig:additional_validation_18}
\end{figure}

\clearpage
\subsection{\rev{NACA4412 and NACA0012 boundary layers}}\label{sec:additionalvalidation19}

The numerical wing databases of Vinuesa et al.~\cite{vinuesa2019wing} and Tanarro et al.~\cite{tanarro2020effect} extend the comparison to curved-wall boundary layers on NACA4412 and NACA0012 airfoils at angles of attack $5^\circ$ and $0^\circ$, respectively. After excluding transitional regions and regions near the trip, the comparison contains 271 velocity profiles, comprising 9,625 samples, and 215 wall-shear values. Edge quantities and integral scales are computed consistently from the local mean-flow records, and Fig.~\ref{fig:additional_validation_19} compares mean velocity for both airfoils and wall shear for the NACA4412 stations only.

\begin{figure}[htbp]
\centering
\includegraphics[width=\textwidth]{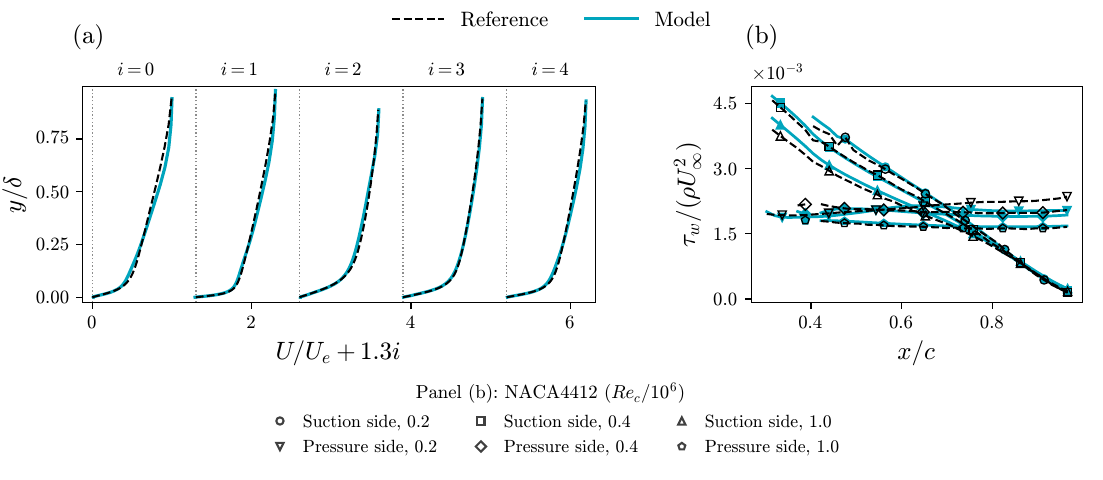}
\caption{\rev{(a) Selected mean velocity profiles, ordered by increasing $i$ as follows: the NACA4412 suction side at $(\Rey_c,x/c)=(2\times10^5,0.687)$ and $(10^6,0.652)$, followed by its pressure side at $(2\times10^5,0.630)$ and $(4\times10^5,0.630)$, and finally the NACA0012 upper surface at $(4\times10^5,0.650)$. Here $c$ is the airfoil chord and $\Rey_c=U_\infty c/\nu$. (b) Streamwise wall shear, $\tau_w/(\rho U_\infty^2)$, with symbols distinguishing the six NACA4412 suction- and pressure-side configurations.}}
\label{fig:additional_validation_19}
\end{figure}

\clearpage
\subsection{\rev{Incoming boundary layer of the NASA hump experiment}}\label{sec:additionalvalidation23}

The NASA wall-mounted hump experiments~\cite{greenblatt2006separation,naughton2006wall} provide an incoming turbulent profile for the baseline configuration without flow control. The comparison in Fig.~\ref{fig:additional_validation_23} uses 46 measured velocity samples at the flat upstream station $x/c=-2.14$, where $c=0.420$~m is the hump reference length. Edge and integral scales are obtained from this incoming profile using the velocity threshold specified in the original dataset to define the boundary-layer edge. The downstream separated region and surface shear are excluded from the comparison because the required matched outer-flow and edge-pressure information is unavailable.

\begin{figure}[htbp]
\centering
\includegraphics[width=.60\textwidth]{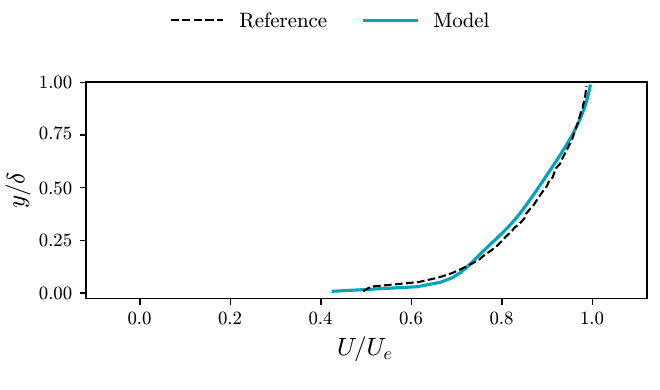}
\caption{\rev{Incoming mean velocity profile at $x/c=-2.14$, where $c=0.420$~m is the hump reference length.}}
\label{fig:additional_validation_23}
\end{figure}

\clearpage
\subsection{\rev{Upstream profiles of the Boeing speed-bump experiment}}\label{sec:additionalvalidation25}

The speed-bump experiments of Gray et al.~\cite{gray2023experimental} provide velocity profiles upstream of two bump configurations, with the bump apex located one and two tunnel widths downstream of the splitter-plate leading edge in configurations A and B, respectively. The comparison includes five profiles and 162 velocity samples at Mach $0.1$ in the nearly planar upstream region, with selected profiles shown in Fig.~\ref{fig:additional_validation_25}. Edge and integral scales are computed from the measured profiles after converting the coordinates as specified in the accompanying documentation.

\begin{figure}[htbp]
\centering
\includegraphics[width=.60\textwidth]{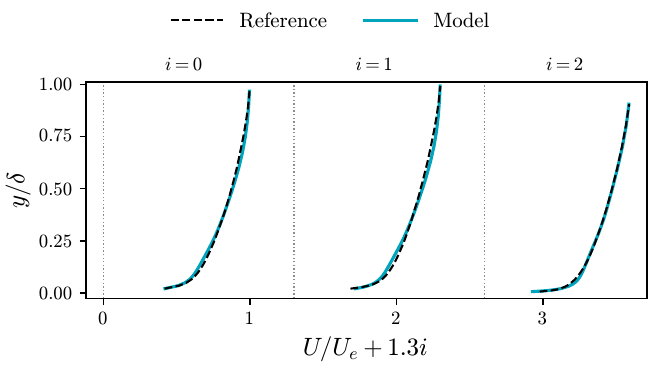}
\caption{\rev{Selected mean velocity profiles, with increasing $i$ corresponding to configuration A at $x/L=-0.683$ and configuration B at $x/L=-1.683,-0.639$. Here $L=0.9144$~m is the tunnel width, and $x$ is measured from the bump apex.}}
\label{fig:additional_validation_25}
\end{figure}

\clearpage
\subsection{\rev{Matched flat- and concave-wall DNS}}\label{sec:additionalvalidation33}

Appelbaum et al.~\cite{appelbaum2025} provide a matched pair of pressure-gradient DNS over flat and concave walls at inflow Mach $0.5$. The comparison uses 87 stations from each geometry, yielding 27,887 velocity samples and 174 wall-shear values. We retain the boundary-layer edge definitions used in the original datasets and use the local edge density and viscosity in the model inputs. Figure~\ref{fig:additional_validation_33} tests the predictions for this pair under moderate compressibility and concave curvature.

\begin{figure}[htbp]
\centering
\includegraphics[width=\textwidth]{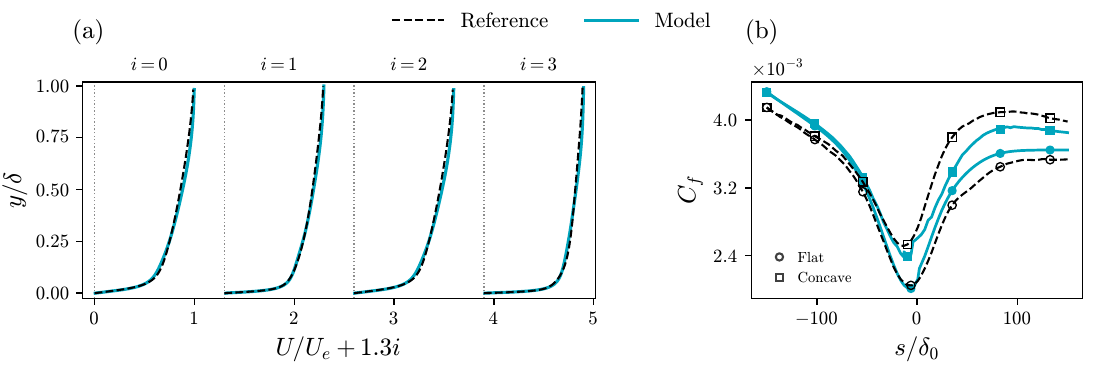}
\caption{\rev{(a) Selected mean velocity profiles, with increasing $i$ corresponding to the flat wall at $s/\delta_0=-150.01,150.04$, followed by the concave wall at the same locations. Here $s$ and $\delta_0$ are the streamwise coordinate and reference thickness used in the simulations. (b) Streamwise wall shear, with $C_f=2\tau_w/(\rho_\infty U_\infty^2)$ and symbols distinguishing the flat and concave walls.}}
\label{fig:additional_validation_33}
\end{figure}

\clearpage
\subsection{\rev{Stationary upstream cylinder boundary layers}}\label{sec:additionalvalidation36}

The cylinder experiments of Bissonnette and Mellor~\cite{bissonnette1974three} and Lohmann~\cite{lohmann1976measurement} include a stationary upstream section followed by a rotating afterbody. The comparison in Fig.~\ref{fig:additional_validation_36} uses one stationary upstream profile from each experiment, comprising 66 velocity samples in total, and tests the mean velocity in the presence of transverse curvature. For these axially developing boundary layers, we retain the edge scales supplied with the data and reconstruct the integral thicknesses consistently.

\begin{figure}[htbp]
\centering
\includegraphics[width=.60\textwidth]{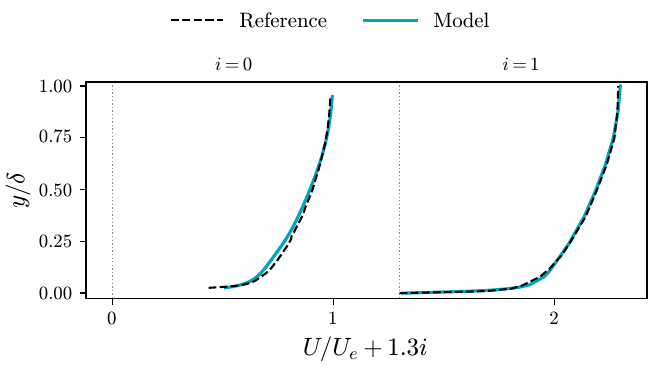}
\caption{\rev{Selected mean velocity profiles, with increasing $i$ corresponding to the Bissonnette--Mellor and Lohmann stationary upstream profiles, respectively.}}
\label{fig:additional_validation_36}
\end{figure}

\clearpage
\subsection{\rev{Adverse-pressure-gradient DNS of Rkein and Laval}}\label{sec:additionalvalidationNC01}

The adverse-pressure-gradient DNS of Rkein and Laval~\cite{rkein2023numerical} provides profiles at thirteen selected stations in a planar turbulent boundary layer. The time- and span-averaged profiles supplied at these stations provide 7,636 velocity samples for the comparison, with selected profiles shown in Fig.~\ref{fig:additional_validation_NC01}. We apply consistent definitions of the edge and integral scales to these profiles without combining them with the separately released data smoothed in the streamwise direction. Only mean velocity is evaluated, as the available records do not establish the corresponding edge pressure gradients needed for a matched wall-shear comparison.

\begin{figure}[htbp]
\centering
\includegraphics[width=.60\textwidth]{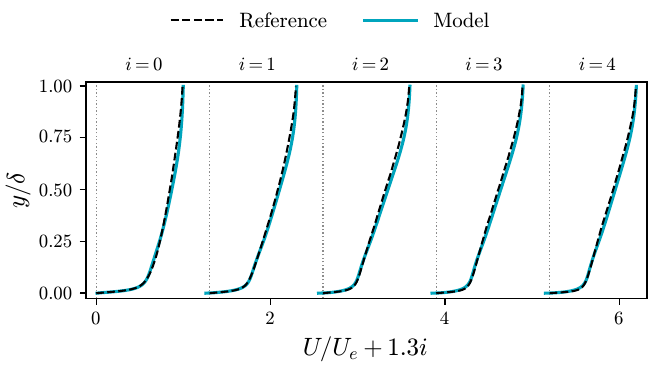}
\caption{\rev{Selected mean velocity profiles, with increasing $i$ corresponding to $x=18.25,124,310,518.875,697.875$ in the units used in the original simulation.}}
\label{fig:additional_validation_NC01}
\end{figure}

\clearpage
\subsection{\rev{Pressure-gradient DNS of Spalart and Watmuff}}\label{sec:additionalvalidationNC04}

The pressure-gradient DNS of Spalart and Watmuff~\cite{spalart1993experimental} provides ten profiles between $x=0.55$ and $1$~m, downstream of the reported inflow transient, comprising 593 velocity samples. Edge quantities and integral scales are reconstructed consistently from the corresponding mean-flow records, accounting for the nonuniform outer flow. The comparison is restricted to these DNS velocity profiles, with selected examples shown in Fig.~\ref{fig:additional_validation_NC04}, and does not include the companion experiment or wall shear.

\begin{figure}[htbp]
\centering
\includegraphics[width=.60\textwidth]{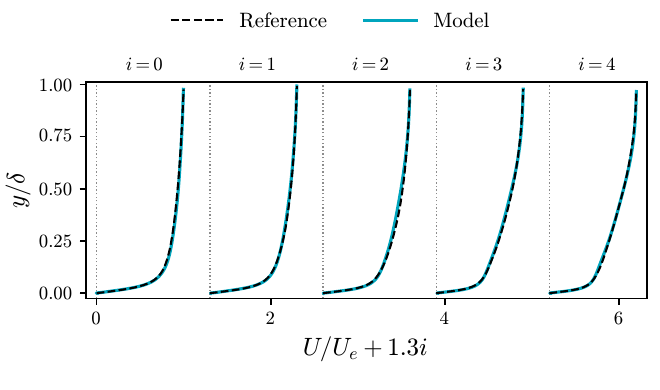}
\caption{\rev{Selected mean velocity profiles, with increasing $i$ corresponding to $x=0.55,0.65,0.75,0.90,1.00$~m.}}
\label{fig:additional_validation_NC04}
\end{figure}

\clearpage
\subsection{\rev{Pressure-history Pitot measurements of Fritsch et al.}}\label{sec:additionalvalidationNC10}

Fritsch et al.~\cite{fritsch2022effects} impose pressure histories on a planar wall using a remote airfoil, with the smooth-wall measurements covering two Reynolds numbers, twelve airfoil angles, and seven stations. The comparison uses the resulting 168 Pitot-rake profiles, comprising 3,621 velocity samples, with selected profiles shown in Fig.~\ref{fig:additional_validation_NC10}. Although the supplied integral scales include a reconstructed near-wall contribution, only measured velocities enter the comparison. Wall shear is not evaluated independently because the released shear estimates are obtained by fitting velocity profiles.

\begin{figure}[htbp]
\centering
\includegraphics[width=.60\textwidth]{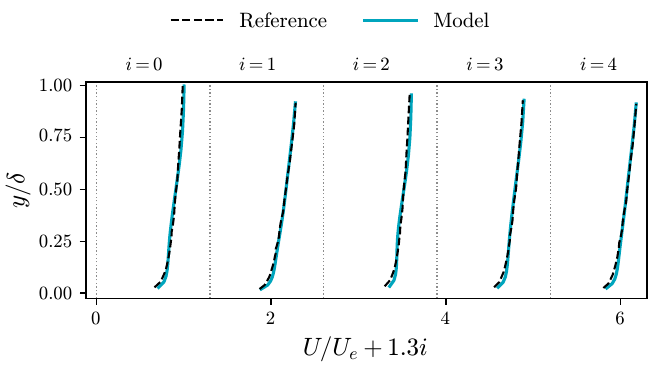}
\caption{\rev{Selected mean velocity profiles, with increasing $i$ corresponding to $(\Rey_c,\alpha)=(2\times10^6,-10^\circ),(2\times10^6,10^\circ),(3.5\times10^6,-10^\circ),(3.5\times10^6,0^\circ),(3.5\times10^6,8^\circ)$, all at $x=3.08$~m. Here $\Rey_c$ and $\alpha$ denote the Reynolds number and angle of attack of the remote airfoil.}}
\label{fig:additional_validation_NC10}
\end{figure}

\clearpage
\subsection{\rev{Companion PIV measurements of Vishwanathan et al.}}\label{sec:additionalvalidationNC11}

The particle-image-velocimetry measurements of Vishwanathan et al.~\cite{vishwanathan2023effects} complement the Pitot measurements from the same pressure-history campaign. The comparison in Fig.~\ref{fig:additional_validation_NC11} uses three consistent profiles at airfoil angles of $-10^\circ$, $0^\circ$, and $12^\circ$, comprising 87 velocity samples, and retains the edge and integral scales supplied with the data. These integral scales include a near-wall contribution obtained from the companion Pitot-rake measurements. The selected files identify the station as $x=3.36$~m, although the accompanying documentation gives $3.31$~m. The profiles provide a further comparison using a different measurement technique within the same campaign, without an independent wall-shear comparison.

\begin{figure}[htbp]
\centering
\includegraphics[width=.60\textwidth]{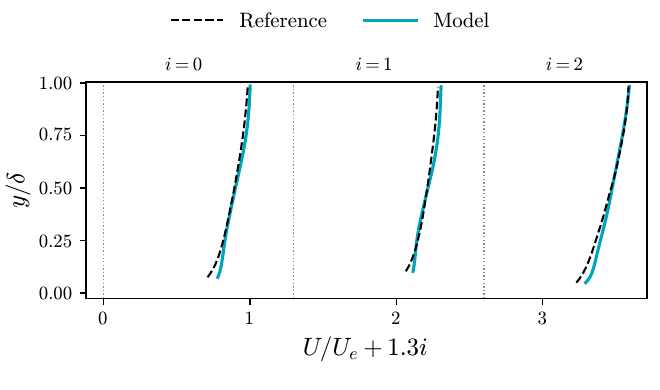}
\caption{\rev{Selected mean velocity profiles, with increasing $i$ corresponding to airfoil angles $0^\circ,-10^\circ,12^\circ$, respectively, at the station labeled $x=3.36$~m in the measurement files.}}
\label{fig:additional_validation_NC11}
\end{figure}

\clearpage
\endgroup

\begingroup
\color{revisioncolor}
\hypersetup{citecolor=revisioncolor,linkcolor=revisioncolor}
\captionsetup{font={color=revisioncolor},labelfont={color=revisioncolor}}
\section[\rev{Interpolation and extrapolation of models in input space}]{\rev{Interpolation and extrapolation of models in input space}}
\label{sec:input-coverage}

We ask how often the unseen cases in the Letter and Sec.~\ref{sec:additionalvalidation} take the model beyond the combinations of
inputs sampled during training. For wall shear, each station gives one point
in the $(\Pi_1^\tau,\Pi_2^\tau)$ plane. For mean velocity, each height in a
profile gives a point in $(\Pi_1^U,\Pi_2^U,\Pi_3^U)$. Different parts of the
same profile can therefore fall inside or outside the training envelope.

We use the same training envelope for every comparison. We include all available stations and profile samples used to assemble the training data. The envelope may therefore include some
regions sampled only by points set aside during training.

To draw the envelope, we divide the range of each input into equal intervals.
These intervals form rectangles in the wall-shear plane and boxes in the
velocity space. A test point counts as \emph{interpolation} if its rectangle
or box contains at least one point from the training data. A point beyond the range of any input or in an empty cell counts as
\emph{extrapolation}. The empty cells matter because a test point can fall
in a gap between sampled regions even when each input, taken separately,
lies within the training range.

We chose the cell widths using the training data alone. We set aside
blocks of neighboring stations in turn, keeping each velocity profile
together, and checked how often the remaining stations covered the points
we had set aside. We used the finest grid that still covered nearly all of
these points on average, giving each simulation equal weight. We then used
all the training data to draw the envelope in Fig.~\ref{fig:input-coverage}.

\begin{figure}[!htbp]
\color{revisioncolor}
\centering
\includegraphics[width=\linewidth]{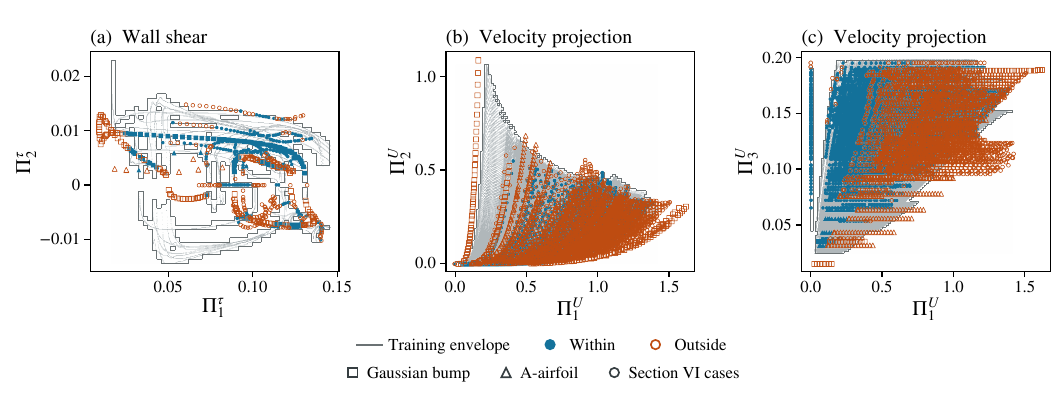}
\caption{The test cases in the input space of the final model. Gray shows the combinations of inputs covered by the training data. Panel (a) shows wall shear. Panels
(b) and (c) show the same velocity samples from two views. Filled blue
symbols fall inside the envelope and open orange symbols fall outside.
An orange symbol can appear inside the gray outline in a velocity panel
because the input omitted from that view places it outside the envelope.
For clarity, only a subset of the test points is drawn.}
\label{fig:input-coverage}
\end{figure}

\begin{table}[!htbp]
\color{revisioncolor}
\centering
\caption{Percentage of points from the unseen cases inside and outside the training envelope.
Each pair of columns sums to 100\% within a row. Wall shear is counted by
station and velocity by wall-normal sample. The last row pools the points
from all cases. A station with known shear inputs can be located in the
figure even if an independent shear measurement is unavailable.}
\label{tab:input-coverage}
\begin{tabular}{@{}lrrrr@{}}
\toprule
& \multicolumn{2}{c}{Wall shear (\%)} & \multicolumn{2}{c}{Mean velocity (\%)} \\
\cmidrule(lr){2-3}\cmidrule(l){4-5}
Test cases & Inside & Outside & Inside & Outside \\
\midrule
Gaussian bump & 35 & 65 & 48 & 52 \\
Near-stall A-airfoil & 33 & 67 & 17 & 83 \\
Additional cases (Sec.~\ref{sec:additionalvalidation}) & 52 & 48 & 72 & 28 \\
\midrule
All cases combined & 45 & 55 & 69 & 31 \\
\bottomrule
\end{tabular}
\end{table}

Across all cases, 55\% of the stations with known shear inputs and
31\% of the velocity samples fall outside the training envelope
(Table~\ref{tab:input-coverage}). The comparisons in Sec.~\ref{sec:additionalvalidation} show
that the model can predict both mean velocity and wall shear with good
accuracy beyond the region sampled by its training data.

\FloatBarrier
\clearpage
\endgroup

\bibliography{references,additional_validation_references}